\documentclass[pdflatex,sn-mathphys-num]{sn-jnl}

\usepackage{booktabs}  
\usepackage{graphicx}%
\usepackage{caption}%
\usepackage{subcaption}%
\usepackage{multirow}%
\usepackage{amsmath,amssymb,amsfonts}%
\usepackage[export]{adjustbox}
\usepackage{amsthm}%
\usepackage{mathrsfs}%
\usepackage[title]{appendix}%
\usepackage{xcolor}%
\usepackage{textcomp}%
\usepackage{manyfoot}%
\usepackage{booktabs}%
\usepackage{algorithm}%
\usepackage{algorithmicx}%
\usepackage{algpseudocode}%
\usepackage{listings}%
\usepackage{comment}
\usepackage{longtable}
\usepackage{tikz}
\usetikzlibrary{arrows.meta,patterns,calc,decorations.markings, shapes.geometric, positioning, fit, backgrounds, calc}

\graphicspath{{figures/}}

\theoremstyle{thmstyleone}%
\theoremstyle{thmstyletwo}%

\theoremstyle{thmstylethree}%

\providecommand{\bfpage}[1]{#1}
\providecommand{\blpage}[1]{#1}

\DeclareMathOperator*{\minimize}{minimize}
\DeclareMathOperator*{\subjectto}{subject~to}

\newcommand{\diff}{\mathrm{d}}

\newcommand{\mat}[1]{\mathbf{#1}}
\newcommand{\R}{\mathbb{R}}
\newcommand{\D}{\mathcal{D}}

\begin{document}

\title[Allen--Cahn Residual-Constrained TO]{Allen--Cahn Residual-Constrained Topology Optimization via an Augmented Lagrangian Method}


\author*[1,2,3]{\fnm{Run} \sur{Du}}\email{rdu@swjtu.edu.cn}

\author[1]{\fnm{Xuan} \sur{Xiao}}

\author[1,2,3]{\fnm{Wenming} \sur{Cheng}}

\author[1,2,3]{\fnm{Zeqiang} \sur{Zhang}}
\affil*[1]{
    \orgdiv{School of Mechanical Engineering}, 
    \orgname{Southwest Jiaotong University}, 
    \orgaddress{    \city{Chengdu}, \postcode{610031}, \state{Sichuan}, \country{China}}}

\affil[2]{
    \orgname{Technology and Equipment of Rail Transit Operation and Maintenance Key Laboratory of Sichuan}, 
    \orgaddress{    \city{Chengdu}, \postcode{610031}, \state{Sichuan}, \country{China}}}
\affil[3]{
    \orgdiv{State Key Laboratory of Bridge Intelligent and Green Construction}, 
    \orgname{Southwest Jiaotong University}, 
    \orgaddress{
    \city{Chengdu}, \postcode{611756}, \state{Sichuan}, \country{China}}}


\abstract{This paper presents a residual-constrained topology optimization framework 
that controls interface transition bandwidth without requiring external density filters. 
The steady conserved Allen--Cahn (AC) equation is formulated as an equality residual 
constraint and solved via a mass-weighted Augmented Lagrangian Method (ALM) coupled 
with the Method of Moving Asymptotes (MMA). 
A closed-form chemical potential projection eliminates the constant residual mode, 
ensuring algebraic compatibility with external volume targets. 
Variational sensitivity analysis reveals a symmetric-plus-rank-one Jacobian that yields 
matrix-free sensitivities without additional adjoint PDE solves, reducing residual 
sensitivity overhead to only $0.11\%$ of the state elasticity solve time in 3D. 
Systematic 2D benchmarks confirm that interface bandwidth is governed intrinsically 
by reaction--diffusion parameters across mesh refinements. 
While diffuse interfaces incur a continuous compliance elevation ($1.4\%$--$12.3\%$), 
a volume-preserving $0/1$ post-processing audit across 48 benchmark cases demonstrates 
that this gap stems primarily from intermediate-density penalization: binarized AC-ALM 
designs achieve equal or superior stiffness in 45 configurations (averaging $1.2\%$ 
stiffer than SIMP). 
Simultaneously, single-node hinges and strut fragmentation are substantially suppressed 
compared to unregularized baselines. 
Three-dimensional benchmarks further demonstrate practical spatial viability 
for complex structures.}

\keywords{Topology Optimization, Conserved Allen--Cahn Equation, Chemical Potential, Residual Constraint, Augmented Lagrangian Method, Diffuse-Interface Profile Control}



\maketitle

\section{Introduction}\label{sec:intro}

Topology optimization (TO) seeks to identify optimal material distributions 
within a design domain $\Omega$ to minimize an objective functional, 
subject to prescribed constraints \citep{Bendsoe2003}. 
In structural mechanics, the canonical problem minimizes mean compliance 
(maximizing overall structural stiffness) 
subject to a global material volume limit \citep{Bendsoe1988}.

Without regularization, continuum compliance minimization is ill-posed. 
Direct finite element discretizations suffer from numerical instabilities, 
including checker-boarding and mesh-dependency \citep{Sigmund2007,Bendsoe2003}. 
To ensure well-posedness, standard regularization techniques are required. 
Established approaches include density filters \citep{Bourdin2001a}, 
perimeter constraints \citep{Haber1996a}, 
and conventional phase-field methods \citep{Bourdin2006a,Burger2006a, 
Takezawa2010a,WangYu2004a, WangYu2004c,ZhouShiwei2007a}.

Alongside level-set methods \citep{WangYu2003,Allaire2004a}, 
phase-field methods provide an implicit interface representation 
for topology optimization \citep{Almi2025a, JinBangti2024a, LiJiajie2025a, Sugimoto2024a}. 
They regularize the ill-posed TO problem by introducing a diffuse interface, 
replacing the discrete 0--1 design space with a continuous design field 
$\phi \in [0,1]$. 
The existing literature predominantly employs phase-field models 
as an energy regularizer. 
The standard approach adds the Ginzburg--Landau free energy \citep{Ginzburg2009} 
as a penalty term to the objective function 
\citep{Bourdin2006a,ZhouShiwei2007a,Blank2014a}. 
The optimization problem then balances structural compliance and interface 
regularization under a prescribed volume fraction:
\begin{equation*}
\minimize_{\phi} \quad 
J_\text{total}(\phi) = \mathcal{J}(\phi) + \gamma J_\text{GL}(\phi).
\end{equation*}
Here, $\mathcal{J}(\phi)$ denotes the structural compliance, 
$\gamma > 0$ is an empirical weighting coefficient, 
and $J_\text{GL}(\phi)$ is the Ginzburg--Landau regularization energy, 
which penalizes diffuse interfaces and implicitly controls the perimeter.

Consequently, $J_\text{total}$ represents a regularized surrogate functional 
rather than the physical structural performance itself. 
The optimizer searches for a mathematical compromise dictated by the penalty term; 
structural stiffness is inevitably compromised to enforce interface regularity. 
Alternatively, conventional phase-field frameworks often evolve the design 
via pseudo-time marching of the Allen--Cahn (AC) \citep{Allen1979} 
or Cahn--Hilliard (CH) \citep{Cahn1958phase} gradient flow dynamics 
\citep{Blank2014a, Takezawa2010a, GaoJiawen2020a, Tavakoli2014b, WangYu2004a}. 
These approaches rely on a heuristic balance between compliance sensitivity 
and interfacial diffusion. 

Related works have derived stabilized evolution equations from energy arguments 
\citep{YuQian2021a, YuQian2022a, YuQian2025a} or neglected curvature effects 
\citep{Takaki2017a, XieWenxuan2023a} to accelerate convergence. 
More rigorous variational treatments have advanced the theoretical foundation 
of phase-field optimization, including Banach-space projected-gradient methods 
\citep{Blank2017a} and multi-phase sharp-interface analyses \citep{Garcke2024a}. 
Across these conventional formulations, however, structural performance 
and interfacial regularization remain coupled through empirical weighting parameters. 
The optimizer must navigate a delicate trade-off. 
An improper parameter choice can result in overly diffuse interfaces 
(if regularization is too weak), or premature convergence to sub-optimal topologies 
with collapsed slender members (if regularization dominates prematurely).

To address these limitations, this work introduces a residual-constrained 
optimization perspective for imposing phase-field interface regularity. 
Instead of appending the Ginzburg--Landau energy as an objective penalty term 
or advancing heuristic gradient flows in pseudo-time, 
we formulate the steady-state \emph{conserved} Allen--Cahn equation as a residual 
equality constraint on the design field. 
Because the conserved residual possesses an intrinsic zero-mean property, 
it is compatible with the independent global volume constraint. 
This formulation decouples structural compliance minimization from interface 
regularization at the solver level, controls transition bandwidth directly through 
physical reaction--diffusion parameters, and eliminates empirical objective weighting. 

While density filtering and projection methods \citep{Bourdin2001a,Guest2004a,Sigmund2007} 
control geometric length scales through spatial convolution, and perimeter regularization 
\citep{Haber1996a} penalizes total interface area, the present framework imposes 
interface regularity through a steady-state reaction--diffusion boundary-value condition. 
The resulting constrained problem is resolved via an inexact Augmented Lagrangian 
Method (ALM) coupled with the Method of Moving Asymptotes (MMA), 
establishing the AC-ALM framework.

In summary, the primary novelties of this work include:
\begin{enumerate}
    \item \emph{Conserved residual formulation:} 
    We formulate diffuse-interface regularization as a steady conserved Allen--Cahn 
    residual constraint. 
    A closed-form mean-zero projection removes the constant-mode compatibility 
    obstruction, ensuring algebraic volume compatibility and eliminating empirical 
    weighting parameters from the objective functional.
    \item \emph{AC-ALM architecture with closed-form sensitivity:} 
    We establish an optimization architecture coupling the Method of Moving 
    Asymptotes (MMA) with ALM. 
    Variational sensitivity analysis reveals a symmetric-plus-rank-one 
    residual Jacobian whose transpose action requires no additional 
    adjoint partial differential equation (PDE) solve for the residual constraint.
    \item \emph{Intrinsic transition-scale control and 3D consistency:} 
    Systematic 2D refinement benchmarks demonstrate predictable interface bandwidths 
    alongside substantial suppression of single-node hinges and mesh-dependent oscillations. 
    Furthermore, 3D benchmarks confirm direct extension to spatial structures, 
    capturing structural curvature while preserving load-carrying connectivity.
\end{enumerate}

The remainder of this paper is organized as follows. 
Section \ref{sec:framework} describes the proposed AC-ALM framework, 
including the problem formulation, the augmented Lagrangian strategy, 
sensitivity analysis, and numerical implementation. 
Section \ref{num_examples} presents numerical tests in 2D and 3D 
to demonstrate the effectiveness, robustness, 
and computational efficiency of the proposed framework. 
Section \ref{discussion} discusses several important aspects of the proposed method.
Finally, concluding remarks are provided in Section \ref{sec:conclusions}.

\section{Proposed AC-ALM Framework}\label{sec:framework}

\subsection{Problem Formulation}

\begin{figure}[ht]
    \centering
\begin{tikzpicture}[
    >=Latex,
    lab/.style={font=\small},
    lead/.style={thin},
    force/.style={-Latex, thick},
    line cap=round,
    line join=round
]

\path[fill=white, draw=black, thick]
(0.2,2.0)
.. controls (0.0,3.3) and (0.9,4.4) .. (2.2,4.3)
.. controls (3.3,4.2) and (4.4,4.8) .. (5.8,4.2)
.. controls (7.2,3.6) and (7.2,2.3) .. (6.3,1.4)
.. controls (5.4,0.5) and (3.9,0.6) .. (2.8,0.8)
.. controls (1.6,1.0) and (0.6,1.2) .. cycle;

\node[lab] at (3.6,2.8) {$\Omega$};

\path (0.4,3.6)
.. controls (0.05,3.1) and (0.05,2.5) .. (0.04,2.0)
coordinate[pos=0.5] (GuMid);

\draw[thick] (0.4,3.6)
.. controls (0.05,3.1) and (0.05,2.5) .. (0.04,2.0);

\draw[
  postaction={decorate},
  decoration={
    markings,
    mark=between positions 0.08 and 0.92 step 6pt with {
      \draw[thin] (0,0) -- ++(-135:4pt);
    }
  }
]
(0.4,3.6) .. controls (0.05,3.1) and (0.05,2.5) .. (0.04,2.0);

\node[label] at (0.5, 2.75) {$\Gamma_D$};


\coordinate (GtA) at (6.5,3.7);
\coordinate (GtB) at (6.8,3.3);

\def\tDX{0.6}
\def\tDY{0.4}

\node[lab, anchor=east] at (6.7,3.4) {$\Gamma_t\subset\Gamma_N$};
\node[label] at (7.4, 4.1) {$\mathbf{t}$};

\foreach \s in {0,0.25,0.50,0.75,1}{
    \coordinate (P) at ($(GtA)!\s!(GtB)$);
    \draw[force, black] ($(P)+(\tDX,\tDY)$) -- (P);
}


\coordinate (PfTop) at (3.60,1.55);
\coordinate (PfBot) at (3.60,0.85);
\draw[force] (PfTop) -- (PfBot);
\node[lab] at (3.4,1.15) {$\mathbf{f}$};

\fill[black]
(5.05,2.9)
.. controls (5.35,3.35) and (6.05,3.25) .. (6.05,2.75)
.. controls (6.05,2.25) and (5.55,2.05) .. (5.15,2.25)
.. controls (4.85,2.40) and (4.75,2.65) .. cycle;

\draw[lead] (5.55,2.75) -- (6.95,1.75);
\node[lab,anchor=west] at (7.05,1.70) {Solid $(\phi=1)$};

\draw[thin] (2.25,3.55) circle (0.10);
\draw[lead] (2.25,3.55) -- (1.05,4.60);
\node[lab,anchor=west] at (0.55,4.65) {A design point ($x$)};

\draw[thin] (2.05,1.55) circle (0.30);
\draw[lead] (2.05,1.55) -- (1.35,0.55);
\node[lab,anchor=west] at (0.85,0.4) {Void $(\phi=0)$};

\end{tikzpicture}
\vspace{25pt}
    \caption{Design domain and boundary conditions.}\label{fig:domain}
\end{figure}
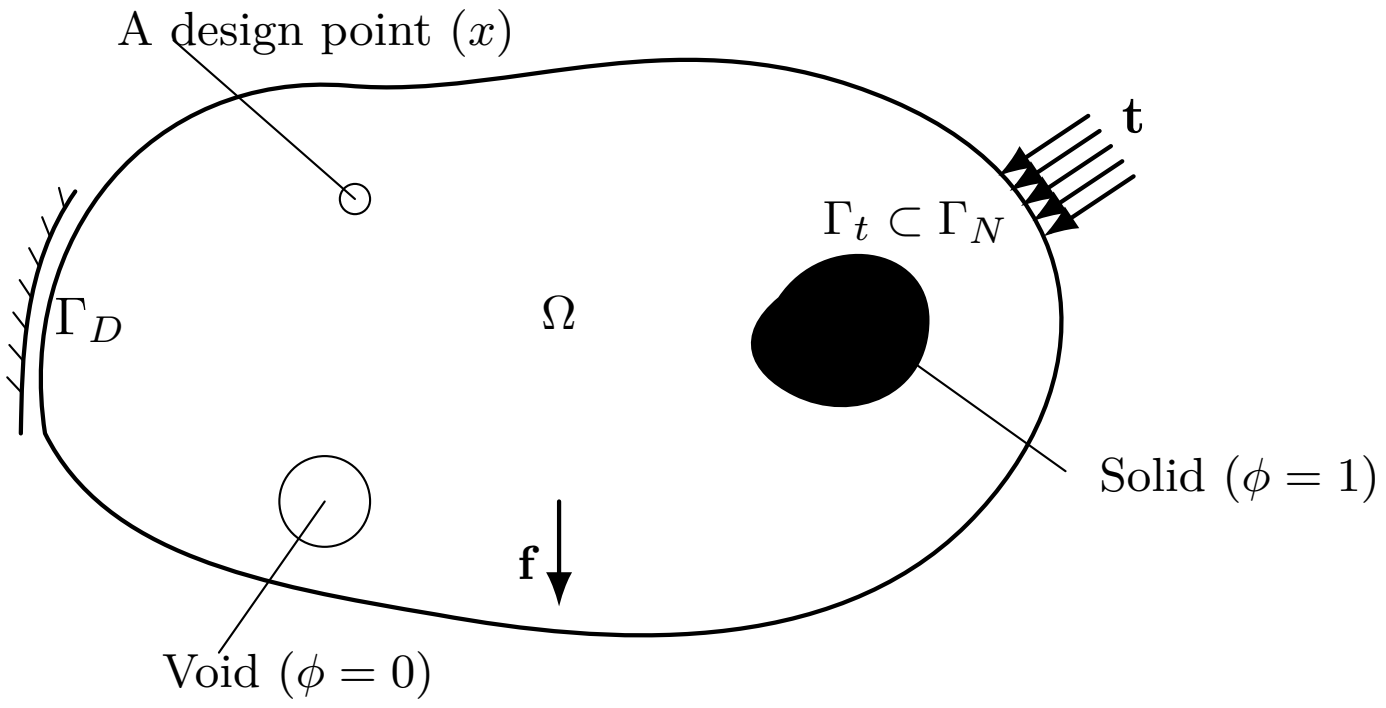

Let $\Omega \subset \R^d$ ($d \in \{2, 3\}$) be a bounded design domain with boundary 
$\partial\Omega = \Gamma_D \cup \Gamma_N$ ($\Gamma_D \cap \Gamma_N = \emptyset$), 
as illustrated in Fig.~\ref{fig:domain}.
Prescribed displacements are imposed on $\Gamma_D$.
Surface tractions $\mathbf{t}$ act on the loaded boundary segment $\Gamma_t \subset \Gamma_N$, 
with the remainder of the natural boundary $\Gamma_N \setminus \Gamma_t$ being traction-free.
The material distribution is described by a continuous design field 
$\phi(\mathbf{x}) : \Omega \to [0,1]$, 
where $\phi=1$ and $\phi=0$ represent solid material and void, respectively.
In the discretized setting, $\phi$ is represented as a nodal field 
co-located with the finite element nodes.

\subsubsection{State Equation and Structural Compliance}\label{sec:state_compliance}
Let $\mathcal{U} = \{\mathbf{u} \in [H^1(\Omega)]^d : \mathbf{u} = \bar{\mathbf{u}} \text{ on } \Gamma_D\}$ 
and $\mathcal{U}_0 = \{\mathbf{w} \in [H^1(\Omega)]^d : \mathbf{w} = \mathbf{0} \text{ on } \Gamma_D\}$ 
denote the spaces of kinematically admissible displacements 
and virtual test displacements, respectively.
For a given design field $\phi$, the equilibrium displacement $\mathbf{u} \in \mathcal{U}$ 
satisfies the weak form of linear elasticity:
\begin{equation} \label{eq:elasticity}
    a(\phi; \mathbf{u}, \mathbf{w}) = l(\mathbf{w}), \quad \forall \mathbf{w} \in \mathcal{U}_0,
\end{equation}
with the bilinear form $a(\cdot; \cdot, \cdot)$ and load functional $l(\cdot)$ defined as:
\begin{align}
    a(\phi; \mathbf{u}, \mathbf{w}) & = \int_{\Omega} \D(\mathbf{w}) : \mat{C}(\phi) : \D(\mathbf{u}) \, \diff\Omega, \\
    l(\mathbf{w}) & = \int_{\Omega} \mathbf{f} \cdot \mathbf{w} \, \diff\Omega 
    + \int_{\Gamma_t} \mathbf{t} \cdot \mathbf{w} \, \diff S.
\end{align}
Here, $\D(\mathbf{u}) = \frac{1}{2}(\nabla \mathbf{u} + (\nabla \mathbf{u})^{\top})$ 
is the symmetric strain tensor, 
$\mathbf{f}$ is the body force (set to zero in the subsequent numerical benchmarks), 
and $\mat{C}(\phi)$ is the effective elasticity tensor interpolated via 
the Solid Isotropic Material with Penalization (SIMP) model 
\citep{Bendsoe1989a,Bendsoe1999a,Bendsoe2003}:
\begin{equation} \label{eq:simp}
    \mat{C}(\phi) = E(\phi) \mat{C}_0, \quad E(\phi) = E_{\min} + \phi^p (E_0 - E_{\min}),
\end{equation}
where $E_0$ is the solid Young's modulus, $E_{\min} > 0$ is a small residual stiffness 
to maintain non-singularity of the global stiffness matrix, 
and $p > 1$ (nominally $p=3$) penalizes intermediate densities.

For an equilibrium displacement state $\mathbf{u}(\phi) \in \mathcal{U}$, 
the overall structural stiffness is quantified by the continuous mean compliance 
(twice the elastic strain energy or total work of external loads):
\begin{equation} \label{eq:compliance_cont}
    \mathcal{J}(\phi) = l(\mathbf{u}(\phi)) = a(\phi; \mathbf{u}(\phi), \mathbf{u}(\phi)) 
    = \int_{\Omega} \D(\mathbf{u}(\phi)) : \mat{C}(\phi) : \D(\mathbf{u}(\phi)) \, \diff\Omega.
\end{equation}
Upon finite element discretization, the continuous state equation \eqref{eq:elasticity} 
yields the linear system $\mat{K}(\boldsymbol{\phi})\mathbf{U} = \mathbf{F}$. 
Here, $\mat{K}(\boldsymbol{\phi})$ is the global stiffness matrix, 
$\mathbf{U}(\boldsymbol{\phi}) \in \R^{N_\text{dof}}$ is the nodal displacement vector, 
and $\mathbf{F} \in \R^{N_\text{dof}}$ is the equivalent nodal load vector. 
Correspondingly, by the principle of virtual work, the continuous compliance 
\eqref{eq:compliance_cont} discretizes to the bilinear algebraic form:
\begin{equation} \label{eq:compliance_disc}
    \mathcal{J}(\boldsymbol{\phi}) = \mathbf{F}^{\top} \mathbf{U}(\boldsymbol{\phi}) 
    = \mathbf{U}(\boldsymbol{\phi})^{\top} \mat{K}(\boldsymbol{\phi}) \mathbf{U}(\boldsymbol{\phi}).
\end{equation}

\subsubsection{Conserved Allen--Cahn Constraint}\label{sec:ac_constraint}

To enforce interface regularity and govern the transition bandwidth, 
we require the design field $\phi$ to satisfy the stationarity condition 
of the volume-conserving Allen--Cahn equation \citep{Rubinstein1992a,Blank2014a}.
In weak form, this requirement seeks $\phi \in H^1(\Omega)$ 
and a constant chemical potential $\mu \in \R$ such that:
\begin{equation} \label{eq:ac_weak}
    G_{\text{AC}}(\phi, \psi) = \int_{\Omega} \left( \epsilon^2 \nabla \phi \cdot \nabla \psi 
    + W^\prime(\phi) \psi \right) \diff\Omega - \mu \int_{\Omega} \psi \, \diff\Omega = 0, 
    \quad \forall \psi \in H^1(\Omega),
\end{equation}
where $\psi$ is an arbitrary test function, $\epsilon$ is the interface-width parameter, 
and $W(\phi)$ is the standard double-well potential:
\begin{equation} \label{eq:double_well}
    W(\phi) = \beta \phi^2 (1 - \phi)^2,
\end{equation}
with $\beta > 0$ denoting the potential-well depth. 
Its derivative, promoting separation between void ($\phi=0$) and solid ($\phi=1$), is given by:
\begin{equation} \label{eq:double_well_prime}
    W^\prime(\phi) = \beta (4\phi^3 - 6\phi^2 + 2\phi).
\end{equation}
Testing Eq.~\eqref{eq:ac_weak} with the constant function $\psi \equiv 1$ 
(for which the gradient term vanishes under homogeneous Neumann boundary conditions) 
determines the chemical potential in closed form:
\begin{equation} \label{eq:mu_def}
    \mu = \frac{1}{|\Omega|} \int_{\Omega} W^\prime(\phi) \, \diff\Omega.
\end{equation}
Thus, $\mu$ is evaluated directly as the domain spatial average of $W^\prime(\phi)$, 
rather than adjusted iteratively.
Applying Green's identity yields the corresponding strong-form condition:
\begin{equation} \label{eq:ac_strong}
    \tilde{h}(\phi) := W^\prime(\phi) - \epsilon^2 \Delta \phi - \mu = 0 \quad \text{in } \Omega, 
    \qquad \nabla \phi \cdot \mathbf{n} = 0 \quad \text{on } \partial\Omega.
\end{equation}

Retaining $\mu$ in Eq.~\eqref{eq:ac_strong} is essential for mathematical compatibility 
with the volume constraint. 
The unprojected condition $W^\prime(\phi) - \epsilon^2\Delta\phi = 0$ requires 
$\int_\Omega W^\prime(\phi)\,\diff\Omega = 0$, which cannot hold for an arbitrary 
prescribed volume fraction. 
Subtracting $\mu$ absorbs this constant mode, eliminating the solvability obstruction 
via a vanishing spatial mean ($\int_\Omega \tilde{h}\,\diff\Omega \equiv 0$). 
Consequently, algebraic compatibility with the volume constraint is ensured, 
leaving the material volume to be satisfied independently by the optimizer.

Furthermore, rewriting the double-well potential as $W = \beta w$ 
with $w(\phi) = \phi^2(1-\phi)^2$, the residual factorizes as:
\begin{equation} \label{eq:h_factorized}
    \tilde{h}(\phi) = \beta \left[ w^\prime(\phi) - \ell_\text{AC}^{2}\, \Delta\phi - \frac{\mu}{\beta} \right],
    \qquad \ell_\text{AC} := \frac{\epsilon}{\sqrt{\beta}},
\end{equation}
where $\mu/\beta$ is the domain average of $w^\prime(\phi)$.
For the zero-residual condition $\{\tilde{h} = 0\}$,
the nominal transition scale is primarily set by $\ell_\text{AC}$, 
with secondary profile corrections from $\mu/\beta$, box bounds,
and finite residual tolerance.
For the one-dimensional stationary interface (with $\tanh$ profile 
$\phi(x) = \frac{1}{2}[1 + \tanh((x-x_0)/(\sqrt{2}\ell_\text{AC}))]$), 
the interface width can be quantified by standard threshold conventions.
The core transition width is
$\xi_{0.1\to 0.9} = \sqrt{2}\ln 9\,\ell_\text{AC} \approx 3.11\,\epsilon/\sqrt{\beta}$, 
which measures the steepest transition zone free from asymptotic tail truncation.
The wider diffuse transition band is
$\xi_{0.05\to 0.95} = \sqrt{2}\ln 19\,\ell_\text{AC} \approx 4.16\,\epsilon/\sqrt{\beta}$, 
encompassing virtually all intermediate grayscale densities. 
Both metrics confirm that $\epsilon$ and $\beta$ prescribe the local interface thickness 
without penalizing the total perimeter or interface length.

In the asymptotic sharp-interface limit ($\epsilon \to 0$) for smooth, unconstrained 
phase boundaries, Eq.~\eqref{eq:ac_strong} formally corresponds to constant-mean-curvature 
(CMC) interfaces via the Gibbs--Thomson relation $|\kappa| = |\mu|/\sigma$ 
\citep{Modica1987a,Luckhaus1989a},
where $\sigma = \epsilon\sqrt{2\beta}/6$ denotes the surface energy. 
The natural homogeneous Neumann boundary condition $\nabla\phi \cdot \mathbf{n} = 0$ 
is applied along all external domain boundaries (including the re-entrant corner of the L-beam), 
naturally avoiding spurious artificial boundary residuals.
Moreover, while Eq.~\eqref{eq:ac_strong} provides an intuitive continuous stationarity condition, 
its direct pointwise enforcement is ill-posed in $H^1(\Omega)$.
We therefore discretize the condition in weak form within the finite element space, 
as formulated in the optimization problem below.

\subsubsection{The Residual-Constrained Optimization Problem}\label{sec:problem_P} 

Via the control-to-state mapping $\boldsymbol{\phi} \mapsto \mathbf{U}(\boldsymbol{\phi})$
established in Section~\ref{sec:state_compliance}, 
the structural compliance $\mathcal{J}(\boldsymbol{\phi})$ defined in Eq.~\eqref{eq:compliance_disc}
is evaluated in reduced form.
The topology optimization problem $(\mathcal{P})$ is formulated to minimize this compliance 
subject to the discrete conserved Allen--Cahn residual constraint and a global volume constraint:
\begin{equation} \label{eq:problem_P}
\begin{aligned}
    \minimize_{\boldsymbol{\phi}} \quad & \mathcal{J}(\boldsymbol{\phi}) = \mathbf{F}^{\top} \mathbf{U}(\boldsymbol{\phi}) \\
    \subjectto \quad & \mathbf{h}(\boldsymbol{\phi}) = \mathbf{0} \\
    \quad & G_\text{v}(\boldsymbol{\phi}) = \mathbf{v}^{\top} \boldsymbol{\phi} - V \leqslant 0 \\
    \quad & 0 \leqslant \phi_i \leqslant 1, \quad i = 1, \dots, N
\end{aligned}
\end{equation}
where $\boldsymbol{\phi} = (\phi_1, \dots, \phi_N)^{\top} \in [0, 1]^N$ collects
the design variables at the $N$ mesh nodes with index $i \in \{1, \dots, N\}$,
$\mathbf{v} \in \R^N$ is the nodal volume-weight vector satisfying $\mathbf{v}^{\top}\mathbf{1} = 1$,
$V \in (0, 1]$ is the prescribed material volume fraction, 
and $\mathbf{h}(\boldsymbol{\phi}) \in \R^N$ is the discrete conserved Allen--Cahn residual vector 
discretizing the weak-form condition Eq.~\eqref{eq:ac_weak}.

In computation, the equality $\mathbf{h}(\boldsymbol{\phi}) = \mathbf{0}$ is enforced 
to a prescribed tolerance $\delta > 0$ via a dimensionless intrinsic norm:
\begin{equation} \label{eq:tol_sense}
    \|\mathbf{h}(\boldsymbol{\phi})\|_{\ast} := \frac{1}{\beta\sqrt{|\Omega|}}
    \left( \mathbf{h}(\boldsymbol{\phi})^{\top} \mat{M}^{-1} \mathbf{h}(\boldsymbol{\phi}) \right)^{1/2} \leqslant \delta,
\end{equation}
where $\mat{M} = \operatorname{diag}(\mathbf{m})$ is the lumped mass matrix
with nodal weights $m_i = \int_\Omega S_i \,\diff\Omega$
satisfying $\mathbf{1}^\top\mathbf{m} = |\Omega|$.
Because the interior weak nodal residual satisfies $h_i \approx m_i \tilde{h}(\mathbf{x}_i)$, 
the discrete metric $\|\mathbf{h}\|_{\ast}$ asymptotically approximates
the root-mean-square of the continuous residual $\tilde{h}$,
normalized by $\beta\sqrt{|\Omega|}$. 
This provides a dimensionless measure normalized against domain volume
and potential-well scaling. 
Enforcing Eq.~\eqref{eq:tol_sense} defines a tolerance-admissible set of regularized designs 
$\mathcal{S}_\delta = \{\boldsymbol{\phi} : \|\mathbf{h}(\boldsymbol{\phi})\|_{\ast} \leqslant \delta\}$ 
without introducing heuristic weighting parameters into the objective functional.

While exact stationarity $\mathbf{h} = \mathbf{0}$ formally implies constant-mean-curvature 
(CMC) interfaces, load-bearing structures fundamentally require variable-curvature 
transitions and branching member junctions. 
These features inevitably introduce localized deviations from ideal phase-field stationarity. 
Furthermore, in saturated bulk regions where $\Delta\phi = 0$, exact stationarity with $\mu \ne 0$ 
induces a slight shift in the unconstrained bulk roots ($W'(\phi_b) = \mu$). 
Because the design variables are subject to box bounds $[0,1]$,
nodes in pure-phase regions and truncated exponential tails of the diffuse interface
carry localized residual contributions. 
From a variational perspective, enforcing stationarity under explicit box bounds 
corresponds to an obstacle-type variational inequality \citep{Blank2013a}. 
Rather than employing non-smooth variational inequality solvers,
the present framework evaluates $\mathbf{h}(\boldsymbol{\phi})$ uniformly across all nodes. 
Adopting the tolerance-regularized admissible set 
$\mathcal{S}_\delta = \{\boldsymbol{\phi} : \|\mathbf{h}(\boldsymbol{\phi})\|_\ast \leqslant \delta\}$ 
serves as a smooth surrogate feasibility metric. 
This cleanly accommodates member-junction, box-truncation, and bulk-shift residuals 
(quantified in Section~\ref{num_boxresidual}) while strictly preserving 
the characteristic transition bandwidth $\ell_\text{AC}$.

\subsection{Augmented Lagrangian Strategy}\label{sec:alm}

The optimization problem $(\mathcal{P})$ couples a global volume constraint 
with a local, nonlinear Allen--Cahn residual field constraint. 
We solve this problem within the AC-ALM framework: 
the Method of Moving Asymptotes (MMA) directly enforces 
the volume inequality constraint ($G_\text{v}(\boldsymbol{\phi}) \leqslant 0$), 
while the spatially distributed Allen--Cahn residual is enforced
via the Augmented Lagrangian Method (ALM). 
Under monotonic SIMP interpolation ($0 < V < 1$), compliance decreases monotonically
with added material.
Thus, the volume constraint remains strictly active at convergence 
($\mathbf{v}^{\top}\boldsymbol{\phi} = V$) for all reported feasible solutions. 
Conceptually, this formulation seeks an optimized topology within the prescribed volume budget, 
while systematically driving the design into the tolerance-admissible
set $\mathcal{S}_\delta$. 
To enforce the nodal Allen--Cahn residual constraint within problem $(\mathcal{P})$ 
and obtain a numerically admissible solution in the tolerance set $\mathcal{S}_\delta$, 
we formulate the Augmented Lagrangian functional:
\begin{equation} \label{eq:alm_functional}
    \begin{split}
    \mathcal{L}_\text{A}(\boldsymbol{\phi}; \boldsymbol{\lambda}_\text{h}) 
    & = \mathcal{J}(\boldsymbol{\phi}) + w_\text{h} \mathcal{L}_\text{AC}(\boldsymbol{\phi}) \\
    & = \mathcal{J}(\boldsymbol{\phi}) 
    + w_\text{h} \left( \boldsymbol{\lambda}_\text{h}^{\top} \mathbf{h}(\boldsymbol{\phi}) 
    + \frac{\rho_\text{h}}{2}\, \mathbf{h}(\boldsymbol{\phi})^{\top}\mat{M}^{-1}\mathbf{h}(\boldsymbol{\phi}) \right),
    \end{split}
\end{equation}
where $\boldsymbol{\lambda}_\text{h} \in \R^N$ is the vector of Lagrange multipliers, 
$\rho_\text{h} > 0$ is the penalty parameter, 
and $w_\text{h} \in (0, 1]$ is the scalar continuation (homotopy) factor
scaling the Allen--Cahn contribution.
The optimization algorithm operates as an inexact continuation ALM scheme while $w_\text{h} < 1$. 
The final feasibility stopping criterion ($\|\mathbf{h}\|_\ast \leqslant \delta$) 
is certified strictly in the final phase where $w_\text{h} = 1$.
Consistent with the tolerance metric in Eq.~\eqref{eq:tol_sense}, 
the quadratic penalty approximates the continuous $L^2$ norm
$\frac{\rho_\text{h}}{2}\int_\Omega \tilde{h}^2\,\diff\Omega$.
This renders $\rho_\text{h}$ mesh-independent across spatial resolutions.
We refer to this form as the \emph{mass-weighted} penalty.
In contrast, an unweighted Euclidean penalty $\frac{\rho_\text{E}}{2}\|\mathbf{h}\|^2$ 
of the assembled residual vector corresponds to an effective continuous penalty 
scaling with element volume $v_e = (\Delta x)^d$:
\begin{equation} \label{eq:penalty_equiv}
    \frac{\rho_\text{E}}{2} \|\mathbf{h}\|^2 \approx \frac{\rho_\text{E} v_e}{2}\, \mathbf{h}^{\top}\mat{M}^{-1}\mathbf{h}
    \approx \frac{\rho_\text{E} v_e}{2} \int_\Omega \tilde{h}^2 \,\diff\Omega,
\end{equation}
which diminishes under mesh refinement ($\rho_\text{E} v_e \propto (\Delta x)^d$ at fixed algorithmic $\rho_\text{E}$).
All numerical benchmarks in Section~\ref{num_examples} employ the mass-weighted formulation.
Their comparative mesh convergence is examined in Section~\ref{num_meshpenalty}.

During early iterations, a small continuation weight $w_\text{h}$ allows the optimizer to establish 
the primary load-bearing topology unhindered by interface constraints.
As iterations progress, $w_\text{h}$ is ramped toward unity 
to drive the nodal residual toward zero, enforcing the feasibility criterion 
$\|\mathbf{h}\|_\ast \leqslant \delta$.
Unlike phase-field continuation strategies that progressively diminish 
the interface thickness $\epsilon$ \citep{Garcke2023c,Garcke2024a}, 
the physical length scale $\ell_\text{AC} = \epsilon/\sqrt{\beta}$ remains fixed here.
Instead, the enforcement weight of the stationarity condition is progressively escalated.

This discrete ALM construction offers three distinct computational advantages:
\begin{enumerate}
    \item \emph{Mitigation of penalty ill-conditioning}: 
    By accumulating constraint violations into the multiplier vector $\boldsymbol{\lambda}_\text{h}$, 
    the framework achieves tolerance feasibility at moderate values of $\rho_\text{h}$.
    This circumvents the severe Hessian ill-conditioning inherent to pure penalty formulations.
    \item \emph{Gradient conditioning}: 
    The mass-weighted quadratic term 
    $\frac{\rho_\text{h}}{2}\mathbf{h}(\boldsymbol{\phi})^{\top}\mat{M}^{-1}\mathbf{h}(\boldsymbol{\phi})$ 
    approximates the continuous $L^2$ penalty $\frac{\rho_\text{h}}{2}\int_\Omega \tilde{h}^2\,\diff\Omega$.
    It supplies a mesh-independent restoring gradient that regularizes the sensitivity field 
    and stably steers intermediate designs toward the residual-feasible manifold.
    \item \emph{Discrete well-posedness}: 
    Formulating the functional on discrete nodal residuals yields a well-defined scalar objective in $\R^N$.
    This avoids demanding $H^2(\Omega)$ regularity of intermediate design fields during optimization.
\end{enumerate}

\subsection{Sensitivity Analysis}

Solving the MMA subproblem requires the gradient of the Augmented Lagrangian
functional $\mathcal{L}_\text{A}$ with respect to the nodal design vector $\boldsymbol{\phi}$. 
By linearity, the total gradient decomposes into compliance and residual contributions:
\begin{equation} \label{eq:total_grad}
    \nabla_{\boldsymbol{\phi}} \mathcal{L}_\text{A} = 
    \nabla_{\boldsymbol{\phi}} \mathcal{J}(\boldsymbol{\phi}) 
    + w_\text{h} \nabla_{\boldsymbol{\phi}} \mathcal{L}_\text{AC}(\boldsymbol{\phi}).
\end{equation}
The gradient of the explicit volume constraint, $\nabla_{\boldsymbol{\phi}} G_\text{v}$, 
is supplied separately to the MMA solver.

\subsubsection{Sensitivity of Compliance}
The sensitivity analysis for the compliance functional $\mathcal{J}$ 
is standard in topology optimization \citep{Bendsoe2003}.
Exploiting the self-adjointness of linear elasticity ($\boldsymbol{\lambda} = \mathbf{u}$), 
the continuous compliance gradient in tensor form is:
\begin{equation} \label{eq:grad_comp_cont}
    \nabla_\phi \mathcal{J}(\phi) = - \D(\mathbf{u}) : \frac{\partial \mat{C}(\phi)}{\partial \phi} : \D(\mathbf{u}).
\end{equation}
In the discrete setting, differentiating the compliance 
$\mathcal{J}(\boldsymbol{\phi}) = \mathbf{U}^{\top} \mat{K}(\boldsymbol{\phi}) \mathbf{U}$ 
with respect to the $j$-th nodal design variable $\phi_j$ 
via the discrete adjoint method \citep{Bendsoe2003} yields:
\begin{equation} \label{eq:grad_comp_disc}
    \frac{\partial \mathcal{J}}{\partial \phi_j} 
    = - \mathbf{U}^{\top} \frac{\partial \mat{K}(\boldsymbol{\phi})}{\partial \phi_j} \mathbf{U}.
\end{equation}

\subsubsection{Sensitivity of the Allen--Cahn Constraint Term}\label{sec:sens_ac}
Differentiating the discrete Allen--Cahn contribution $\mathcal{L}_\text{AC}(\boldsymbol{\phi})$
from Eq.~\eqref{eq:alm_functional} with respect to the design vector $\boldsymbol{\phi}$ yields:
\begin{equation} \label{eq:grad_ac_chain}
    \nabla_{\boldsymbol{\phi}} \mathcal{L}_\text{AC} = \left[ \frac{\partial \mathbf{h}}{\partial \boldsymbol{\phi}} \right]^{\top} 
    (\boldsymbol{\lambda}_\text{h} + \rho_\text{h} \mat{M}^{-1}\mathbf{h}(\boldsymbol{\phi})) = \mat{H}^{\top} \mathbf{z},
\end{equation}
where $\mathbf{z} = \boldsymbol{\lambda}_\text{h} + \rho_\text{h} \mat{M}^{-1}\mathbf{h}(\boldsymbol{\phi})$ 
is an auxiliary vector, and $\mat{H} = \partial \mathbf{h} / \partial \boldsymbol{\phi}$
is the Jacobian matrix of the discrete residual $\mathbf{h}(\boldsymbol{\phi})$.

The linearization of the unprojected residual functional is defined by the bilinear form:
\begin{equation} \label{eq:jacobian_bilinear}
    a_\text{AC}(\phi_h; \eta_h,\psi_h)
    = \int_\Omega W''(\phi_h)\eta_h\psi_h\,\diff\Omega
    + \epsilon^2\int_\Omega \nabla \eta_h\cdot\nabla\psi_h\,\diff\Omega,
\end{equation}
where $\phi_h=\sum_j S_j\phi_j$ is the finite-element design field, 
and $\eta_h$ and $\psi_h$ denote trial and test increments expanded in the nodal basis $\{S_i\}$.
Setting $\eta_h=S_j$ and $\psi_h=S_i$, the leading part is
$H^{\text{sym}}_{ij} = a_\text{AC}(\phi_h; S_j, S_i)$, which is manifestly symmetric.
Accounting for the design dependence of the chemical potential 
through $\partial\mu/\partial\phi_j = w_j / |\Omega|$
with $w_j = \int_\Omega W''(\phi_h) S_j \,\diff\Omega$, 
the complete discrete Jacobian entries evaluate to:
\begin{equation} \label{eq:jacobian_H}
    H_{ij} = \underbrace{\int_\Omega W''(\phi_h)S_iS_j\,\diff\Omega
    + \epsilon^2\int_\Omega \nabla S_i\cdot\nabla S_j\,\diff\Omega}_{\textstyle =:\, H^{\text{sym}}_{ij}}
    \;-\; \frac{1}{|\Omega|}\, m_i\, w_j,
\end{equation}
where $m_i = \int_\Omega S_i\,\diff\Omega$ are the lumped mass weights defined in Eq.~\eqref{eq:tol_sense}. 
The mean-zero projection via $\mu$ thus contributes only the rank-one correction
$-\mathbf{m}\mathbf{w}^{\top}/|\Omega|$, whose transpose action is available in closed form.
Consequently, the transposed Jacobian action required for the gradient reduces to:
\begin{equation} \label{eq:grad_ac}
    \nabla_{\boldsymbol{\phi}} \mathcal{L}_\text{AC} = \mat{H}^{\top} \mathbf{z}
    = \mat{H}^{\text{sym}} \mathbf{z} - \frac{\mathbf{m}^{\top}\mathbf{z}}{|\Omega|}\, \mathbf{w}.
\end{equation}
Crucially, evaluating $\nabla_{\boldsymbol{\phi}} \mathcal{L}_\text{AC}$ via Eq.~\eqref{eq:grad_ac} 
requires only one symmetric matrix-vector product plus one vector inner product, 
with no adjoint PDE solve required. 
The diagonal lumped mass vector $\mathbf{m}$ is utilized exclusively
for the scalar constant-mode projection and the mass-weighted penalty metric $\mat{M}^{-1}$, 
which only requires scalar volume conservation $\mathbf{1}^\top\mathbf{m} = |\Omega|$.

\paragraph{LICQ and Multiplier Gauge Invariance.}
Because $\mathbf{1}^\top \mathbf{h}(\boldsymbol{\phi}) \equiv 0$ holds identically
for all designs $\boldsymbol{\phi}$, 
left-multiplying Eq.~\eqref{eq:jacobian_H} by $\mathbf{1}^\top$ yields:
\begin{equation*}
    \mathbf{1}^\top \mat{H} 
    = \mathbf{1}^\top \mat{H}^\text{sym} - \frac{\mathbf{1}^\top \mathbf{m}}{|\Omega|}\, \mathbf{w}^\top 
    = \mathbf{w}^\top - \frac{|\Omega|}{|\Omega|}\, \mathbf{w}^\top \equiv \mathbf{0}^\top.
\end{equation*}
This confirms that $\mat{H}$ is singular with a permanent left null vector $\mathbf{1}$. 
Consequently, the full $N$-dimensional equality system $\mathbf{h}(\boldsymbol{\phi}) = \mathbf{0}$ 
does not satisfy the Linear Independence Constraint Qualification (LICQ). 
In the dual space, this manifests as a gauge freedom 
$\boldsymbol{\lambda}_\text{h} \mapsto \boldsymbol{\lambda}_\text{h} + c\mathbf{1}$ 
for any constant $c \in \mathbb{R}$. 

Because the algorithm initializes $\boldsymbol{\lambda}_\text{h}^0 = \mathbf{0}$ 
and updates multipliers along $\mat{M}^{-1}\mathbf{h}$ 
(which satisfies $\mathbf{m}^\top \mat{M}^{-1}\mathbf{h} = \mathbf{1}^\top \mathbf{h} \equiv 0$), 
the mass-weighted mean $\mathbf{m}^\top \boldsymbol{\lambda}_\text{h}^k \equiv 0$ 
is conserved automatically at every iteration. 
This uniquely fixes the multiplier gauge without requiring artificial rank reduction. 
If the safeguarding bound $|\lambda_{h,i}| \leqslant \lambda_{\max}$ is enforced, 
gauge preservation is strictly maintained by the projection 
$\boldsymbol{\lambda}_\text{h} \leftarrow \boldsymbol{\lambda}_\text{h}
- \frac{\mathbf{m}^\top \boldsymbol{\lambda}_\text{h}}{|\Omega|}\mathbf{1}$.

\subsection{Numerical Implementation}\label{sec:num_impl}

\subsubsection{Discretization and Compliance Evaluation}\label{sec:discretization_comp}

We discretize the governing equations using conforming finite elements ($Q4$ in 2D, $H8$ in 3D) 
on structured meshes and place the design variables $\boldsymbol{\phi}$ at the nodes.
The $C^0$-continuous nodal representation naturally suppresses element-level two-node checkerboard 
patterns \citep{Guest2004a}, while the intrinsic Allen--Cahn diffusion regularizes 
high-frequency boundary oscillations.
A node-to-element mapping transfers the nodal design $\boldsymbol{\phi}$
to element physical densities $\bar{\phi}_e$ for SIMP stiffness and volume evaluation.
In the intrinsic AC regime ($r_{\min} \leqslant \frac{\sqrt{d}}{2}\Delta x$, typically $r_{\min} = 0.1$), 
the sub-element projection serves purely as a local numerical operator, 
and the mapping reduces to an element-corner average:
\begin{equation}
    \bar{\phi}_e = \frac{1}{2^d} \sum_{i=1}^{2^d} \phi_{e,i},
\end{equation}
where $\phi_{e,i}$ denotes the design value at the $i$-th corner node of element $e$.
In the filter-assisted regime ($r_{\min} > \frac{\sqrt{d}}{2}\Delta x$, e.g., $r_{\min} = 2.0$), 
a conical projection weighting \citep{Guest2004a} is adopted:
\begin{equation}\label{eq:projection}
    \bar{\phi}_e = \dfrac{\displaystyle\sum_{j \in \mathcal{N}_e} w_{ej} \phi_j}{\displaystyle\sum_{j \in \mathcal{N}_e} w_{ej}}, \quad 
    w_{ej} = \max\left(0, \frac{r_{\min} - d_{ej}}{r_{\min}}\right)^{q_f},
\end{equation}
where $\mathcal{N}_e$ denotes the set of nodes within distance $r_{\min}$ from the center of element $e$, 
$d_{ej}$ is the Euclidean distance, and $q_f=2$ is the weight exponent.
For any scalar function $f$ evaluated on the physical element densities
(specifically, the structural compliance $\mathcal{J}$ and the volume constraint $G_\text{v}$),
sensitivities with respect to nodal design variables follow via the chain rule:
\begin{equation}\label{eq:chain_rule}
    \frac{\partial f}{\partial \phi_j} = \sum_{e \in \mathcal{E}_j} 
    \frac{\partial \bar{\phi}_e}{\partial \phi_j} \frac{\partial f}{\partial \bar{\phi}_e}
    = \sum_{e \in \mathcal{E}_j} 
    \left( \dfrac{w_{ej}}{\displaystyle\sum_{i \in \mathcal{N}_e} w_{ei}} \right)
    \frac{\partial f}{\partial \bar{\phi}_e},
\end{equation}
where $\mathcal{E}_j$ denotes the set of elements within influence radius $r_{\min}$ of node $j$
(reducing to $\partial \bar{\phi}_e / \partial \phi_j = 1/2^d$
for element corner nodes when $r_{\min} \leqslant \frac{\sqrt{d}}{2}\Delta x$).
Under the SIMP interpolation \eqref{eq:simp}, the elemental compliance sensitivity evaluates to:
\begin{equation} \label{eq:grad_comp_elem}
    \frac{\partial \mathcal{J}}{\partial \bar{\phi}_e} 
    = - \mathbf{u}_e^{\top} \frac{\partial \mathbf{k}_e}{\partial \bar{\phi}_e} \mathbf{u}_e 
    = - p \bar{\phi}_e^{p-1} (E_0 - E_{\min})\, \mathbf{u}_e^{\top} \mathbf{k}_0 \mathbf{u}_e,
\end{equation}
where $\mathbf{u}_e$ is the element displacement vector and $\mathbf{k}_0$ is the solid element stiffness matrix. 
The displacement state vector $\mathbf{U}$ is likewise discretized at the nodes.
Similarly, the volume constraint in Eq.~\eqref{eq:problem_P} evaluates over elements as
$G_\text{v}(\boldsymbol{\phi}) = \frac{1}{|\Omega|}\sum_e \bar{\phi}_e v_e - V \equiv \mathbf{v}^\top \boldsymbol{\phi} - V$
with element volume $v_e$, whose gradient $\mathbf{v} = \nabla_{\boldsymbol{\phi}} G_\text{v}$
follows directly from Eq.~\eqref{eq:chain_rule} with $\partial G_\text{v}/\partial \bar{\phi}_e = v_e / |\Omega|$
and satisfies $\mathbf{v}^\top\mathbf{1} = 1$.

\subsubsection{Conserved Allen--Cahn Residual and Sensitivity Evaluation}\label{sec:ac_impl}
Unlike the compliance and volume constraints, which are evaluated on element physical densities, 
the conserved Allen--Cahn residual and its sensitivity are formulated directly on the continuous 
nodal design field $\phi_h(\mathbf{x}) = \sum_j S_j(\mathbf{x}) \phi_j$ without requiring 
node-to-element mapping.
Substituting this finite element approximation and the nodal test functions $\psi_i = S_i$ 
into weak form Eq.~\eqref{eq:ac_weak} yields the discrete nodal residual vector:
\begin{equation} \label{eq:h_discrete}
    \mathbf{h}(\boldsymbol{\phi}) = \mathbf{r}_W(\boldsymbol{\phi}) 
    + \epsilon^2 \mat{K}_\text{lap} \boldsymbol{\phi} - \mu(\boldsymbol{\phi})\, \mathbf{m},
    \qquad
    \mu(\boldsymbol{\phi}) = \frac{\mathbf{1}^{\top}\mathbf{r}_W(\boldsymbol{\phi})}{|\Omega|},
\end{equation}
where $\mat{K}_\text{lap}$ is the standard Laplacian stiffness matrix with entries
$(\mat{K}_\text{lap})_{ij} = \int_\Omega \nabla S_i \cdot \nabla S_j \,\diff\Omega$,
$\mathbf{m}$ is the lumped mass vector introduced in Eq.~\eqref{eq:tol_sense}, 
and the nonlinear reaction vector has entries
$r_{W,i}(\boldsymbol{\phi}) = \int_\Omega W'(\phi_h) S_i \,\diff\Omega$.
Because $\mat{K}_\text{lap}$ annihilates constant fields ($\mathbf{1}^{\top}\mat{K}_\text{lap} = \mathbf{0}^{\top}$), 
the mean-zero projection via $\mu$ guarantees strict volume compatibility via
$\mathbf{1}^{\top}\mathbf{h}(\boldsymbol{\phi}) = 0$ 
identically at every iteration at the cost of a single dot product.
Since $\mat{K}_\text{lap}$ discretizes $-\Delta$ via integration by parts,
it enters the residual with a positive sign.
Working with this weak-form residual avoids the mass matrix inversion required
by strong-form collocation, preserving operator sparsity.
At each optimization iteration, the residual $\mathbf{h}(\boldsymbol{\phi})$ monitors 
constraint feasibility ($\|\mathbf{h}\|_* \leqslant \delta$), updates the multipliers 
$\boldsymbol{\lambda}_\text{h}$, and forms the auxiliary vector 
$\mathbf{z} = \boldsymbol{\lambda}_\text{h} + \rho_\text{h} \mat{M}^{-1}\mathbf{h}(\boldsymbol{\phi})$ 
for sensitivity evaluation via Eq.~\eqref{eq:grad_ac}. 
Decomposing the leading symmetric action as $\mat{H}^\text{sym}\mathbf{z} = \mat{H}_W\mathbf{z} + \epsilon^2 \mat{K}_\text{lap}\mathbf{z}$ 
with $(H_W)_{ij} = \int_\Omega W''(\phi_h) S_i S_j \,\diff\Omega$, 
evaluating the residual and its sensitivity shares the same Laplacian matrix $\mat{K}_\text{lap}$ 
and Gauss quadrature points.
Specifically, a $3\times3$ tensor-product Gauss rule in 2D ($3\times3\times3$ in 3D) 
simultaneously evaluates the reaction residual $r_{W,i}$, the rank-one vector 
$w_j = \int_\Omega W''(\phi_h) S_j \,\diff\Omega$, and the matrix-free reaction product 
$(\mat{H}_W\mathbf{z})_i = \int_\Omega W''(\phi_h) S_i (\sum_j S_j z_j)\,\diff\Omega$ 
in a single quadrature pass, without explicit assembly or storage of the Jacobian $\mat{H}$.

\subsubsection{Optimization Subproblem Solver}\label{subsec:opt_mma}
Unlike the pure compliance objective whose sensitivities are strictly negative, 
the Augmented Lagrangian objective $\mathcal{L}_\text{A}$ exhibits indefinite gradient 
signs (with positive and negative components across nodes induced by the double-well 
potential and multipliers) alongside Laplacian spatial coupling. 
This renders standard Optimality Criteria (OC) update schemes \citep{Sigmund2001}, 
which rely on strictly negative sensitivities and separable scaling, mathematically inapplicable. 
We therefore employ the Method of Moving Asymptotes (MMA) \citep{Svanberg1987}, 
which accommodates arbitrary gradient signs through convex separable approximations 
and efficiently handles explicit volume and box bounds.
At each iteration $k$, MMA constructs a strictly convex, separable approximating subproblem:
\begin{equation}
    \tilde{f}^{(k)}(\boldsymbol{\phi}) = r^{(k)} 
    + \sum_{i=1}^N \left( \frac{p_{i}^{(k)}}{U_{i}^{(k)} - \phi_i} + \frac{q_{i}^{(k)}}{ \phi_i - L_{i}^{(k)} } \right),
\end{equation}
where the coefficients $p_i^{(k)}, q_i^{(k)} \geqslant 0$ and $r^{(k)}$ are computed 
from $\mathcal{L}_\text{A}(\boldsymbol{\phi}^k)$ 
and its gradient $\nabla_{\boldsymbol{\phi}}\mathcal{L}_\text{A}(\boldsymbol{\phi}^k)$ (Eq.~\eqref{eq:total_grad}), 
while the moving asymptotes $L_i^{(k)}$ and $U_i^{(k)}$ are updated following 
standard MMA rules \citep{Svanberg1987}.
The subproblem is solved subject to the explicit volume constraint $G_\text{v}(\boldsymbol{\phi}) \leqslant 0$
and box bounds $0 \leqslant \phi_i \leqslant 1$.
The specific asymptote initialization, bound-fraction clamping, and move-limit schedules 
used to stabilize convergence are detailed in Section~\ref{subsec:algo_params}.

\subsection{Overall AC-ALM Optimization Algorithm}\label{sec:overall_algorithm}

The numerical solution under the tolerance-regularized scheme proceeds via
an inexact single-step Augmented Lagrangian iterative formulation. 
In each iteration $k$, the design variables $\boldsymbol{\phi}$ are updated 
by solving the MMA subproblem for the augmented objective $\mathcal{L}_\text{A}$ 
subject to the explicit volume constraint and box bounds.
Subsequently, the Lagrange multipliers $\boldsymbol{\lambda}_\text{h}$ 
and penalty parameter $\rho_\text{h}$ are updated adaptively. 
The overall procedure is summarized in Algorithm~\ref{alg:alm_mma}.

\begin{algorithm}[ht]
\caption{The AC-ALM Algorithm}
\label{alg:alm_mma}
\begin{algorithmic}[1]
\State \textbf{Initialize:} Set $k=0$, uniform design $\boldsymbol{\phi}^0 = V$, 
multipliers $\boldsymbol{\lambda}_\text{h}^0 = \mathbf{0}$, initial penalty $\rho_\text{h}^0$, 
and continuation weight $w_\text{h}^0$.
\Repeat
    \State \textbf{State Solution:} Solve linear elasticity $\mat{K}(\boldsymbol{\phi}^k)\mathbf{U}^k = \mathbf{F}$ 
    for displacement $\mathbf{U}^k$.
    \State \textbf{Residual Evaluation:} Compute $\mu(\boldsymbol{\phi}^k)$, 
    the discrete conserved AC residual $\mathbf{h}(\boldsymbol{\phi}^k)$ via Eq.~\eqref{eq:h_discrete}, 
    and volume $G_\text{v}(\boldsymbol{\phi}^k)$.
    \State \textbf{Continuation:} Update continuation weight $w_\text{h}^k = \min(w_{\max}, w_\text{h}^0 + \alpha_w k)$.
    \State \textbf{MMA Subproblem:} Compute sensitivities via Eq.~\eqref{eq:total_grad} and update design:
    \Statex \hspace{2em} $\boldsymbol{\phi}^{k+1} = \text{MMA}(\boldsymbol{\phi}^k, \nabla_{\boldsymbol{\phi}} \mathcal{L}_{\text{A}}, \nabla_{\boldsymbol{\phi}} G_\text{v})$ 
    \Statex \hspace{2em} subject to $G_\text{v}(\boldsymbol{\phi}) \leqslant 0, \quad 0 \leqslant \phi_i \leqslant 1$.
    \State \textbf{Residual Re-evaluation:} Compute $\mathbf{h}(\boldsymbol{\phi}^{k+1})$ 
    and evaluate tolerance metric $\|\mathbf{h}(\boldsymbol{\phi}^{k+1})\|_{\ast}$ via Eq.~\eqref{eq:tol_sense}.
    \State \textbf{Multiplier Update:} Clamped update with gauge preservation for all $i = 1, \dots, N$:
    \Statex \hspace{2em} $\lambda_{h,i}^{k+1} \leftarrow \max\left(-\lambda_{\max},\, \min\left(\lambda_{\max},\, \lambda_{h,i}^k + \rho_\text{h}^k [\mat{M}^{-1}\mathbf{h}(\boldsymbol{\phi}^{k+1})]_i\right)\right)$,
    \Statex \hspace{2em} followed by zero-mean projection $\boldsymbol{\lambda}_\text{h} \leftarrow \boldsymbol{\lambda}_\text{h} - \frac{\mathbf{m}^\top \boldsymbol{\lambda}_\text{h}}{|\Omega|}\mathbf{1}$ if clamped.
    \State \textbf{Adaptive Penalty:} Update penalty at interval $N_\rho$:
    \Statex \hspace{2em} $\rho_\text{h}^{k+1} \leftarrow \min\left(\rho_\text{h}^{\max},\, \gamma_\rho(h_{\text{error}}^k)\,\rho_\text{h}^k\right)$.
    \State $k \gets k+1$
\Until{($k \geqslant k_{\min}$, $\|\boldsymbol{\phi}^{k} - \boldsymbol{\phi}^{k-1}\|_\infty < \tau_\phi$, 
$|\mathcal{J}^k - \mathcal{J}^{k-1}|/\max(1,|\mathcal{J}^k|) < \tau_{\text{obj}}$, 
$|G_\text{v}| \leqslant \tau_g$, 
and $\|\mathbf{h}\|_\text{norm}^k \leqslant \tau_h$) \textbf{or} $k \geqslant k_{\max}$}
\end{algorithmic}
\end{algorithm}

In Algorithm~\ref{alg:alm_mma}, the primary stopping criterion requires
convergence of design and compliance alongside strict residual tolerance feasibility
($\|\mathbf{h}\|_\text{norm}^k \leqslant \tau_h$, corresponding to $\|\mathbf{h}\|_{\ast}^k \leqslant \delta$). 
A residual stalling diagnostic
($|\|\mathbf{h}\|_\text{norm}^k - \|\mathbf{h}\|_\text{norm}^{k-1}| < 10^{-3}\tau_h$) 
monitors potential numerical plateaus without premature termination.

\section{Numerical Examples}\label{num_examples}

This section evaluates the proposed AC-ALM framework on standard two-
and three-dimensional benchmarks to demonstrate its effectiveness,
mesh robustness, and interface regularity.

\subsection{Algorithm Parameters and Computational Setup}\label{subsec:algo_params}

The algorithm parameters and operational schedules applied across all numerical examples 
are consolidated in Table~\ref{tab:algorithm_parameters}. 
First, for the Allen--Cahn constraint configuration, the gradient parameter $\epsilon$ 
and potential scaling $\beta$ scale with grid spacing as $\epsilon \propto \sqrt{\Delta x}$ 
and $\beta \propto \Delta x$. 
This scaling preserves an invariant physical interface length scale 
$\ell_\text{AC} = \epsilon/\sqrt{\beta}$ across mesh discretizations. 
Simultaneously, adapting $\beta$ ensures that the diffuse interface spans $2$--$4$ elements 
across resolutions, preventing Gauss quadrature under-sampling.

Second, regarding the Augmented Lagrangian parameters, the initial penalty $\rho_\text{h}^0$ 
is calibrated such that during the initial formation of spatial gradients ($k \in [1, 20]$), 
the AC penalty gradient remains two to three orders of magnitude smaller than compliance sensitivity 
($\|\nabla \mathcal{L}_\text{AC}\|_2 / \|\nabla \mathcal{J}\|_2 \sim 10^{-3}$--$10^{-2}$). 
This prevents premature interface pinning and allows primary load paths to self-organize unhindered. 
In parallel, a linear continuation schedule $w_\text{h}^k = \min(w_{\max}, w_\text{h}^0 + \alpha_w k)$ 
permits controlled residual infeasibility during early iterations before progressively 
steering the design toward the admissible manifold.

Third, the adaptive update strategy updates the penalty $\rho_\text{h}$ periodically 
based on the composite relative violation metric 
$h_{\text{error}} = \max(\varepsilon_\text{h,rel},\, \varepsilon_{\text{h,max,rel}})$. 
Here, $\varepsilon_\text{h,rel} = \|\mathbf{h}\|_\text{norm}/\tau_h$ tracks domain-averaged 
residual feasibility, whereas $\varepsilon_{\text{h,max,rel}} = \|\mathbf{h}\|_{\infty,\text{norm}}/\tau_{h,\max}$ 
detects localized residual spikes at structural junctions via 
$\|\mathbf{h}\|_{\infty,\text{norm}} = (\Delta x)^d\|\mathbf{h}\|_\infty/|\Omega|$. 
Tiered growth factors dynamically scale $\rho_\text{h}$ according to violation severity 
up to a safeguard ceiling $\rho_\text{h}^{\max}$, while multipliers are clamped at 
$|\lambda_i| \leqslant \lambda_{\max}$ to preserve numerical conditioning (Table~\ref{tab:algorithm_parameters}). 
Crucially, this peak localized metric steers only the penalty ramping rate without altering the global termination criterion.

Fourth, within the MMA subproblem, conservative asymptote initialization and a two-phase 
bound-fraction clamping (\texttt{albefa}) stabilize subproblem approximations against severe 
non-convex oscillations. 
In late iterations, the design move limit is periodically contracted via 
$\Delta \phi_{\max} \leftarrow \max(\Delta \phi_{\text{floor}},\, \gamma_\phi\,\Delta \phi_{\max})$ 
to suppress high-frequency interface chatter. 
Enforcing a strictly positive lower bound exceeding the design tolerance, $\Delta \phi_{\text{floor}} > \tau_\phi$, 
prevents premature termination and guarantees that convergence is governed by physical stationarity rather than optimizer step starvation.

Finally, convergence requires simultaneous satisfaction of the volume tolerance $\tau_g$, 
maximum nodal design change $\tau_\phi$, relative objective change $\tau_{\text{obj}}$, 
and residual norm tolerance $\tau_h$ within designated iteration bounds $[k_{\min}, k_{\max}]$ 
(Table~\ref{tab:algorithm_parameters}). 
To compute the residual metric efficiently, the diagonal lumped mass matrix $\mat{M}$ 
(which requires trivial $O(N)$ inversion) is evaluated directly to track
the intrinsic norm $\|\mathbf{h}\|_*$ of Eq.~\eqref{eq:tol_sense}. 
For comparison with Euclidean norms, the normalized Euclidean residual norm 
$\|\mathbf{h}\|_\text{norm} := \sqrt{v_e}\,\|\mathbf{h}\|_2 / |\Omega|$
provides an interior nodal approximation:
\begin{equation} \label{eq:delta_equiv}
    \|\mathbf{h}\|_\text{norm} \approx \frac{0.065\,(\Delta x)^{d+1}}{\sqrt{|\Omega|}}\,\|\mathbf{h}\|_*,
    \qquad
    \delta = \frac{\tau_h \sqrt{|\Omega|}}{\beta\,(\Delta x)^d}
    = \frac{\tau_h \sqrt{|\Omega|}}{0.065\,(\Delta x)^{d+1}},
\end{equation}
where the asymptotic scaling holds because interior nodes dominate the domain ($m_i = v_e$), 
with boundary nodes carrying geometric weight fractions ($v_e/2$ and $v_e/4$). 
Because $h_i \approx v_e \tilde{h}(\mathbf{x}_i)$, the Euclidean norm satisfies 
$\|\mathbf{h}\|_2 \approx \sqrt{v_e|\Omega|}\,\tilde{h}_{\text{rms}}$, so that 
$\|\mathbf{h}\|_\text{norm} \approx \frac{v_e}{|\Omega|}\|\tilde{h}\|_{L^2}
= \frac{(\Delta x)^d}{\sqrt{|\Omega|}}\tilde{h}_{\text{rms}}$. 
Because the weak residual scales as $h_i \approx m_i \tilde{h} \propto m_i \beta \propto (\Delta x)^{d+1}$, 
scaling the residual tolerance as $\tau_h \propto (\Delta x)^{d+1}$ preserves an invariant 
continuous admissible tolerance ($\delta \approx 0.075$ in 2D and $0.034$ in 3D) via Eq.~\eqref{eq:delta_equiv}. 
This confirms that the discrete stopping criterion $\|\mathbf{h}\|_\text{norm} \leqslant \tau_h$ 
faithfully enforces feasibility of the tolerance-regularized admissible set $\mathcal{S}_\delta$.

All benchmark simulations were conducted using our in-house MATLAB implementation, 
executed on a workstation featuring an Intel Core i7-11850H processor
and 128 GB of RAM (2933 MT/s), running MATLAB R2023a on Debian 12 Linux.

\begin{table}[!ht]
\centering
\footnotesize
\setlength{\tabcolsep}{2.5pt}
\caption{Consolidated algorithm parameters.}\label{tab:algorithm_parameters}
\begin{tabular}{p{6.0cm}cc}
\toprule
\textbf{Parameter Name} & \textbf{Symbol} & \textbf{Nominal Value} \\
\midrule
\multicolumn{3}{l}{\textit{Allen--Cahn Constraint Configuration}} \\
\quad Phase-field gradient parameter & $\epsilon$ & $0.175\sqrt{\Delta x}$ \\
\quad Double-well potential scaling & $\beta$ & $0.065\,\Delta x$ \\
\midrule
\multicolumn{3}{l}{\textit{Augmented Lagrangian Parameters}} \\
\quad Initial AC residual penalty & $\rho_\text{h}^0$ & 0.1 \\
\quad AC weight ramp intercept & $w_\text{h}^0$ & 0.05 \\
\quad AC weight ramp rate & $\alpha_w$ & 0.06 / iteration (0.04 in 3D) \\
\quad Maximum AC weight factor & $w_{\max}$ & 1.0 \\
\midrule
\multicolumn{3}{l}{\textit{Adaptive Update Strategy}} \\
\quad Penalty trigger metric & $h_{\text{error}} $ & $\max(\varepsilon_\text{h,rel}, \varepsilon_{\text{h,max,rel}})$ \\
\quad Growth factor ($h_{\text{error}} > 5$) & & $1.5 \times$ \\
\quad Growth factor ($h_{\text{error}} > 2$) & & $1.2 \times$ \\
\quad Growth factor ($h_{\text{error}} > 1$) & & $1.1 \times$ \\
\quad Maximum AC penalty value & $\rho_\text{h}^{\max}$ & 10 (50 for L-beam) \\
\quad Penalty update interval & $N_\rho$ & 10 iterations \\
\quad Multiplier clamp bound & $\lambda_{\max}$ & 10 \\
\midrule
\multicolumn{3}{l}{\textit{MMA Optimizer Settings}} \\
\quad Initial move limit scaling & $\Delta \phi_{\max}^0$ & 1.0 \\
\quad Asymptote initialization parameter & $s_0$ & 0.5 \\
\quad Late-phase move limit decay (iter $>$ 100) & $\gamma_\phi,\, \Delta \phi_{\text{floor}}$ & $0.95\times$ / 5 iters (min 0.02) \\
\quad Asymptote bound fraction (\texttt{albefa}) & $\alpha/\beta_{\text{mma}}$ & 0.1 (early), 0.22 (late) \\
\midrule
\multicolumn{3}{l}{\textit{Convergence Criteria}} \\
\quad Volume constraint tolerance & $\tau_g$ & $2 \times 10^{-3}$ \\
\quad Residual norm tolerance $\|\cdot\|_\text{norm}$ & $\tau_h$ & $5 \times 10^{-5}(\Delta x)^3$ (2D), $1 \times 10^{-5}(\Delta x)^4$ (3D) \\
\quad Maximum design variable change & $\tau_\phi$ & $10^{-2}$ \\
\quad Objective function relative change & $\tau_{\text{obj}}$ & $10^{-3}$ \\
\quad Iteration bounds & $k_{\min},\, k_{\max}$ & 30,\, 500 \\
\bottomrule
\end{tabular}
\end{table}

\subsection{Benchmark Geometries and Problem Setup}\label{num_setup}

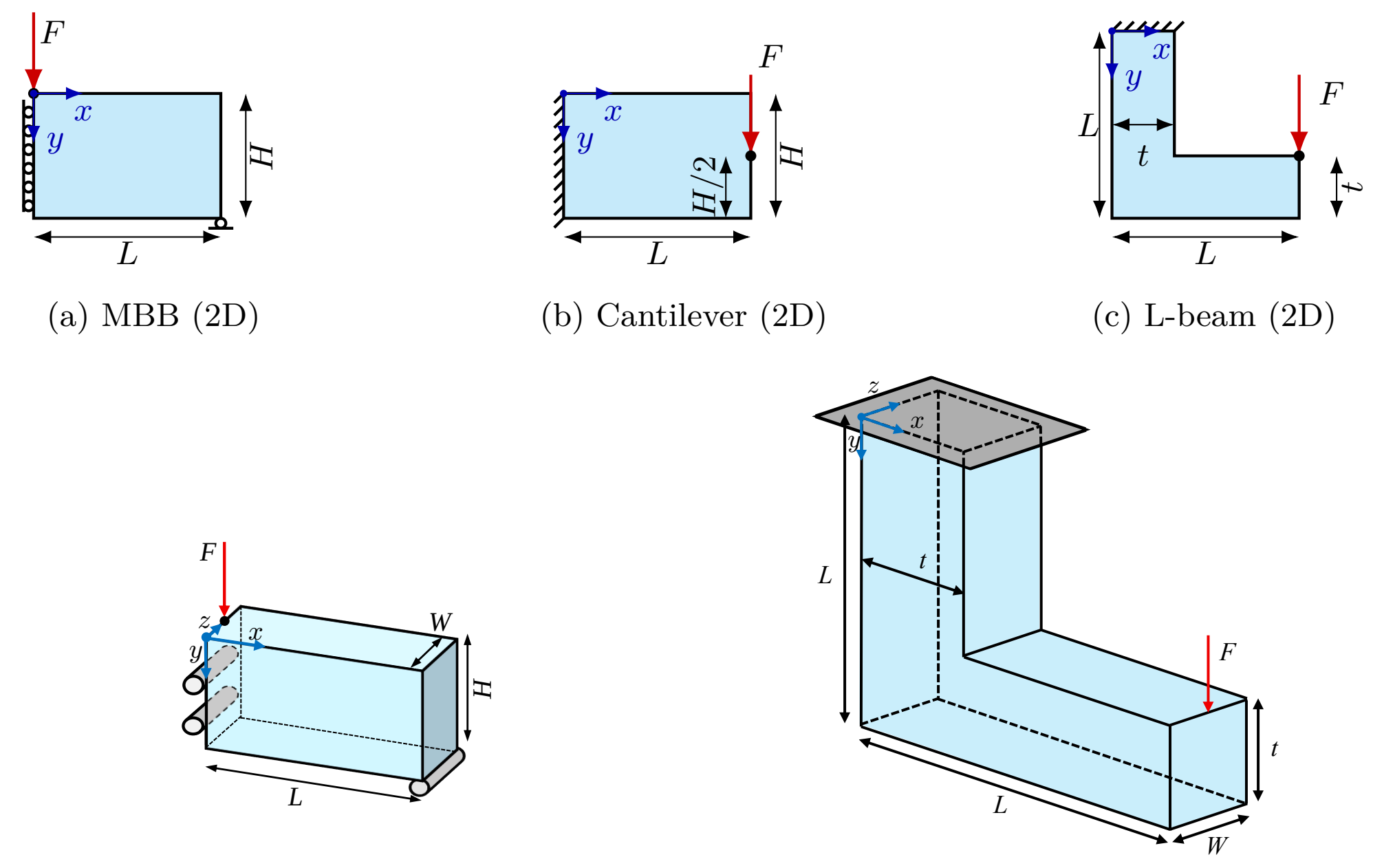
\begin{figure}[ht]
    \begin{subfigure}[b]{0.22\textwidth}
        \centering
\begin{tikzpicture}[
    x=0.06cm, y=0.06cm, 
    >=Latex,
    force/.style={-Latex, thick, red!80!black}, 
    dim/.style={Latex-Latex, thin},
    label/.style={font=\small},
]

\def\L{30}    
\def\H{20}    

\draw[thick, fill=cyan!20] (0,0) rectangle (\L,\H);

\foreach \y in {2,5,...,18}{
  \draw[thick] (-0.8, \y) circle (0.8);
}
\draw[thick] (-1.6, 1) -- (-1.6, 19); 

\draw[thick] (\L, -0.8) circle (0.8);
\draw[thick] (\L-2, -1.6) -- (\L+2, -1.6); 


\draw[force] (0, \H*1.5 + 3) -- (0, \H);
\node[label,anchor=south] at (3, \H +6) {$F$};

\fill (0, \H) circle (1.5pt);

\coordinate (Center) at (0,20);
\draw[->,thin, blue!70!black] (Center) -- ++(8,0) node[label,anchor=north] {$x$};
\draw[->,thin, blue!70!black] (Center) -- ++(0,-8) node[label,anchor=west] {$y$};
\fill[blue!70!black] (Center) circle (1pt);

\draw[dim] (0,-3) -- (\L,-3);
\node[label] at (\L/2,-5.5) {$L$};

\draw[dim] (\L+4,0) -- (\L+4,\H);
\node[label,rotate=90] at (\L+6.5,\H/2) {$H$};

\end{tikzpicture}
        \caption{MBB (2D)}\label{fig:sketch_mbb}
    \end{subfigure}
    \hfill
    \begin{subfigure}[b]{0.22\textwidth}
        \centering
\begin{tikzpicture}[
    x=0.06cm, y=0.06cm, 
    >=Latex,
    force/.style={-Latex, thick, red!80!black}, 
    dim/.style={Latex-Latex, thin},
    label/.style={font=\small},
]

\def\L{30}    
\def\H{20}    

\draw[thick, fill=cyan!20] (0,0) rectangle (\L,\H);


\foreach \y in {0,2,...,20}{
  \draw[thick] (0,\y) -- (-1.5,\y-1.5);
}


\draw[force] (\L, \H+3) -- (\L, \H/2);
\node[label,anchor=west] at (\L-1,\H+6) {$F$};

\fill (\L,\H/2) circle (1.5pt);

\coordinate (Center) at (0,20);
\draw[->,thin, blue!70!black] (Center) -- ++(8,0) node[label,anchor=north] {$x$};
\draw[->,thin, blue!70!black] (Center) -- ++(0,-8) node[label,anchor=west] {$y$};
\fill[blue!70!black] (Center) circle (1pt);

\draw[dim] (0,-3) -- (\L,-3);
\node[label] at (\L/2,-5.5) {$L$};

\draw[dim] (\L+4,0) -- (\L+4,\H);
\node[label,rotate=90] at (\L+6.5,\H/2) {$H$};

\draw[dim] (\L-4,0) -- (\L-4,\H/2);
\node[label, rotate=90] at (\L-7,\H/4) {$H/2$};

\end{tikzpicture}
        \caption{Cantilever (2D)}\label{fig:sketch_cantilever}
    \end{subfigure}
    \hfill
    \begin{subfigure}[b]{0.22\textwidth}
        \centering
\begin{tikzpicture}[
    x=0.06cm, y=0.06cm, 
    >=Latex,
    force/.style={-Latex, thick, red!80!black}, 
    dim/.style={Latex-Latex, thin},
    label/.style={font=\small},
]

\def\W{30}    
\def\H{30}    
\def\t{10}    

\draw[thick, fill=cyan!20] (0,\H) -- (0,0) -- (\W,0) -- (\W,\t) -- (\t,\t) -- (\t,\H) -- cycle;

\foreach \x in {0,2,...,10}{
  \draw[thick] (\x,\H) -- (\x+1.5,\H+1.5);
}


\draw[force] (\W, \t+13) -- (\W, \t);
\node[label,anchor=west] at (\W+1,\t+10) {$F$};

\fill (\W,\t) circle (1.5pt);

\coordinate (Center) at (0,30);
\draw[->,thin, blue!70!black] (Center) -- ++(8,0) node[label,anchor=north] {$x$};
\draw[->,thin, blue!70!black] (Center) -- ++(0,-8) node[label,anchor=west] {$y$};
\fill[blue!70!black] (Center) circle (1pt);

\draw[dim] (-2,0) -- (-2,\H);
\node[label] at (-3.8,\H/2) {$L$};

\draw[dim] (0,-3) -- (\W,-3);
\node[label] at (\W/2,-5.5) {$L$};

\draw[dim] (\W+6,0) -- (\W+6,\t);
\node[label,rotate=90] at (\W+8.5,\t/2) {$t$};

\draw[dim] (0,\H/2) -- (\t,\H/2);
\node[label] at (\t/2,\H/2-5) {$t$};

\end{tikzpicture}

        \caption{L-beam (2D)}\label{fig:sketch_lbeam}
    \end{subfigure}
    \\
    \begin{subfigure}[b]{0.48\textwidth}
        \centering
        \includegraphics[width=0.5\linewidth]{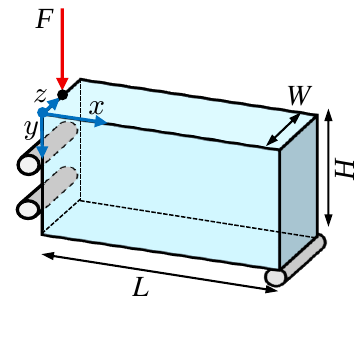}
        \caption{MBB (3D)}\label{fig:sketch_MBB_3d}
    \end{subfigure}
    \hfill
    \begin{subfigure}[b]{0.48\textwidth}
        \centering
        \includegraphics[width=0.75\linewidth]{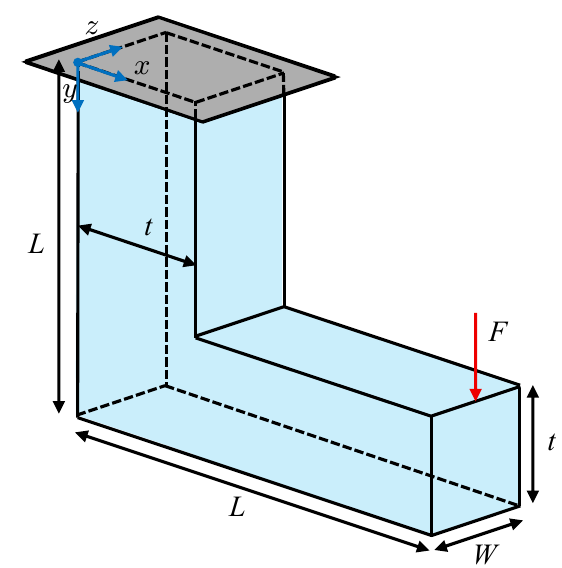}
        \caption{L-beam (3D)}\label{fig:sketch_Lbeam_3d}
    \end{subfigure}
    \caption{Benchmark problem setups.}
    \label{fig:setup}
\end{figure}

All benchmark problems are discretized using uniform meshes
of bilinear quad (Q4) elements in 2D or trilinear hexahedral (H8) elements in 3D. 
The geometric configurations are sketched in Fig.~\ref{fig:setup}:
\begin{itemize}
    \item \textbf{2D MBB beam} (Fig.~\ref{fig:sketch_mbb}): 
    Exploiting bilateral symmetry, we model the right half on a rectangular domain
    of size $L = 120$ by $H = 80$, corresponding to a full beam aspect ratio of $3:1$. 
    Horizontal displacement is fixed along the left symmetry edge ($u_x = 0$), 
    the bottom-right corner is vertically supported ($u_y = 0$), 
    and a vertical unit point load is applied downward at the top-left corner.
    \item \textbf{2D Cantilever beam} (Fig.~\ref{fig:sketch_cantilever}): 
    The domain is a rectangle of dimension $L = 120$ by $H = 80$. 
    The left edge is fully clamped ($\mathbf{u} = \mathbf{0}$), 
    and a downward unit point load is applied at the midpoint of the right edge.
    \item \textbf{2D L-beam} (Fig.~\ref{fig:sketch_lbeam}): 
    The non-convex domain has an overall bounding dimension of $120 \times 120$
    with an arm thickness $t = 48$, forming a re-entrant corner
    with a passive void region of $72 \times 72$. 
    The top edge is fully clamped.
    A downward unit point load is applied at the upper-right corner
    of the horizontal arm (level with the re-entrant corner).
    \item \textbf{3D MBB beam} (Fig.~\ref{fig:sketch_MBB_3d}): 
    Taking advantage of quarter symmetry, the design domain has dimensions
    $L = 60$, $H = 40$, and $W = 20$. 
    Symmetry conditions are imposed on the corresponding cut planes, 
    the outer bottom line is vertically restrained, 
    and a downward unit point load is applied at the midpoint
    of the top-left edge ($z = W/2$).
    \item \textbf{3D L-beam} (Fig.~\ref{fig:sketch_Lbeam_3d}): 
    The domain has bounding dimensions $60 \times 60 \times 20$
    with an arm thickness $t = 24$. 
    The top surface is fully clamped.
    A downward unit point load is applied at the midpoint
    of the upper-right edge of the horizontal arm
    ($z = W/2$, level with the re-entrant edge).
\end{itemize}
In all cases, the isotropic solid phase has Young's modulus $E_0 = 1$
and Poisson's ratio $\nu = 0.3$. 
The artificial void stiffness is set to $E_{\min} = 10^{-9}$, 
and the SIMP interpolation power is $p = 3$. 
All optimizations are initialized from a uniform density field $\phi(\mathbf{x}) = V$.

\subsection{2D Benchmark Problems}\label{subsec:2d_benchmarks}

The two-dimensional study evaluates the AC-ALM framework against a nodal-variable 
SIMP baseline using the identical node-to-element mapping operator (Eq.~\eqref{eq:projection}) 
without the Allen--Cahn residual constraint ($\mathbf{h} = \mathbf{0}$ disabled). 
The comparative test matrix comprises 48 optimization runs for each method. 
It spans three benchmark geometries (MBB beam, cantilever, and L-beam), 
four mesh discretizations ($120\times 80$, $180\times 120$, $240\times 160$, and $300\times 200$ 
for the rectangular domains, and $120\times 120$ to $300\times 300$ for the L-beam domain, 
sharing element sizes $\Delta x \in \{1,\,2/3,\,1/2,\,2/5\}$), and two target volume 
fractions ($V \in \{0.3,\,0.4\}$).

Importantly, the proposed AC-ALM formulation does not apply an external density filter. 
The design field is directly mapped to element densities via standard bilinear interpolation; 
feature regularity is enforced intrinsically by the reaction--diffusion residual constraint. 
To benchmark this mechanism fairly against established practices, the comparative baseline 
incorporates two SIMP regimes: unregularized SIMP ($r_\text{min} = 0.1$, without external 
filtering) and filtered SIMP ($r_\text{min} = 2.0$, with a standard linear density filter).

All benchmark cases are investigated across two distinct regularization regimes. 
The \emph{filter-assisted regime} ($r_{\min} = 2.0$) demonstrates compatibility with 
external manufacturing filters. 
The \emph{intrinsic AC regime} ($r_{\min} = 0.1 \leqslant \frac{\sqrt{d}}{2}\Delta x$) 
disables sub-element filtering; length-scale control and interface regularity rely 
exclusively on the Allen--Cahn residual constraint. 

Performance is evaluated across four criteria: structural compliance $\mathcal{J}$, 
residual feasibility $\|\mathbf{h}\|_* \leqslant \delta$ (monitored via the equivalent 
normalized Euclidean surrogate $\|\mathbf{h}\|_\text{norm} \leqslant \tau_h$), 
topological mesh consistency under refinement, and computational cost. 
An algorithmic ablation isolating the necessity of the Lagrange multiplier against 
a pure penalty formulation is detailed in Section~\ref{subsec:ablation_multiplier}.

\subsubsection{MBB Beam}\label{num_examples_mbb}

\begin{figure}[ht]
\centering
\includegraphics{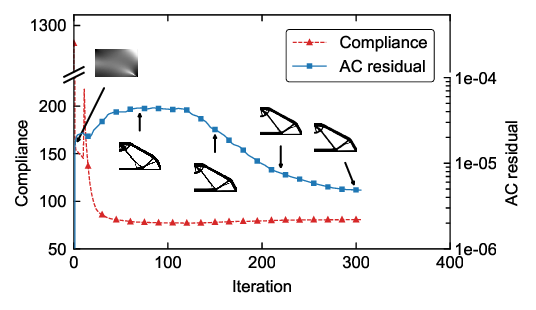}
\caption{Convergence history for MBB under AC-ALM ($180\times 120$, $V=0.3$, $r_{\min}=0.1$).}\label{fig:mbb_conv}
\end{figure}

We first examine the MBB beam problem described in Section~\ref{num_setup}. 
Figure~\ref{fig:mbb_conv} shows the convergence history of the reference case 
($180\times 120$ mesh, $V = 0.3$, $r_{\min} = 0.1$). 
During the early optimization phase, the compliance decreases sharply to a minimum 
of $\mathcal{J}_{\min} = 77.05$ (at iteration 122). 
This establishes the primary load-bearing topology while the AC residual remains elevated. 
As the ALM updates progressively tighten residual enforcement,
the normalized residual drops by a factor of 8.3 (nearly an order of magnitude),
from $\|\mathbf{h}\|_{\text{norm}} = 4.02\times 10^{-5}$
to $4.85\times 10^{-6}$ ($0.33\,\tau_h$), 
successfully satisfying the feasibility tolerance. 
During this residual-reduction stage, the compliance stabilizes
with a modest adjustment of $4.67\%$ to $\mathcal{J}_{\mathrm{final}} = 80.64$.
This reflects the regularizing effect of the finite diffuse interface. 
Concurrently, the conserved AC chemical potential settles to $\mu \approx -5.6\times 10^{-5}$ 
($|\mu|/\beta \approx 1.3\times 10^{-3}$; see its full evolution
compared with the 3D case in Fig.~\ref{fig:mu_history}). 
Because the 2D truss members are bounded predominantly by straight boundaries
with near-zero interface curvature ($\kappa \approx 0$),
the asymptotic Gibbs--Thomson relation of Section~\ref{sec:ac_constraint}
corresponds to an essentially vanishing chemical potential. 
The volume constraint error remains below $2\times 10^{-3}$ throughout, 
confirming seamless compatibility between volume conservation and local residual enforcement.

The optimized topologies under mesh refinement are presented
in Table~\ref{tab:mbb-rmin2} ($r_{\min} = 2.0$) 
and Table~\ref{tab:mbb-rmin01} ($r_{\min} = 0.1$). 
In the filter-assisted regime ($r_{\min} = 2.0$),
both SIMP and AC-ALM yield consistent, well-resolved topologies, 
with AC-ALM suppressing minor secondary branches and producing smoother boundaries. 
The decisive contrast appears in the intrinsic regime ($r_{\min} = 0.1$).
While the unregularized SIMP baseline exhibits severe mesh dependency 
and fragments into increasingly thin members as the mesh is refined, 
AC-ALM maintains an identical, mesh-consistent structural layout
from $120\times 80$ to $300\times 200$. 
This confirms that the intrinsic interface length scale
$\ell_\text{AC} = \epsilon/\sqrt{\beta}$ robustly governs the structural 
feature size independently of filtering.

\begin{table}[!htbp]
    \centering
    \small
    \setlength{\tabcolsep}{2pt}
    \caption{Structure comparison of MBBs ($r_{\min}=2$)} 
    \label{tab:mbb-rmin2}
    \begin{tabular}{cccccc}
        \toprule
        $V$ & Method & $120\times 80$ & $180\times 120$ & $240\times 160$ & $300\times 200$ \\
        \midrule
        \multirow{2}{*}[-11.5pt]{0.3} & SIMP &
        \includegraphics[width=1.5cm,valign=m]{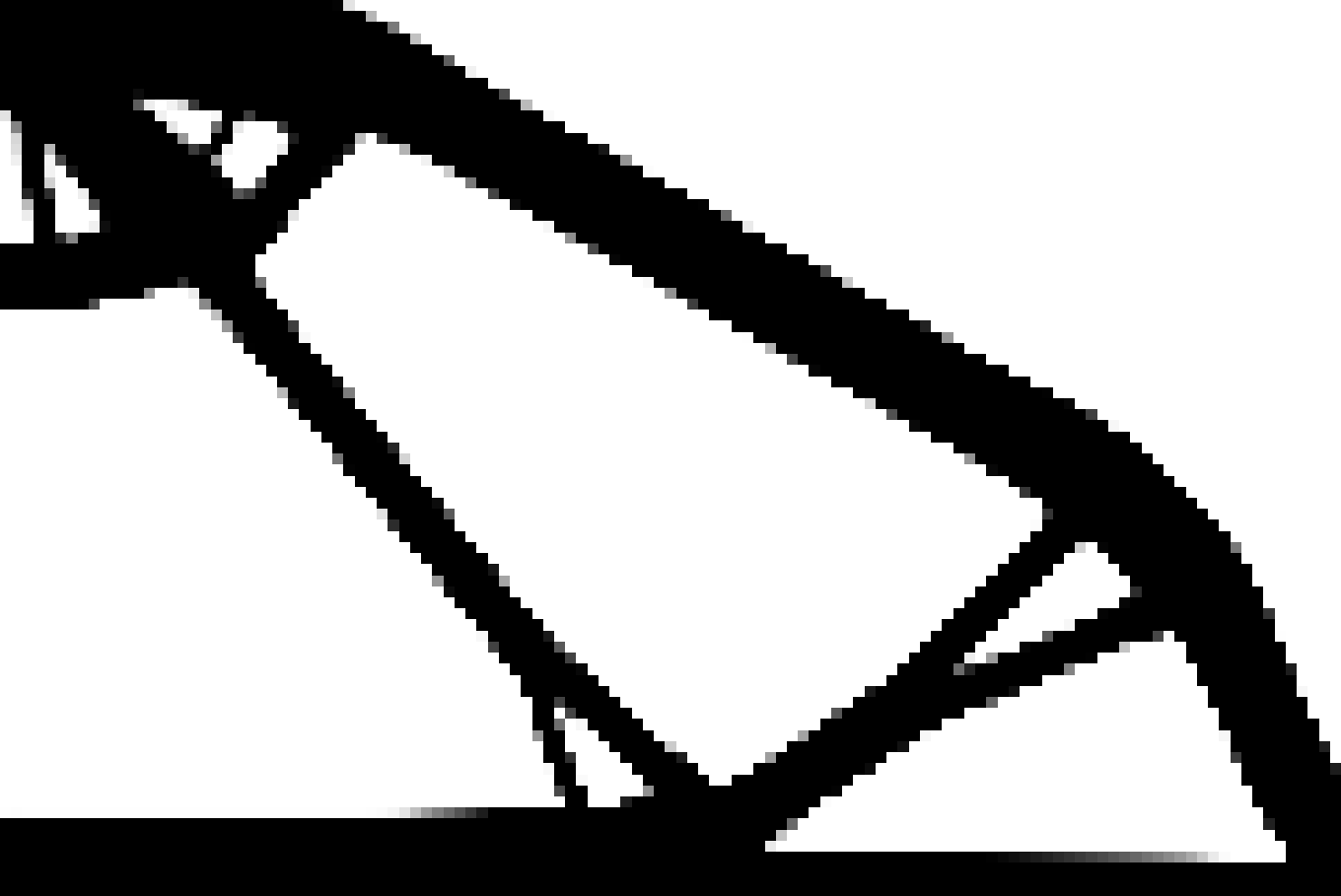} &
        \includegraphics[width=1.5cm,valign=m]{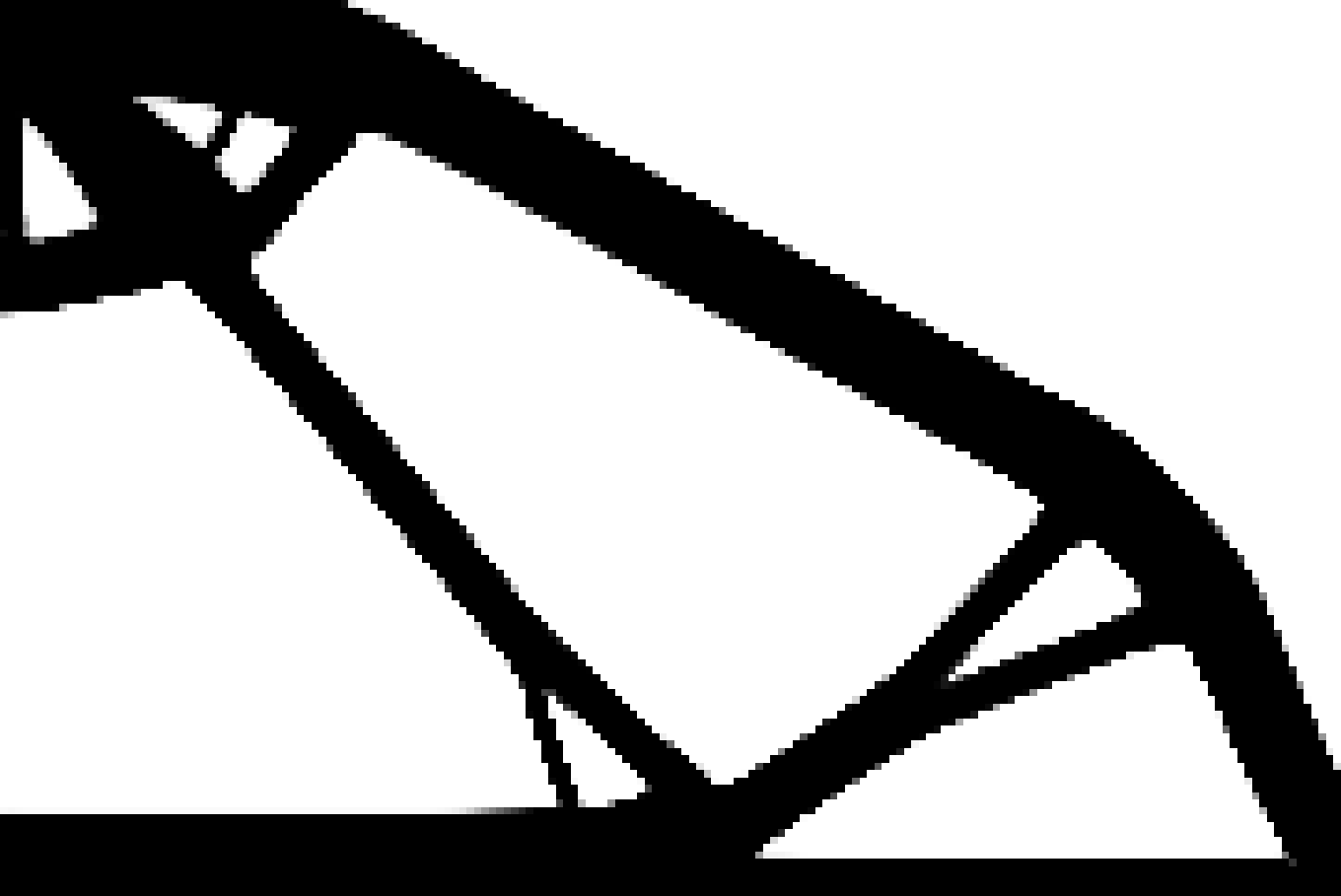} &
        \includegraphics[width=1.5cm,valign=m]{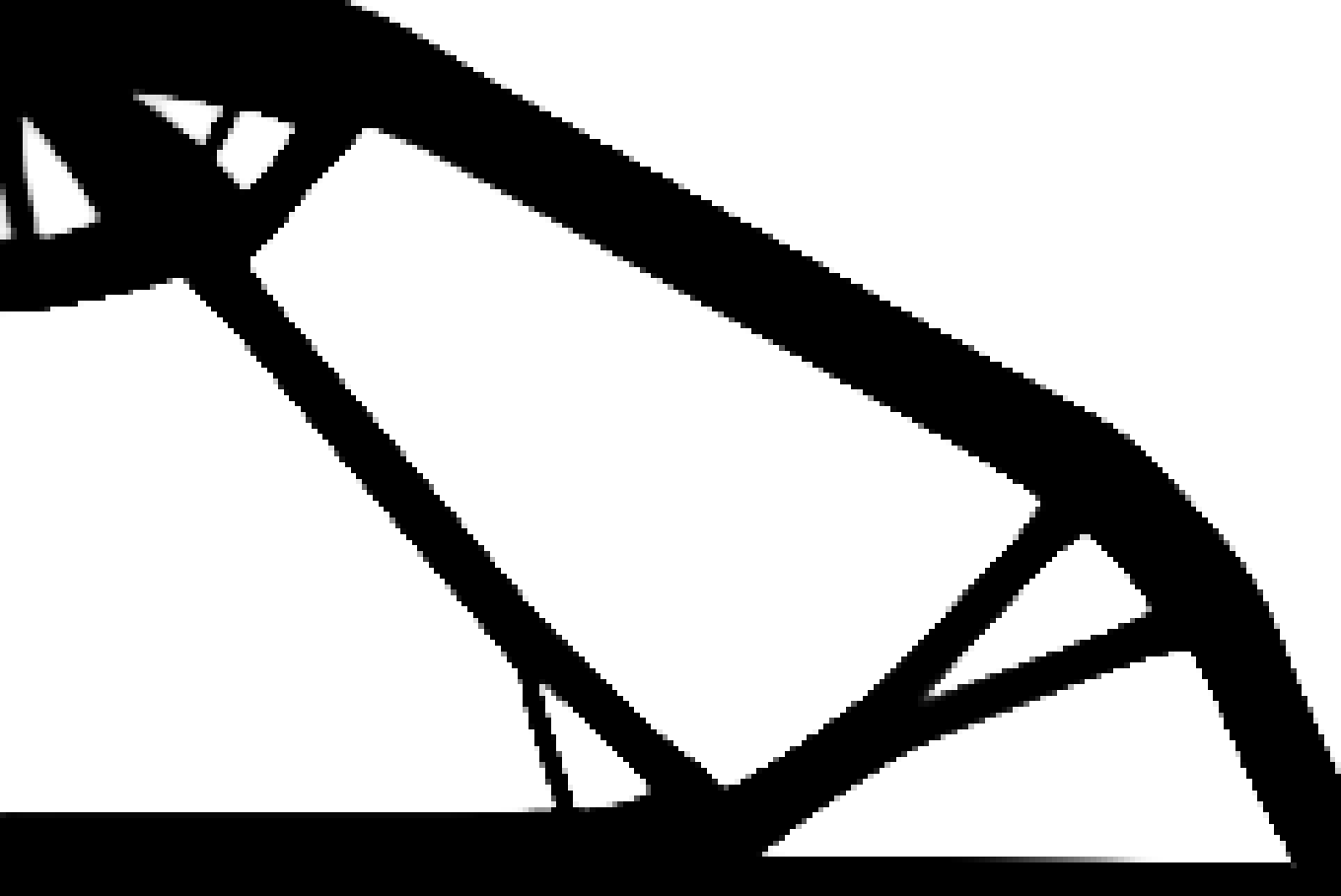} &
        \includegraphics[width=1.5cm,valign=m]{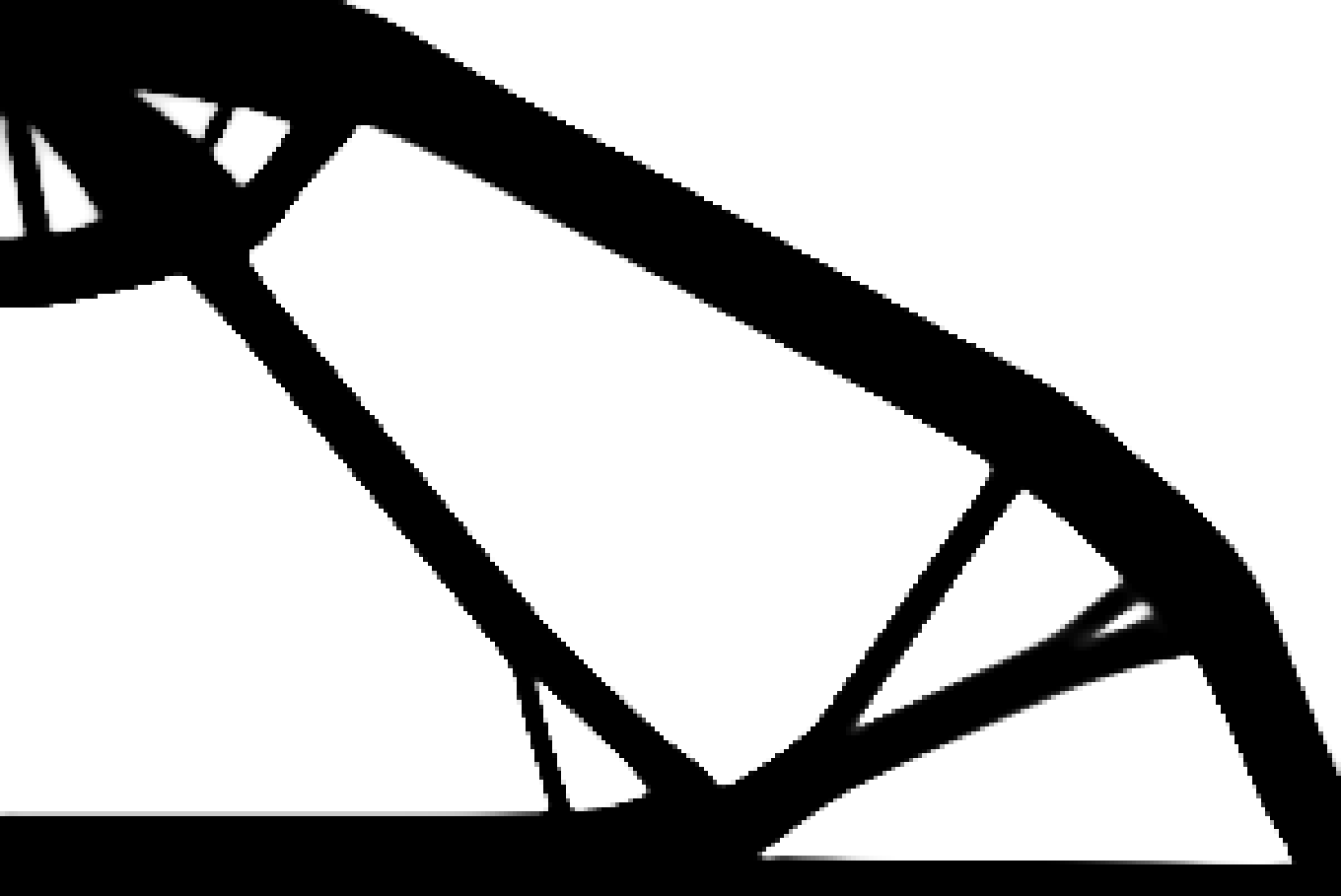} \\ \addlinespace[12pt]
                             & AC-ALM &
        \includegraphics[width=1.5cm,valign=m]{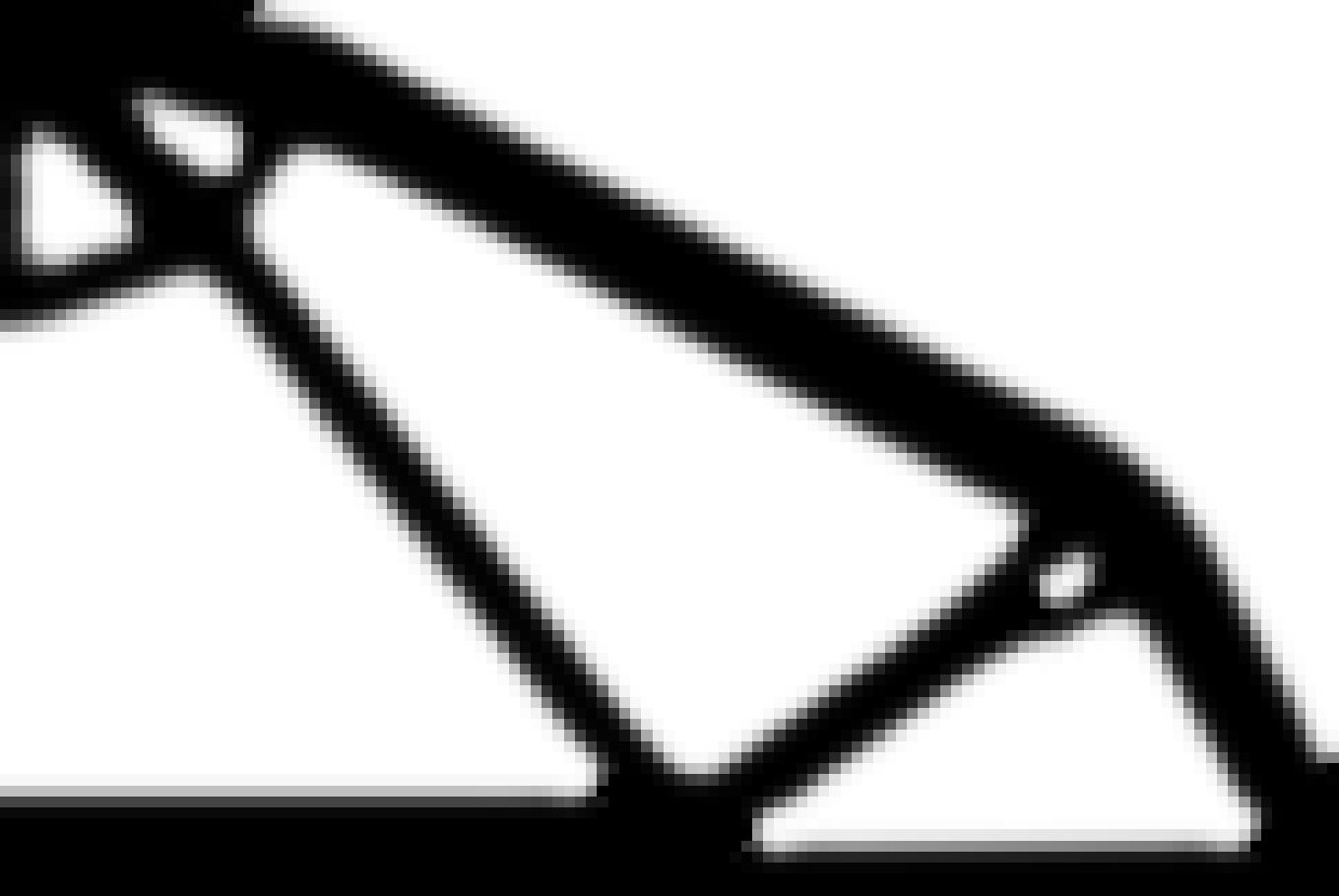} &
        \includegraphics[width=1.5cm,valign=m]{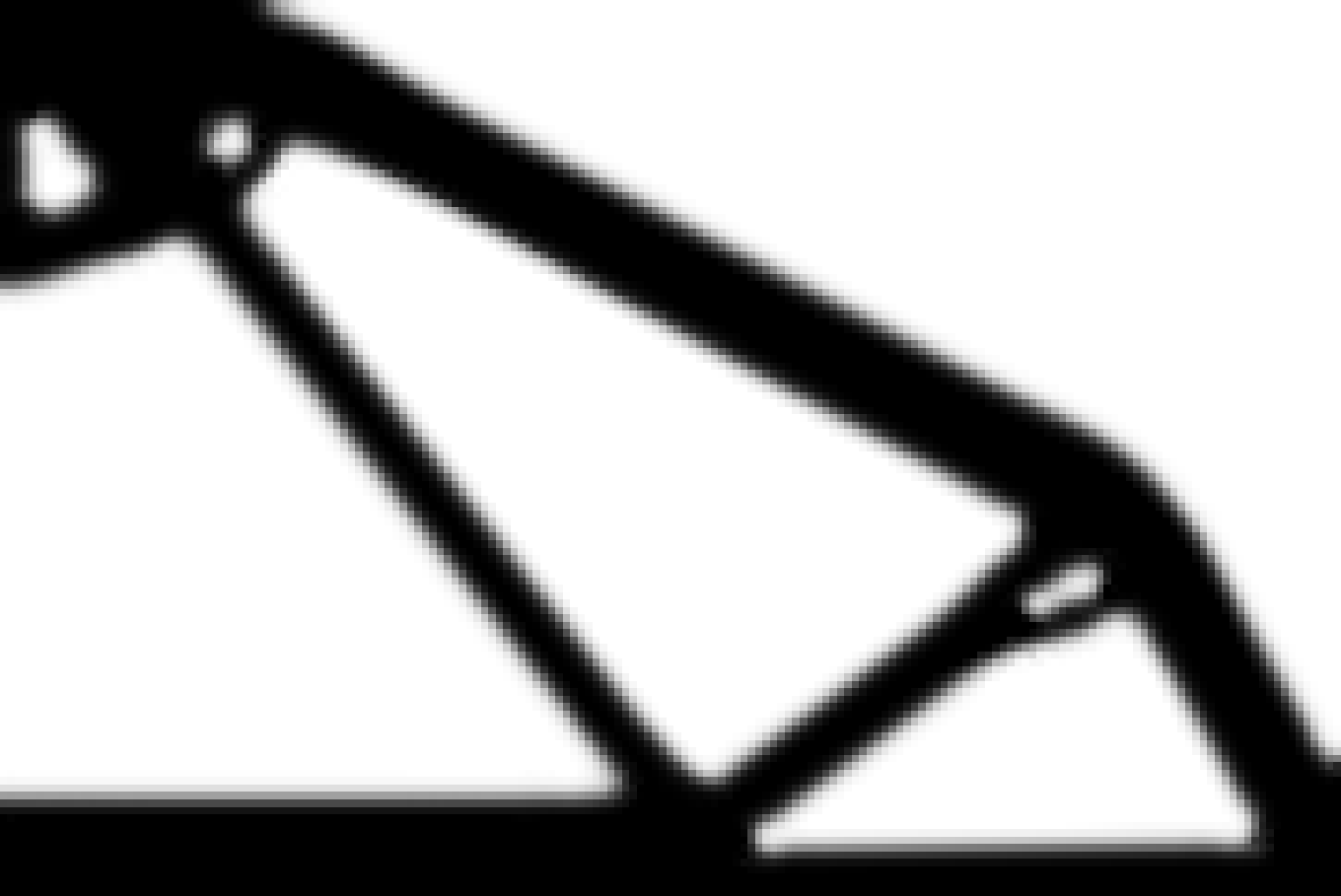} &
        \includegraphics[width=1.5cm,valign=m]{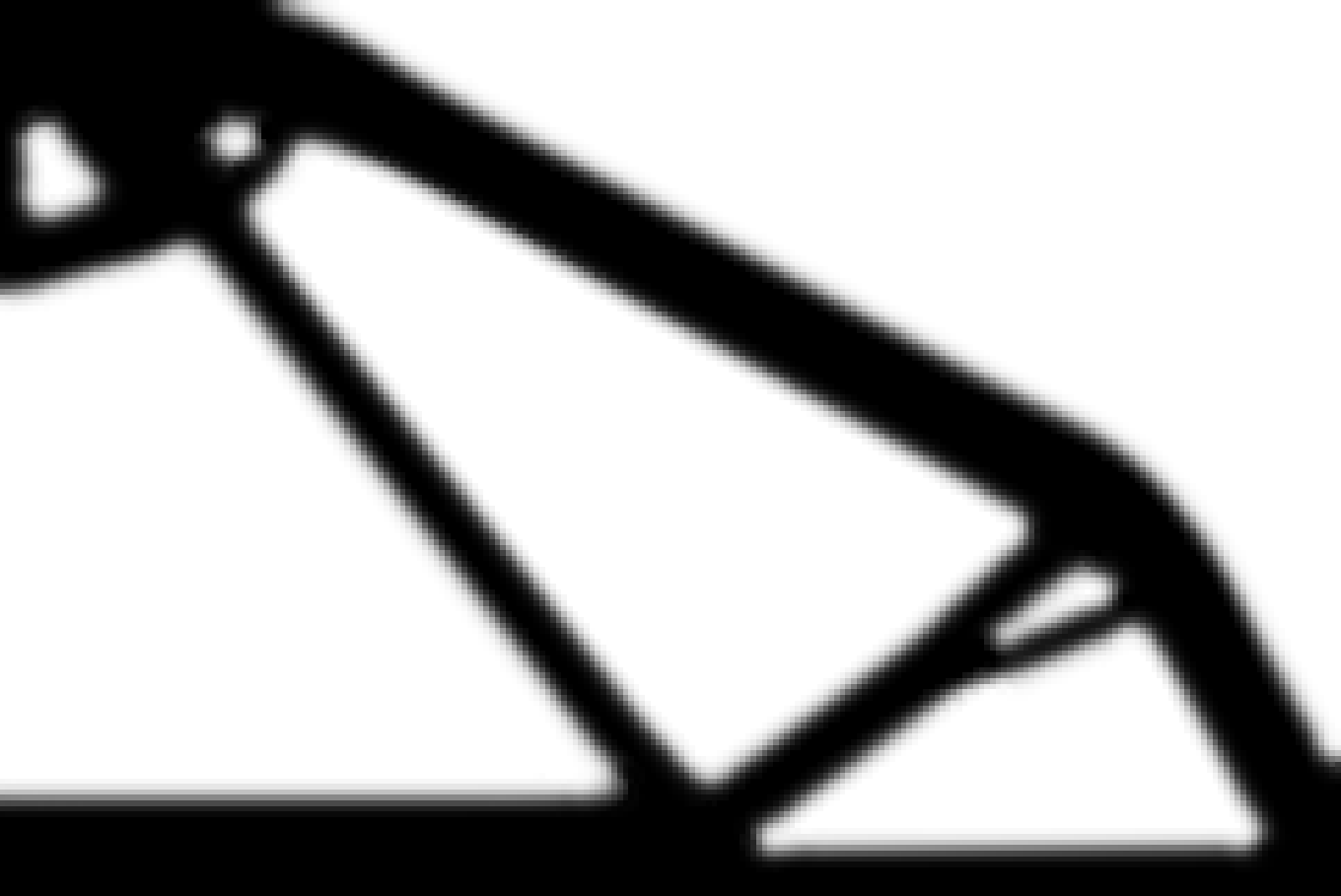} &
        \includegraphics[width=1.5cm,valign=m]{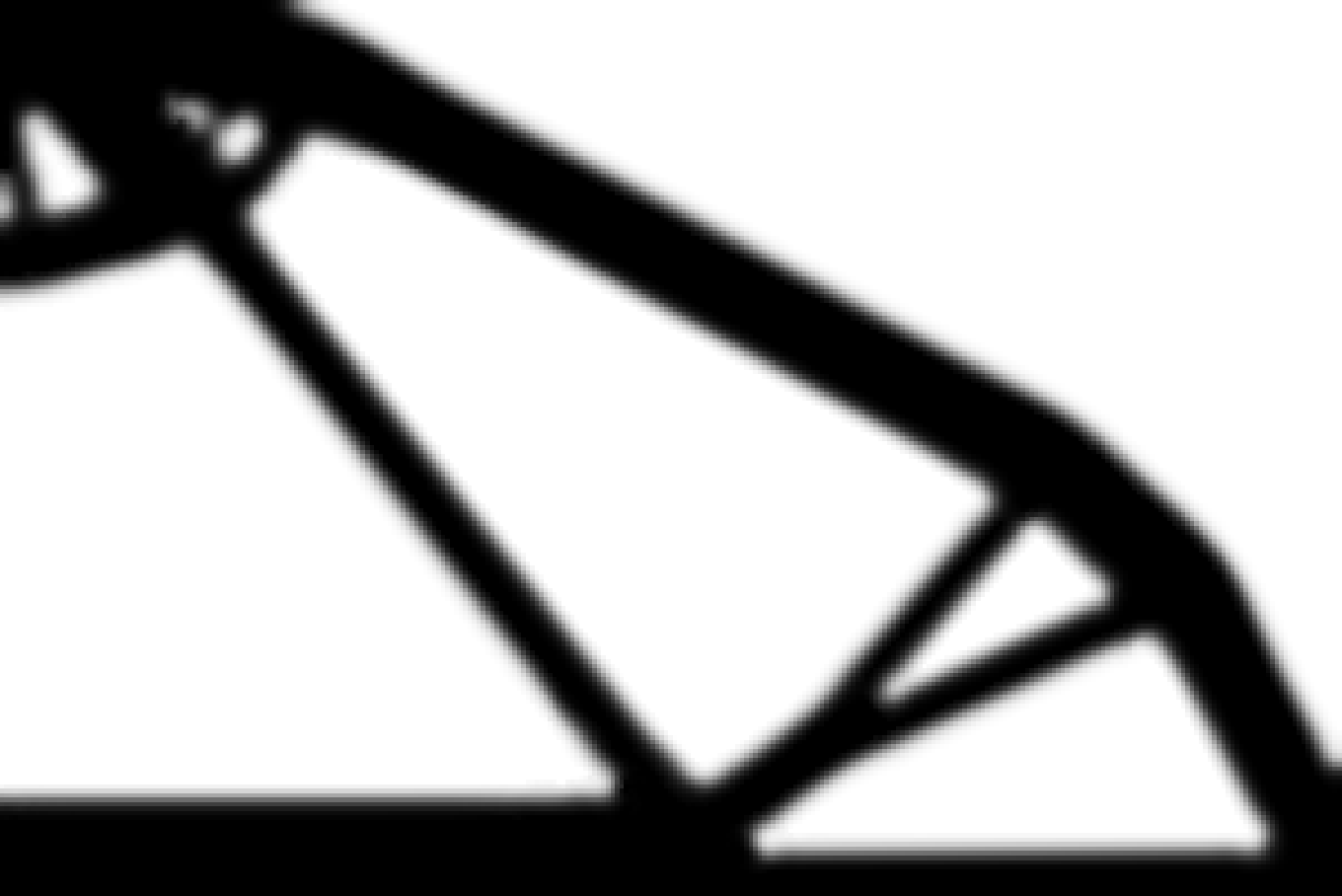} \\
        \midrule
        \multirow{2}{*}[-11.5pt]{0.4} & SIMP &
        \includegraphics[width=1.5cm,valign=m]{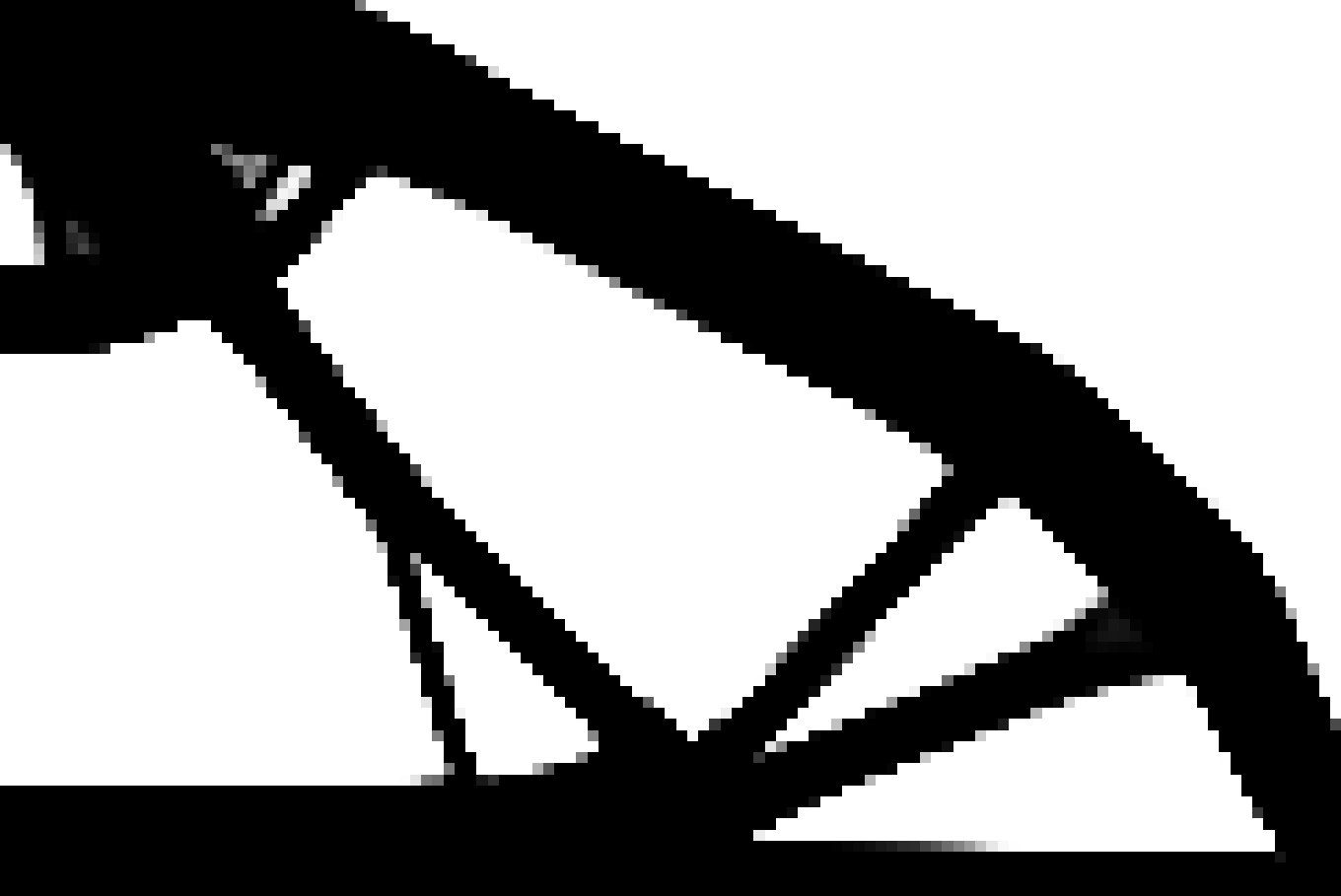} &
        \includegraphics[width=1.5cm,valign=m]{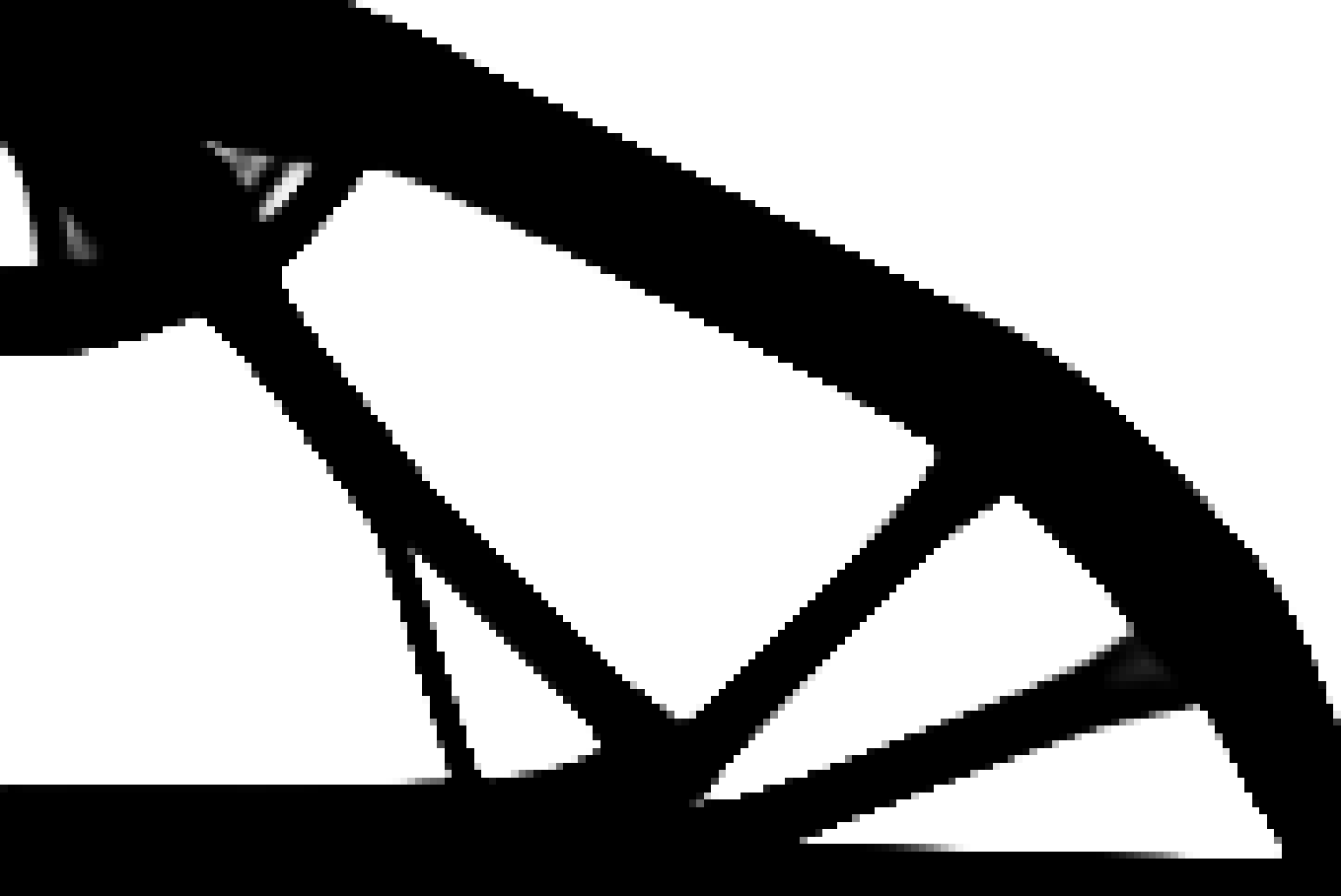} &
        \includegraphics[width=1.5cm,valign=m]{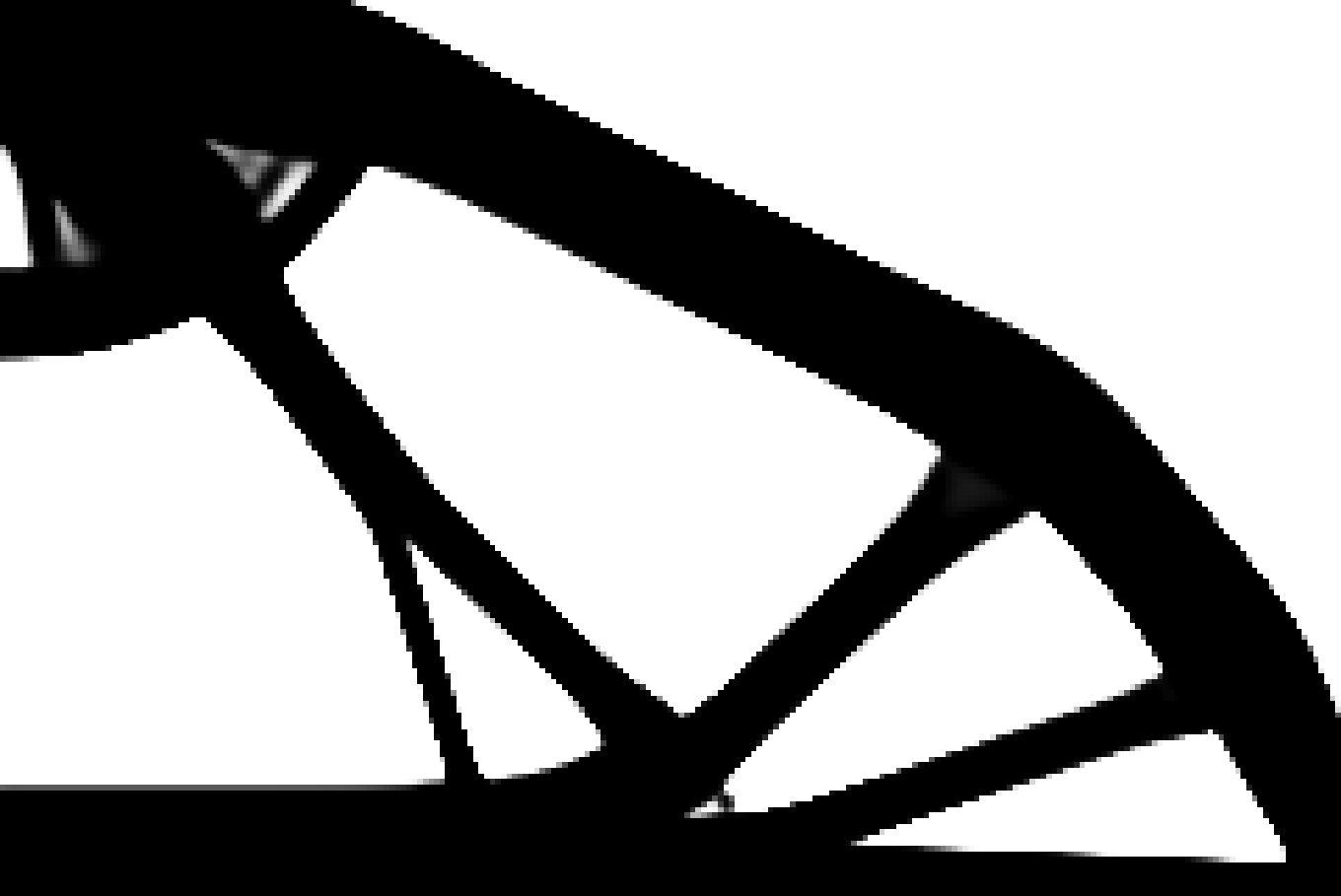} &
        \includegraphics[width=1.5cm,valign=m]{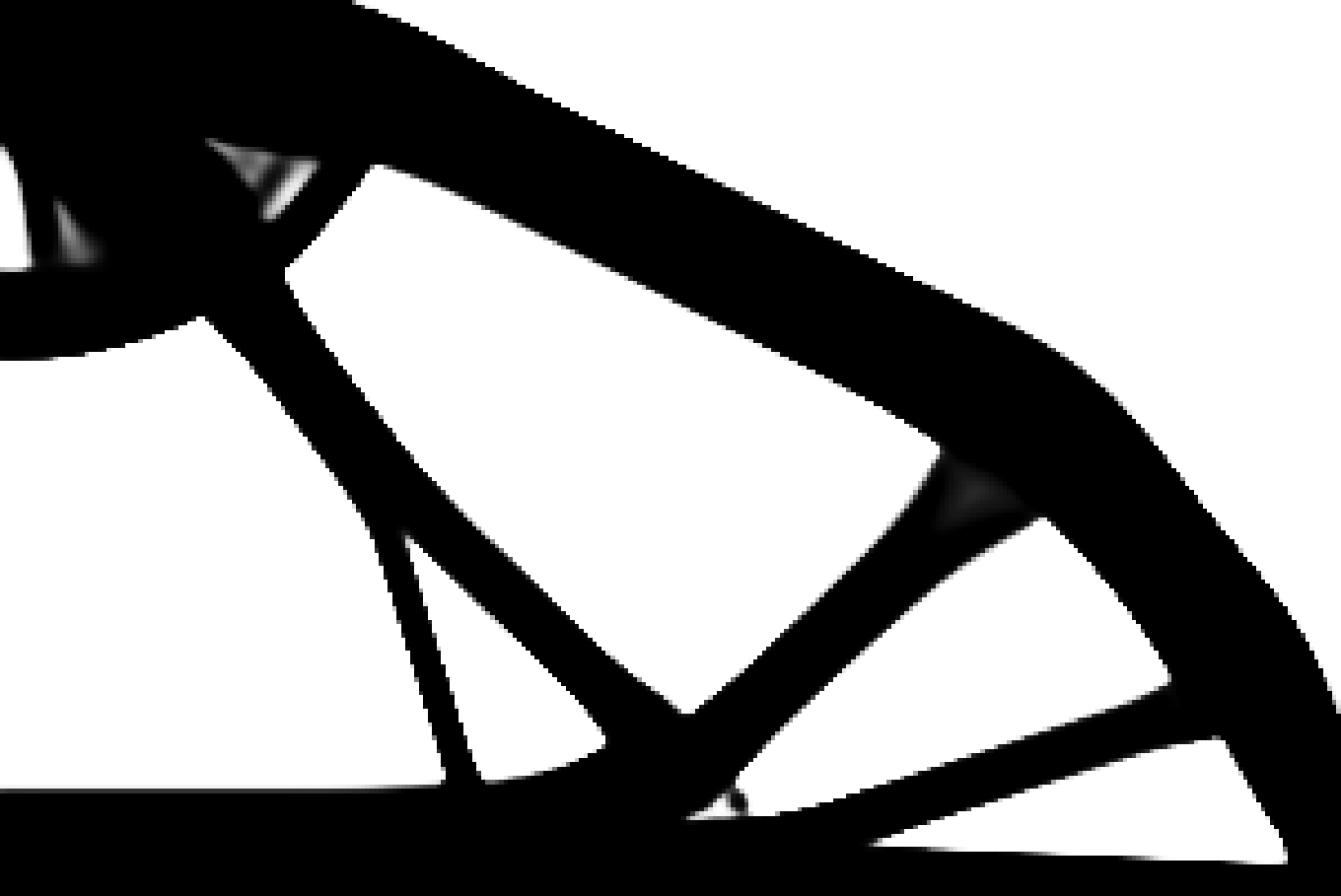} \\ \addlinespace[12pt]
                             & AC-ALM &
        \includegraphics[width=1.5cm,valign=m]{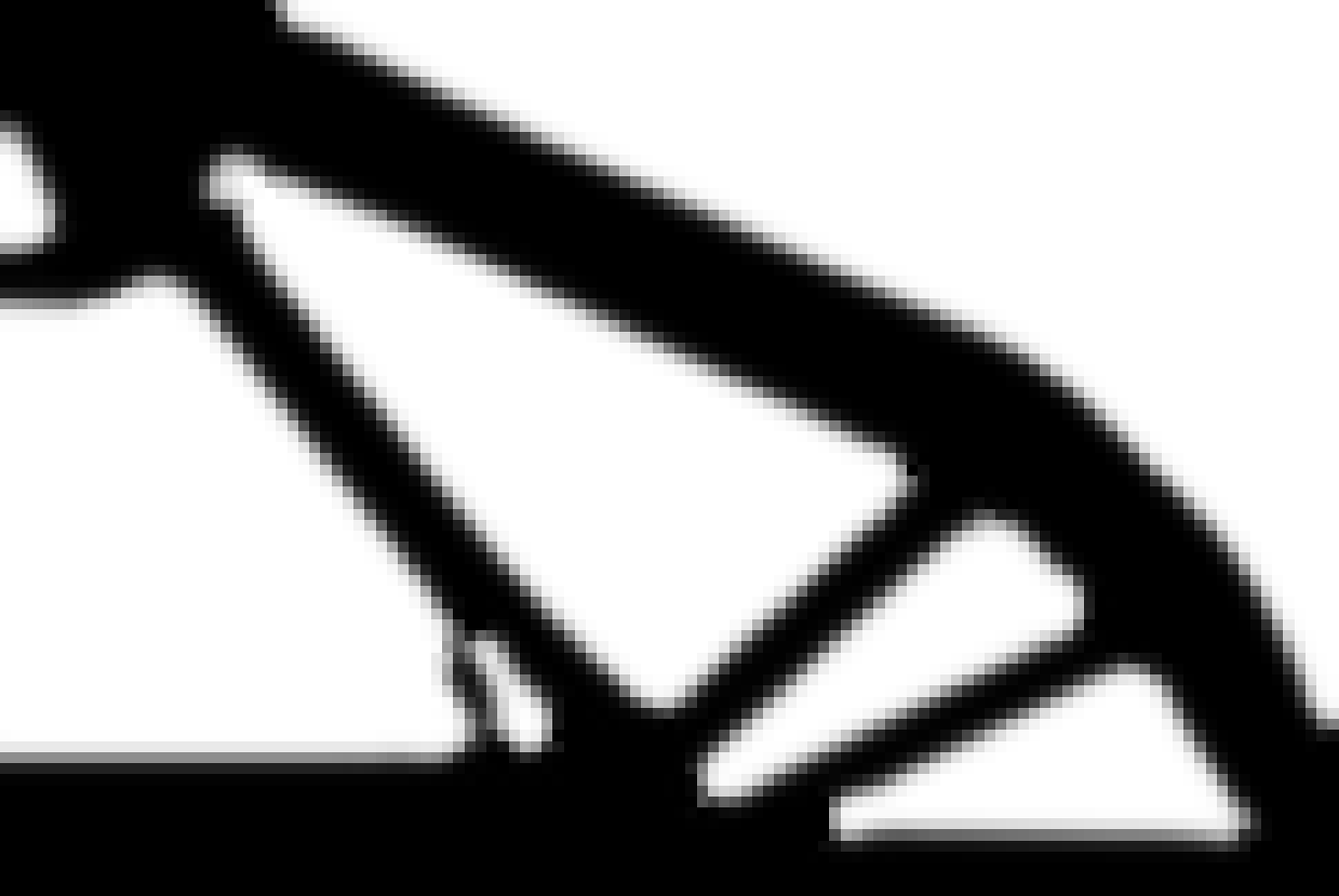} &
        \includegraphics[width=1.5cm,valign=m]{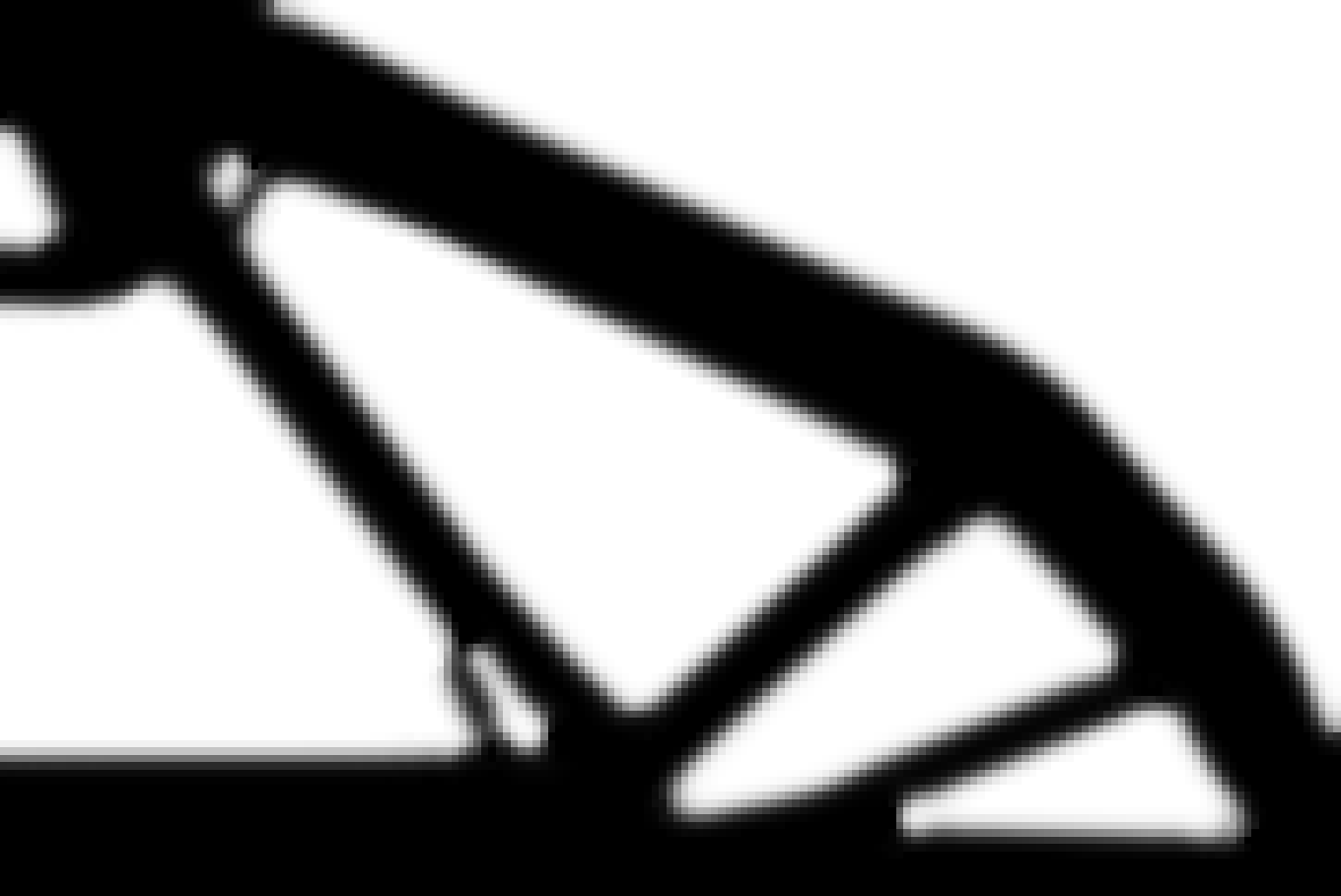} &
        \includegraphics[width=1.5cm,valign=m]{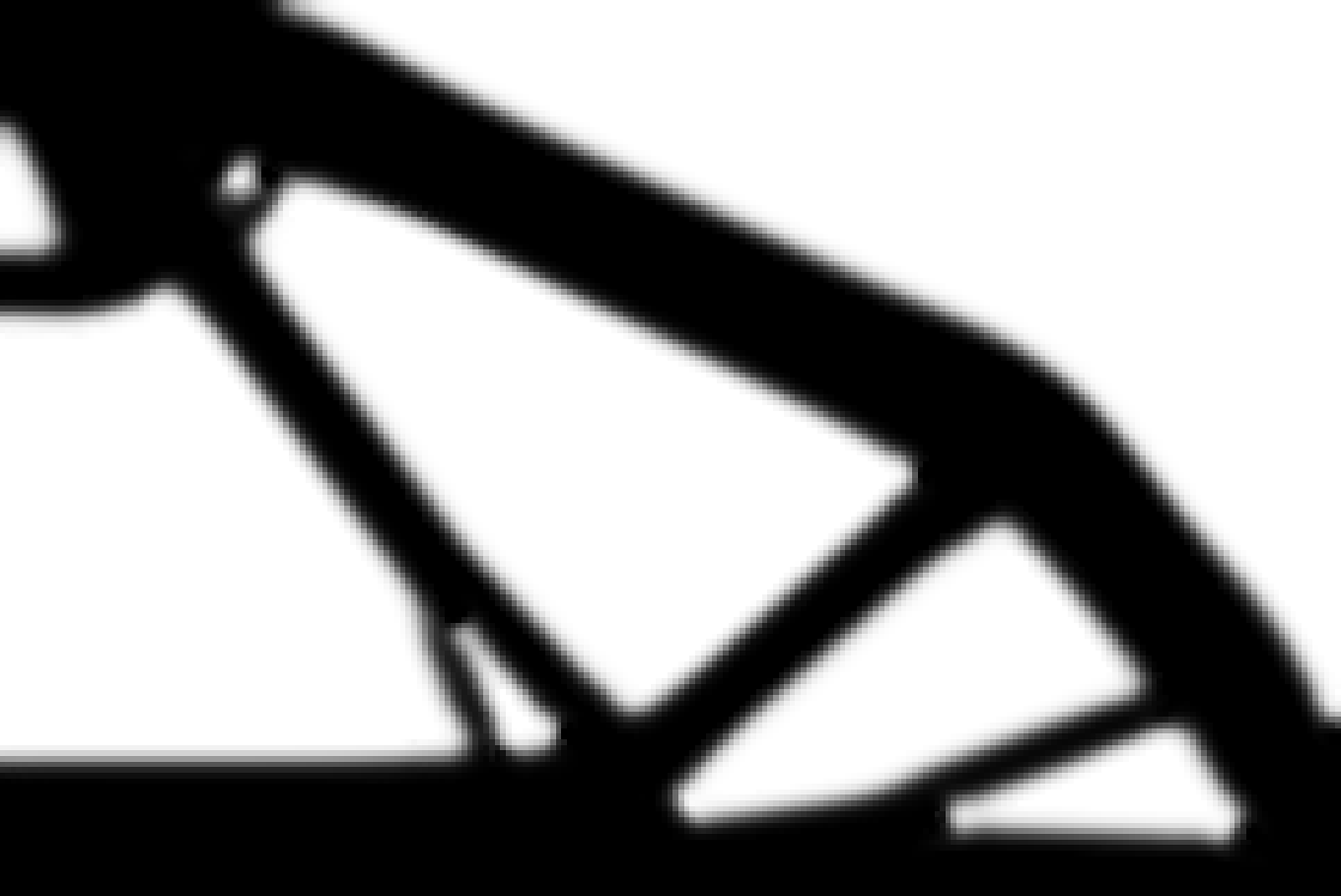} &
        \includegraphics[width=1.5cm,valign=m]{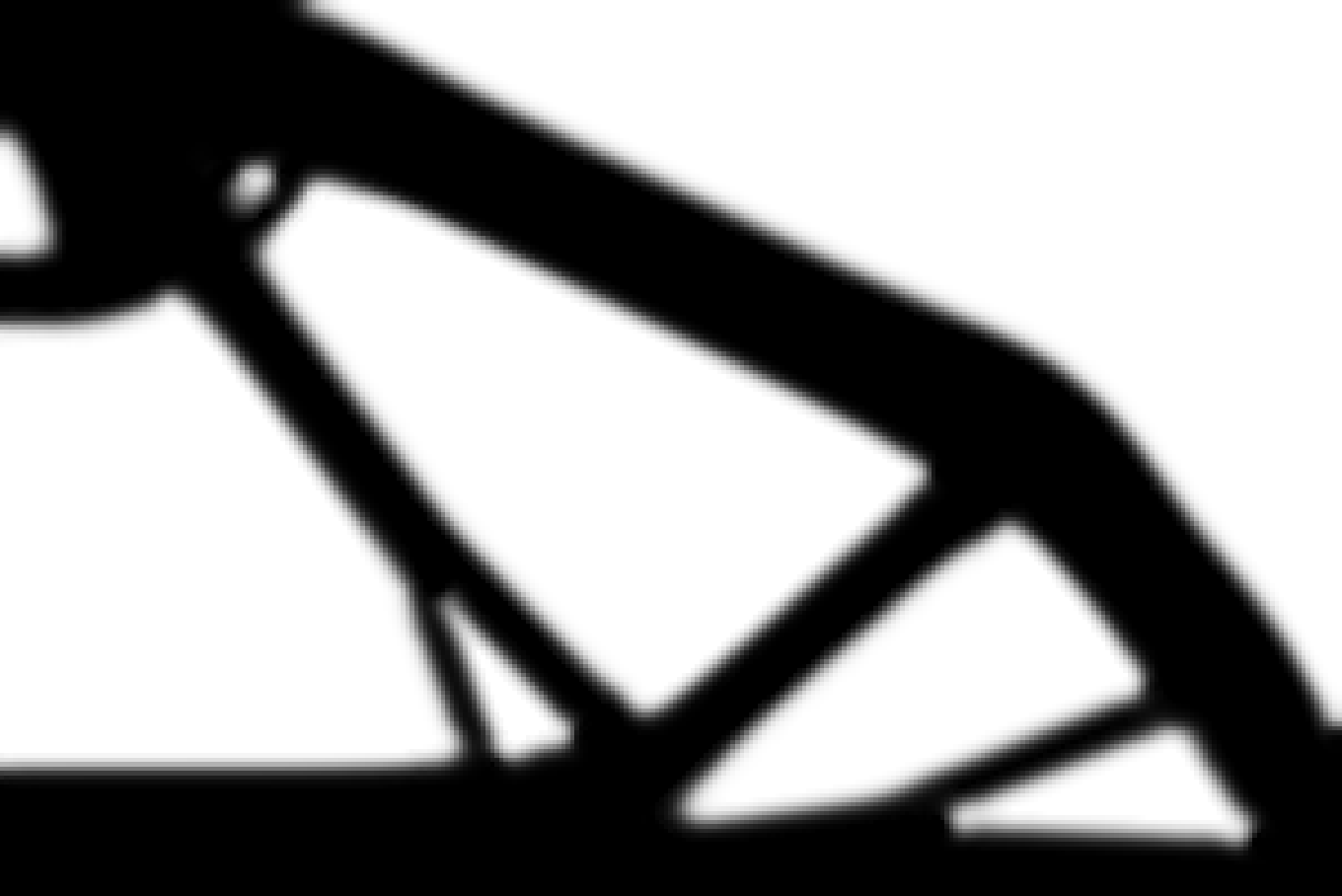} \\
        \bottomrule
    \end{tabular}
\end{table}

\begin{table}[!htbp]
    \centering
    \small
    \setlength{\tabcolsep}{2pt}
    \caption{Structure comparison of MBBs ($r_{\min}=0.1$)} 
    \label{tab:mbb-rmin01}
    \begin{tabular}{cccccc}
        \toprule
        $V$ & Method & $120\times 80$ & $180\times 120$ & $240\times 160$ & $300\times 200$ \\
        \midrule
        \multirow{2}{*}[-11.5pt]{0.3} & SIMP &
        \includegraphics[width=1.5cm,valign=m]{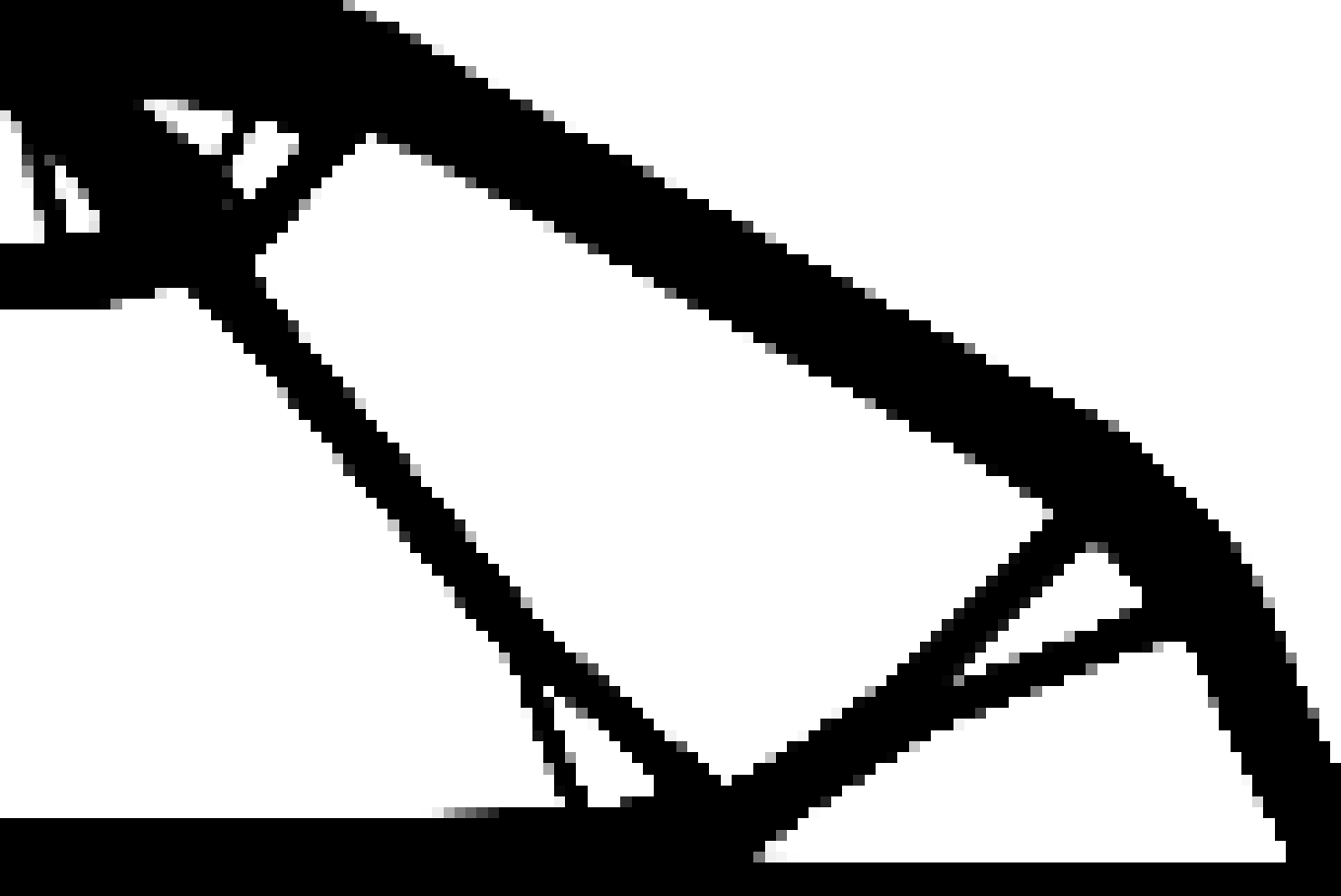} &
        \includegraphics[width=1.5cm,valign=m]{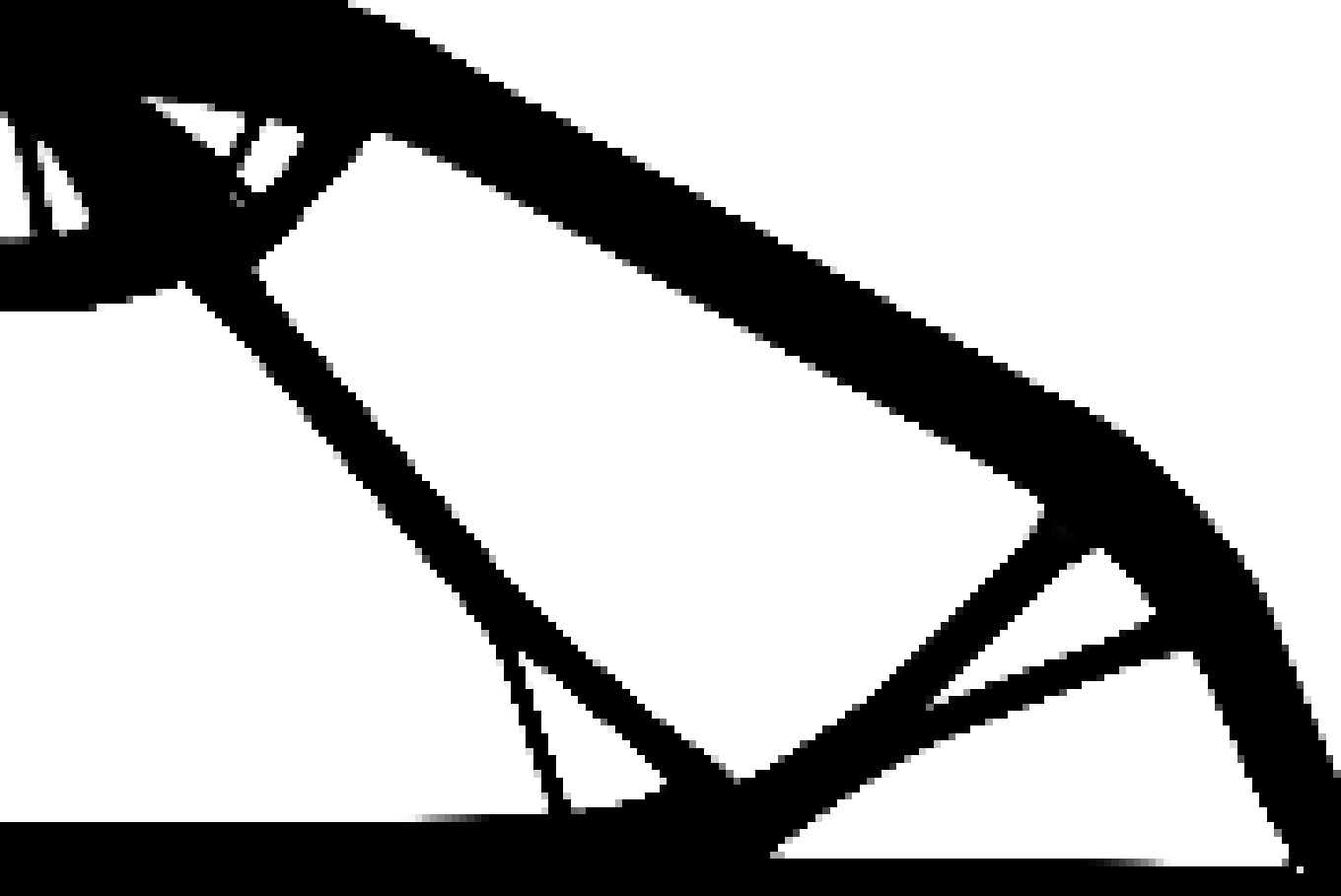} &
        \includegraphics[width=1.5cm,valign=m]{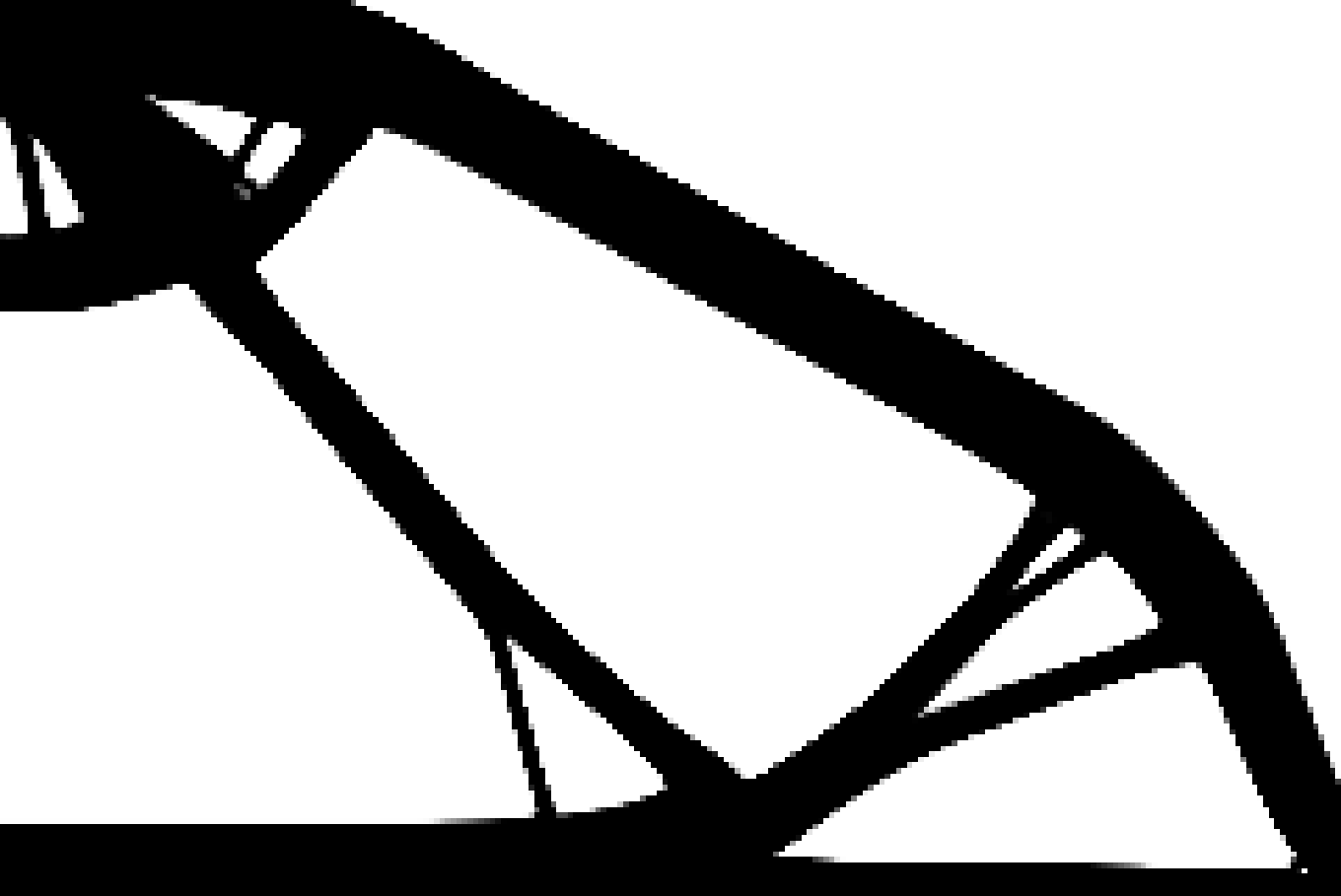} &
        \includegraphics[width=1.5cm,valign=m]{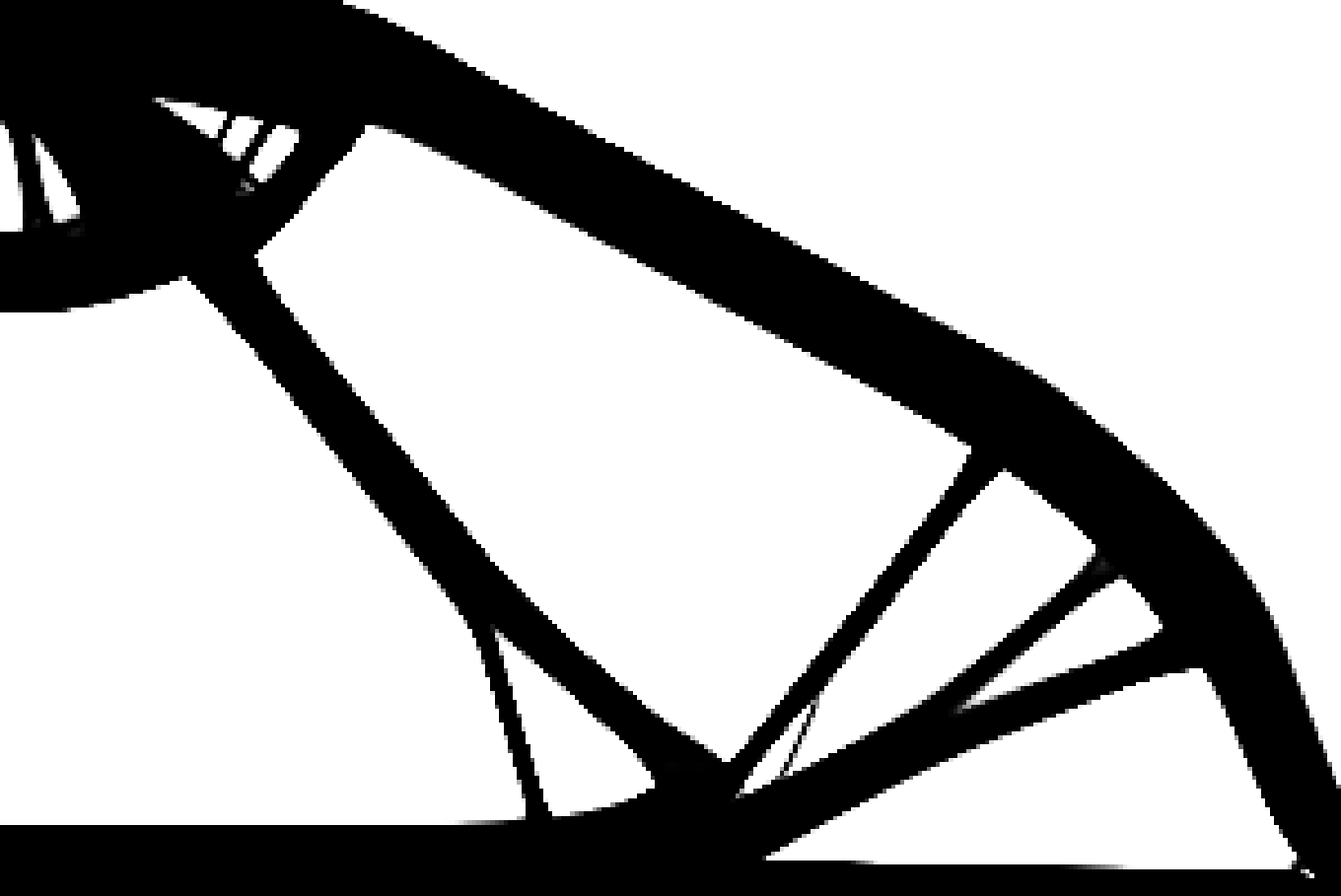} \\ \addlinespace[12pt]
                             & AC-ALM &
        \includegraphics[width=1.5cm,valign=m]{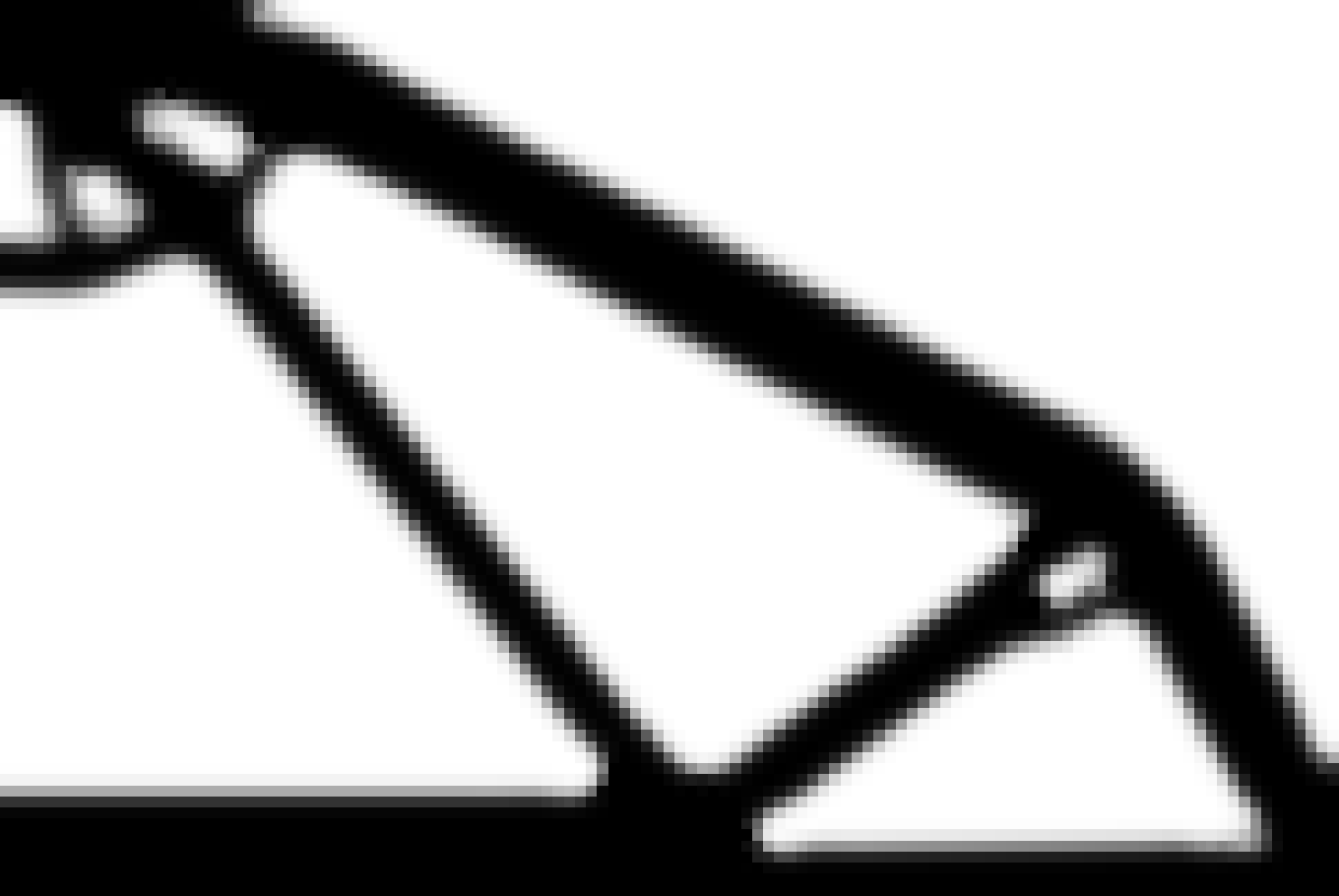} &
        \includegraphics[width=1.5cm,valign=m]{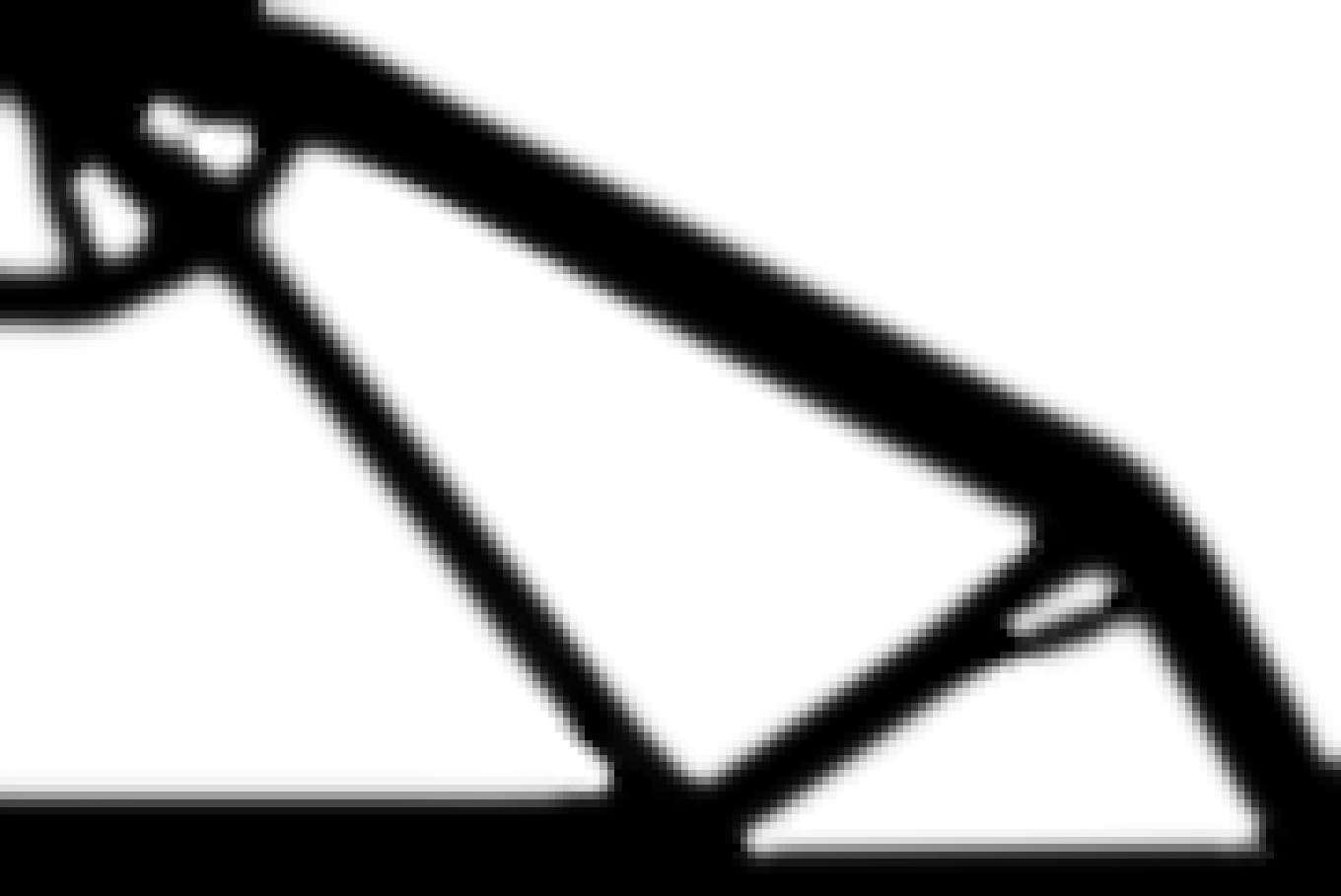} &
        \includegraphics[width=1.5cm,valign=m]{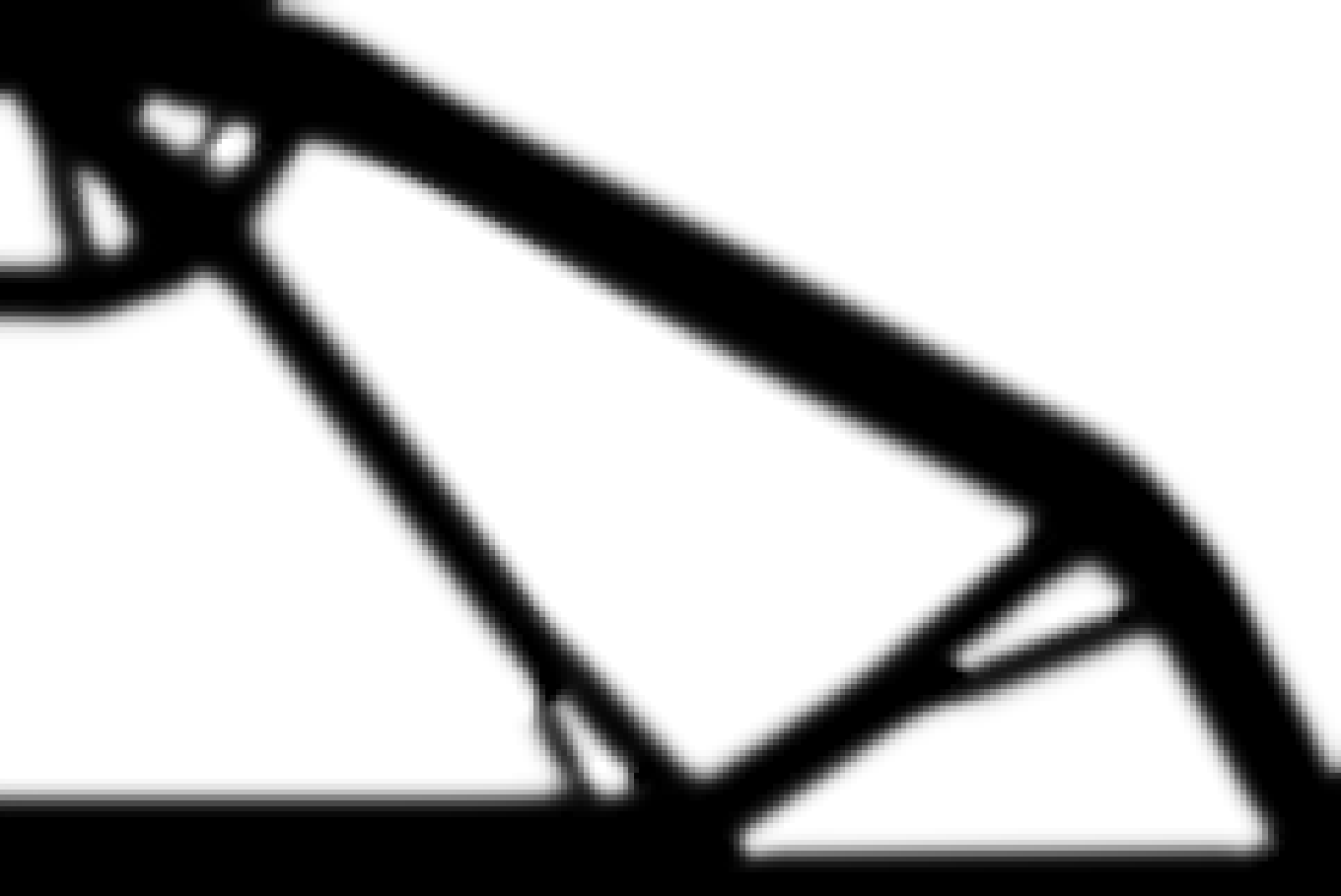} &
        \includegraphics[width=1.5cm,valign=m]{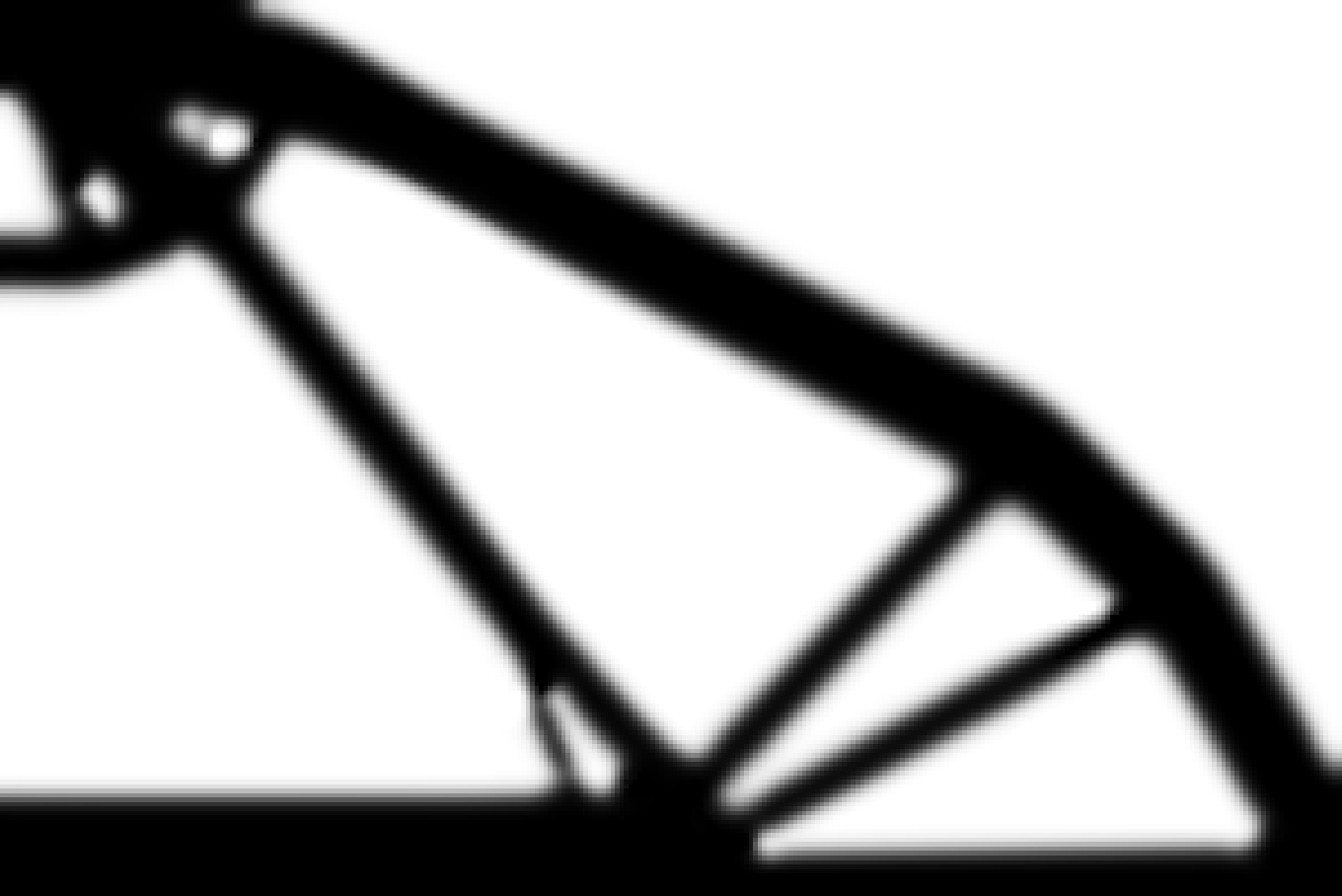} \\
        \midrule
        \multirow{2}{*}[-11.5pt]{0.4} & SIMP &
        \includegraphics[width=1.5cm,valign=m]{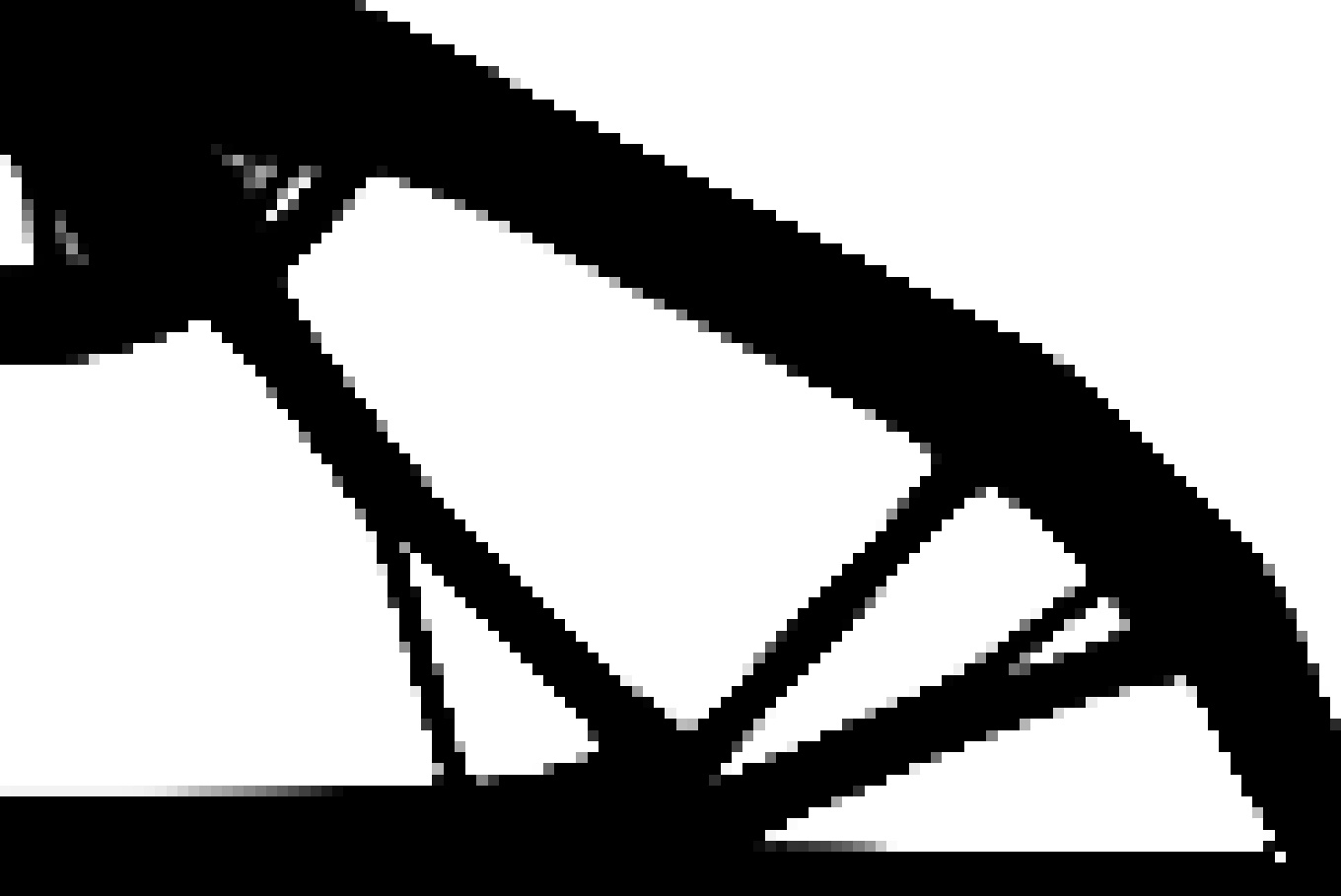} &
        \includegraphics[width=1.5cm,valign=m]{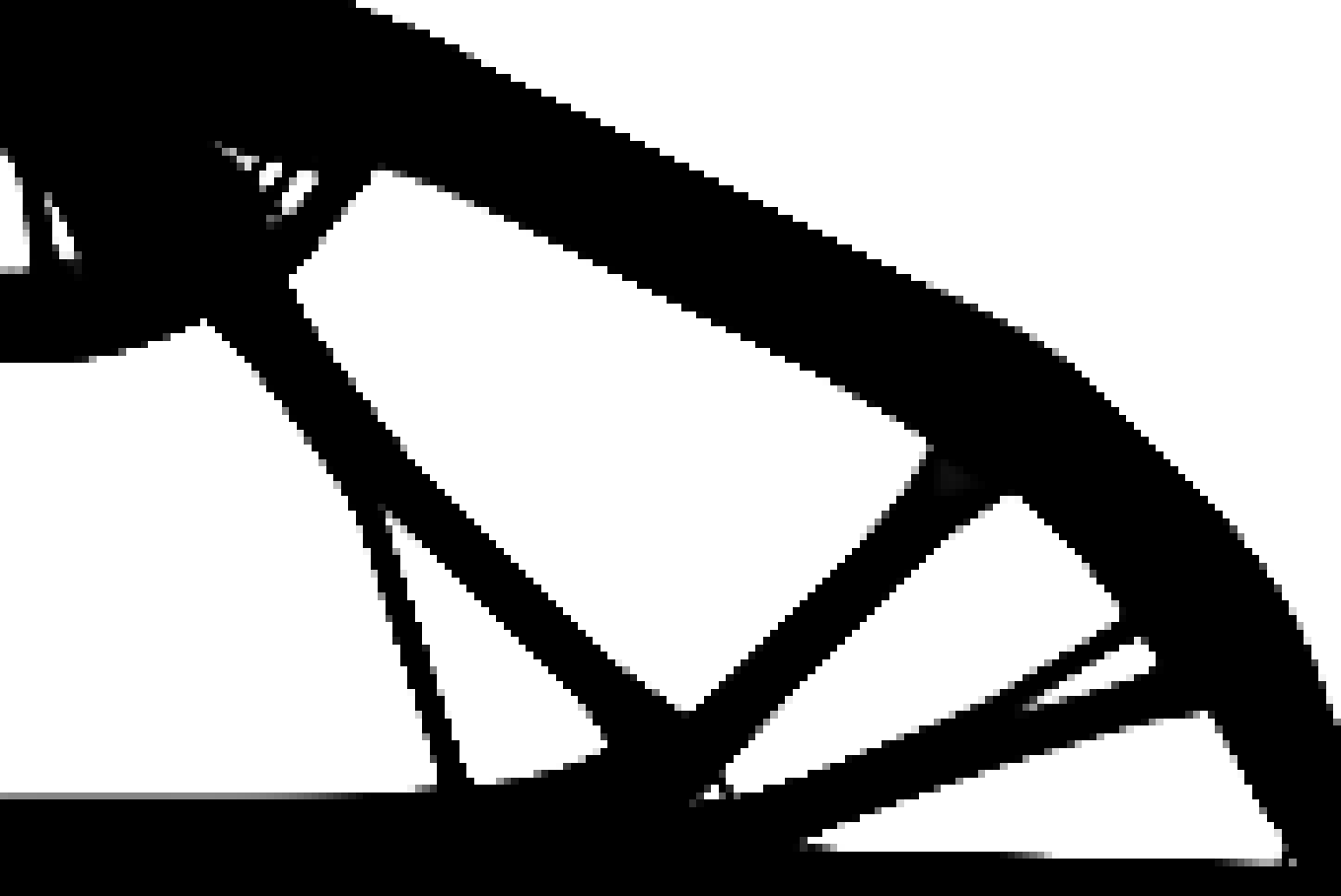} &
        \includegraphics[width=1.5cm,valign=m]{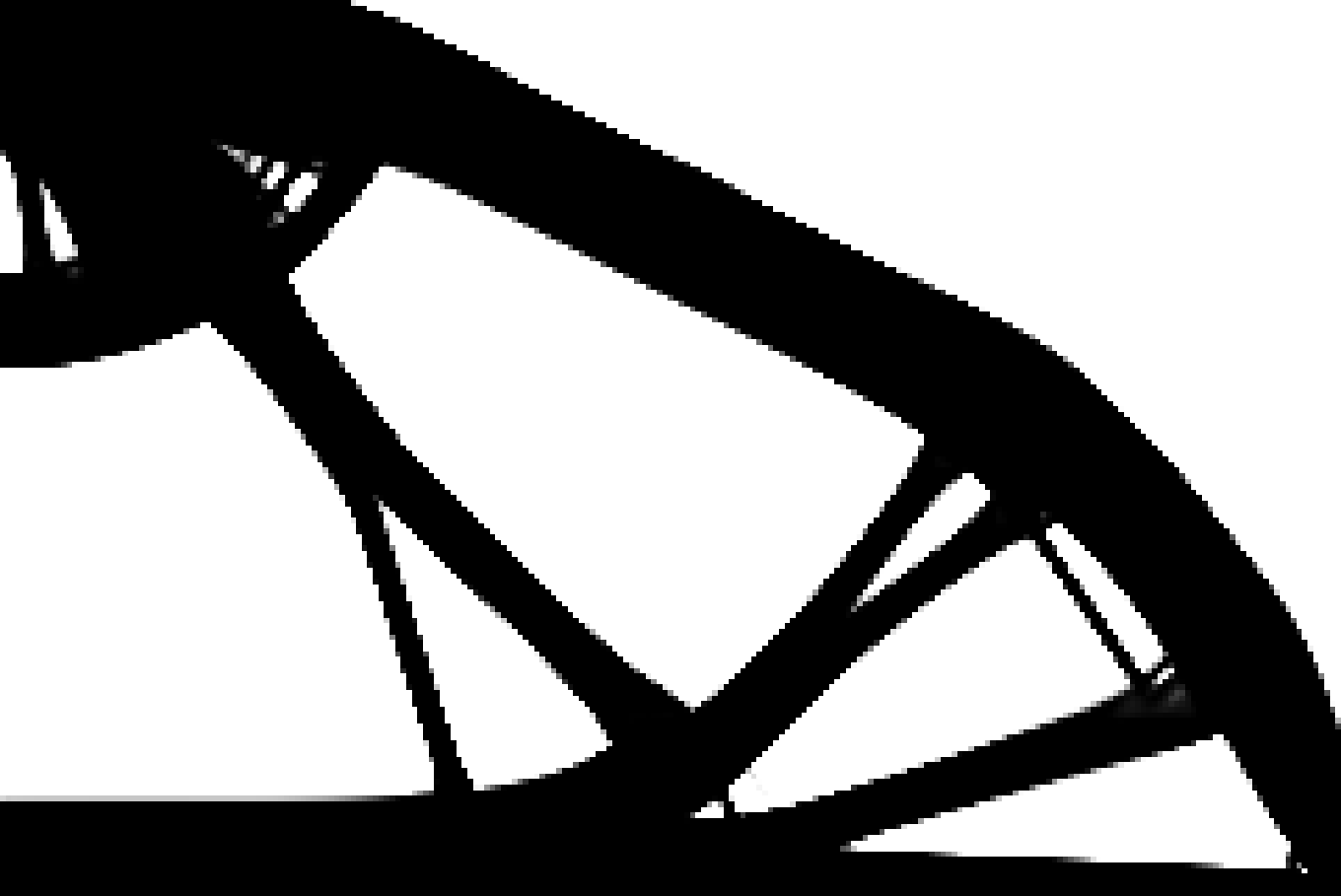} &
        \includegraphics[width=1.5cm,valign=m]{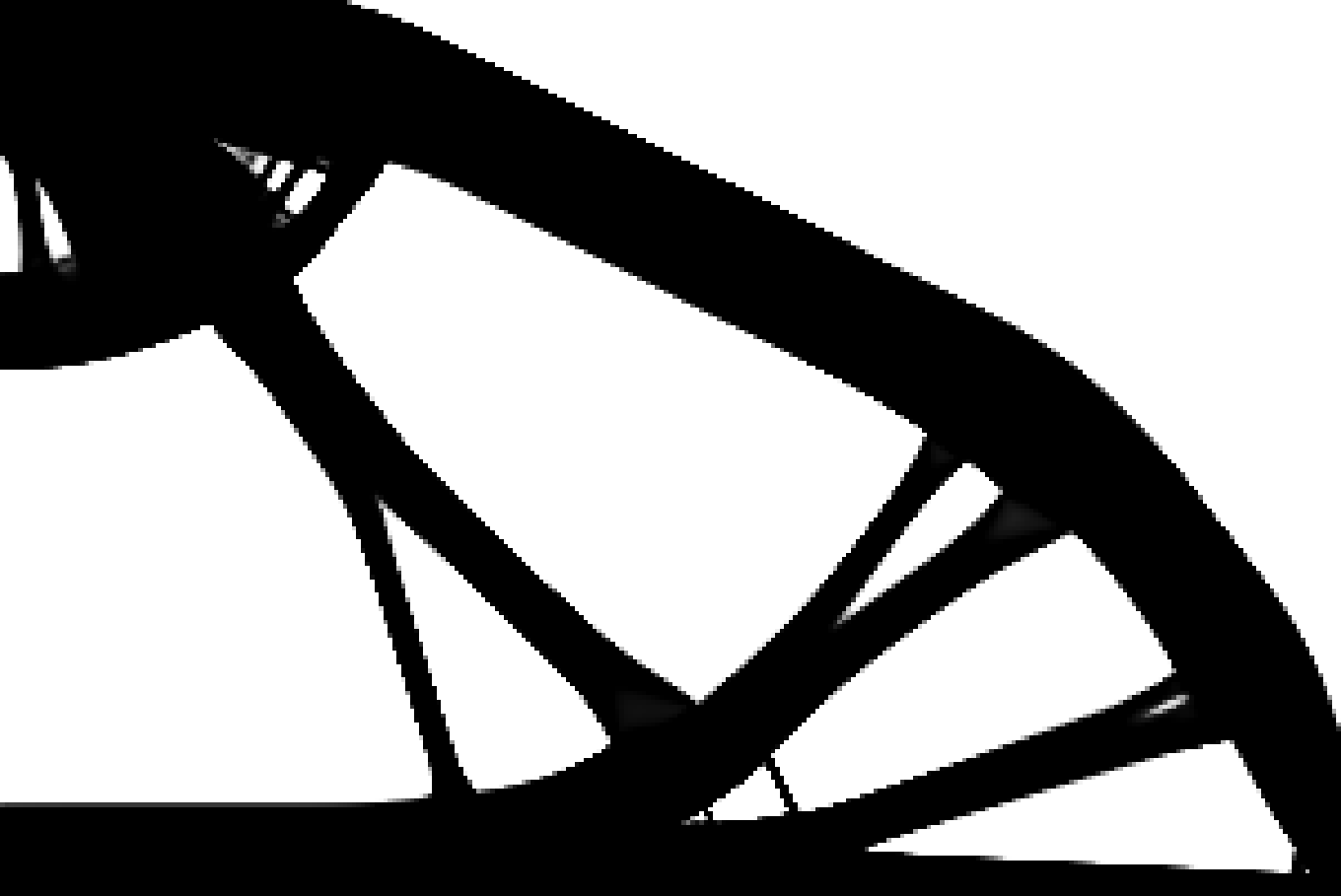} \\ \addlinespace[12pt]
                             & AC-ALM &
        \includegraphics[width=1.5cm,valign=m]{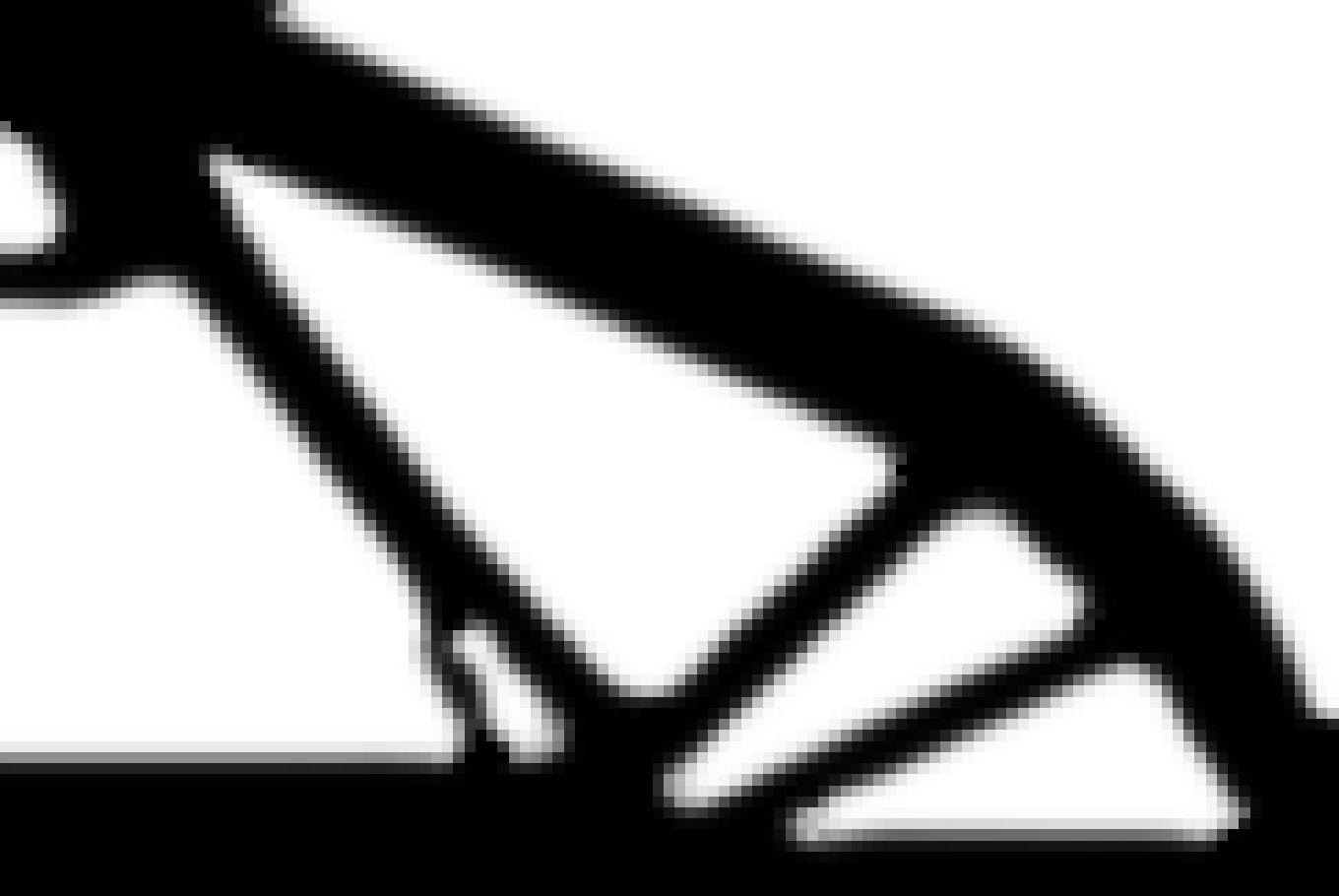} &
        \includegraphics[width=1.5cm,valign=m]{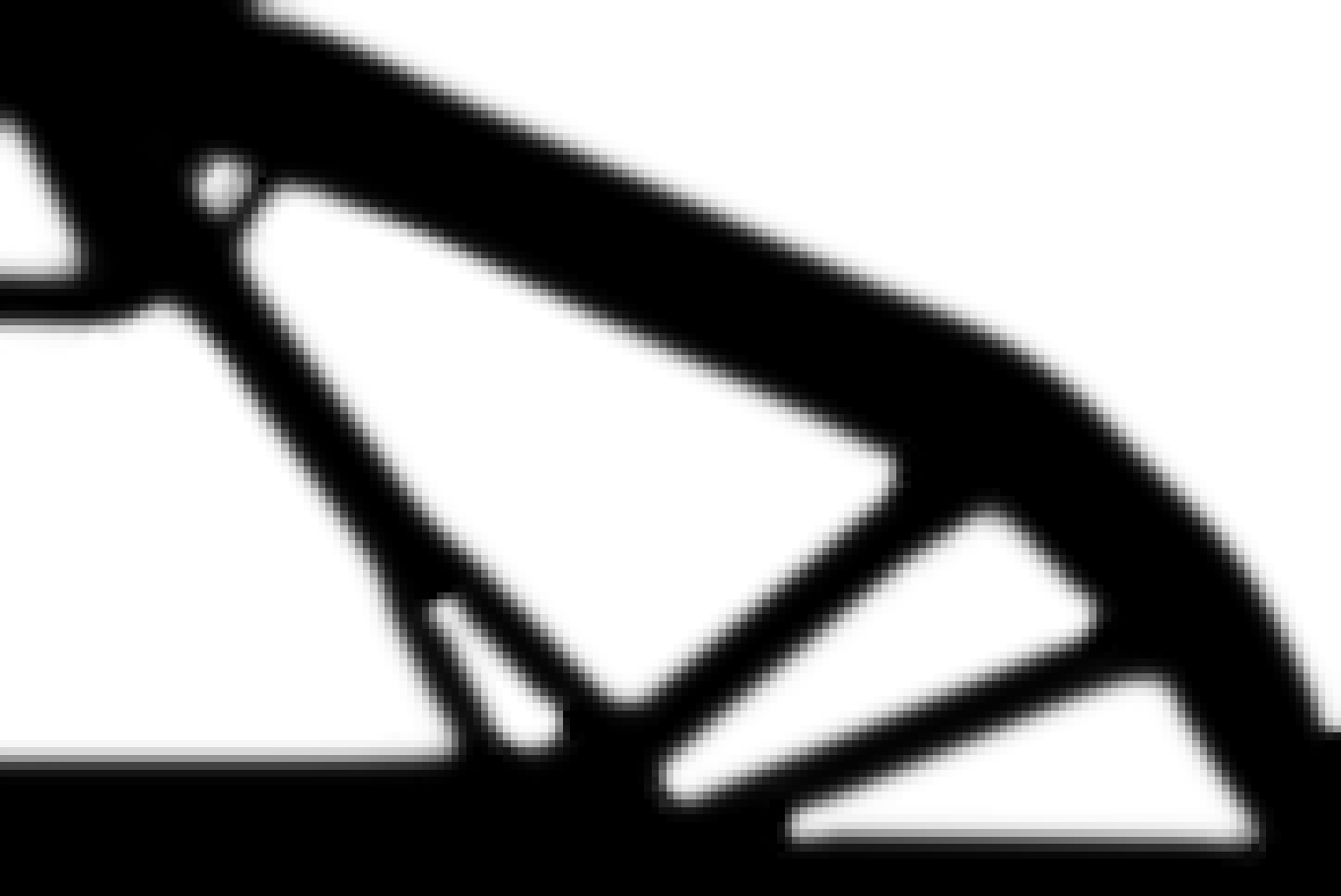} &
        \includegraphics[width=1.5cm,valign=m]{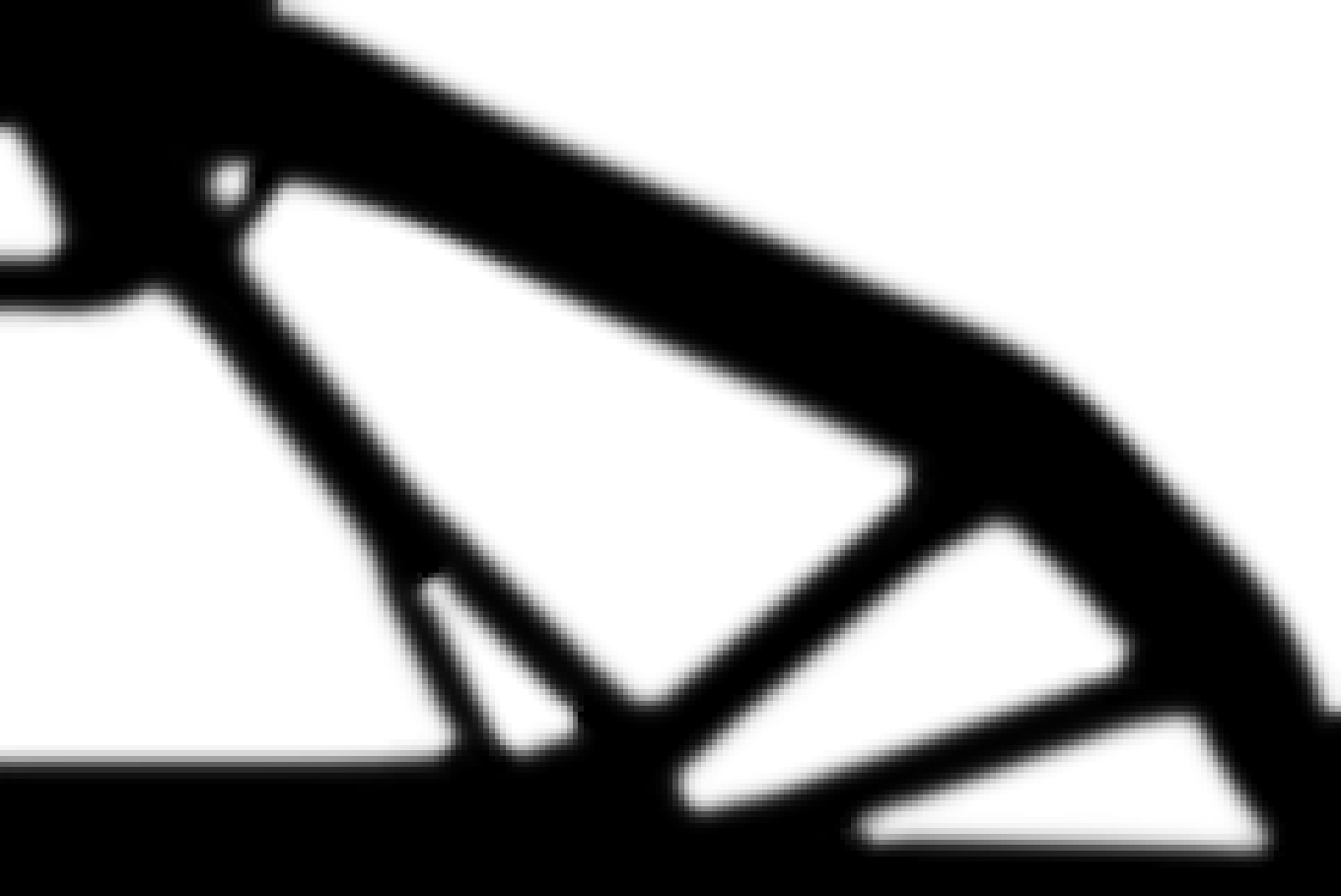} &
        \includegraphics[width=1.5cm,valign=m]{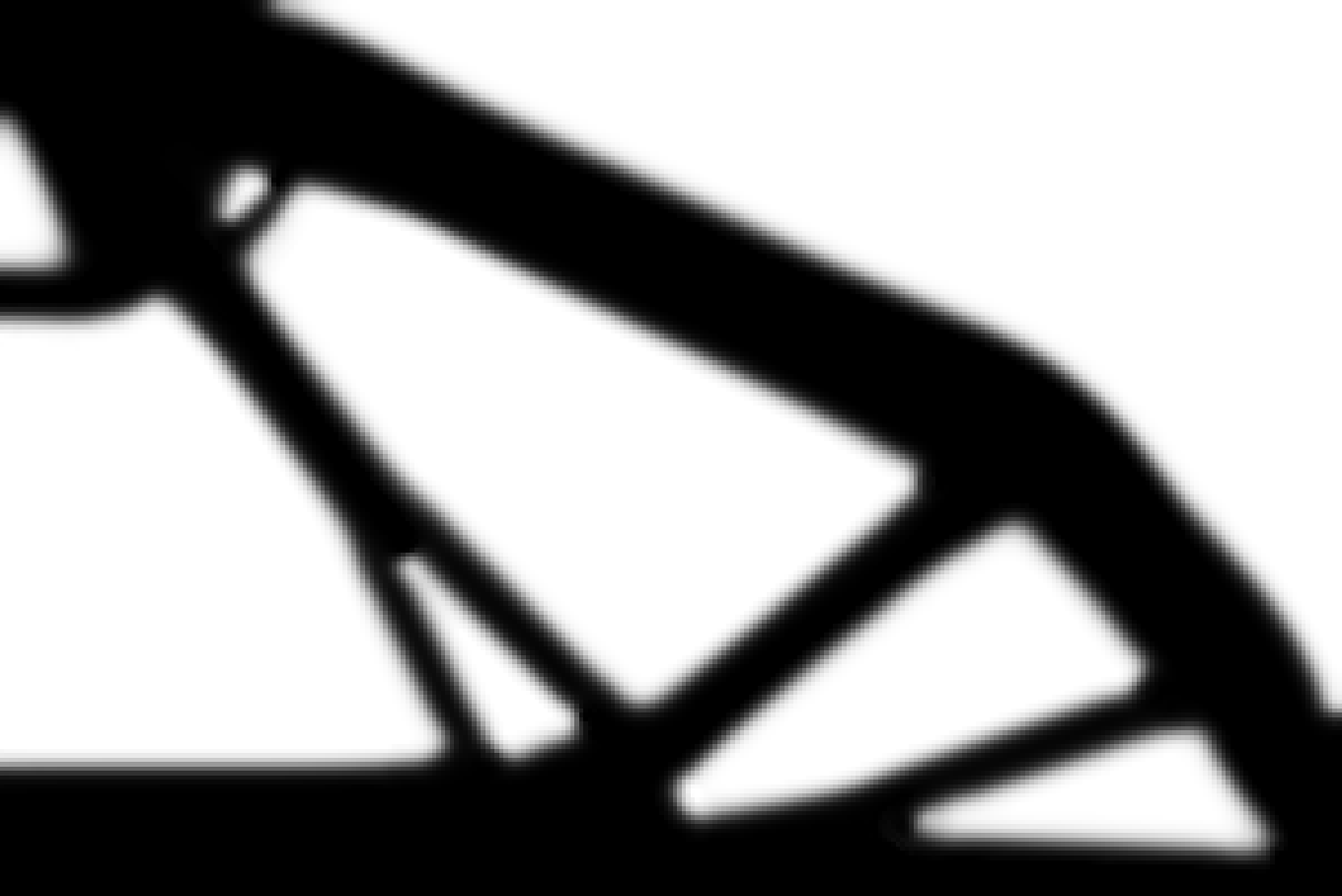} \\
        \bottomrule
    \end{tabular}
\end{table}

\subsubsection{Cantilever Beam}\label{num_examples_cantilever}

\begin{figure}[!ht]
\centering
\includegraphics{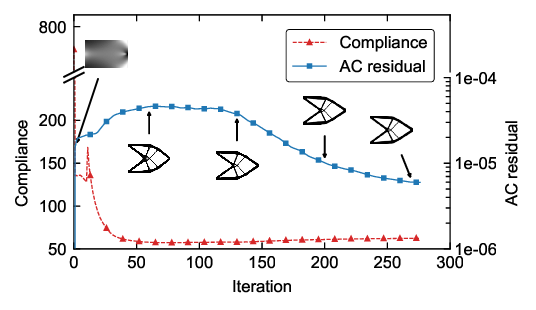}
\caption{Convergence history for cantilever under AC-ALM ($180\times 120$, $V=0.3$, $r_{\min}=0.1$).}\label{fig:cantilever_conv}
\end{figure}

The cantilever beam benchmark evaluates the framework under asymmetric bending
and shear loads. 
As shown in the convergence history of Fig.~\ref{fig:cantilever_conv} 
($180\times 120$ mesh, $V = 0.3$, $r_{\min} = 0.1$), 
the compliance decreases rapidly in the first 40 iterations before stabilizing, 
while the normalized AC residual norm steadily decreases below the prescribed tolerance threshold.

The optimized designs under mesh refinement are presented
in Table~\ref{tab:cantilever-rmin2} ($r_{\min} = 2.0$) 
and Table~\ref{tab:cantilever-rmin01} ($r_{\min} = 0.1$). 
At $r_{\min} = 2.0$, both AC-ALM and SIMP generate consistent primary load paths
with cross-bracing diagonals. 
In the intrinsic regime ($r_{\min} = 0.1$), the SIMP baseline branches into numerous thin,
mesh-dependent trusses. 
By contrast, AC-ALM preserves consistent primary diagonal members across all four mesh resolutions.
This verifies that the residual constraint regularizes complex asymmetric load paths
without filter intervention.

\begin{table}[!htbp]
    \centering
    \small
    \setlength{\tabcolsep}{2pt}
    \caption{Structure comparison of cantilevers ($r_{\min}=2$)} 
    \label{tab:cantilever-rmin2}
    \begin{tabular}{cccccc}
        \toprule
        $V$ & Method & $120\times 80$ & $180\times 120$ & $240\times 160$ & $300\times 200$ \\
        \midrule
        \multirow{2}{*}[-11.5pt]{0.3} & SIMP &
        \includegraphics[width=1.5cm,valign=m]{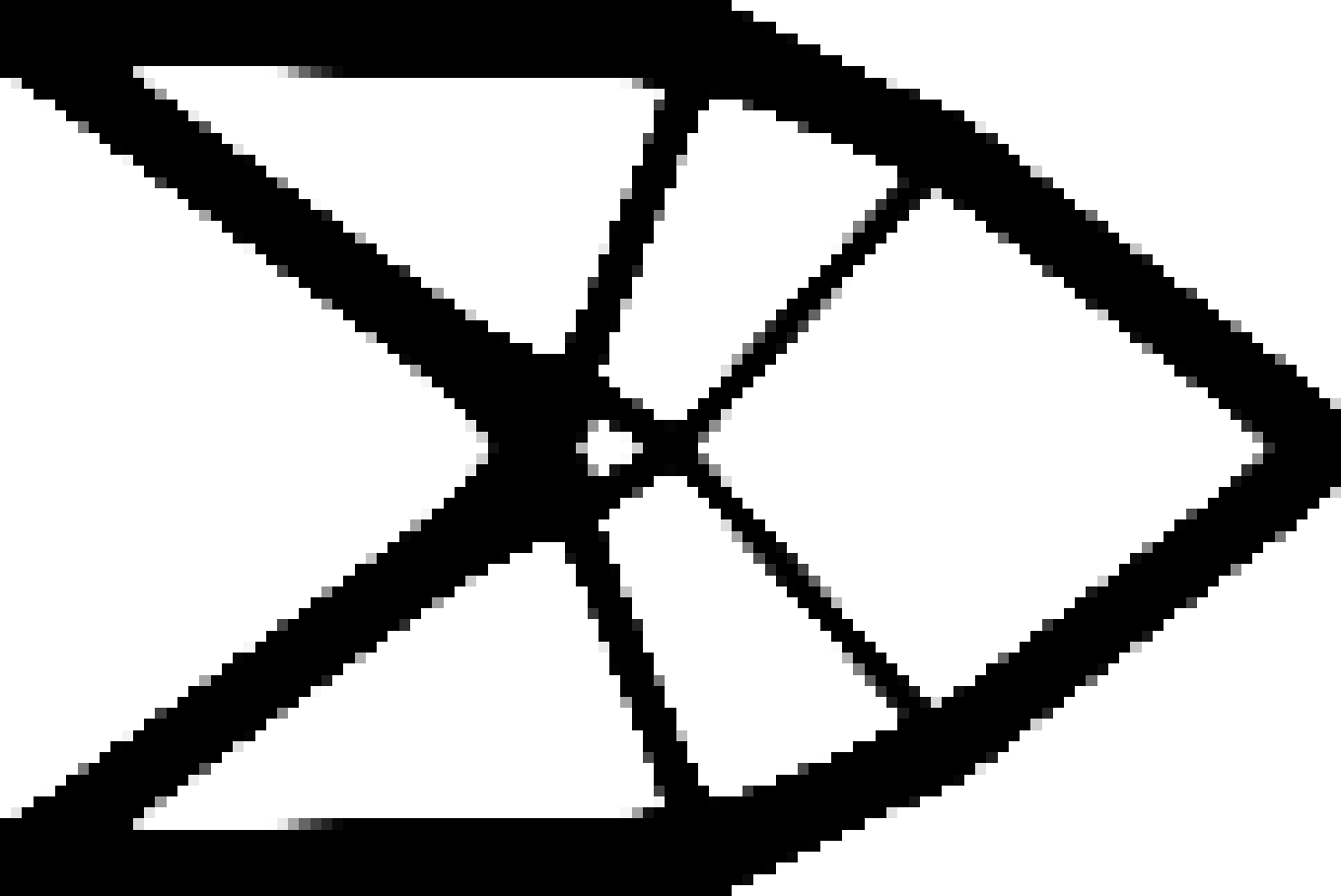} &
        \includegraphics[width=1.5cm,valign=m]{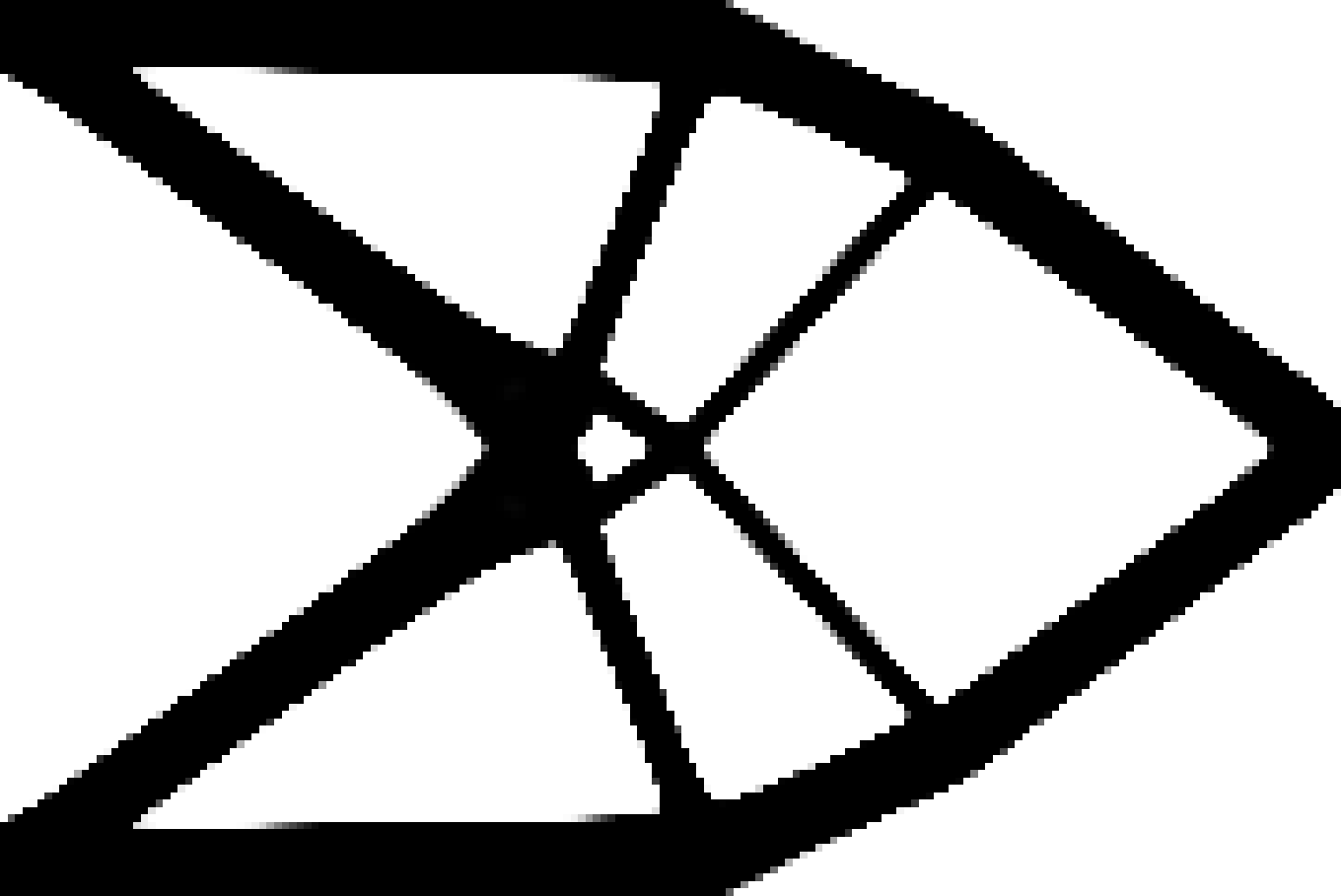} &
        \includegraphics[width=1.5cm,valign=m]{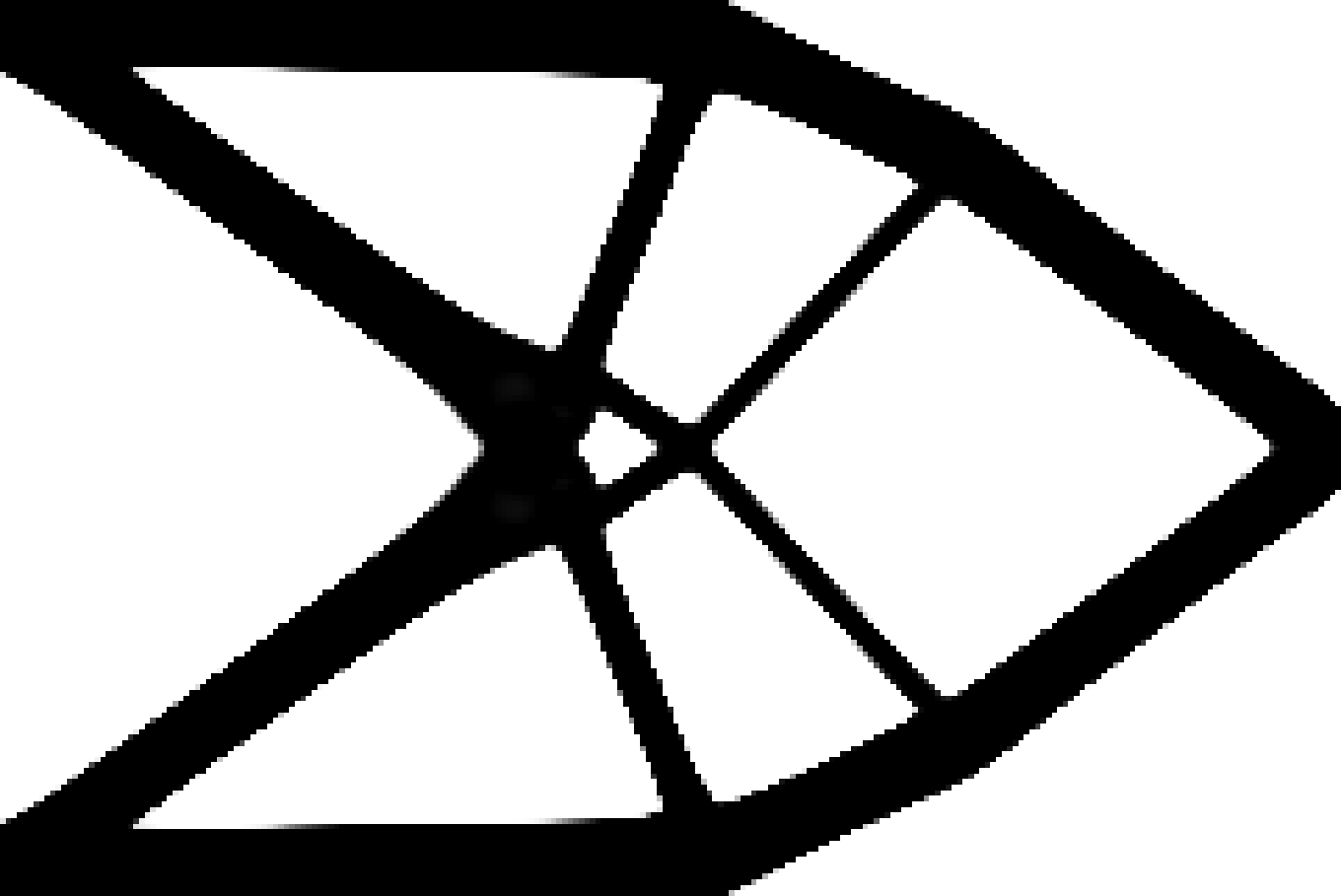} &
        \includegraphics[width=1.5cm,valign=m]{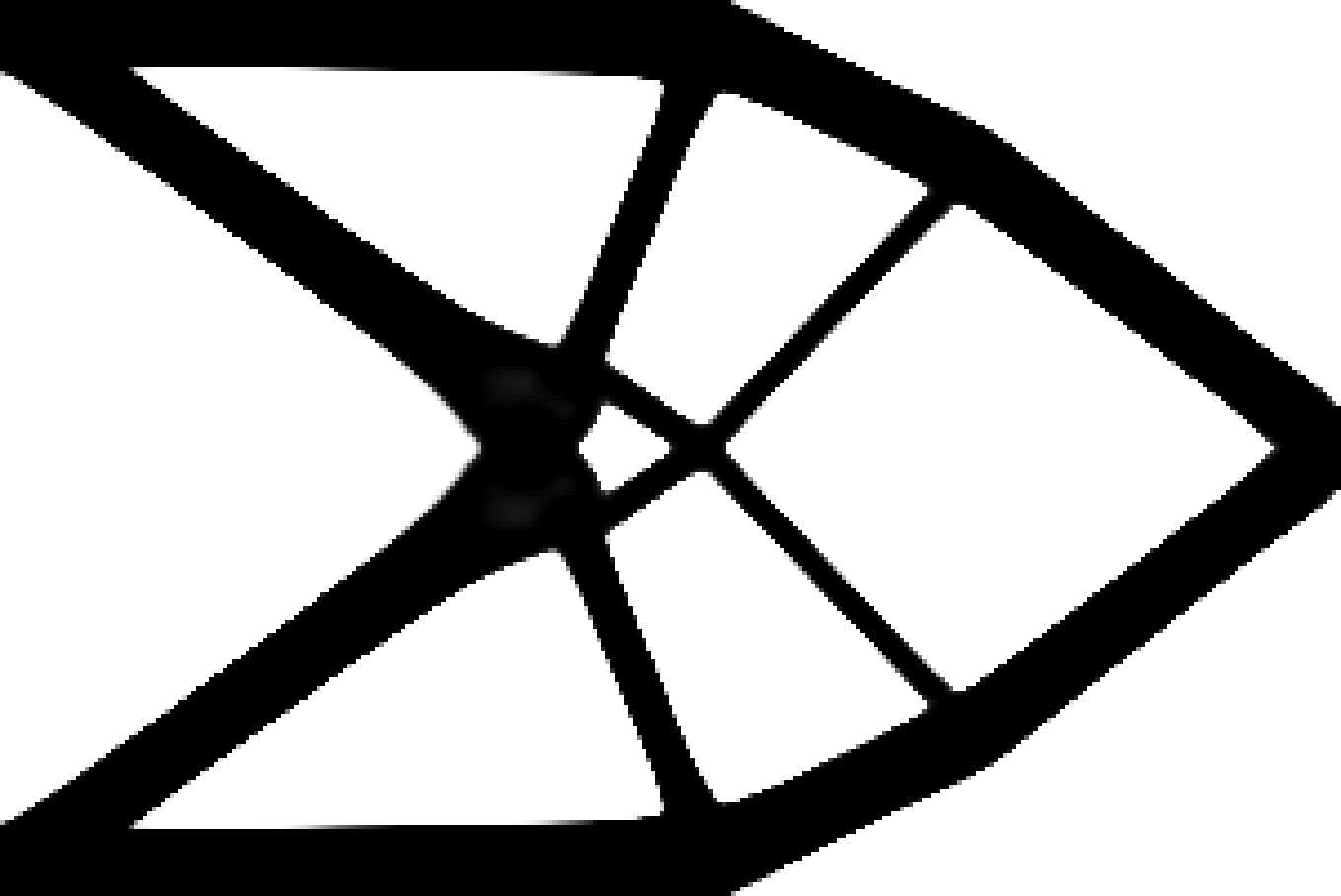} \\ \addlinespace[12pt]
                             & AC-ALM &
        \includegraphics[width=1.5cm,valign=m]{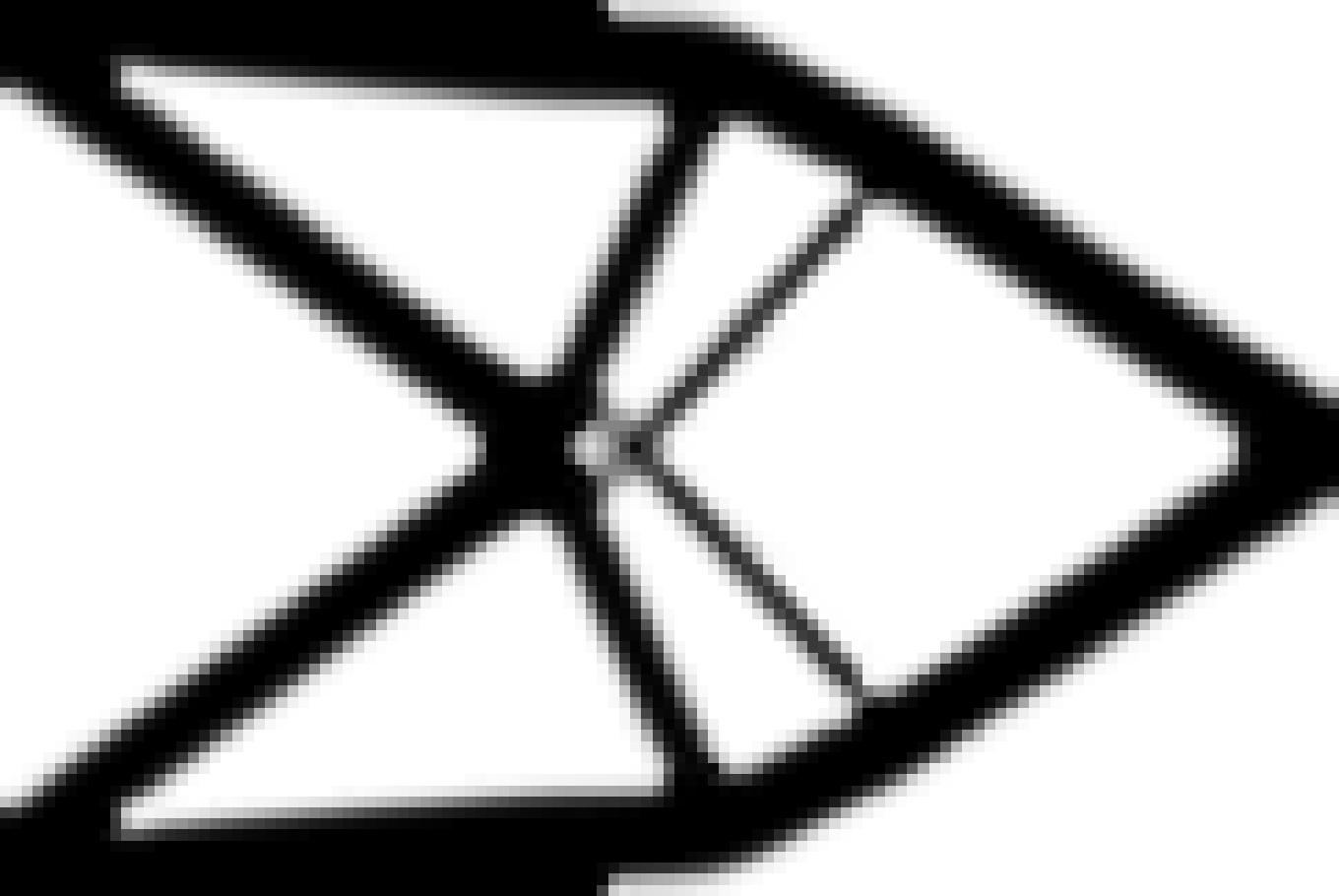} &
        \includegraphics[width=1.5cm,valign=m]{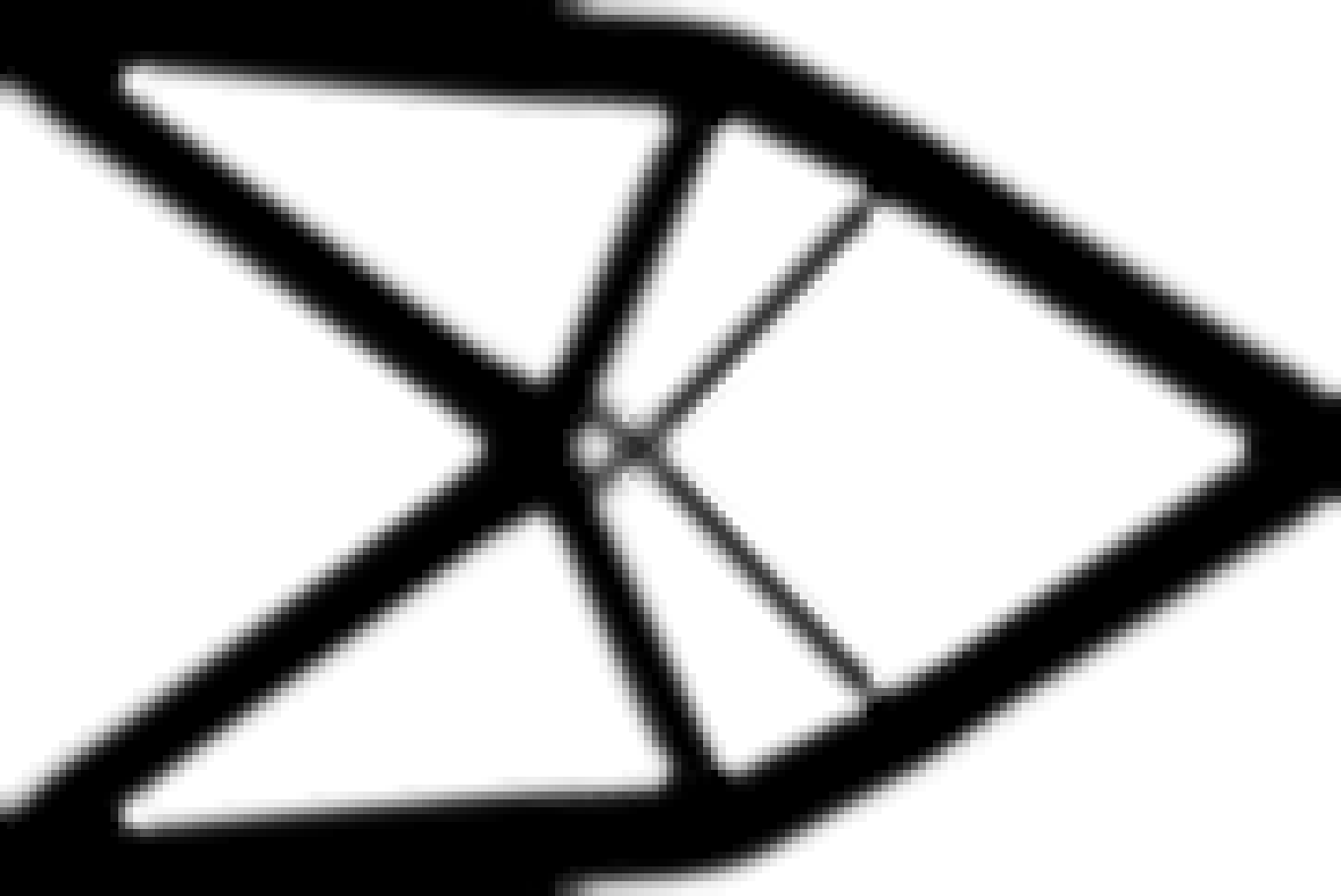} &
        \includegraphics[width=1.5cm,valign=m]{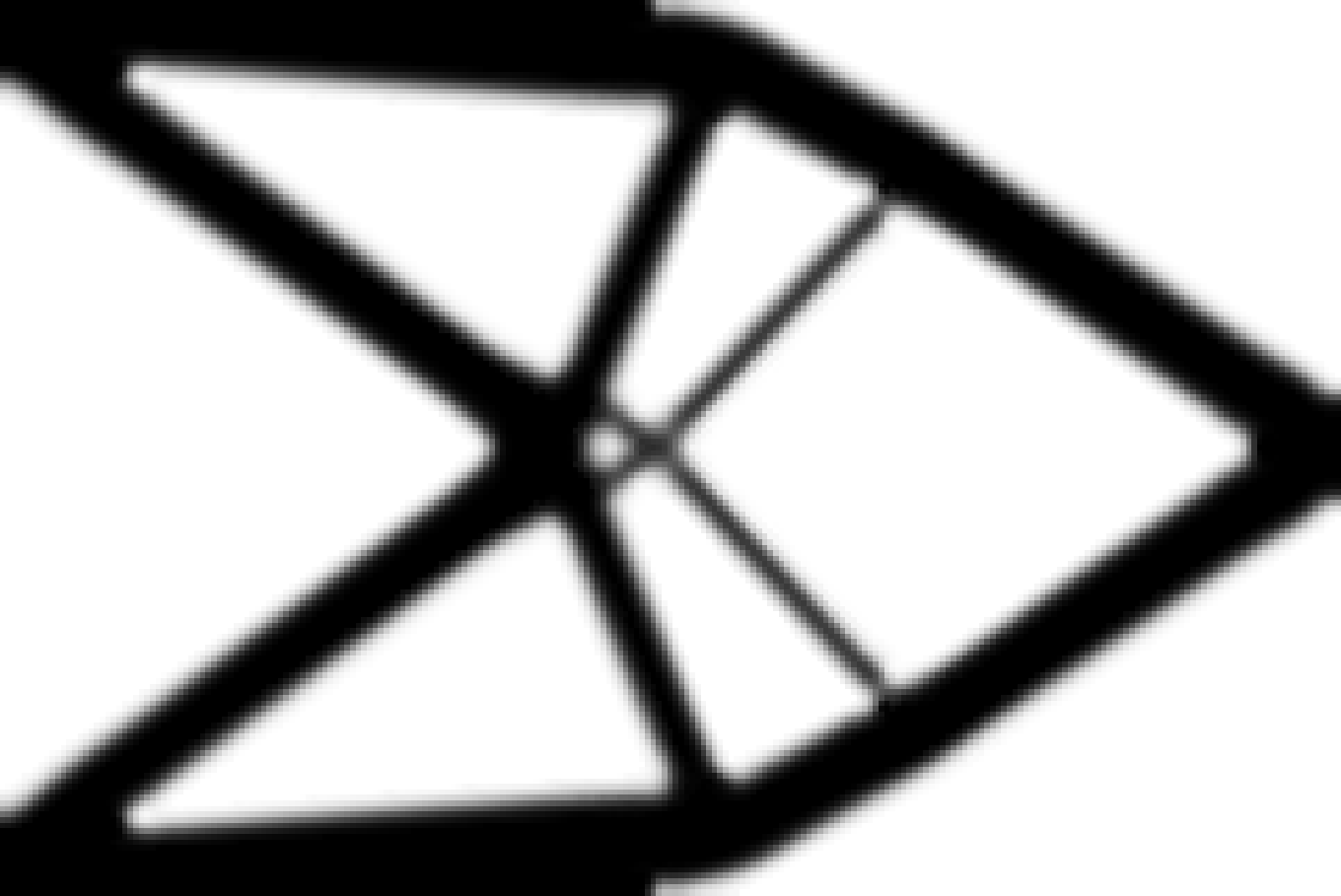} &
        \includegraphics[width=1.5cm,valign=m]{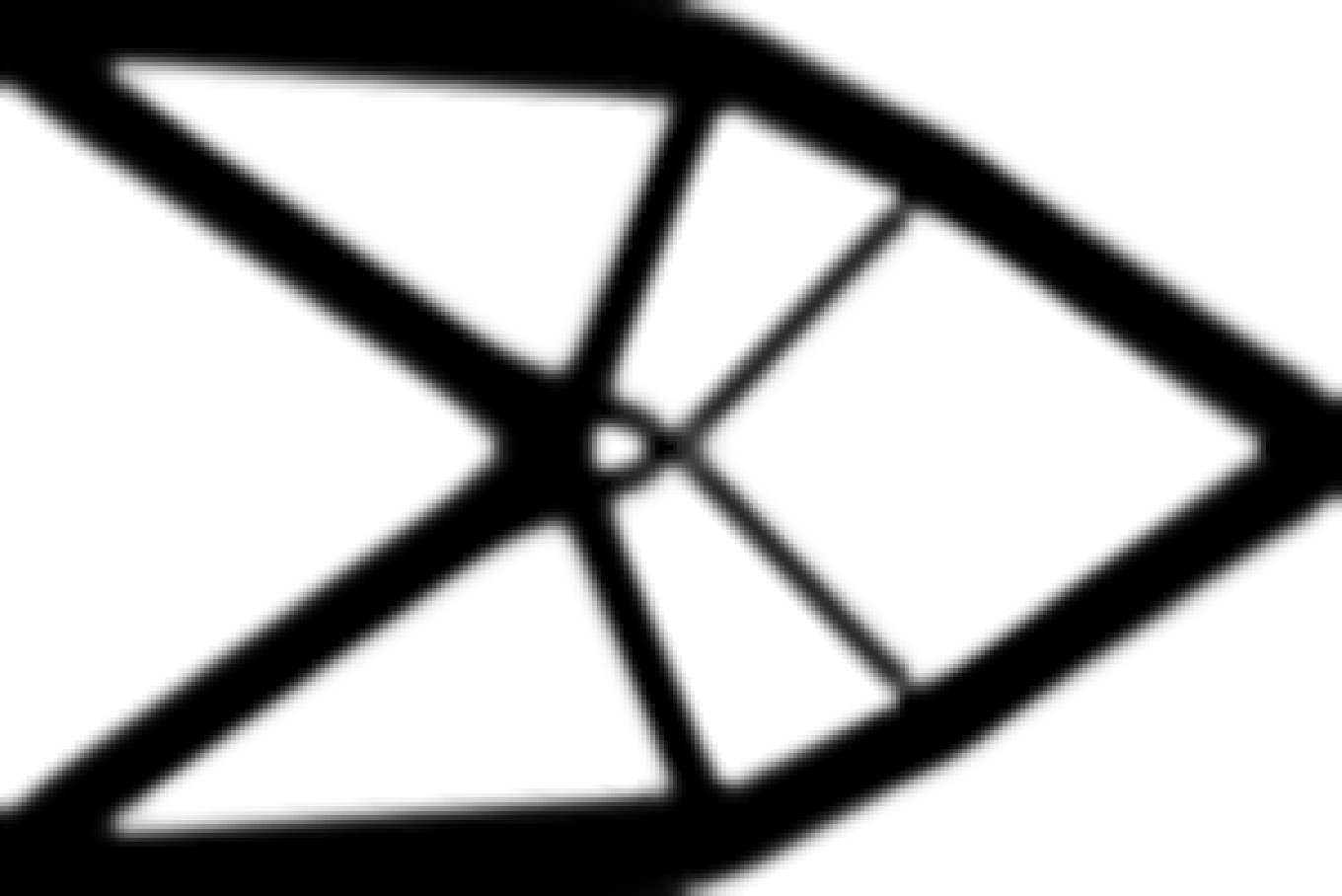} \\
        \midrule
        \multirow{2}{*}[-11.5pt]{0.4} & SIMP &
        \includegraphics[width=1.5cm,valign=m]{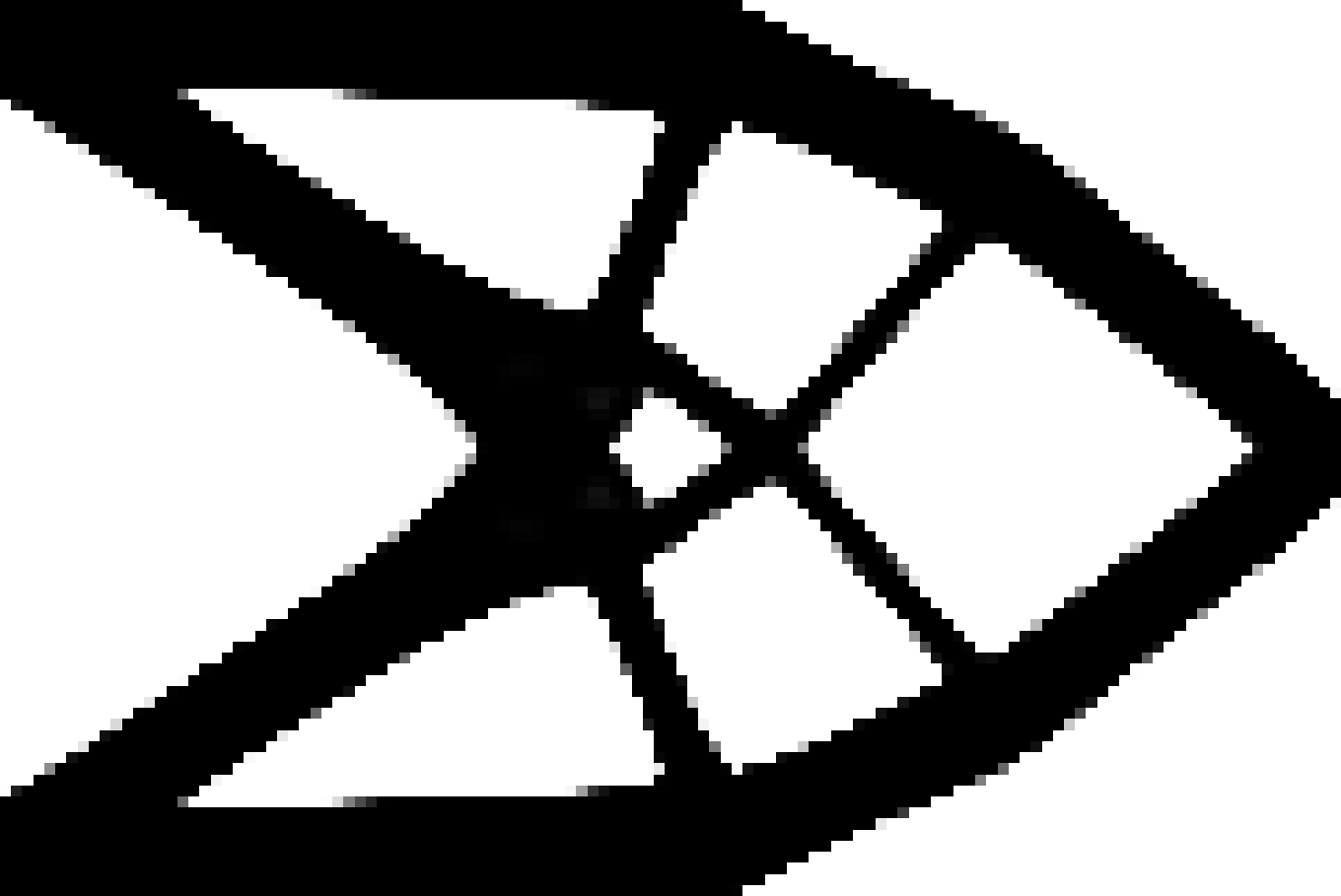} &
        \includegraphics[width=1.5cm,valign=m]{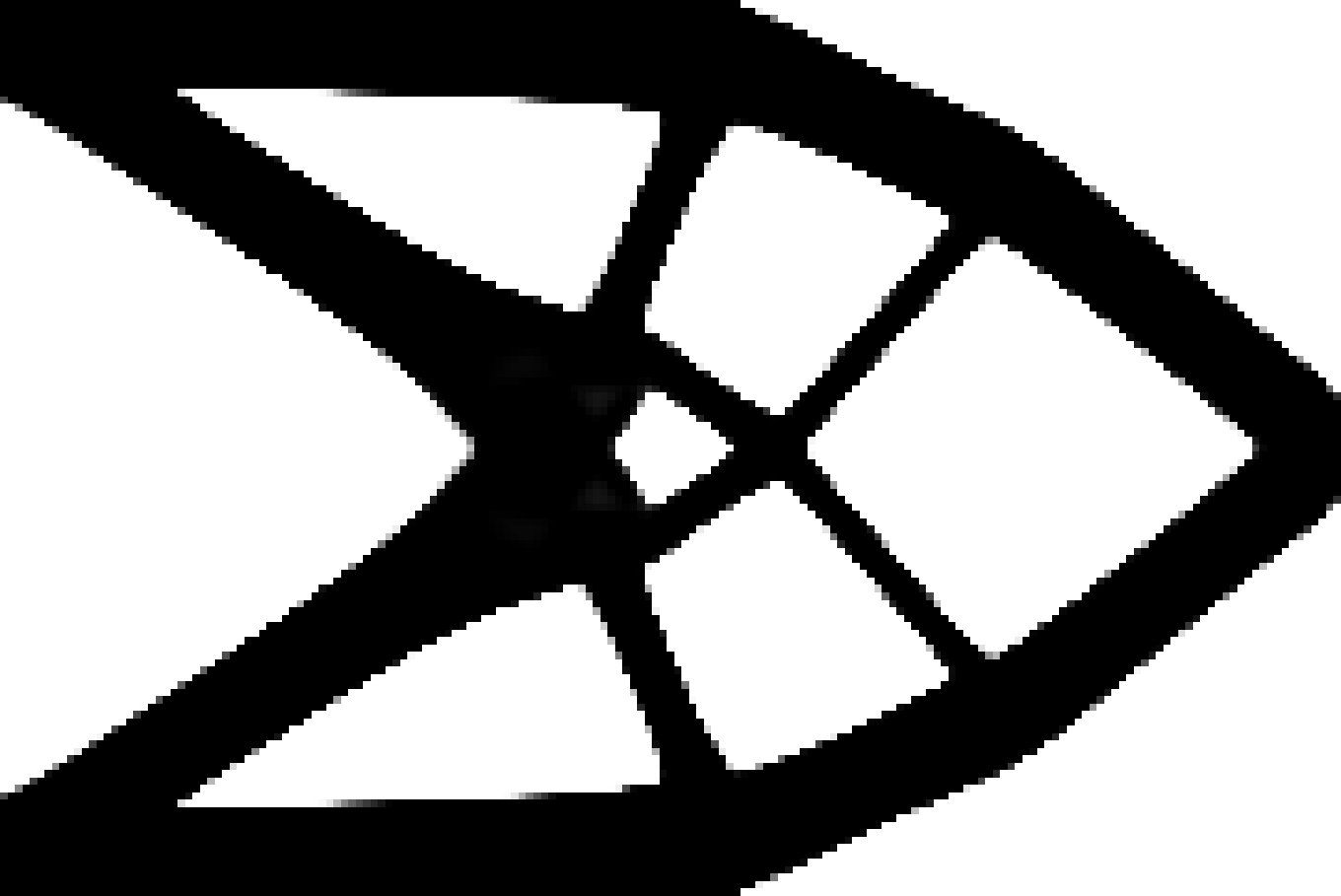} &
        \includegraphics[width=1.5cm,valign=m]{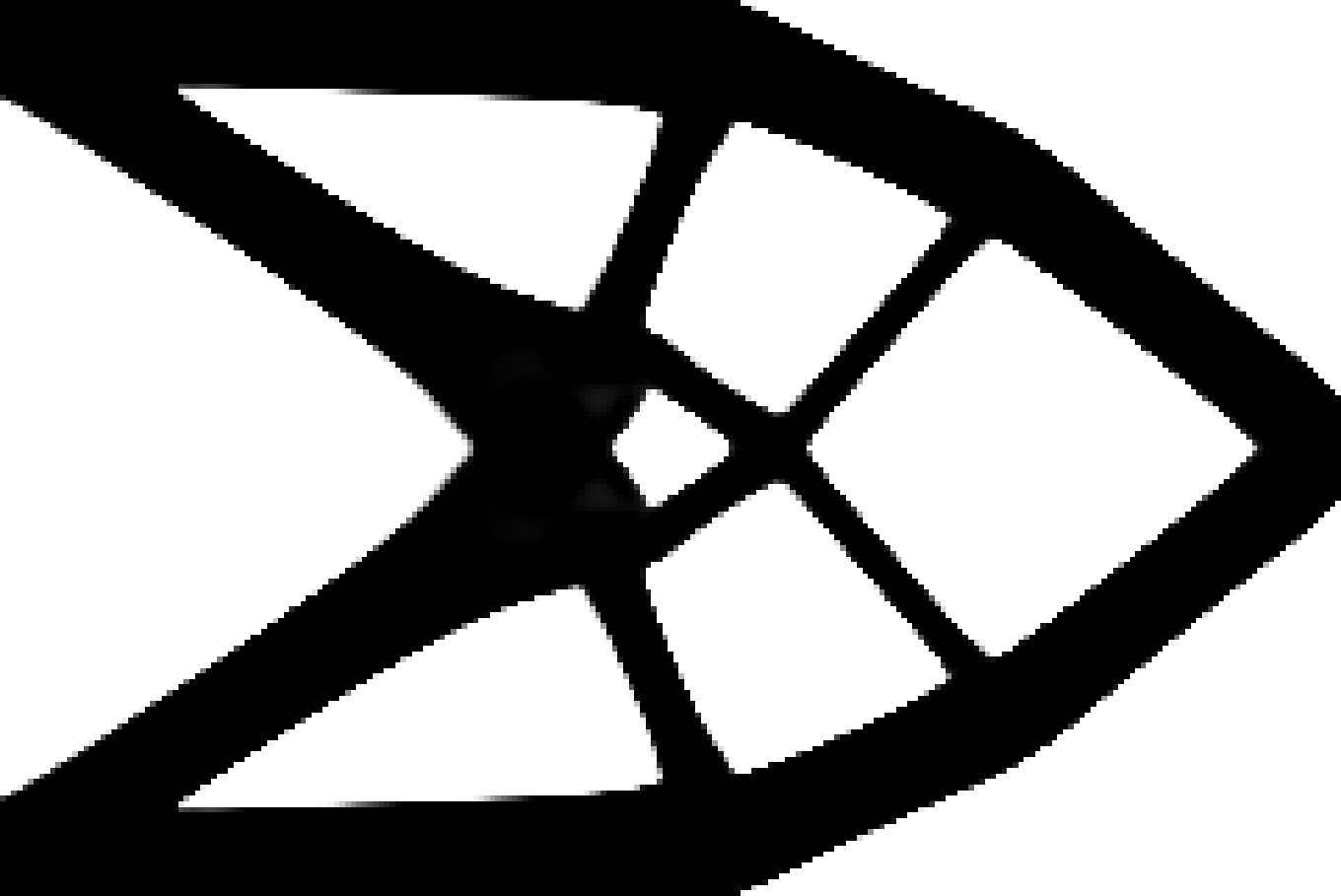} &
        \includegraphics[width=1.5cm,valign=m]{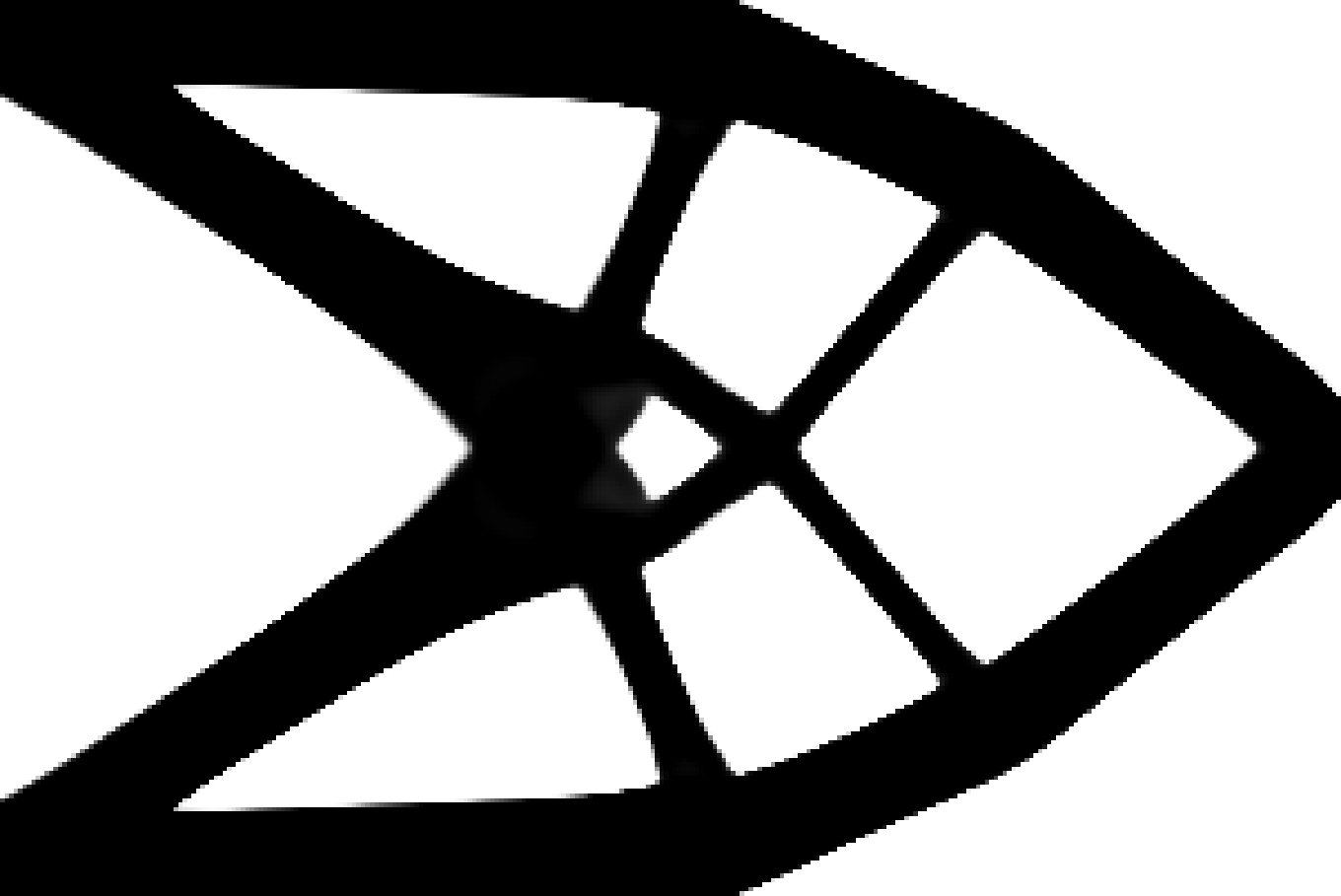} \\ \addlinespace[12pt]
                             & AC-ALM &
        \includegraphics[width=1.5cm,valign=m]{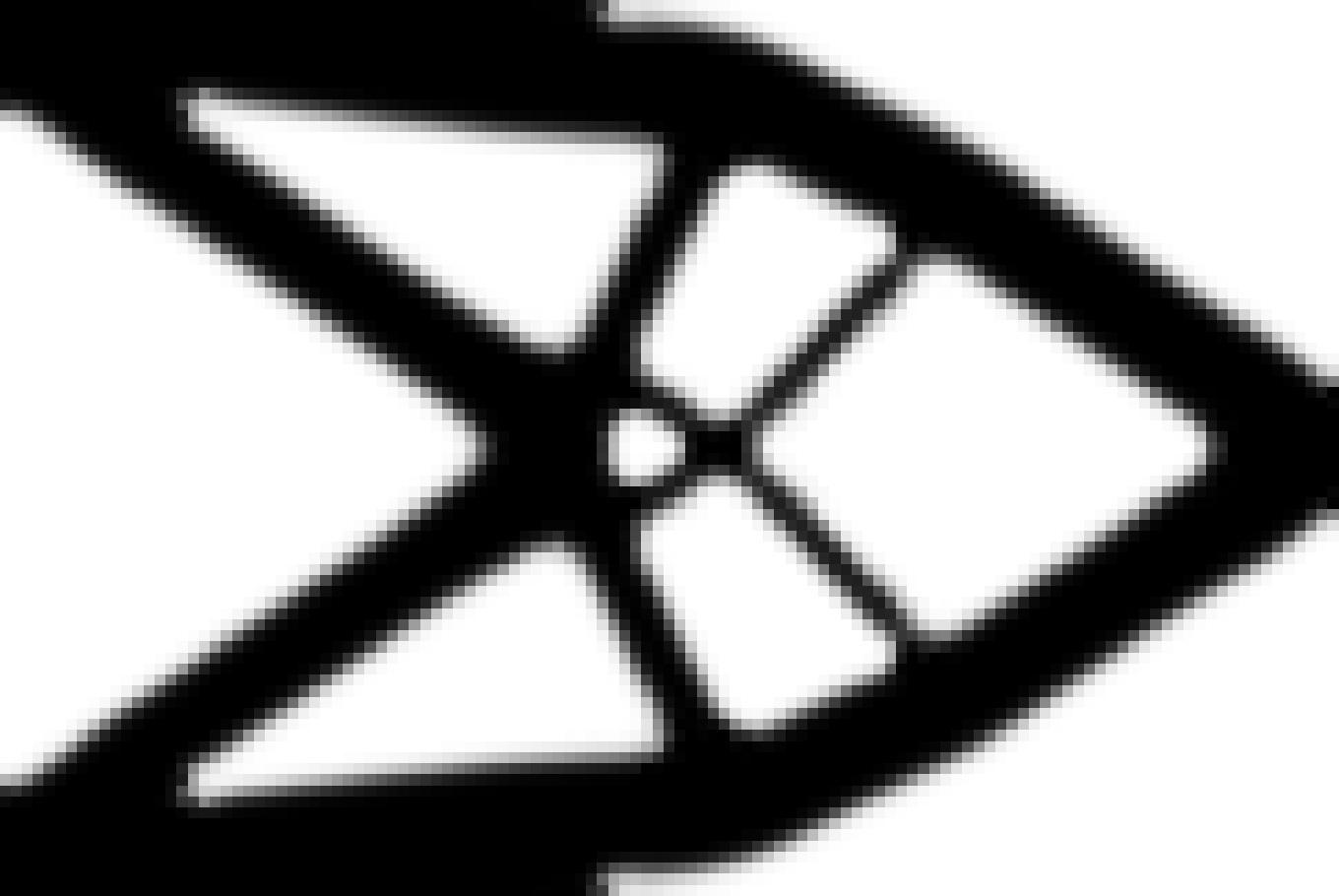} &
        \includegraphics[width=1.5cm,valign=m]{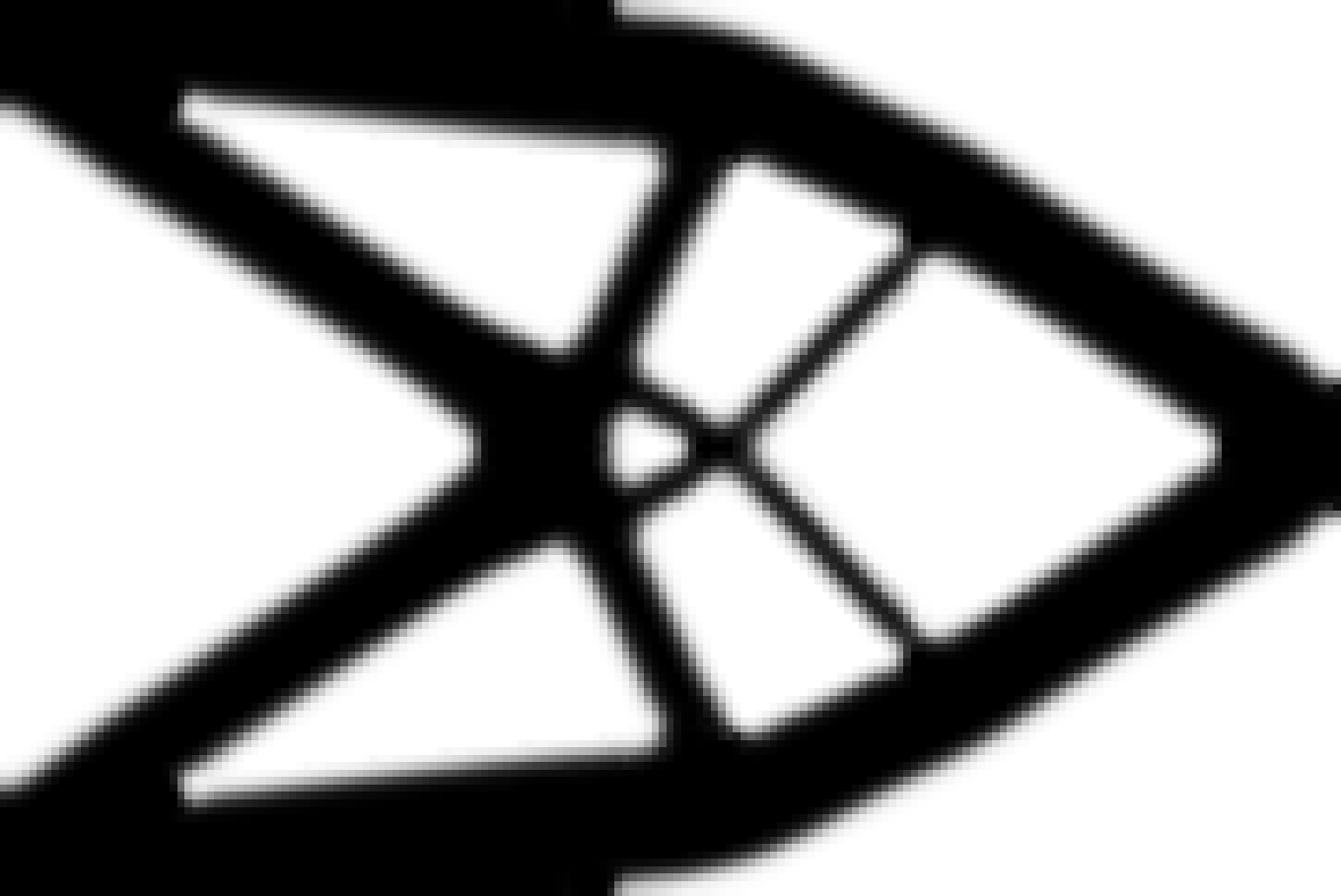} &
        \includegraphics[width=1.5cm,valign=m]{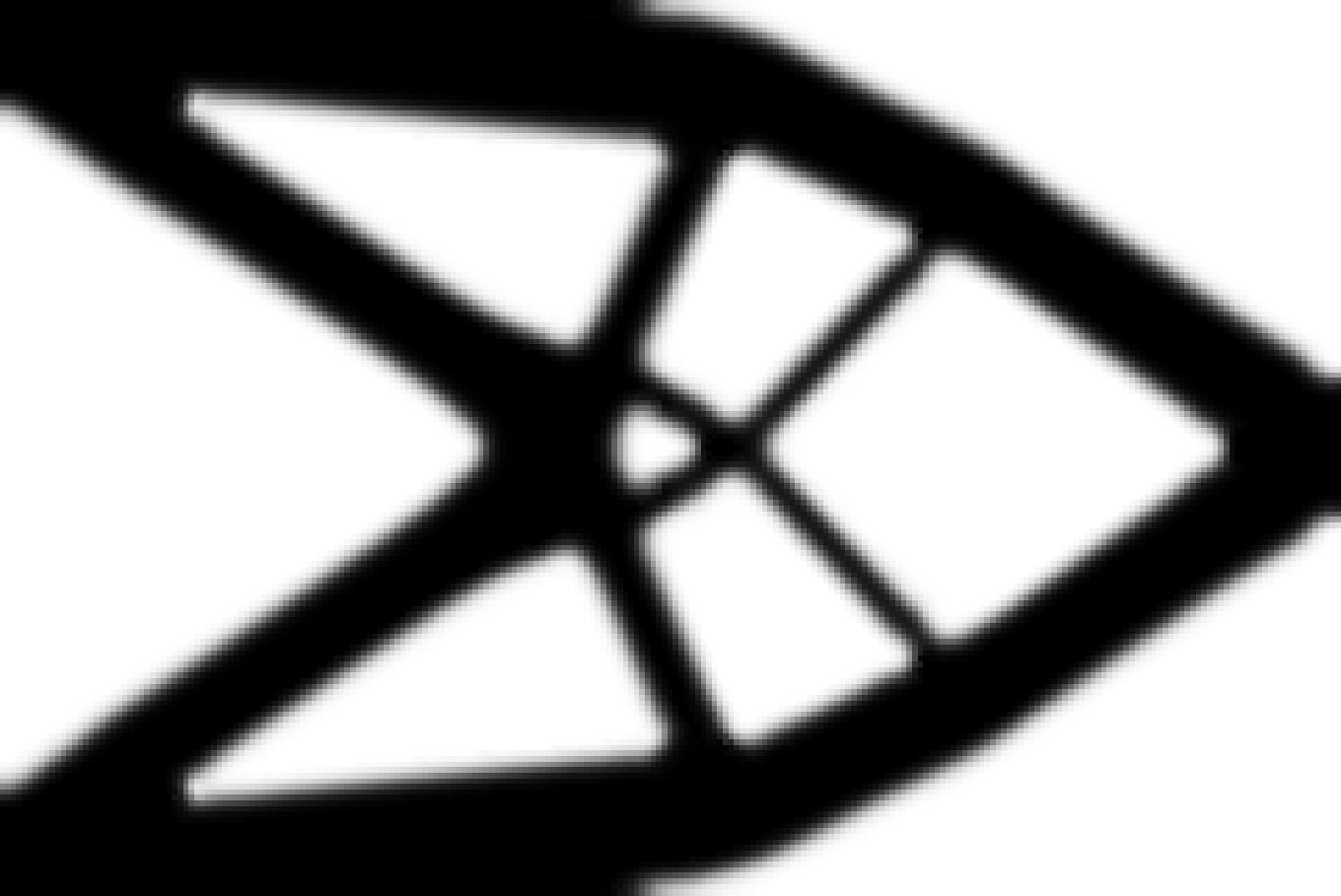} &
        \includegraphics[width=1.5cm,valign=m]{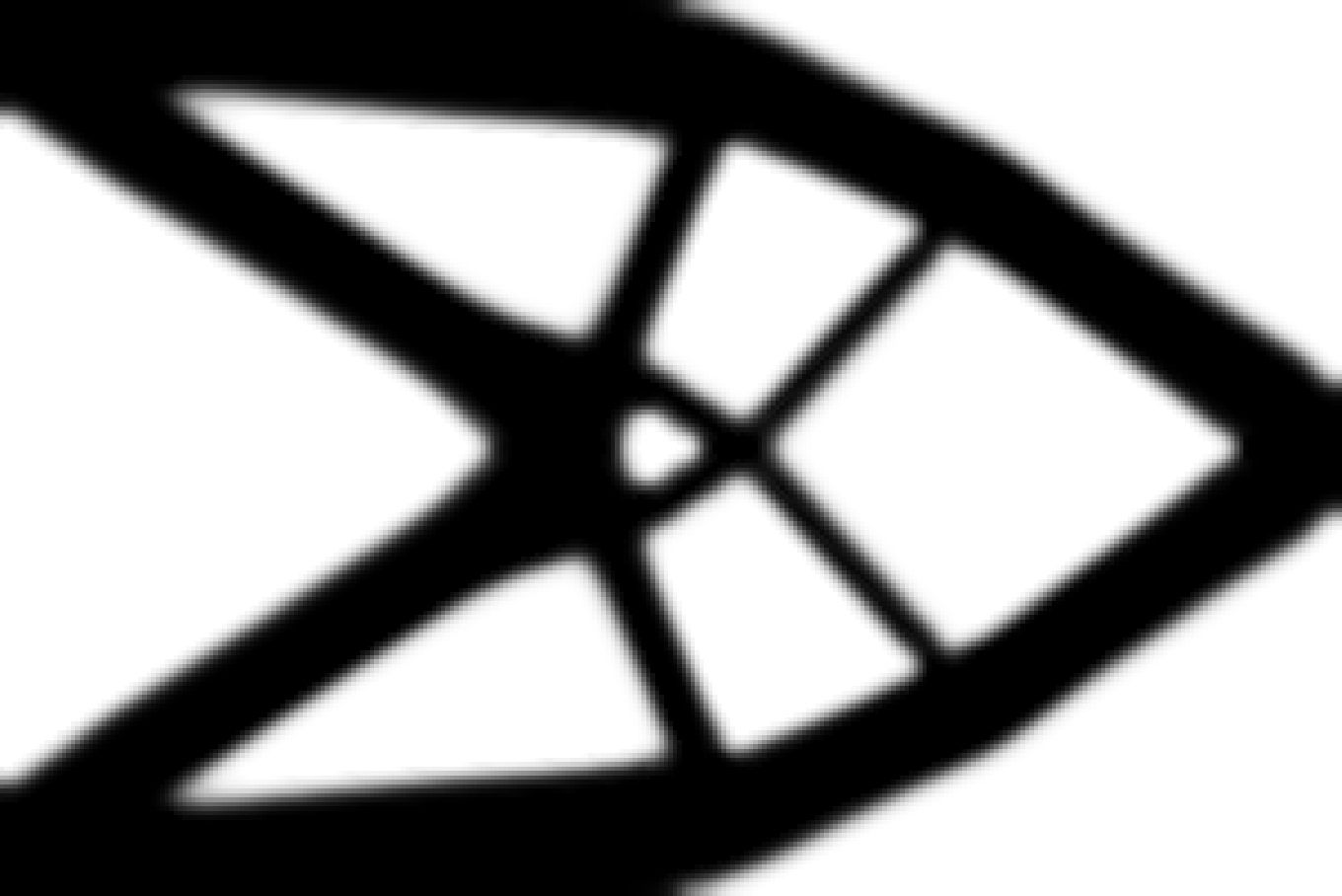} \\
        \bottomrule
    \end{tabular}
\end{table}

\begin{table}[!htbp]
    \centering
    \small
    \setlength{\tabcolsep}{2pt}
    \caption{Structure comparison of cantilevers ($r_{\min}=0.1$)} 
    \label{tab:cantilever-rmin01}
    \begin{tabular}{cccccc}
        \toprule
        $V$ & Method & $120\times 80$ & $180\times 120$ & $240\times 160$ & $300\times 200$ \\
        \midrule
        \multirow{2}{*}[-11.5pt]{0.3} & SIMP &
        \includegraphics[width=1.5cm,valign=m]{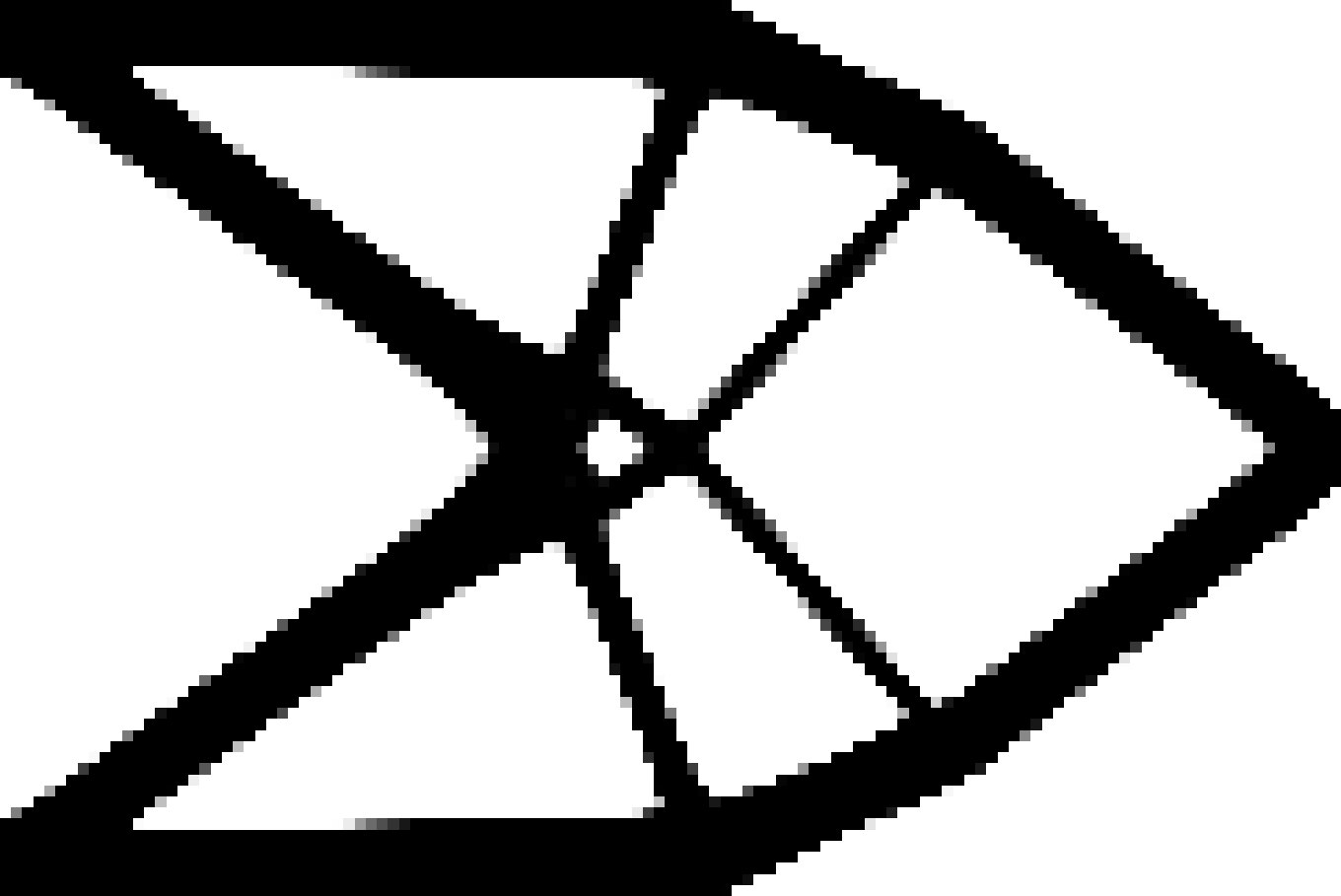} &
        \includegraphics[width=1.5cm,valign=m]{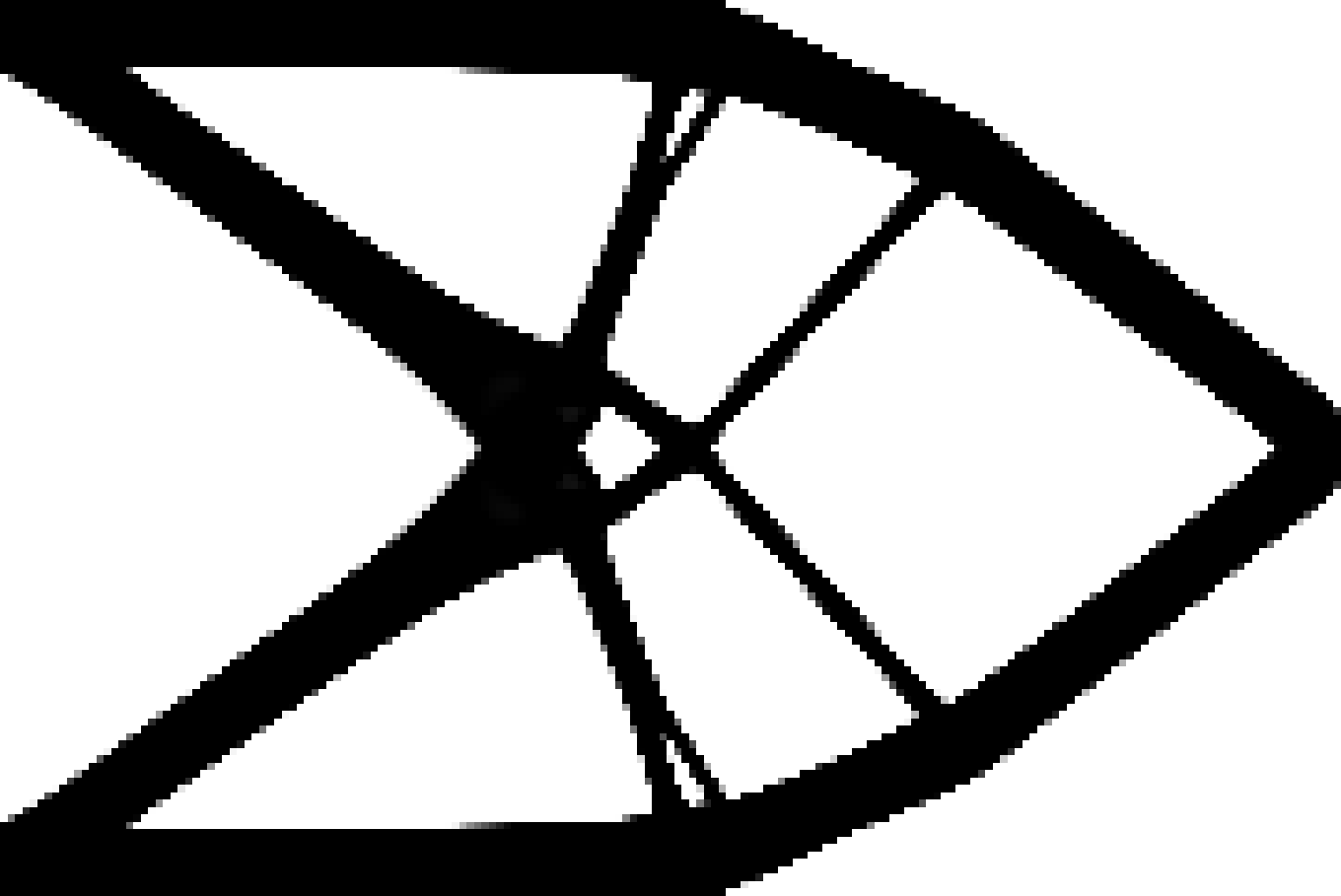} &
        \includegraphics[width=1.5cm,valign=m]{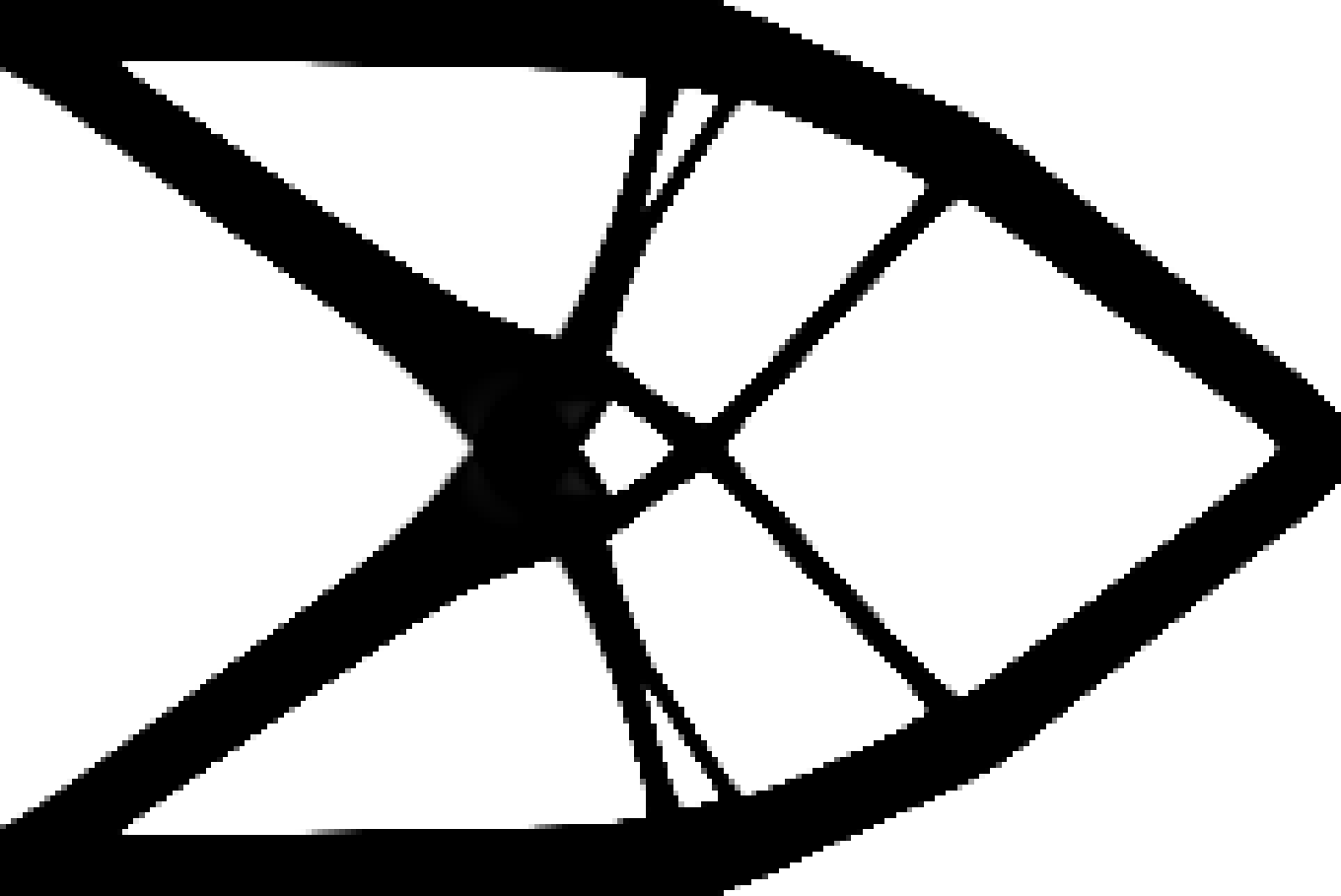} &
        \includegraphics[width=1.5cm,valign=m]{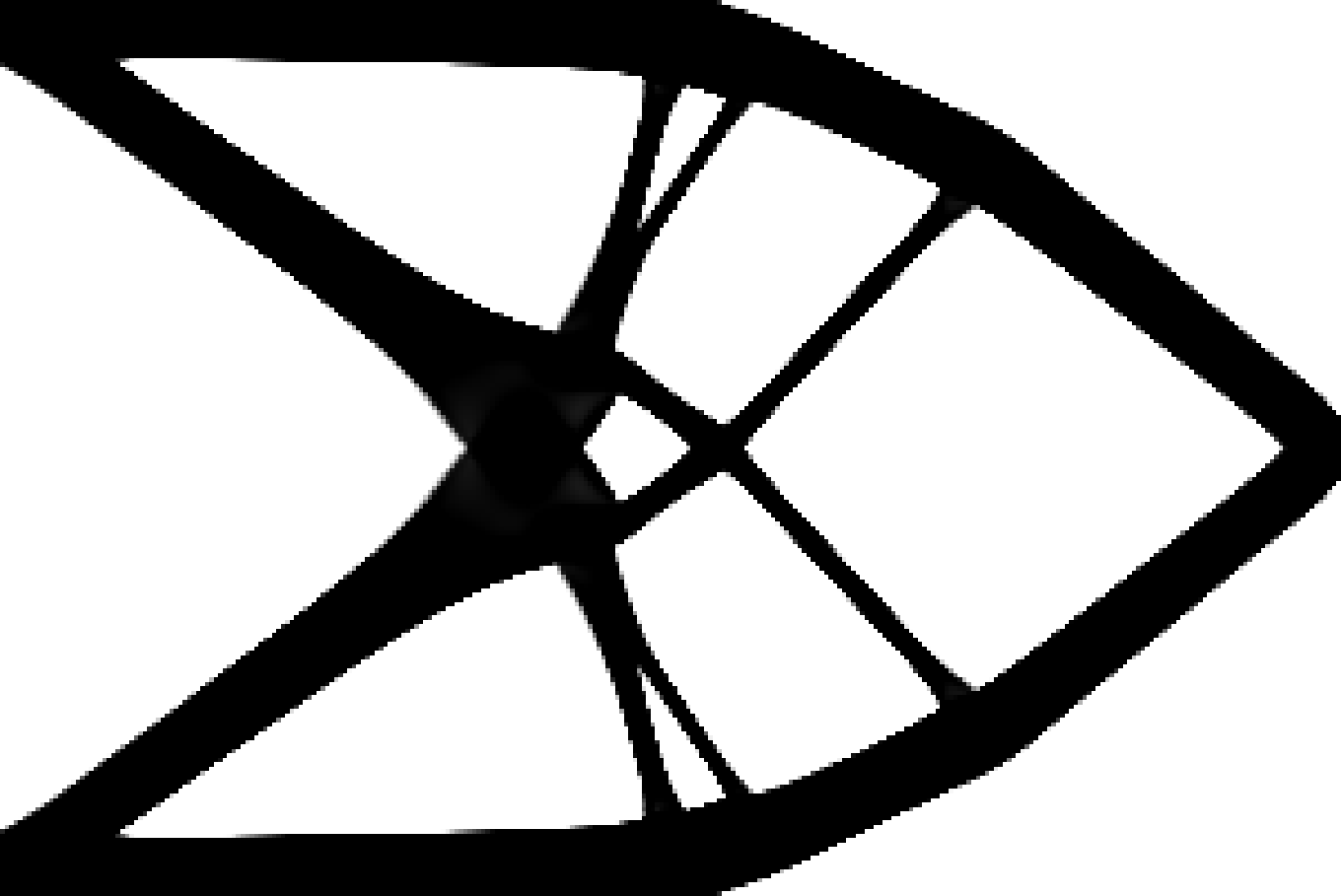} \\ \addlinespace[12pt]
                             & AC-ALM &
        \includegraphics[width=1.5cm,valign=m]{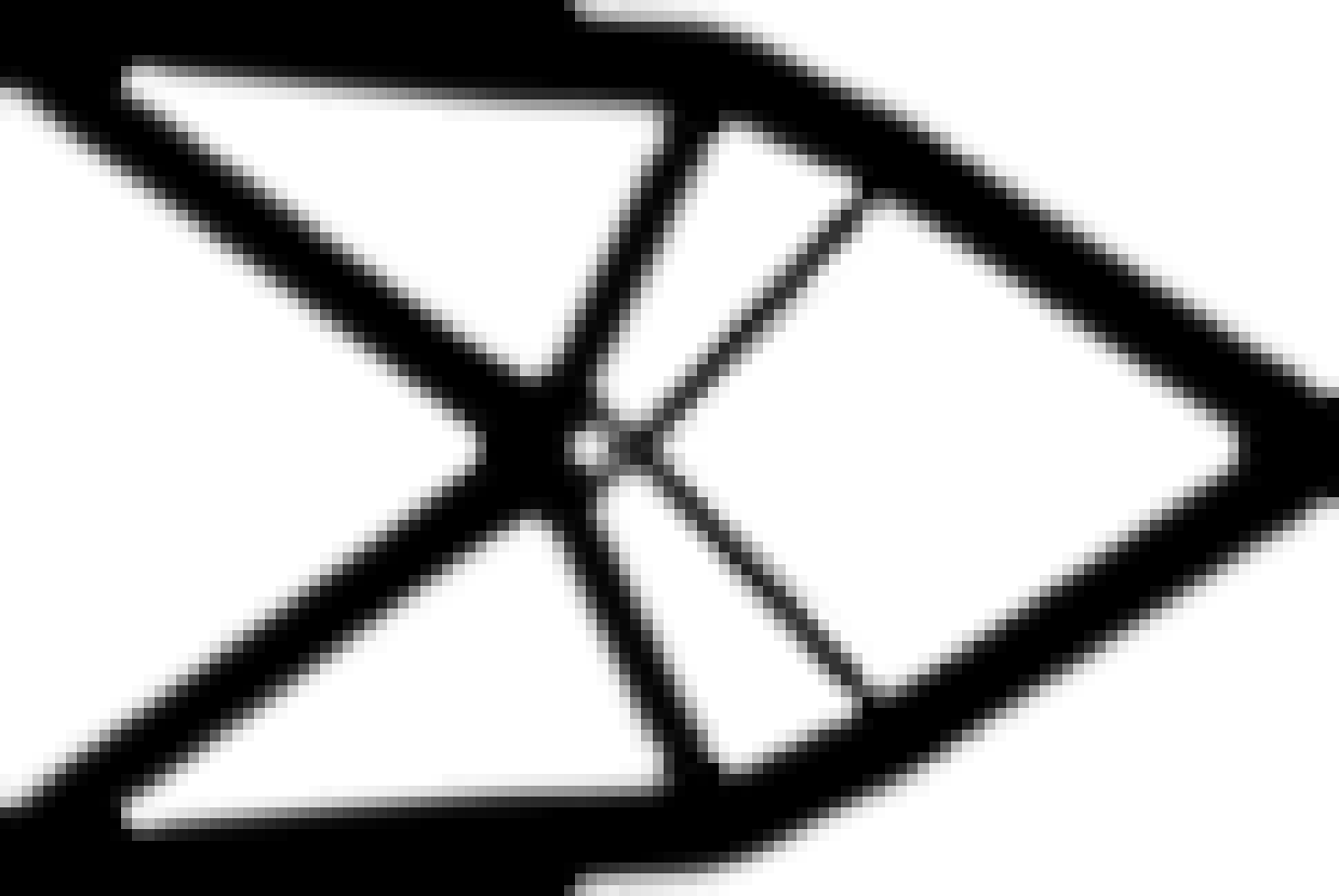} &
        \includegraphics[width=1.5cm,valign=m]{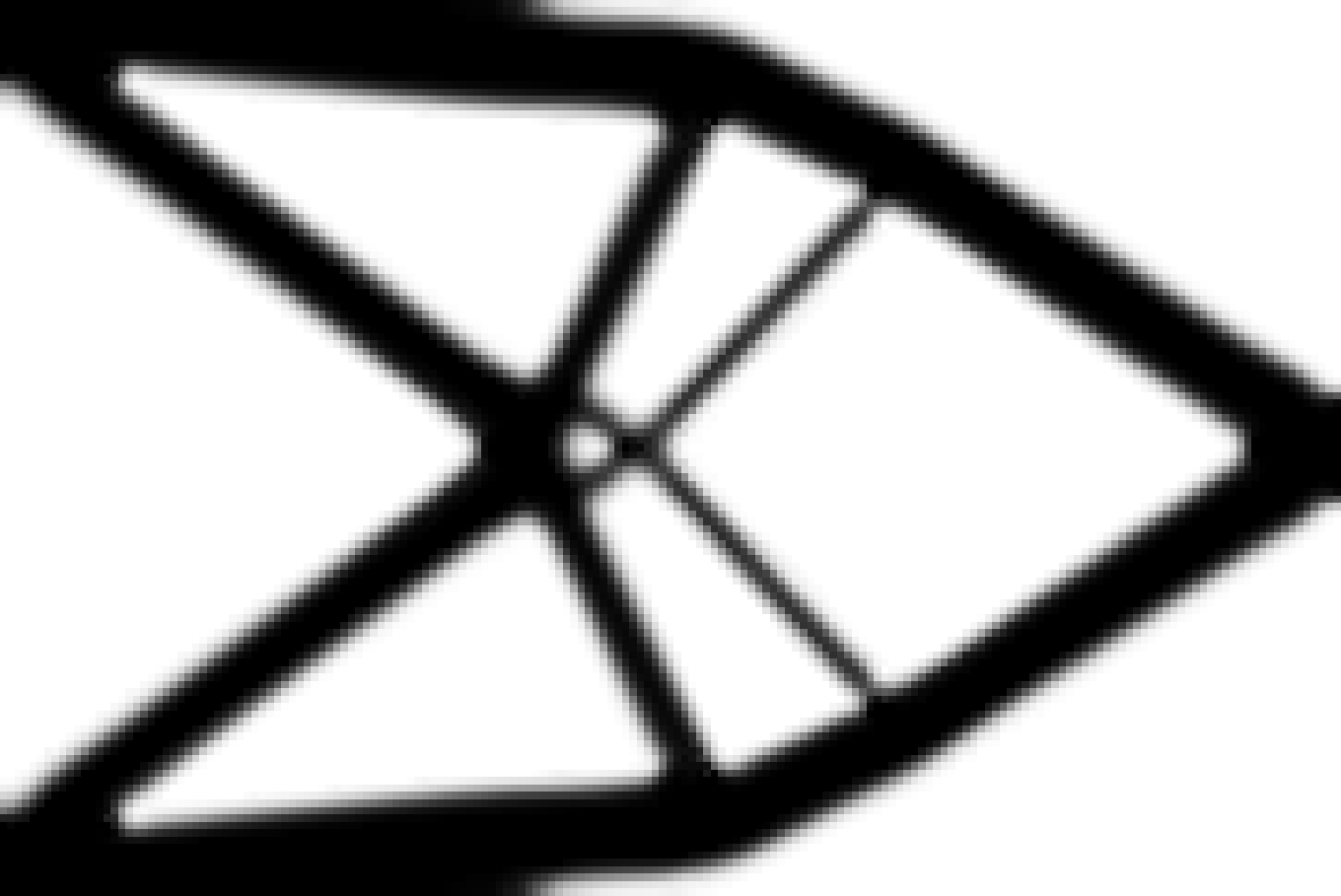} &
        \includegraphics[width=1.5cm,valign=m]{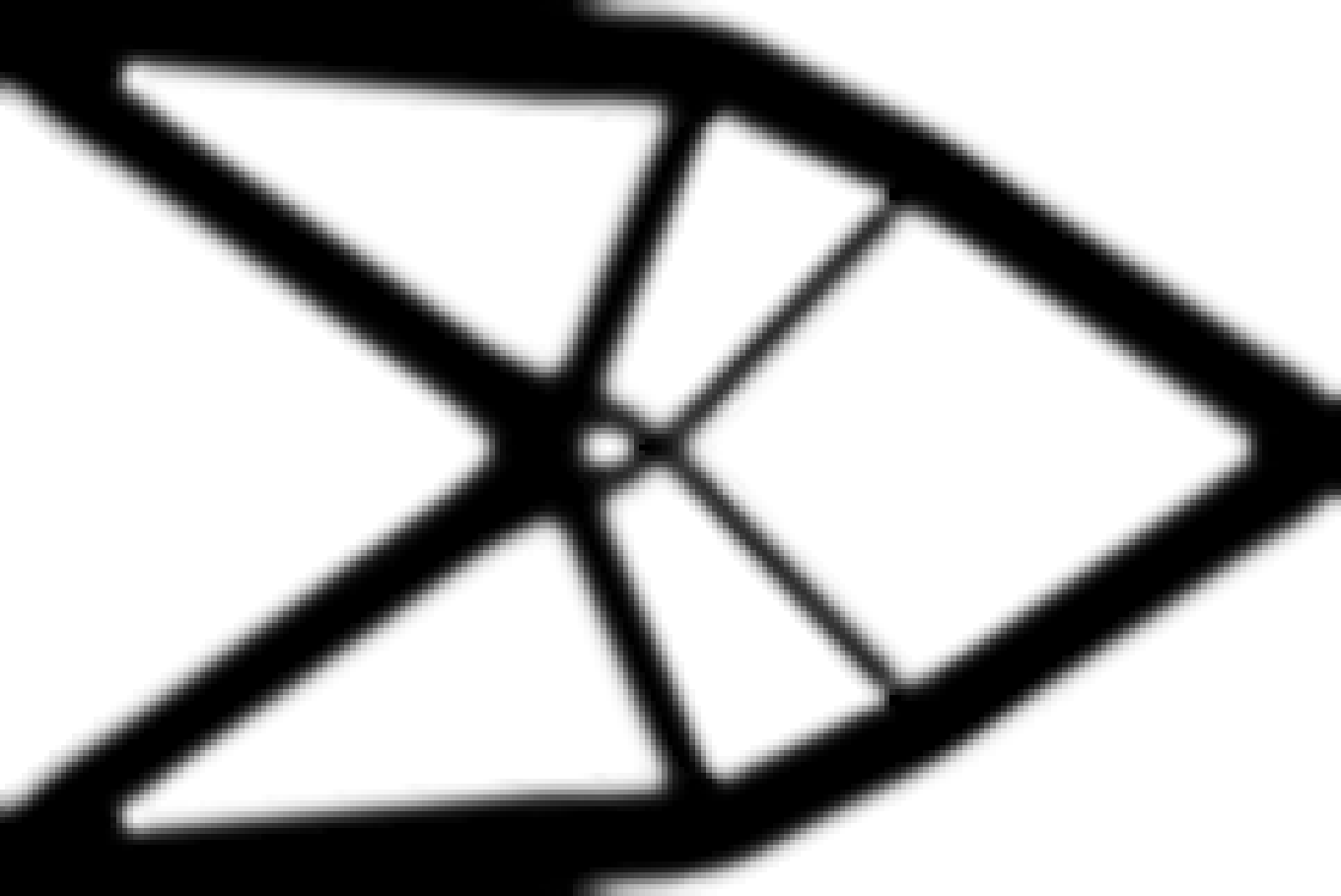} &
        \includegraphics[width=1.5cm,valign=m]{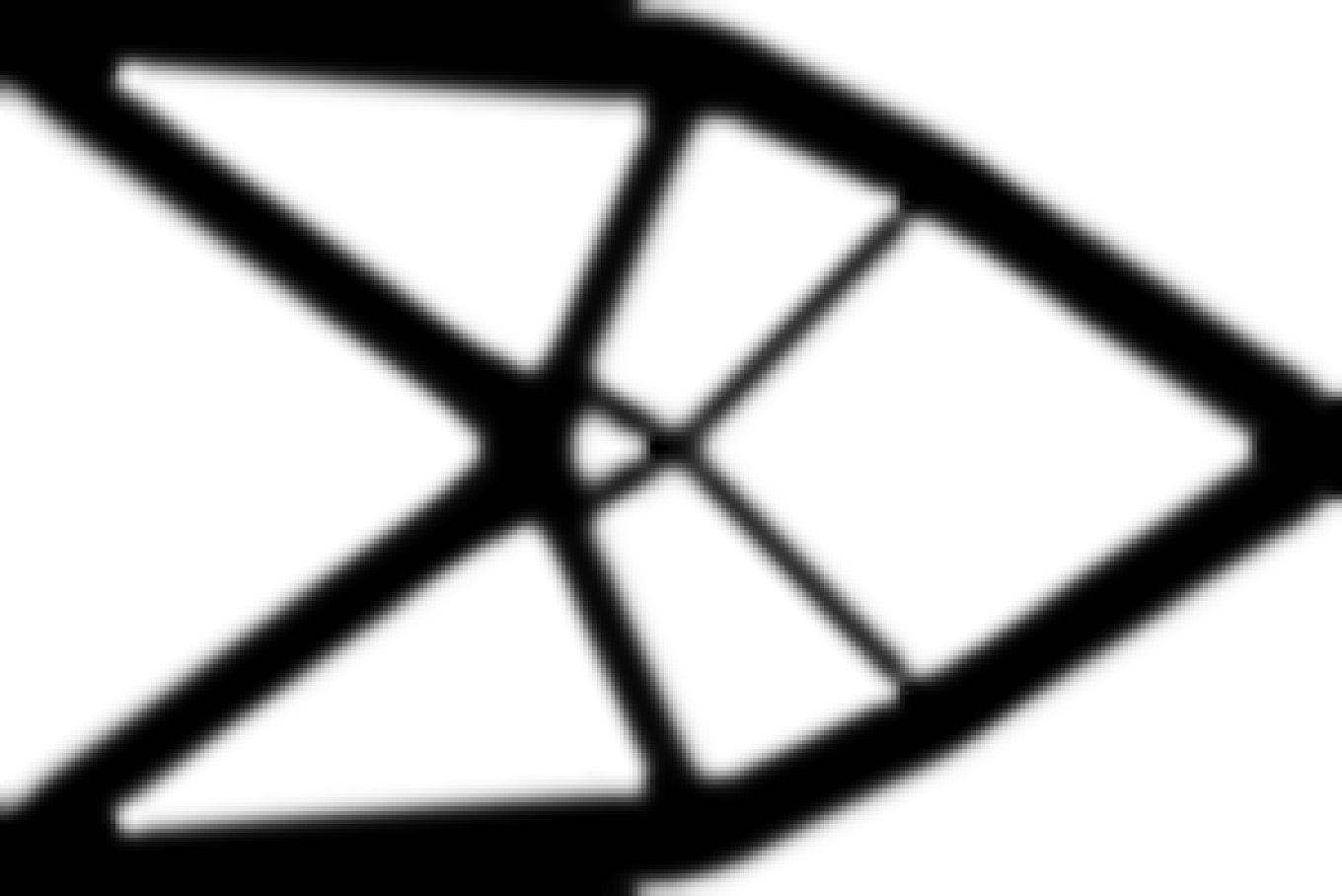} \\
        \midrule
        \multirow{2}{*}[-11.5pt]{0.4} & SIMP &
        \includegraphics[width=1.5cm,valign=m]{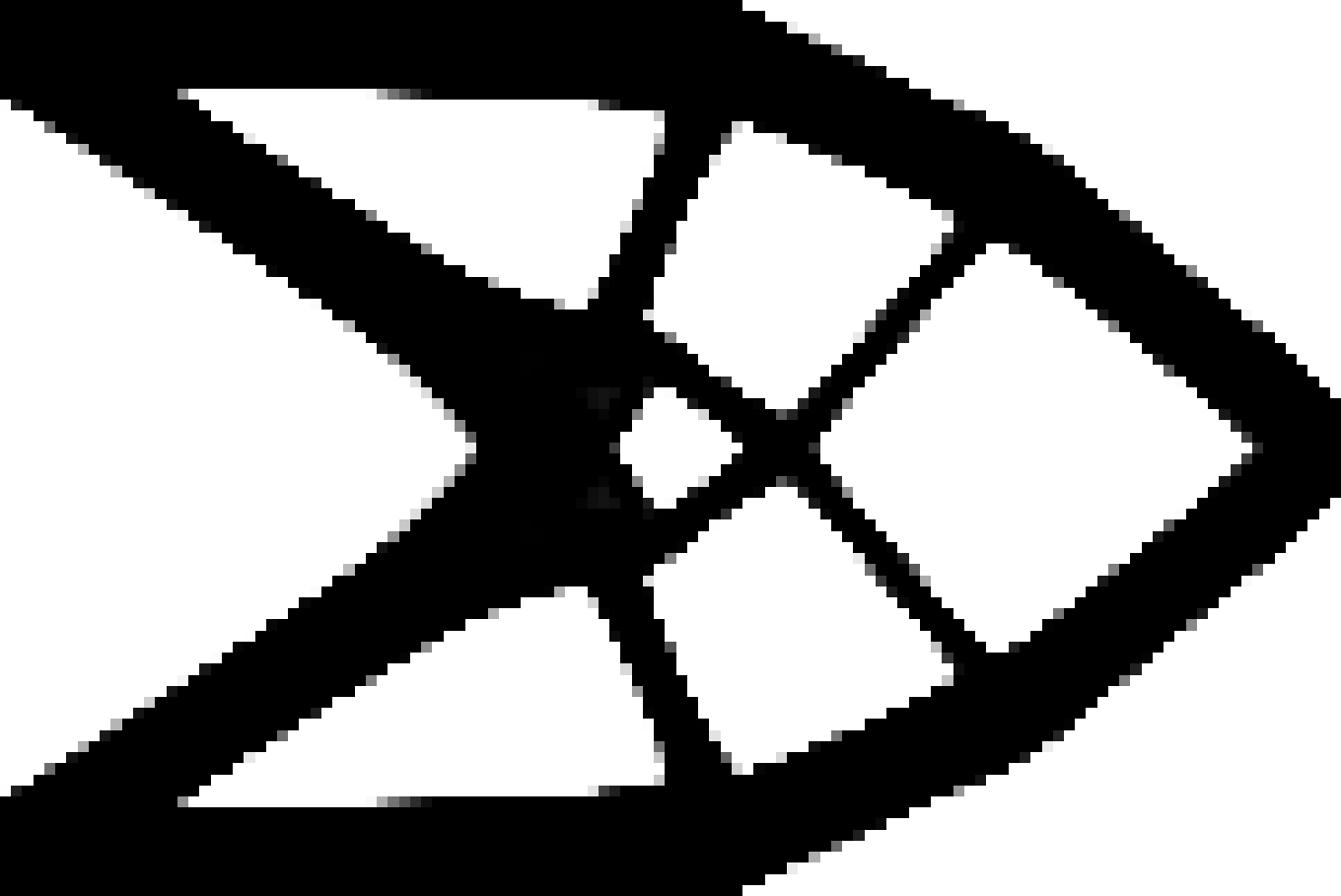} &
        \includegraphics[width=1.5cm,valign=m]{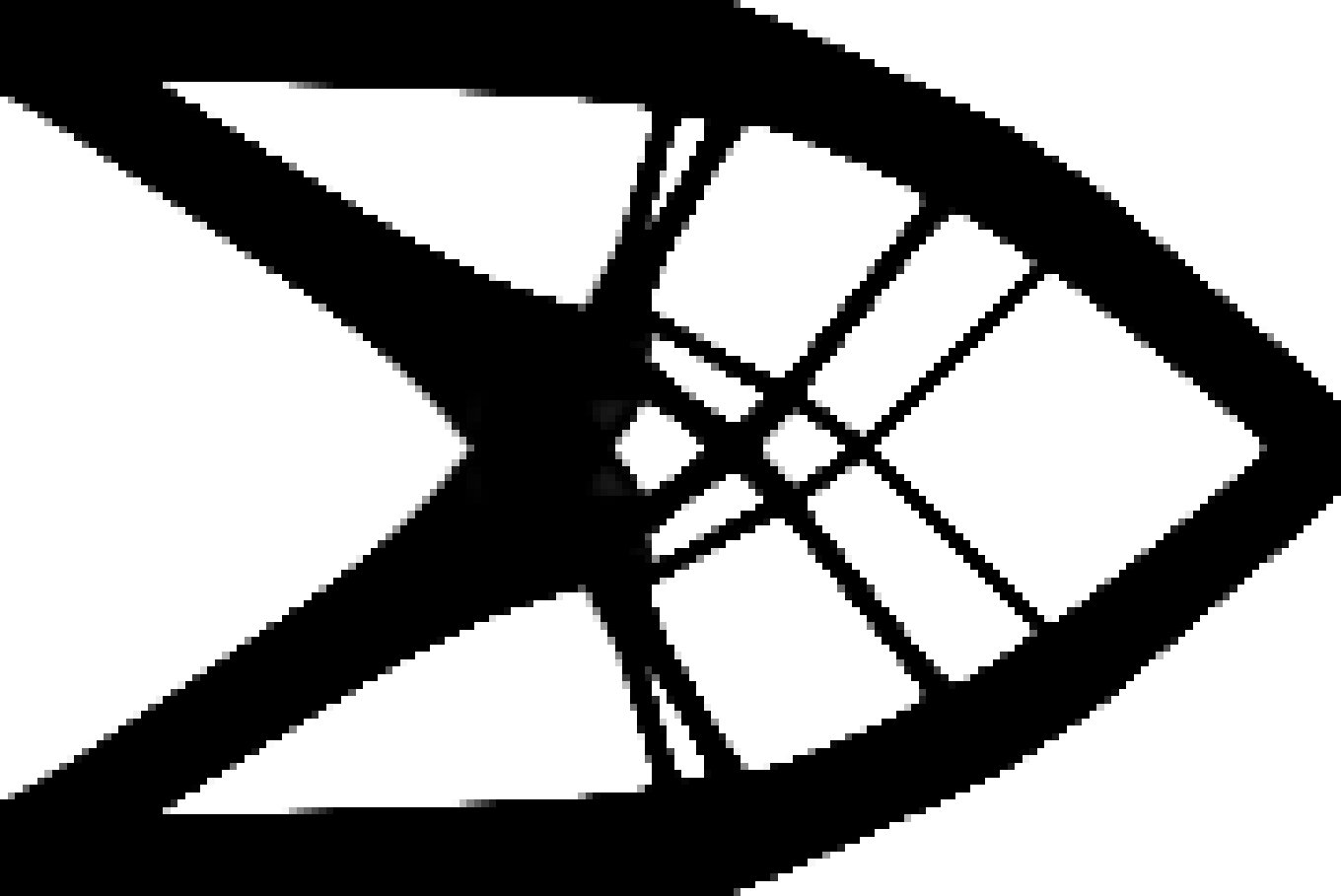} &
        \includegraphics[width=1.5cm,valign=m]{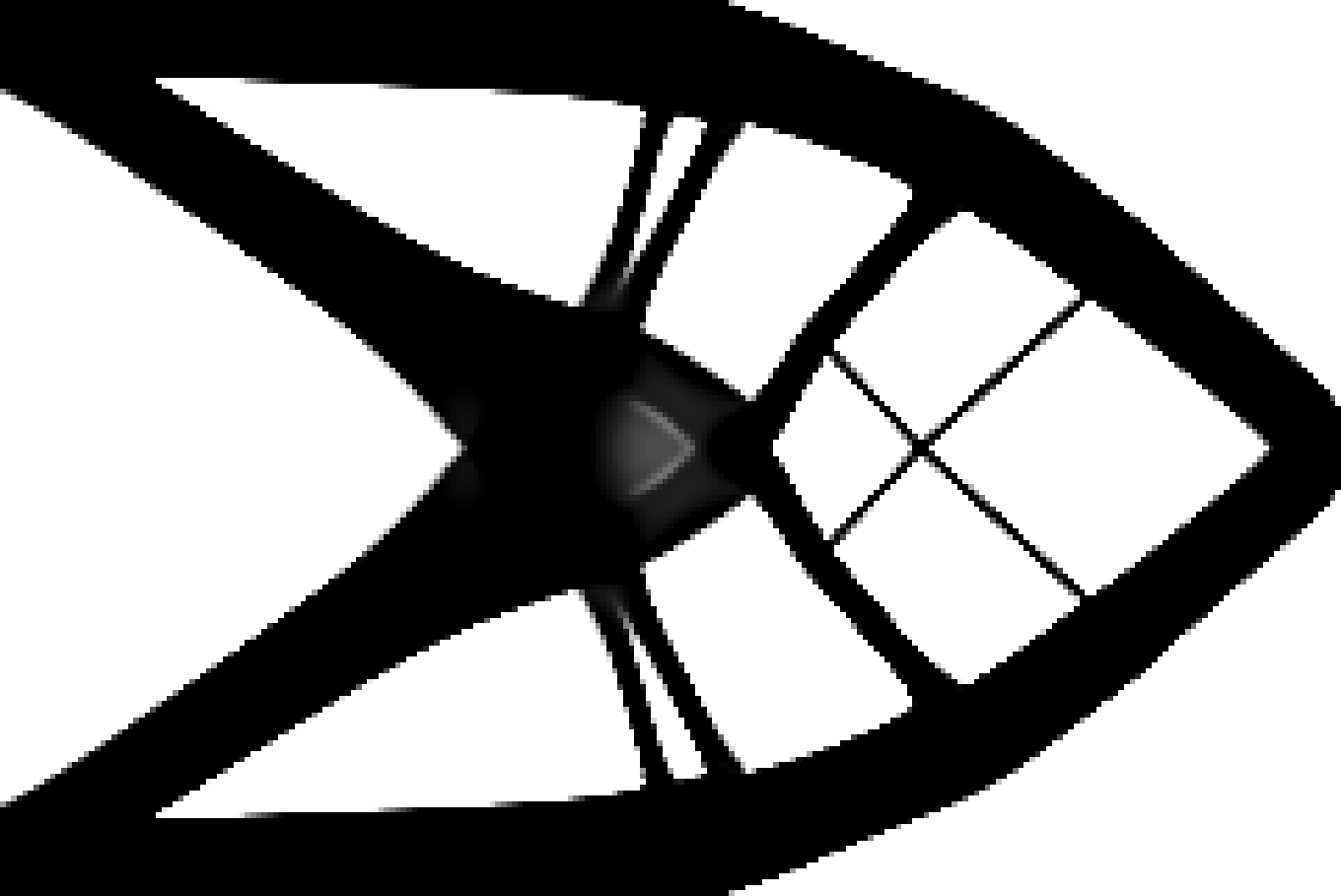} &
        \includegraphics[width=1.5cm,valign=m]{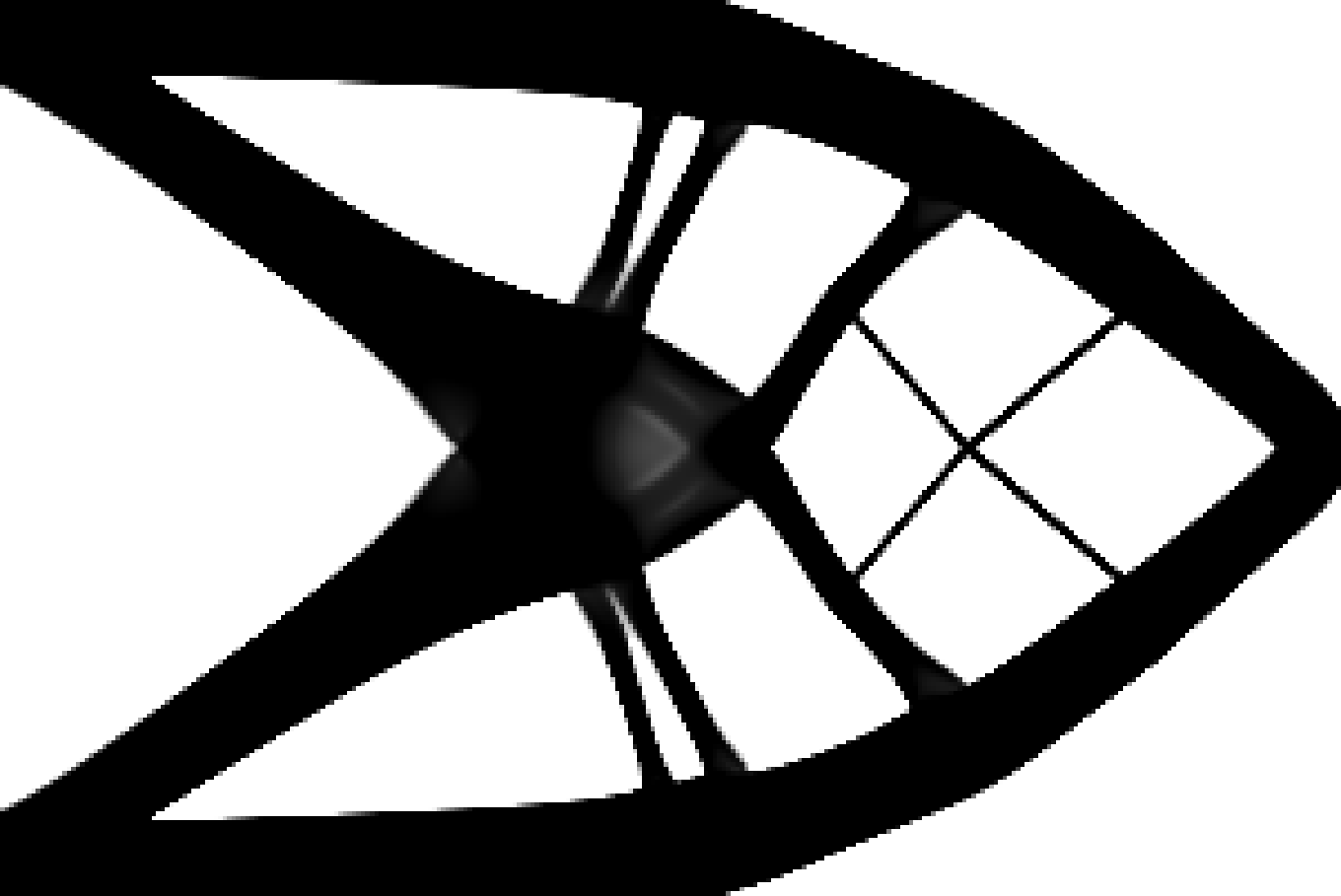} \\ \addlinespace[12pt]
                             & AC-ALM &
        \includegraphics[width=1.5cm,valign=m]{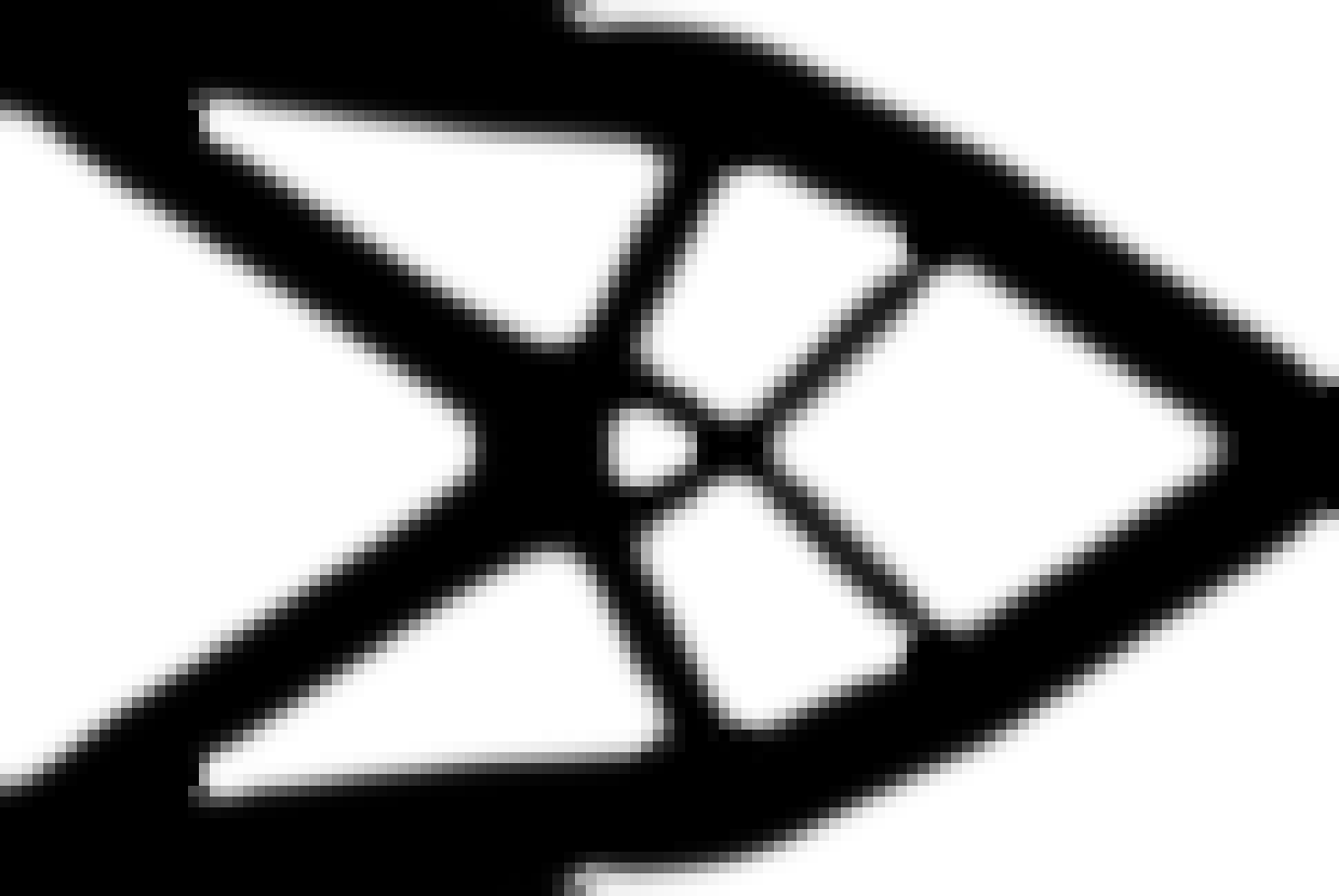} &
        \includegraphics[width=1.5cm,valign=m]{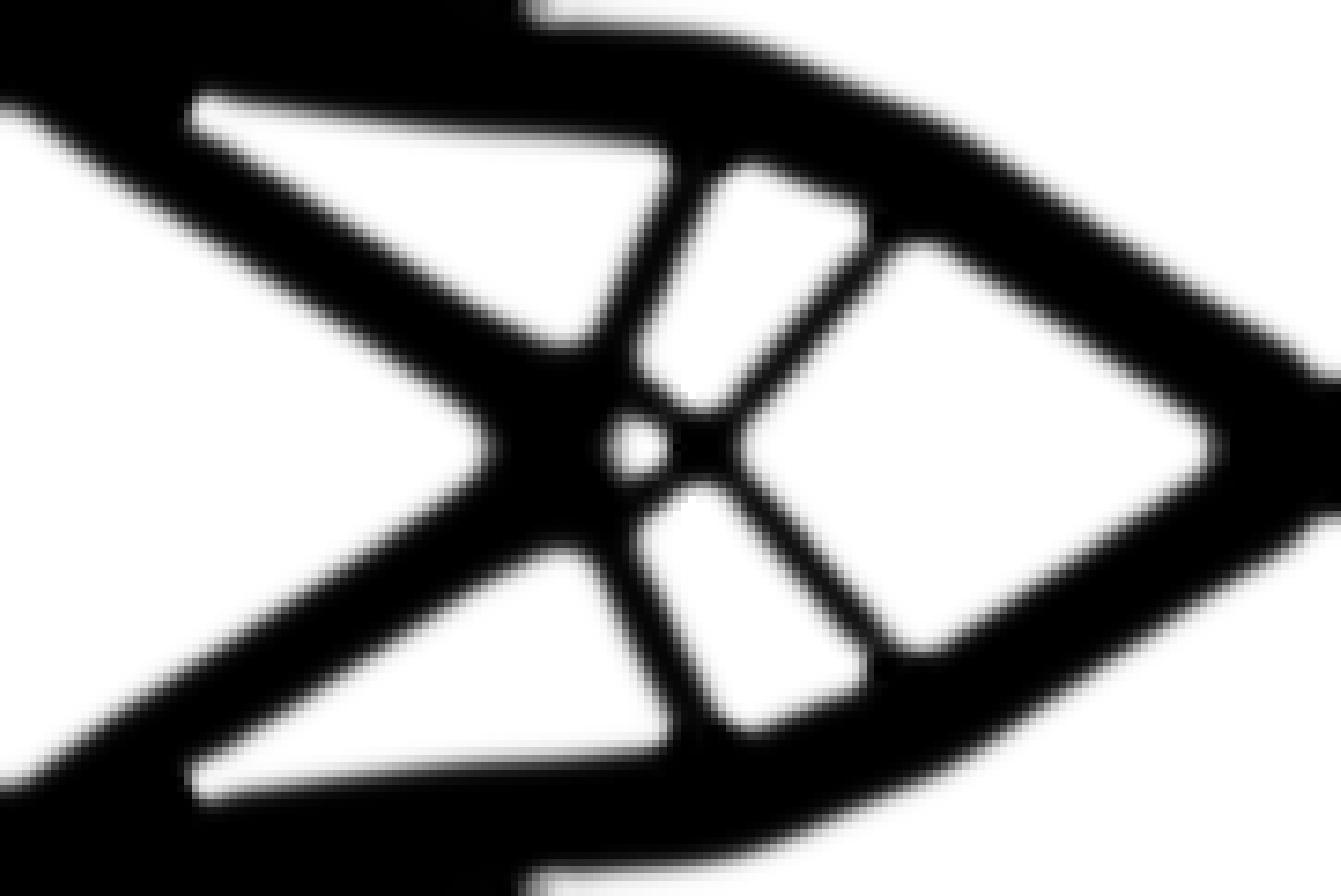} &
        \includegraphics[width=1.5cm,valign=m]{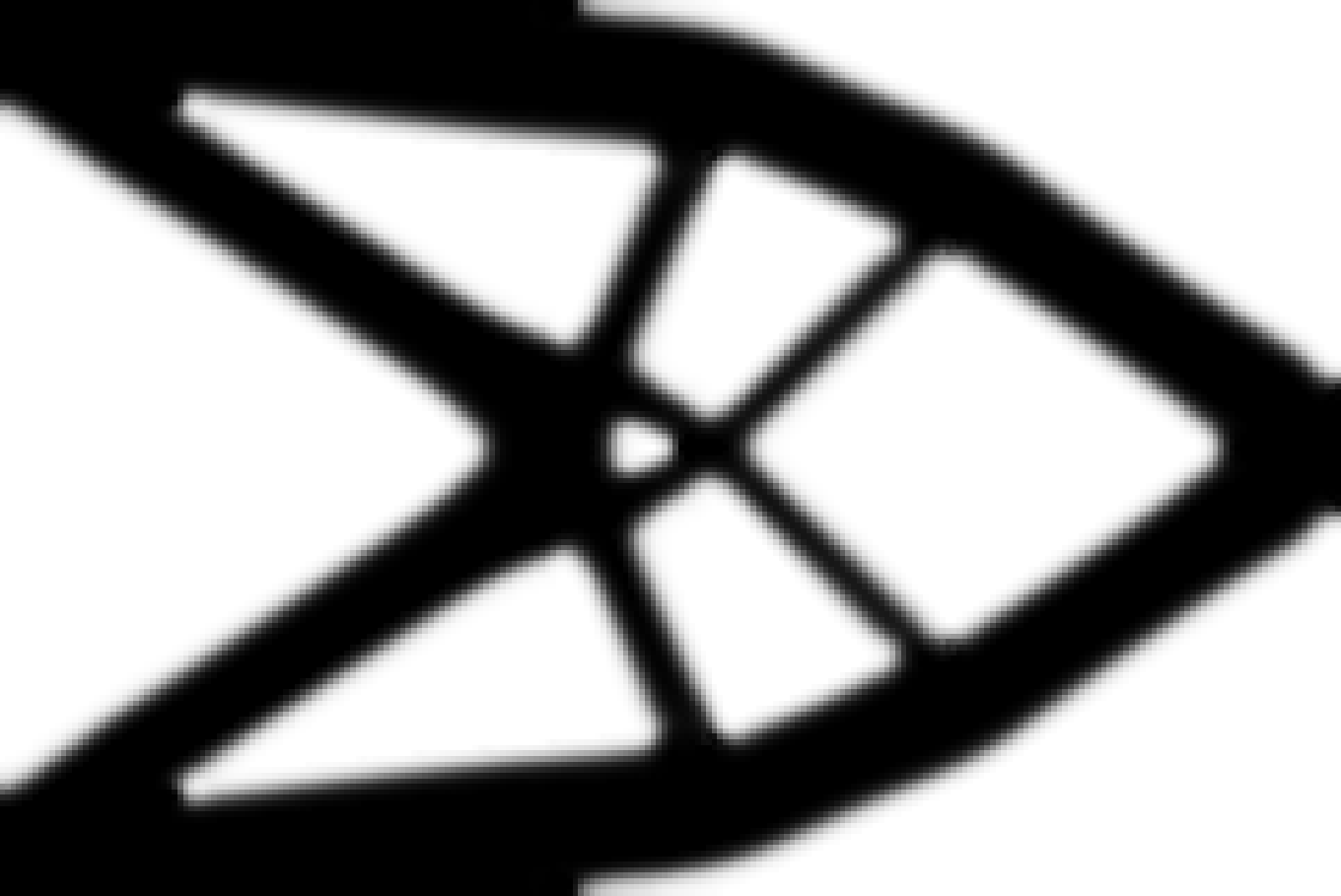} &
        \includegraphics[width=1.5cm,valign=m]{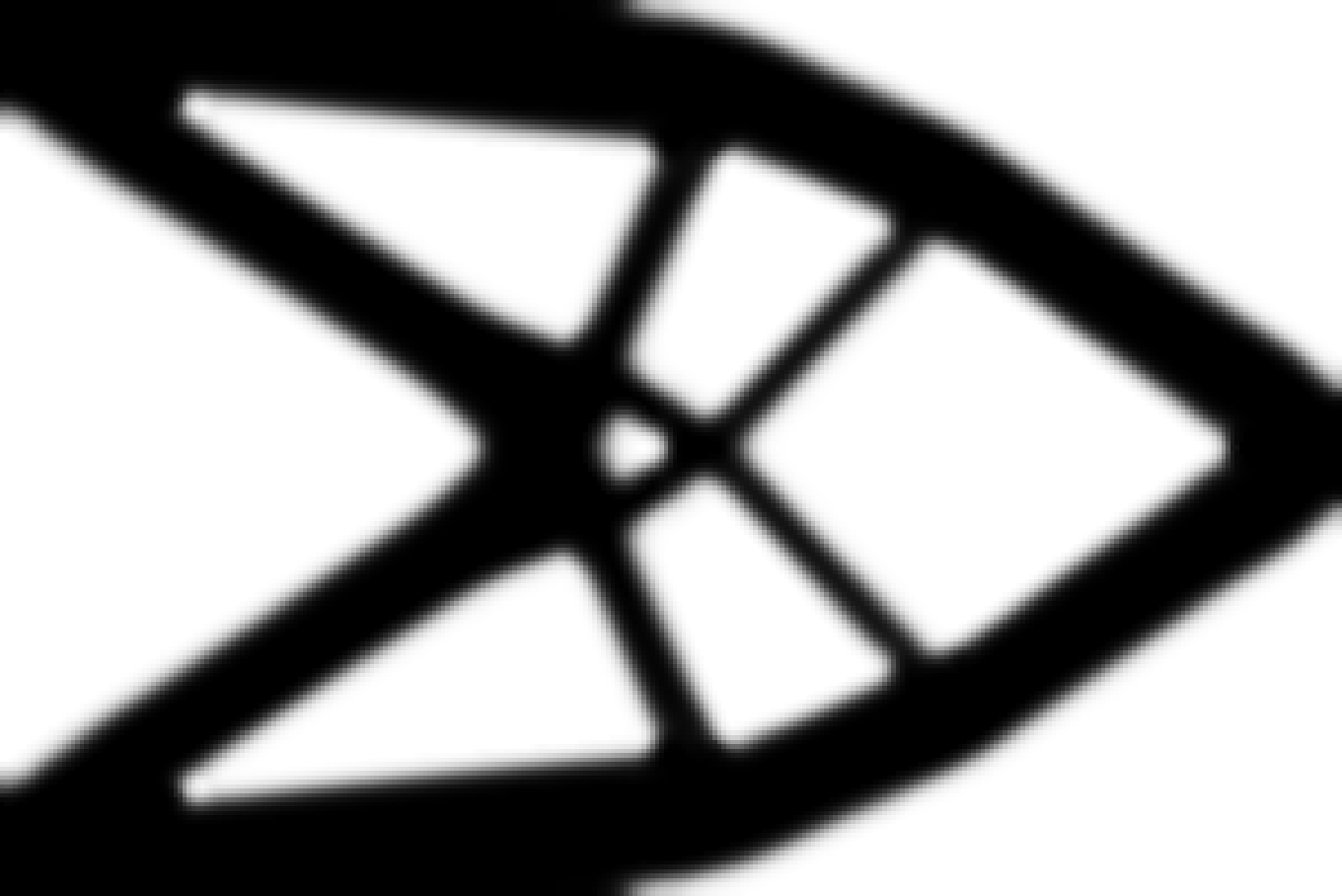} \\
        \bottomrule
    \end{tabular}
\end{table}

\subsubsection{L-Beam}\label{num_examples_lbeam}

The L-beam introduces a non-convex domain with a sharp re-entrant corner 
that induces a severe stress singularity. 
To preserve consistent boundary conditions, the passive void region ($72\times 72$)
is excluded from both the structural elasticity equations
and the AC finite element assembly. 
The nodes along the cut boundary remain active design variables
and satisfy the natural homogeneous Neumann condition $\mathbf{n}\cdot\nabla\phi = 0$. 
This allows material to terminate naturally at the boundary
without generating spurious artificial residuals. 
An identical treatment is applied to the SIMP baseline.

\begin{figure}[!ht]
\centering
\includegraphics{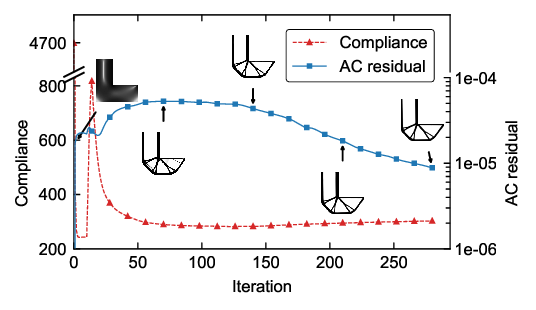}
\caption{Convergence history for L-beam under AC-ALM ($180\times 180$, $V=0.3$, $r_{\min}=0.1$).}\label{fig:lbeam_conv}
\end{figure}

The convergence history in Fig.~\ref{fig:lbeam_conv} confirms stable optimization progress 
around the geometric singularity. 
Across all 16 L-beam runs, the residual norm terminates safely below the tolerance $\tau_h$ 
(with a peak ratio of $\|\mathbf{h}\|_\text{norm}/\tau_h = 0.67$). 
Feasibility is achieved with the adaptive penalty remaining well below the configured ceiling 
$\rho_\text{h}^{\max} = 50$, avoiding cap saturation and preventing numerical ill-conditioning.

Tables~\ref{tab:lbeam-rmin2} and \ref{tab:lbeam-rmin01} display
the optimized L-beam topologies. 
In the filter-assisted regime ($r_{\min} = 2.0$, Table~\ref{tab:lbeam-rmin2}), 
both SIMP and AC-ALM produce closely matched primary load paths consisting of an outer arch 
and radiating web members, with AC-ALM suppressing minor secondary branchlets and faint gray-scale artifacts. 
The decisive contrast appears in the intrinsic regime ($r_{\min} = 0.1$, Table~\ref{tab:lbeam-rmin01}). 
Here, the unregularized baseline exhibits severe mesh-dependent branch proliferation 
and thin member fragmentation near the re-entrant corner. 
In contrast, AC-ALM forms clean, well-connected arches
that route smoothly around the corner, 
preserving topological consistency and an intrinsic length scale governed by
$\ell_\text{AC} = \epsilon/\sqrt{\beta}$ across all meshes.

\begin{table}[!htbp]
    \centering
    \small
    \setlength{\tabcolsep}{2pt}
    \caption{Structure comparison of L-beams ($r_{\min}=2$)} 
    \label{tab:lbeam-rmin2}
    \begin{tabular}{cccccc}
        \toprule
         $V$ & Method & $120\times 120$ & $180\times 180$ & $240\times 240$ & $300\times 300$ \\
        \midrule
        \multirow{2}{*}[-11.5pt]{0.3} & SIMP &
        \includegraphics[width=1.5cm,valign=m]{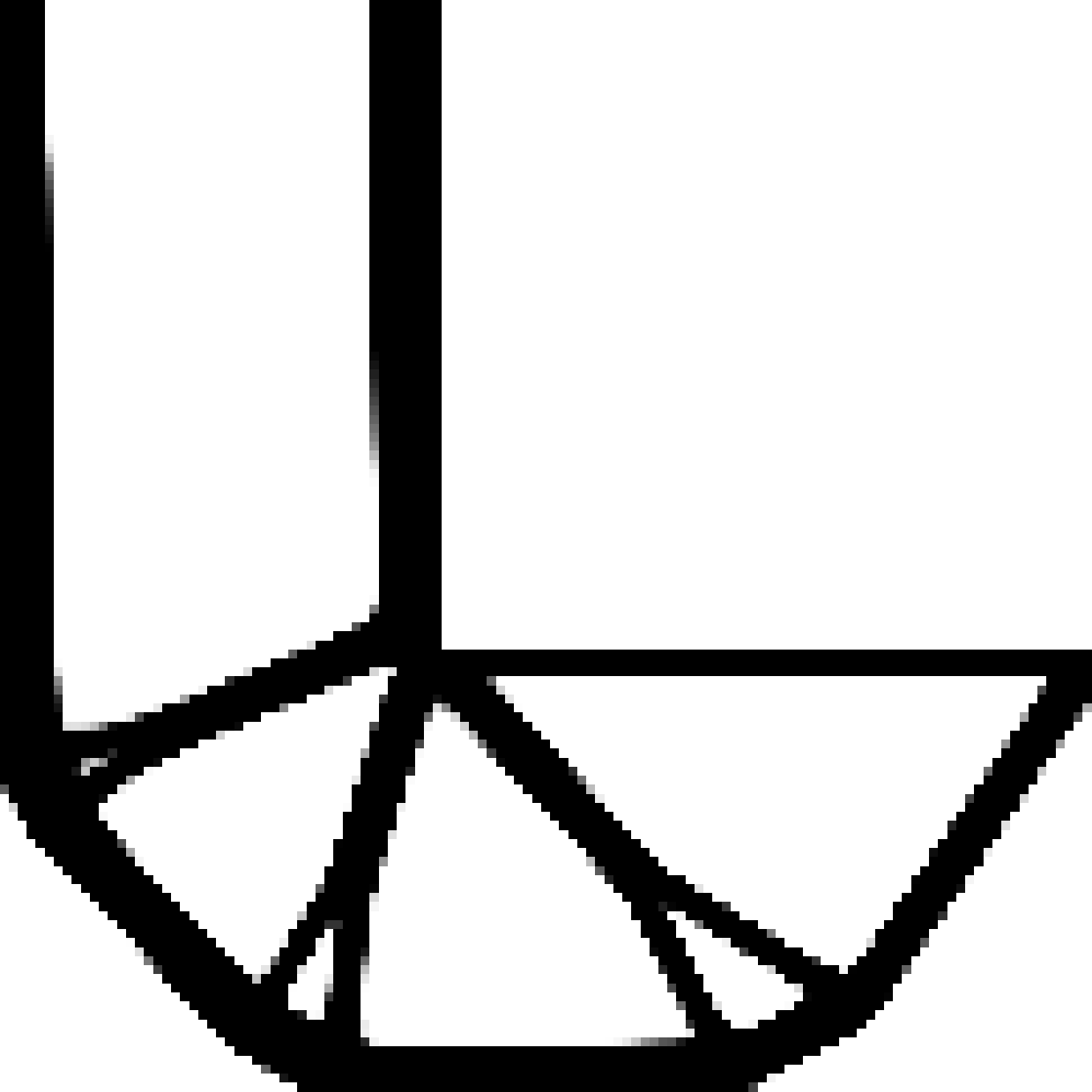} &
        \includegraphics[width=1.5cm,valign=m]{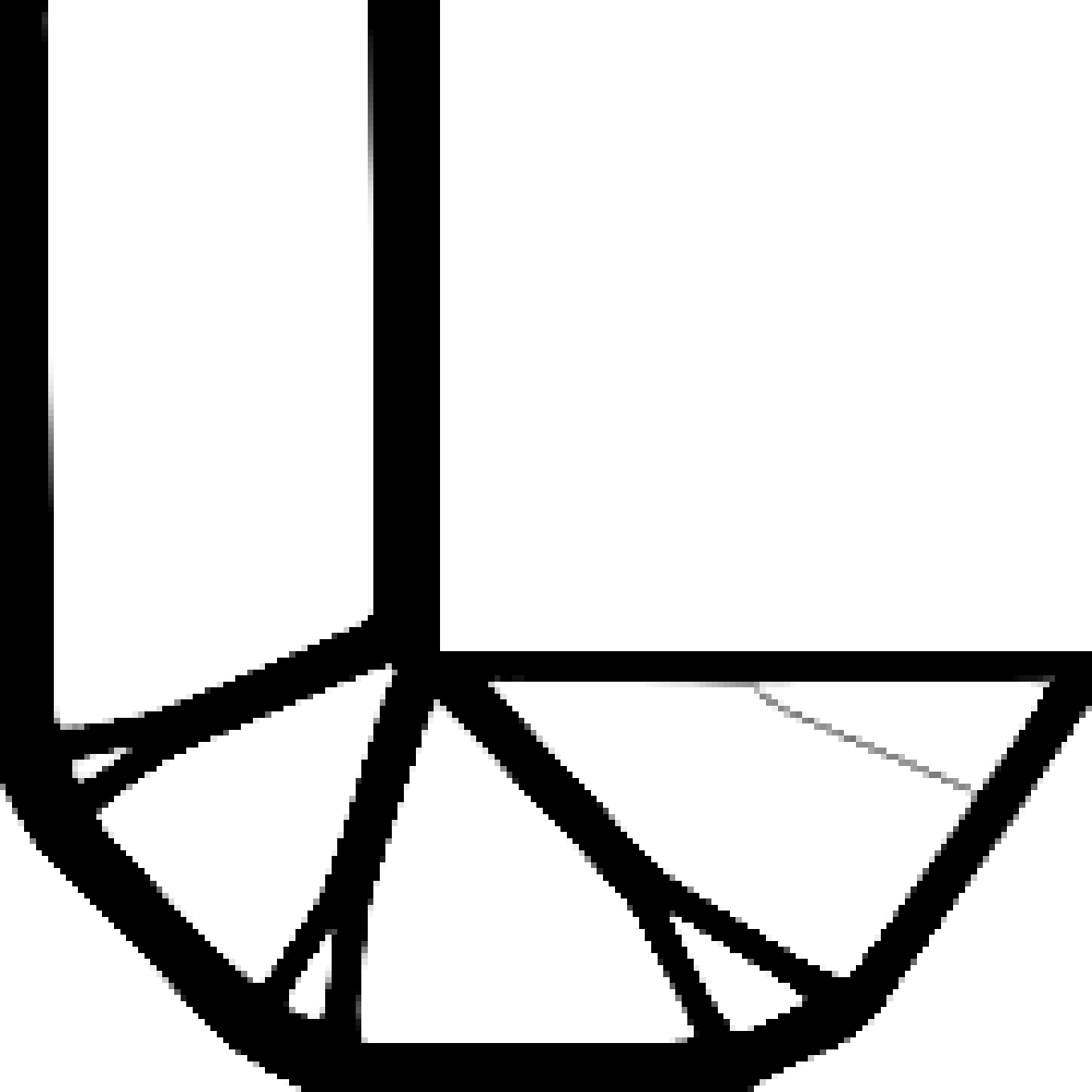} &
        \includegraphics[width=1.5cm,valign=m]{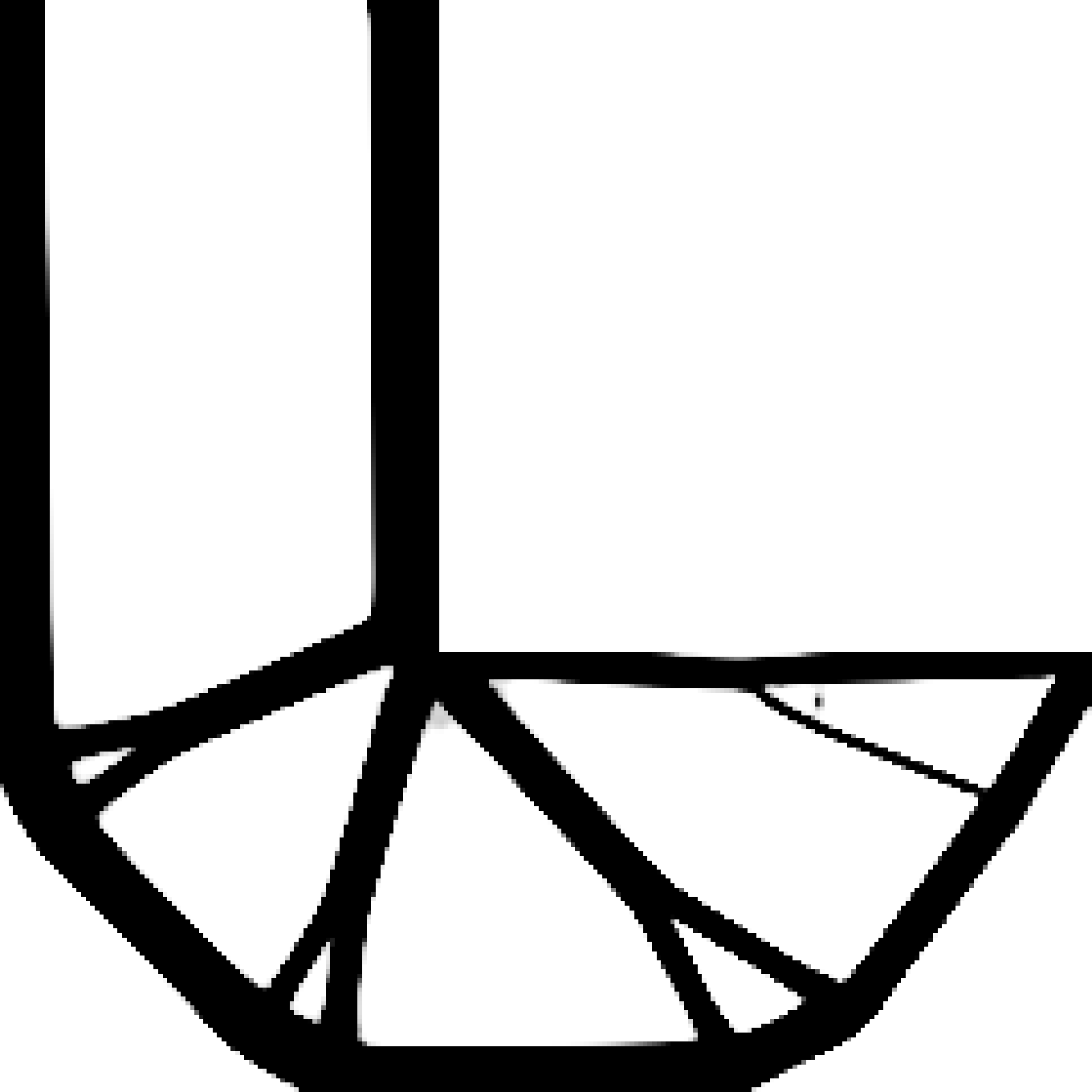} &
        \includegraphics[width=1.5cm,valign=m]{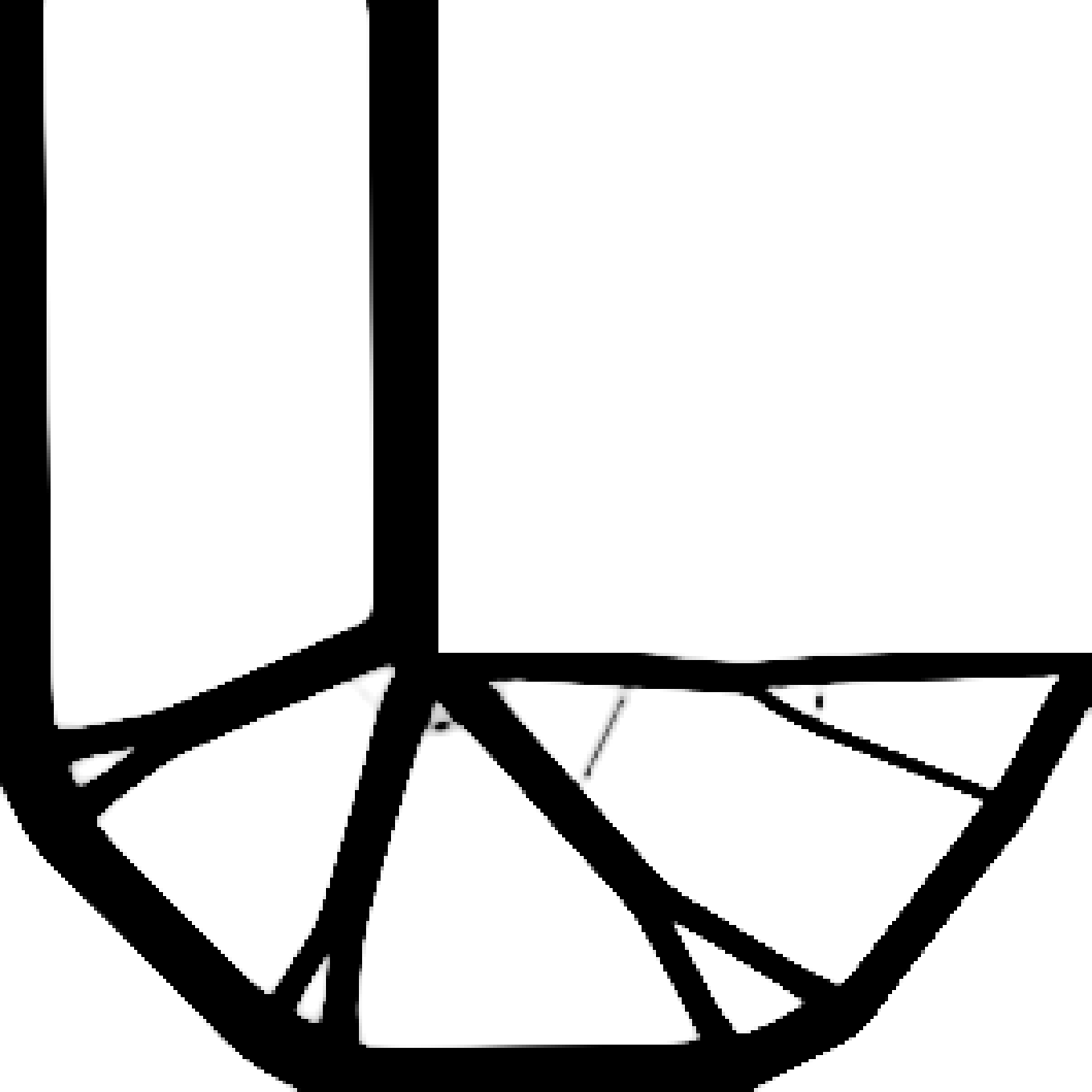} \\ \addlinespace[12pt]
                             & AC-ALM &
        \includegraphics[width=1.5cm,valign=m]{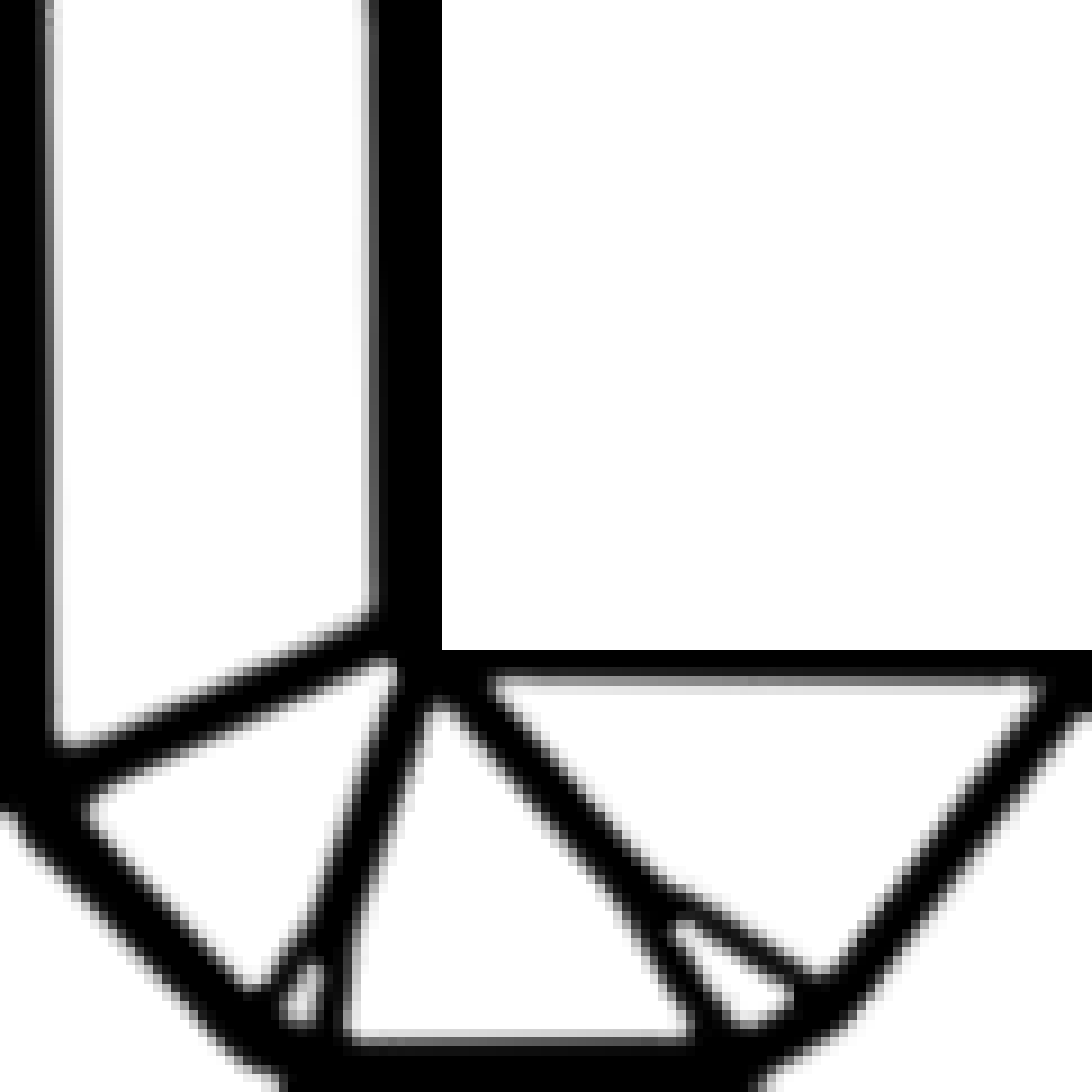} &
        \includegraphics[width=1.5cm,valign=m]{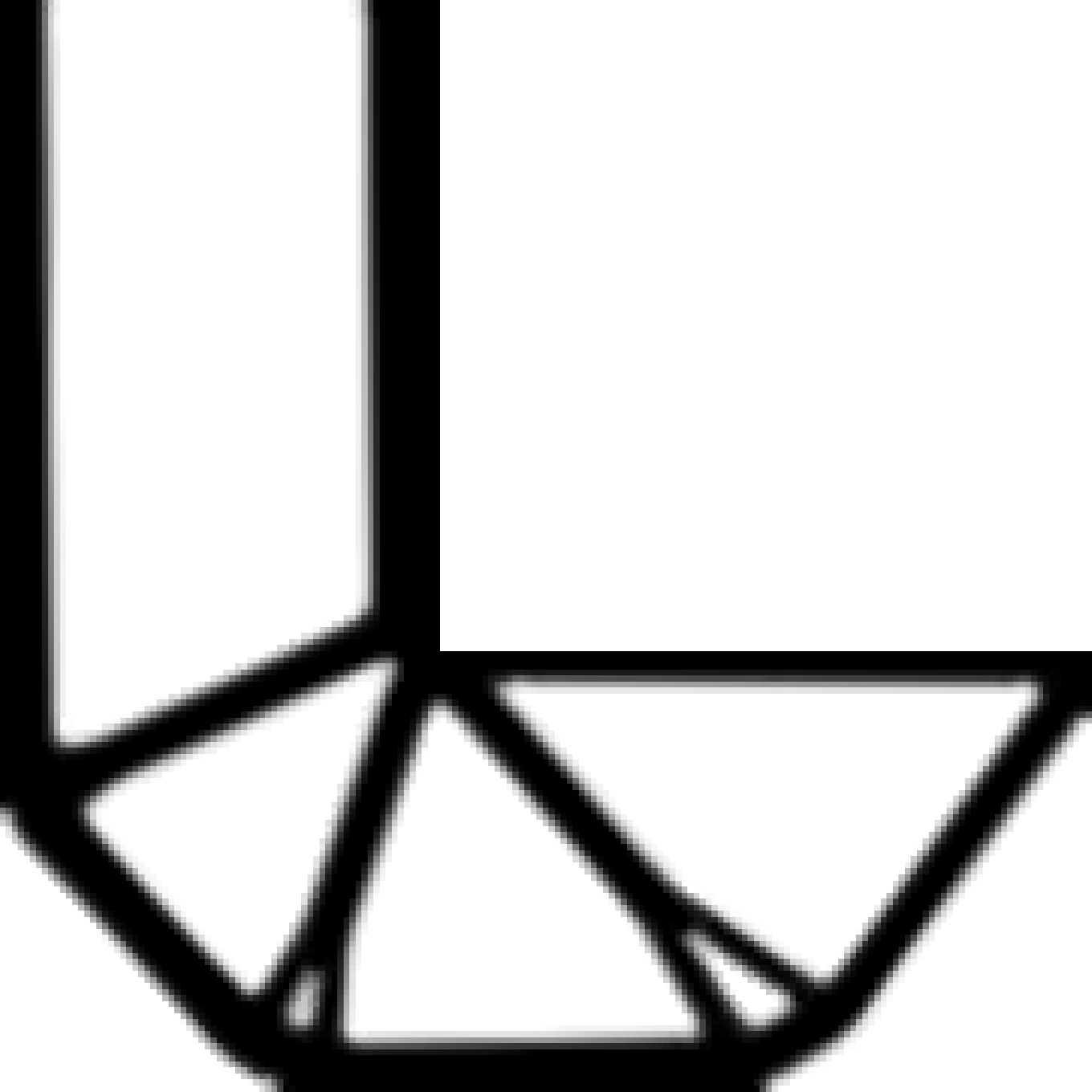} &
        \includegraphics[width=1.5cm,valign=m]{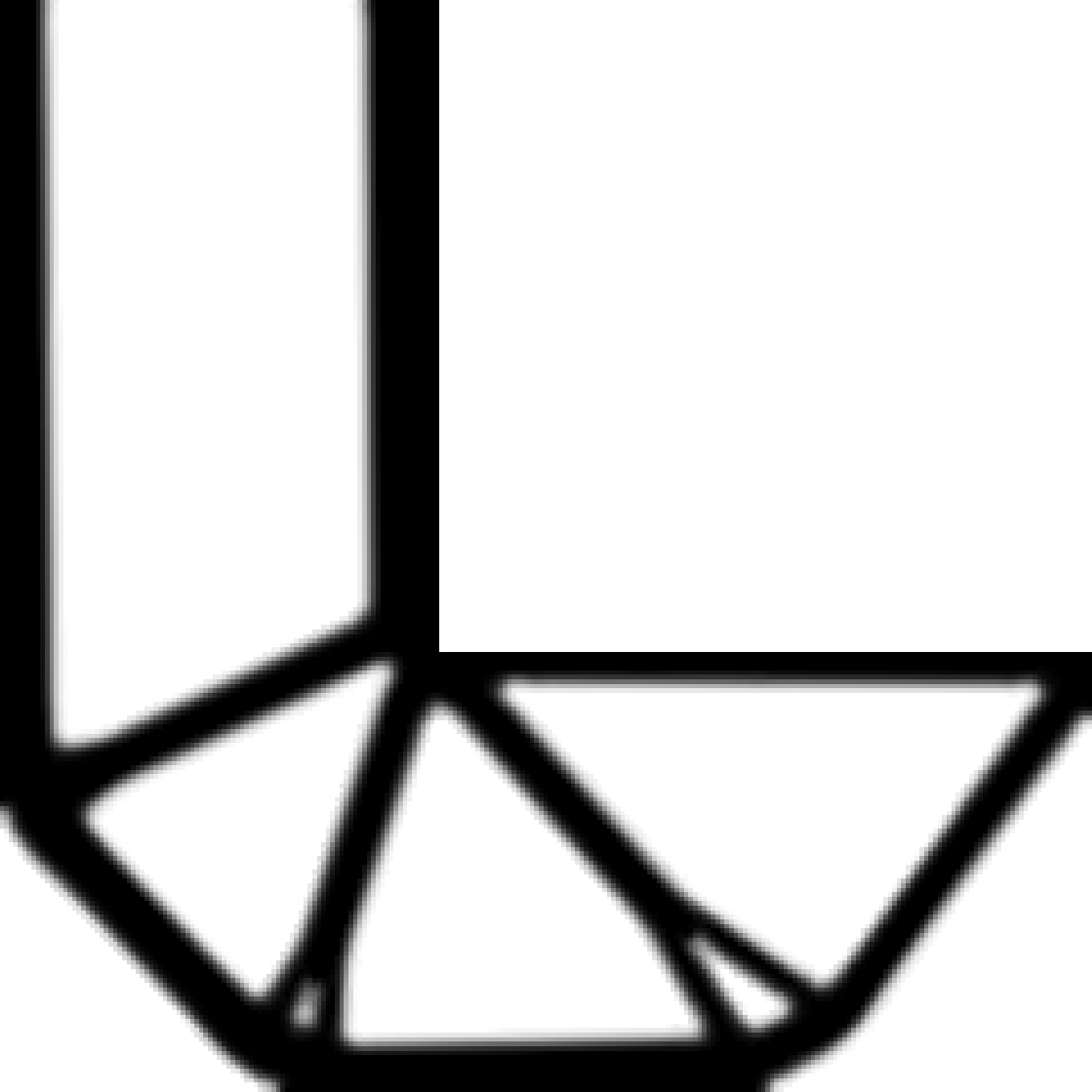} &
        \includegraphics[width=1.5cm,valign=m]{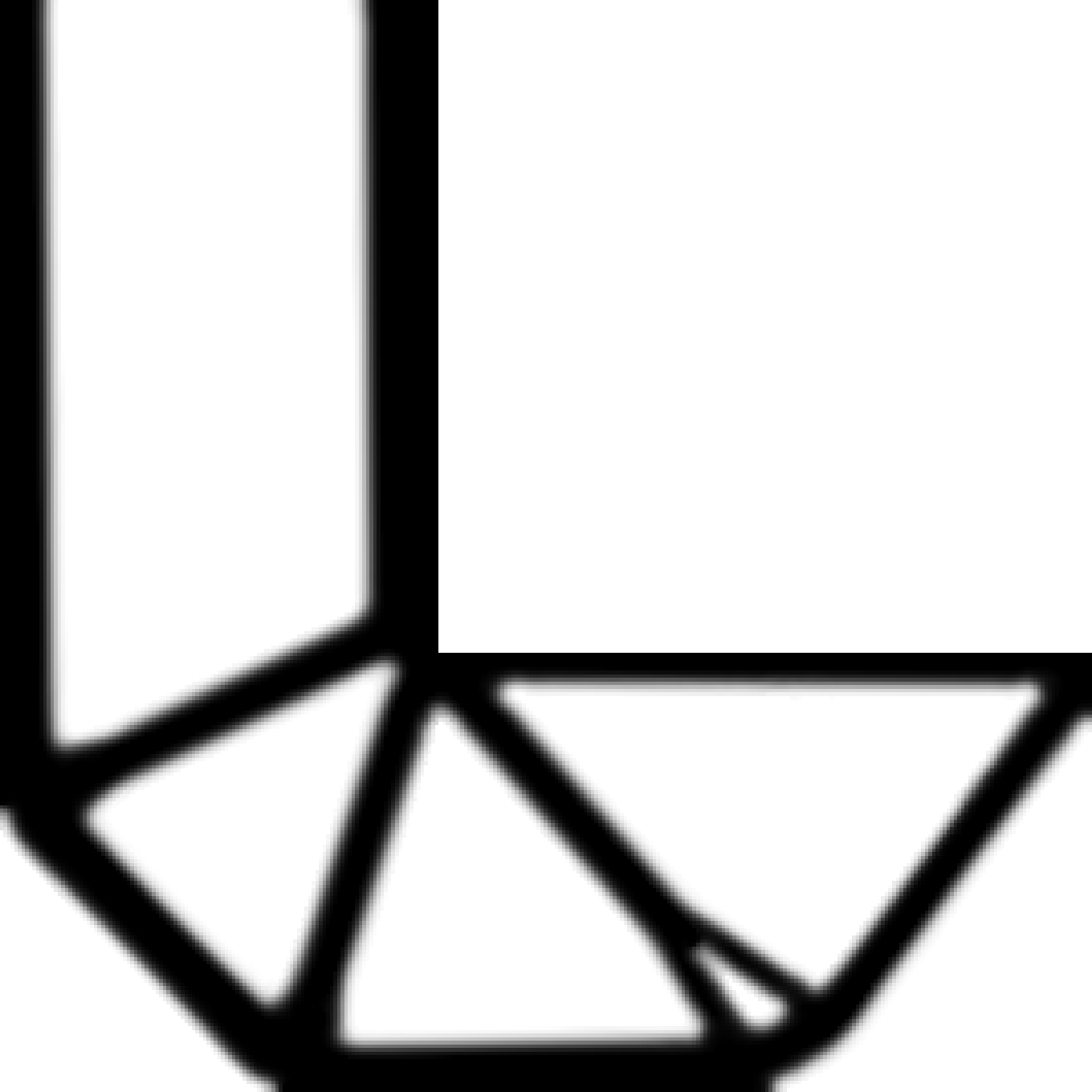} \\
        \midrule
        \multirow{2}{*}[-11.5pt]{0.4} & SIMP &
        \includegraphics[width=1.5cm,valign=m]{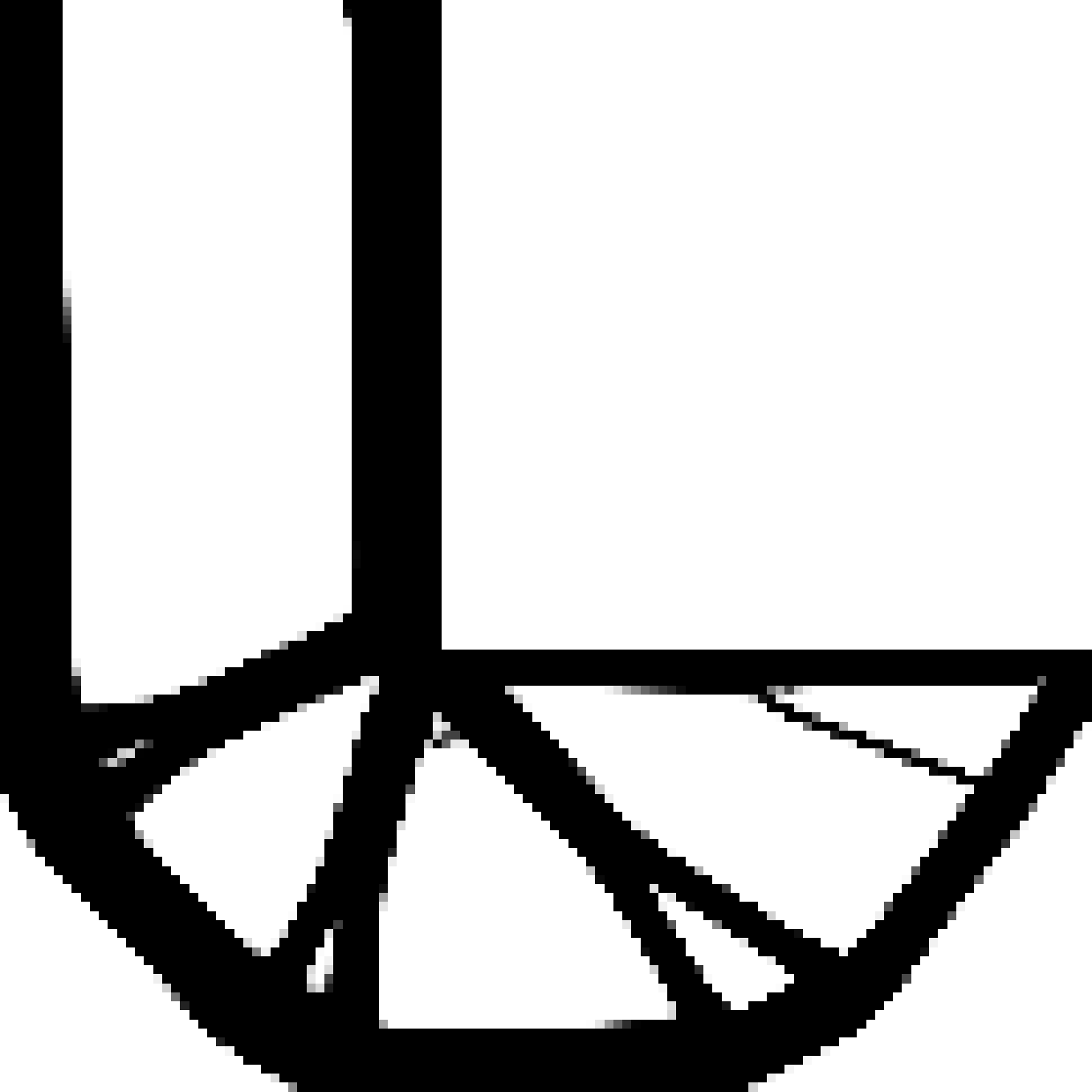} &
        \includegraphics[width=1.5cm,valign=m]{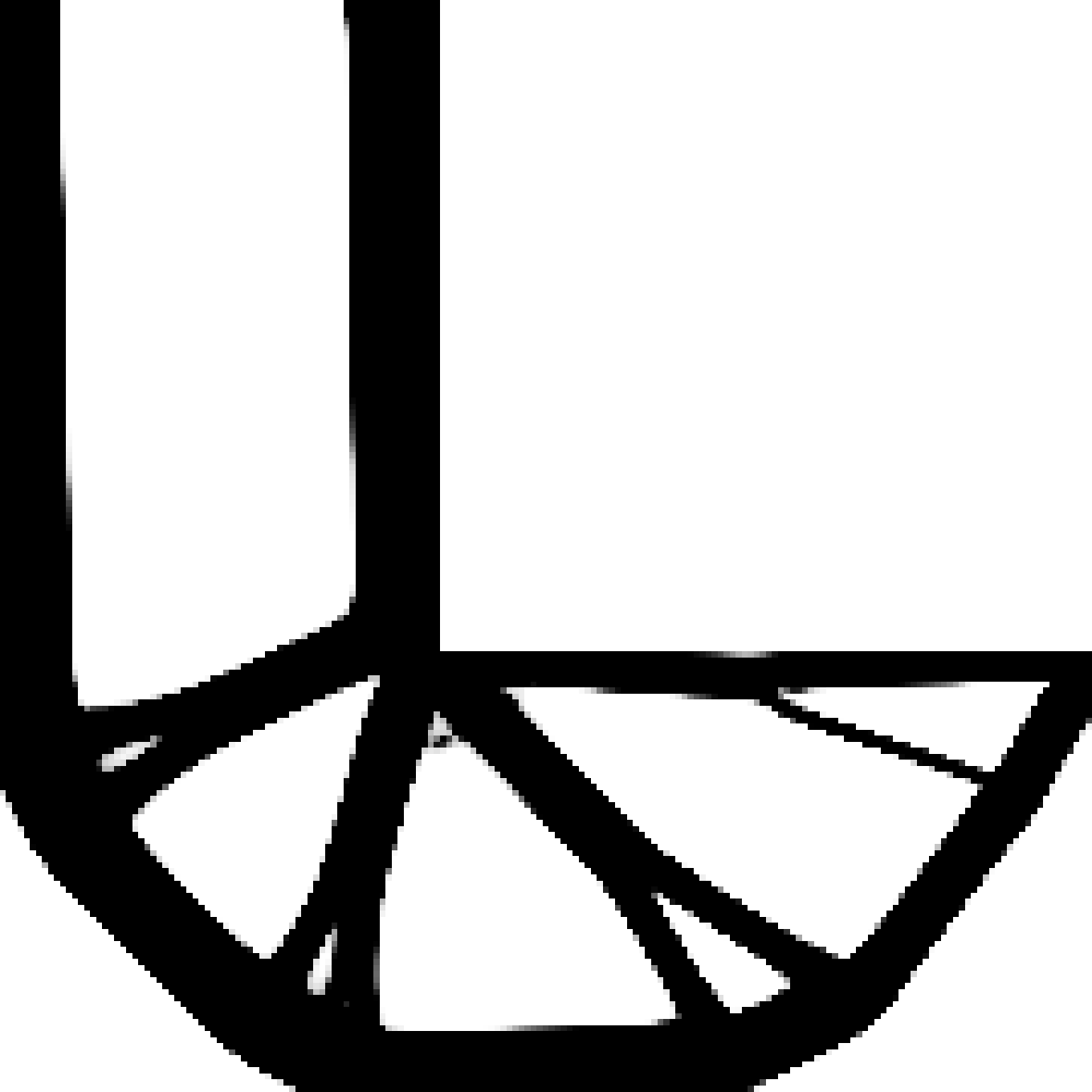} &
        \includegraphics[width=1.5cm,valign=m]{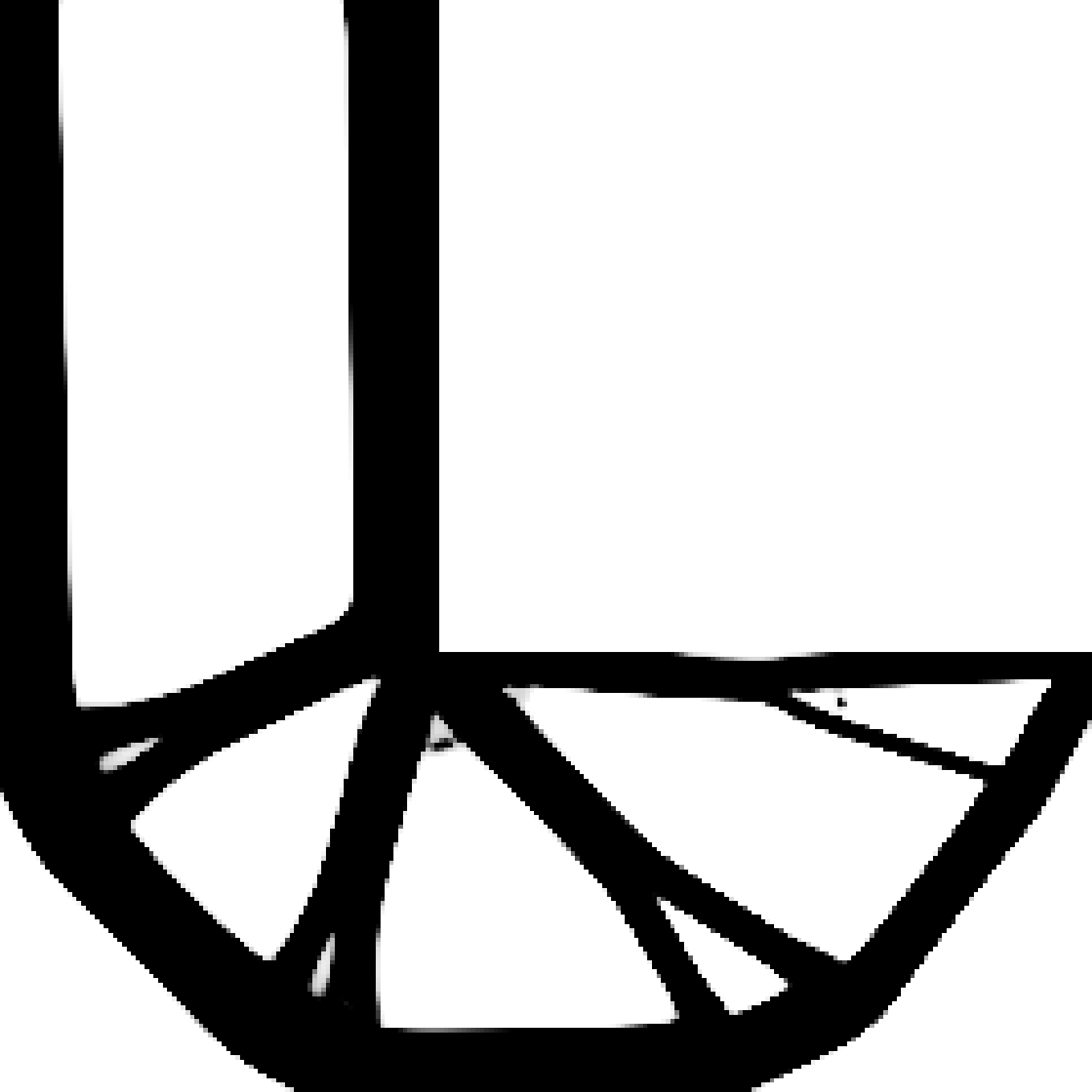} &
        \includegraphics[width=1.5cm,valign=m]{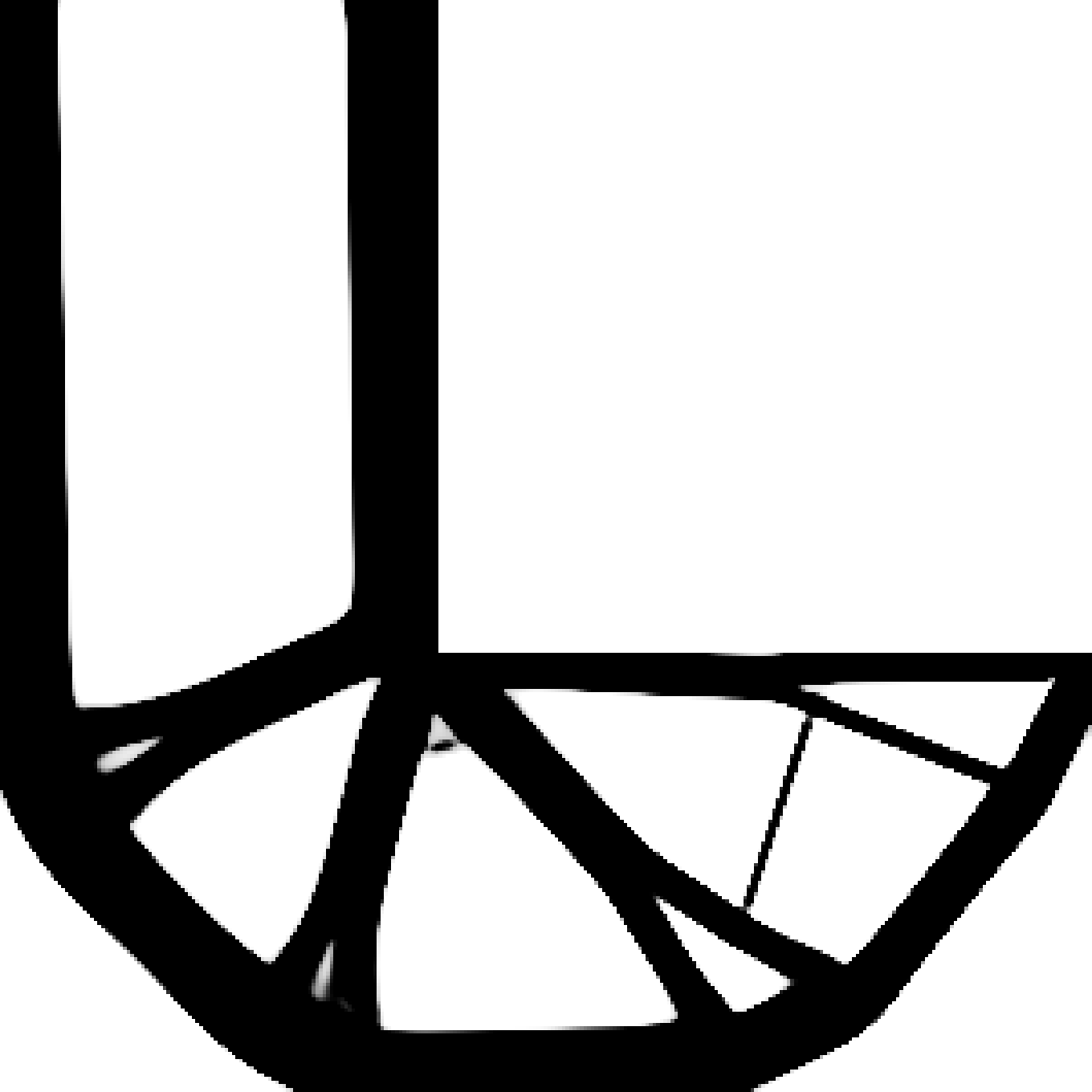} \\ \addlinespace[12pt]
                             & AC-ALM &
        \includegraphics[width=1.5cm,valign=m]{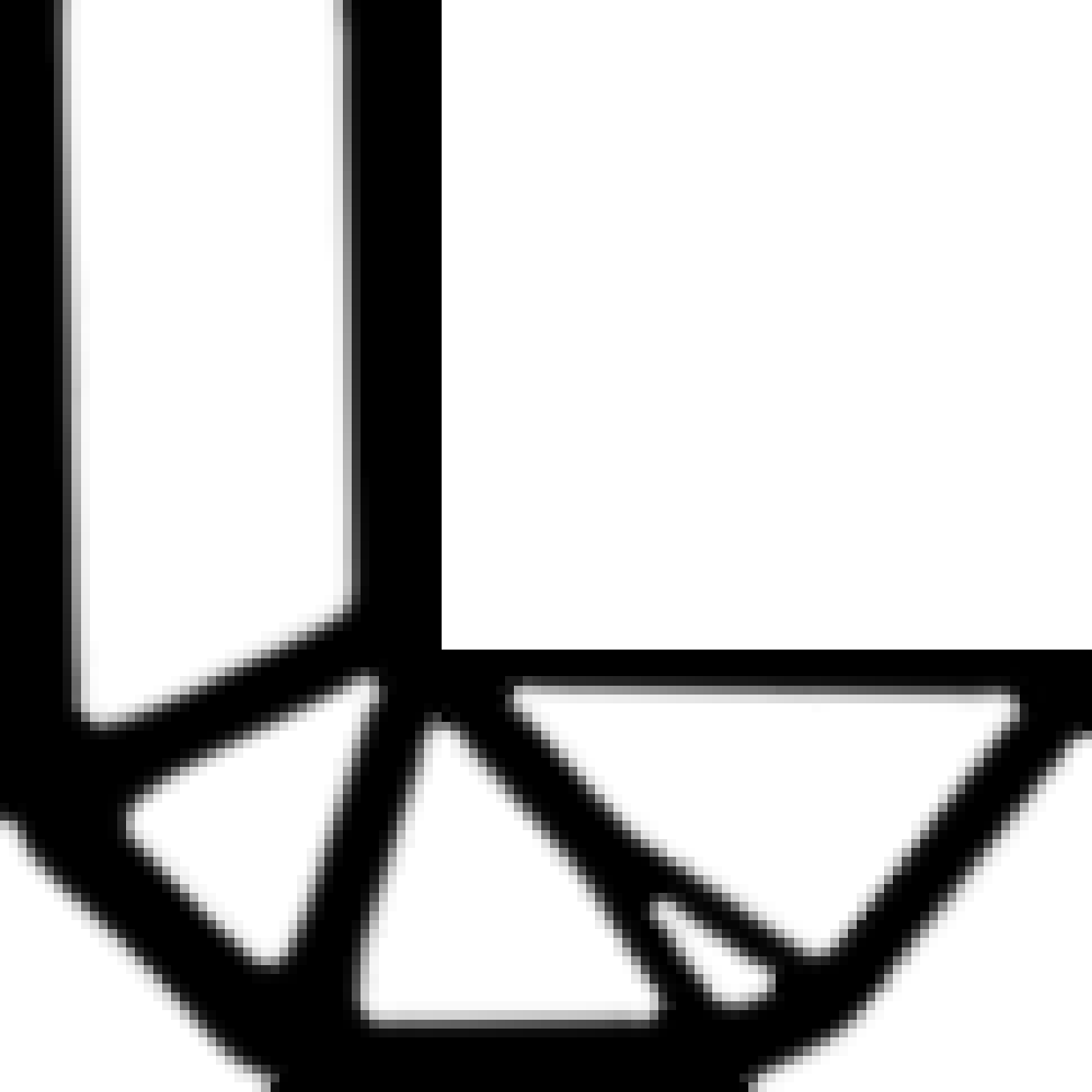} &
        \includegraphics[width=1.5cm,valign=m]{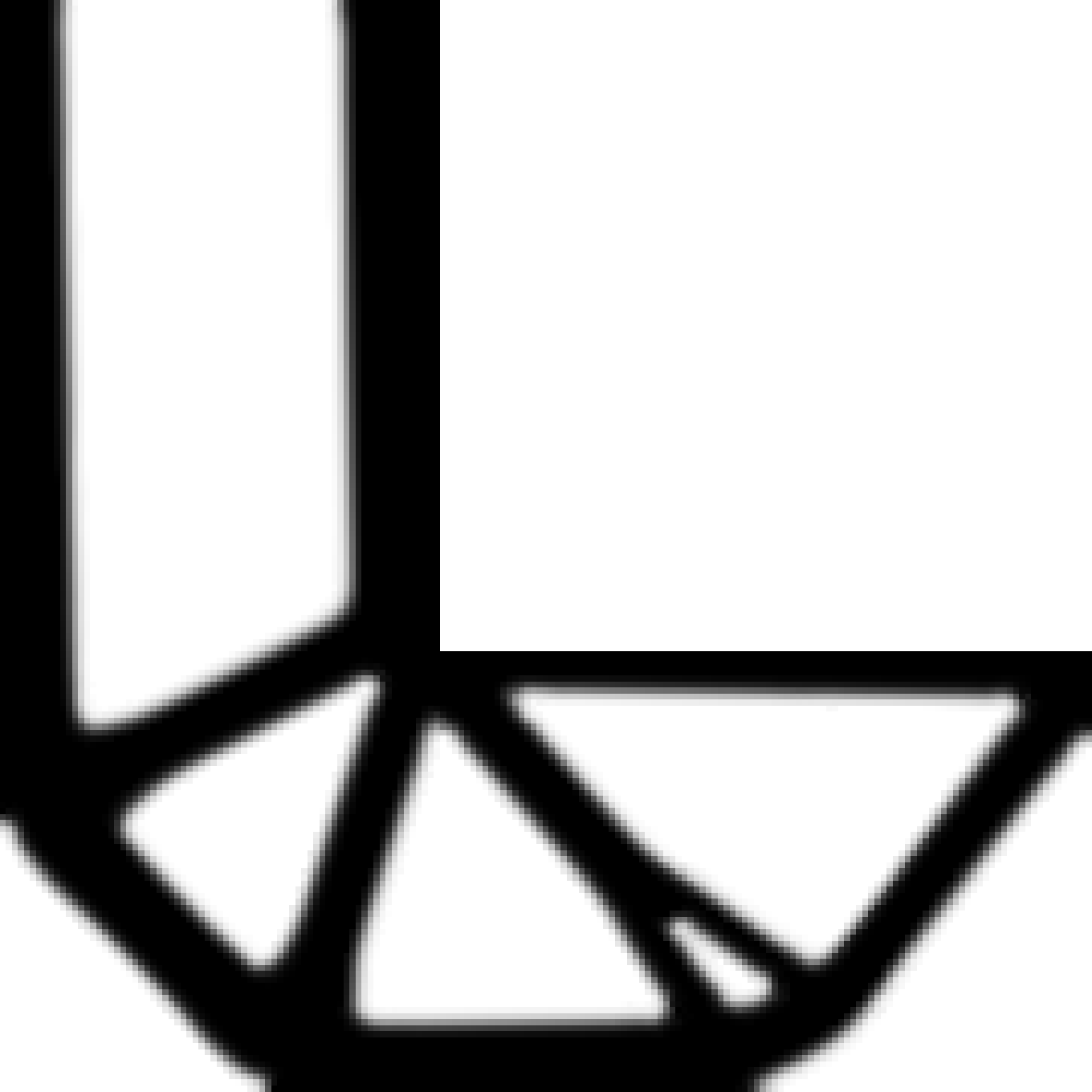} &
        \includegraphics[width=1.5cm,valign=m]{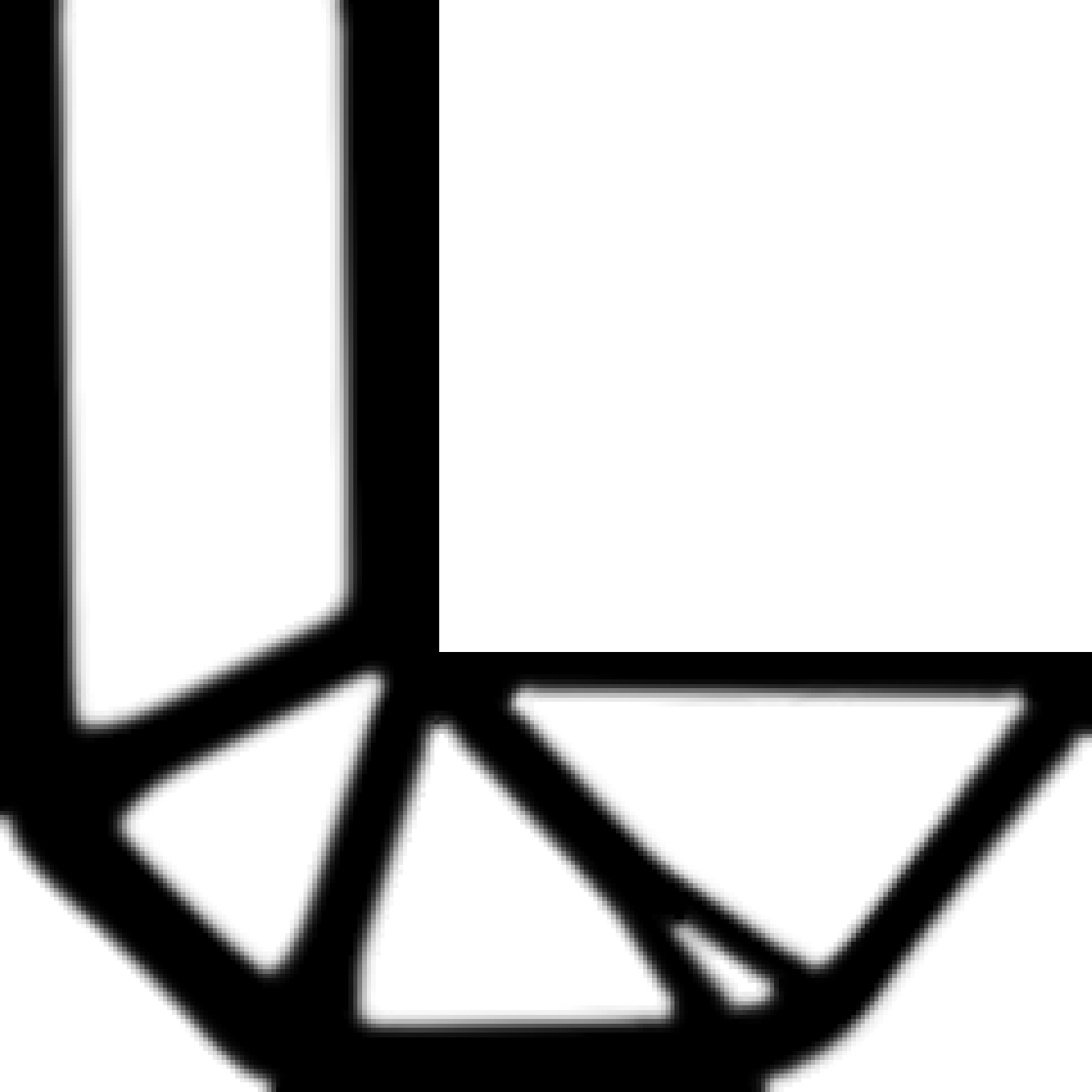} &
        \includegraphics[width=1.5cm,valign=m]{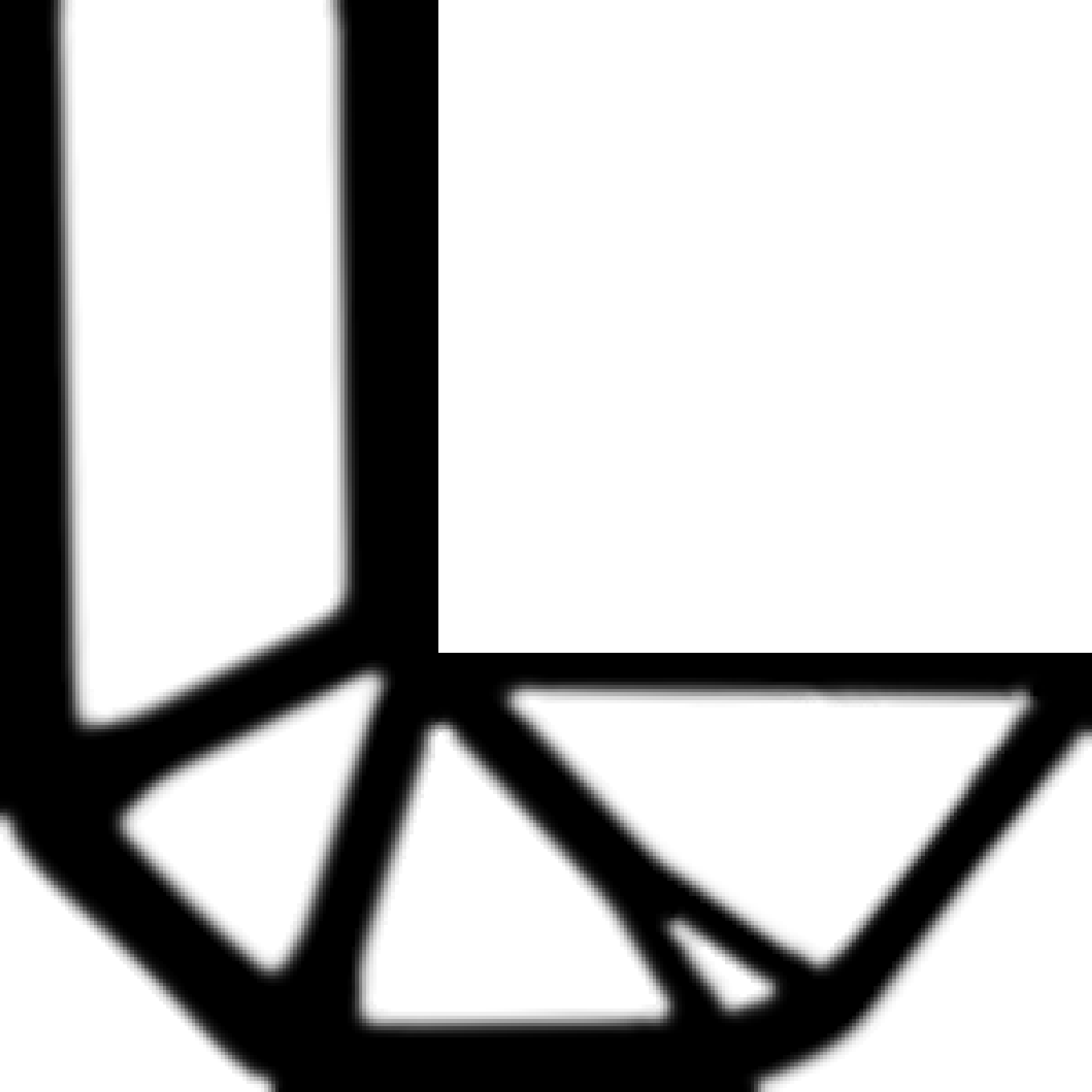} \\
        \bottomrule
    \end{tabular}
\end{table}

\begin{table}[!htbp]
    \centering
    \small
    \setlength{\tabcolsep}{2pt}
    \caption{Structure comparison of L-beams ($r_{\min}=0.1$)} 
    \label{tab:lbeam-rmin01}
    \begin{tabular}{cccccc}
        \toprule
        $V$ & Method & $120\times 120$ & $180\times 180$ & $240\times 240$ & $300\times 300$ \\
        \midrule
        \multirow{2}{*}[-11.5pt]{0.3} & SIMP &
        \includegraphics[width=1.5cm,valign=m]{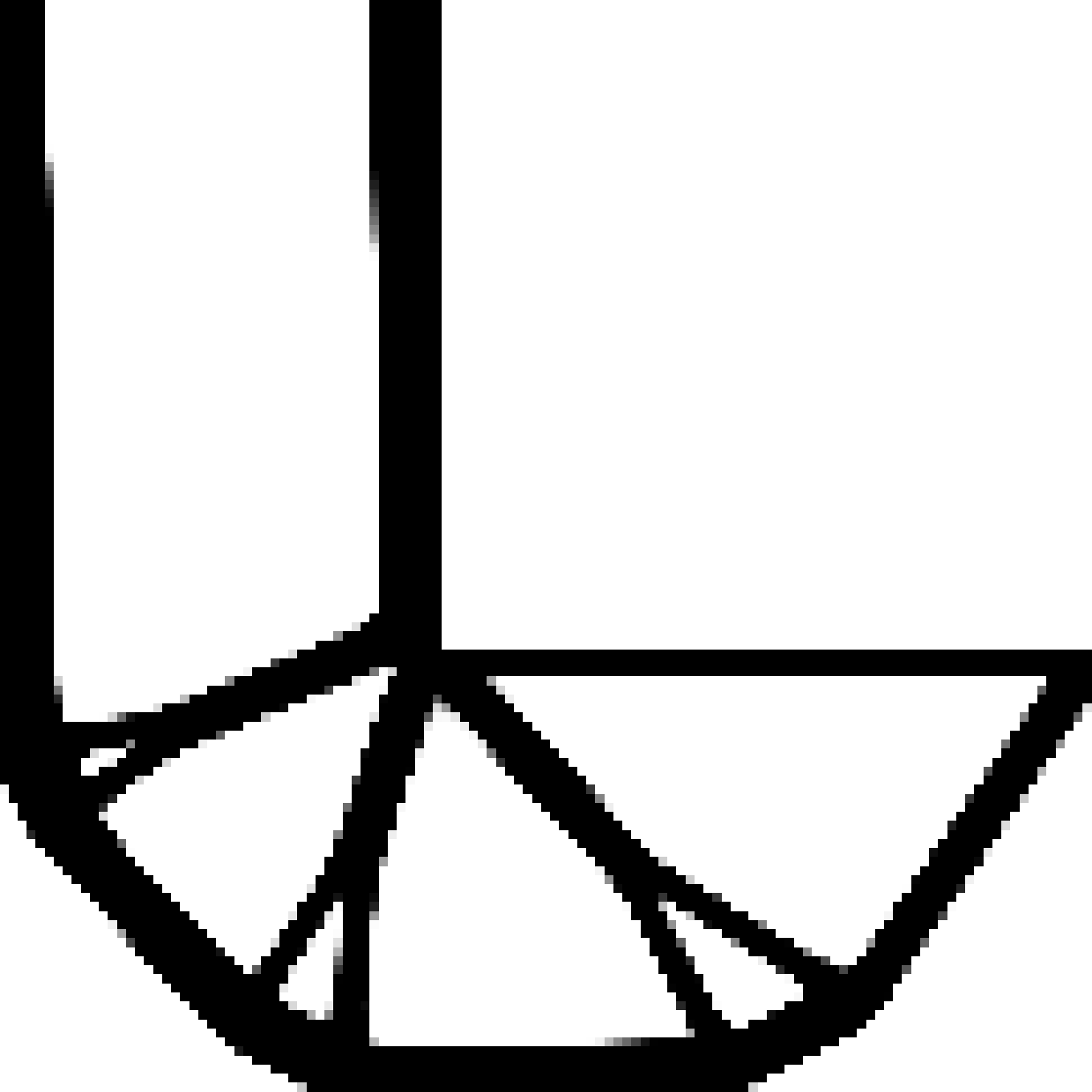} &
        \includegraphics[width=1.5cm,valign=m]{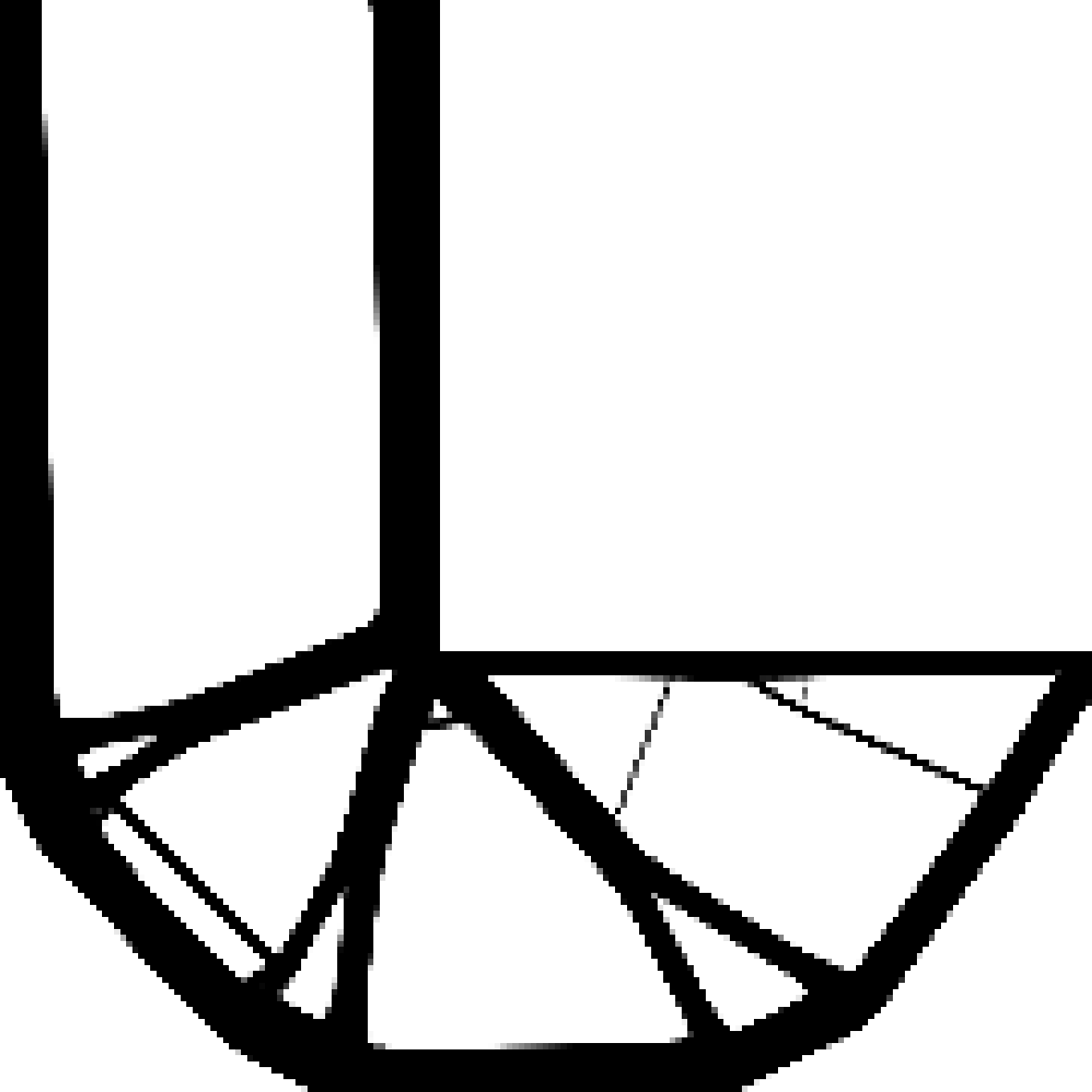} &
        \includegraphics[width=1.5cm,valign=m]{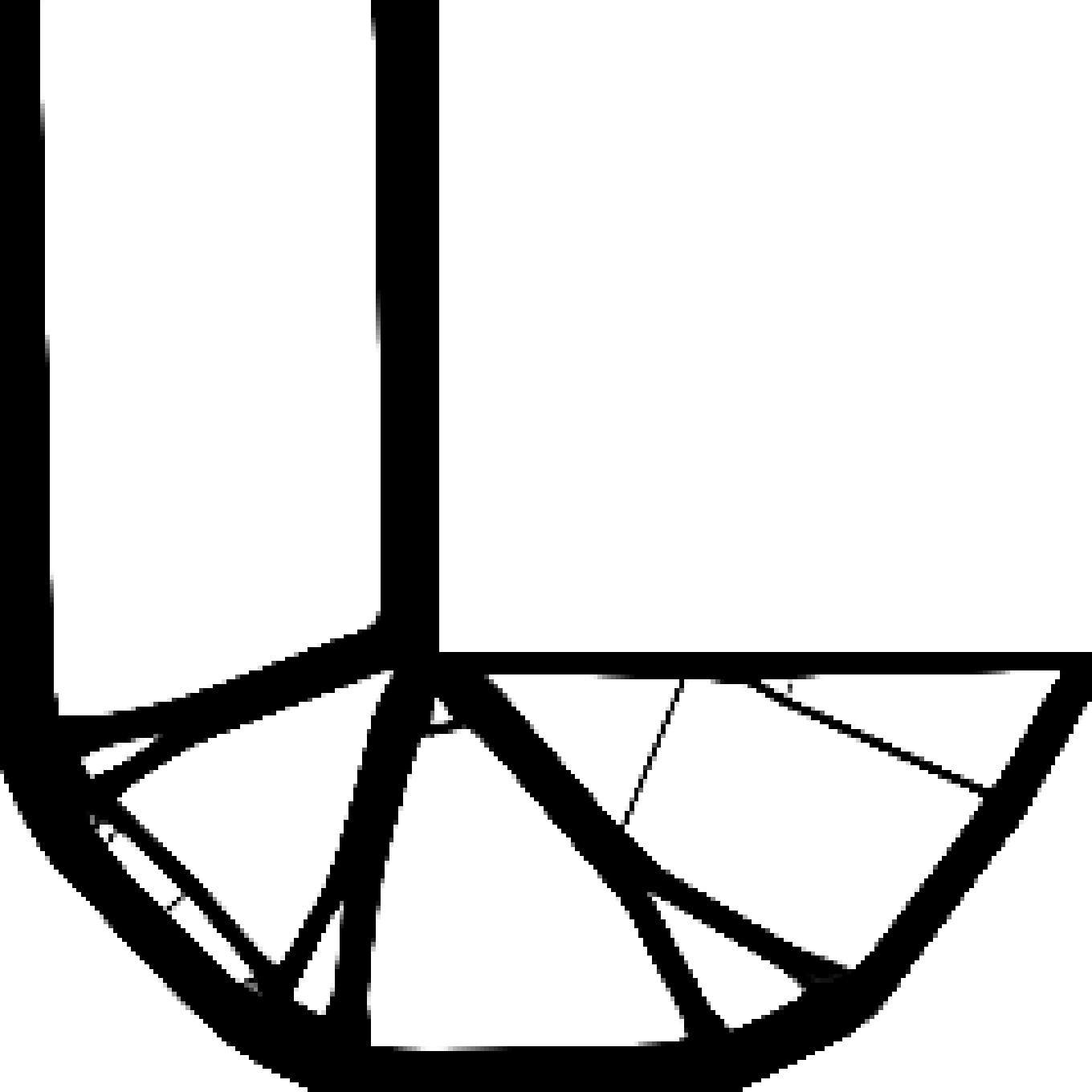} &
        \includegraphics[width=1.5cm,valign=m]{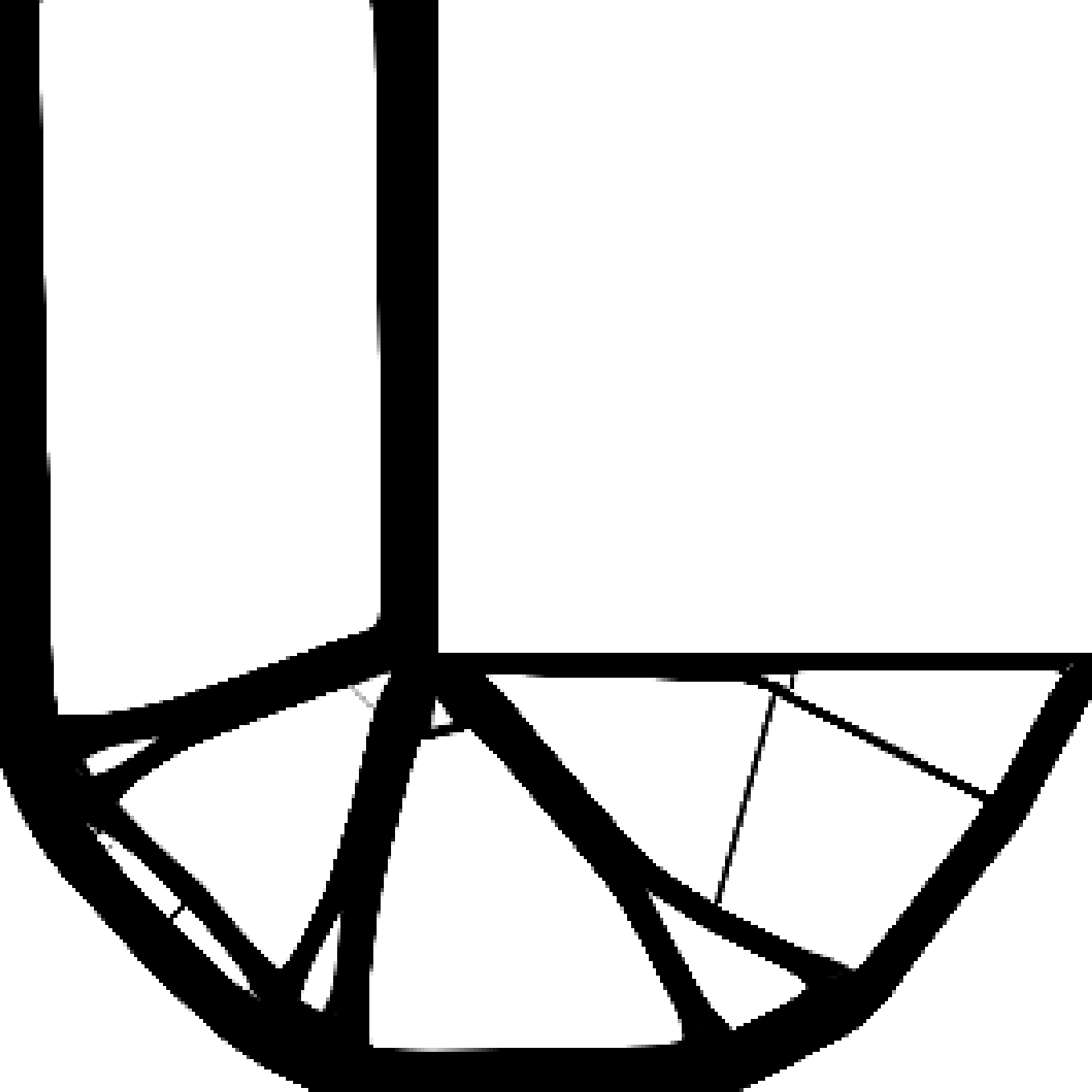} \\ \addlinespace[12pt]
                             & AC-ALM &
        \includegraphics[width=1.5cm,valign=m]{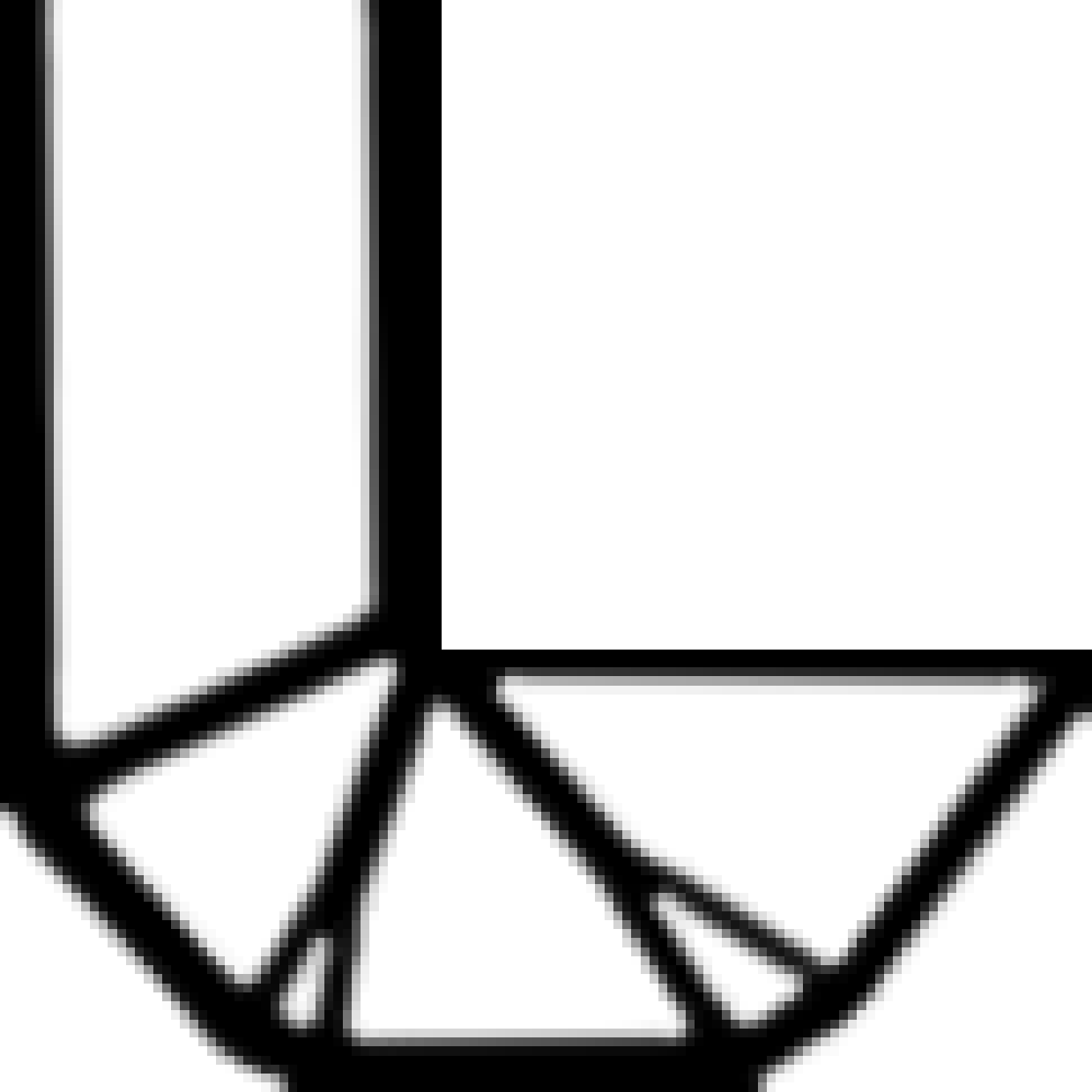} &
        \includegraphics[width=1.5cm,valign=m]{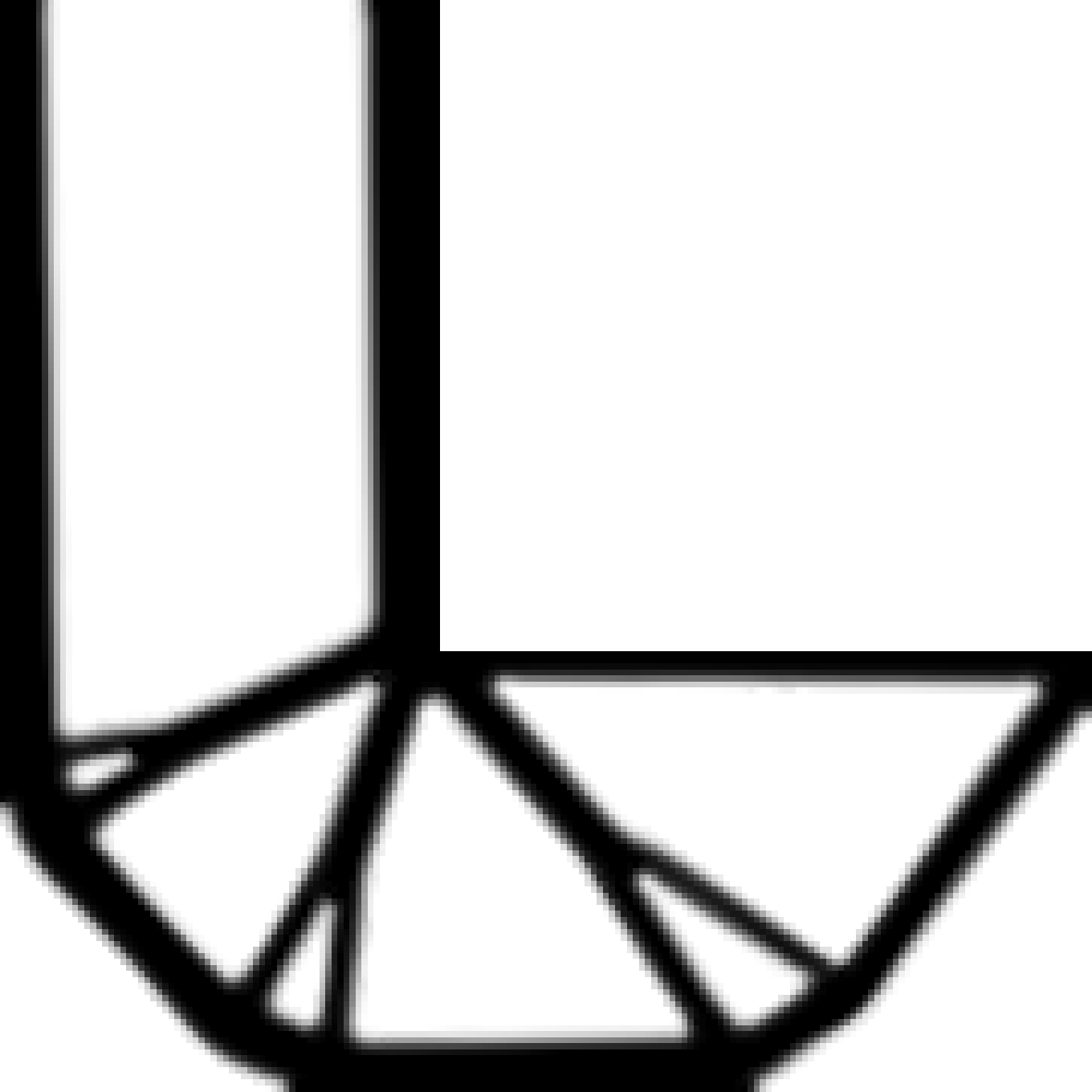} &
        \includegraphics[width=1.5cm,valign=m]{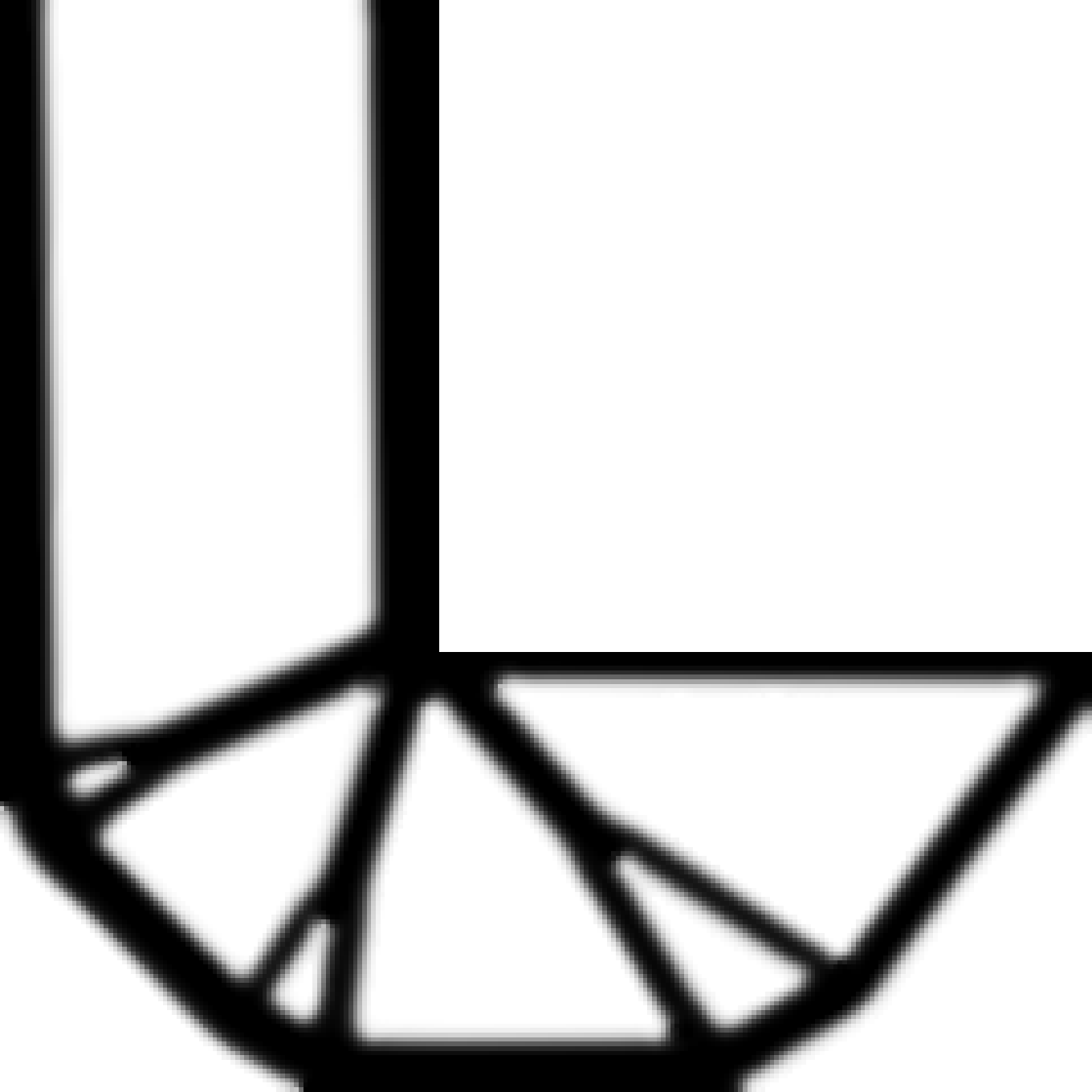} &
        \includegraphics[width=1.5cm,valign=m]{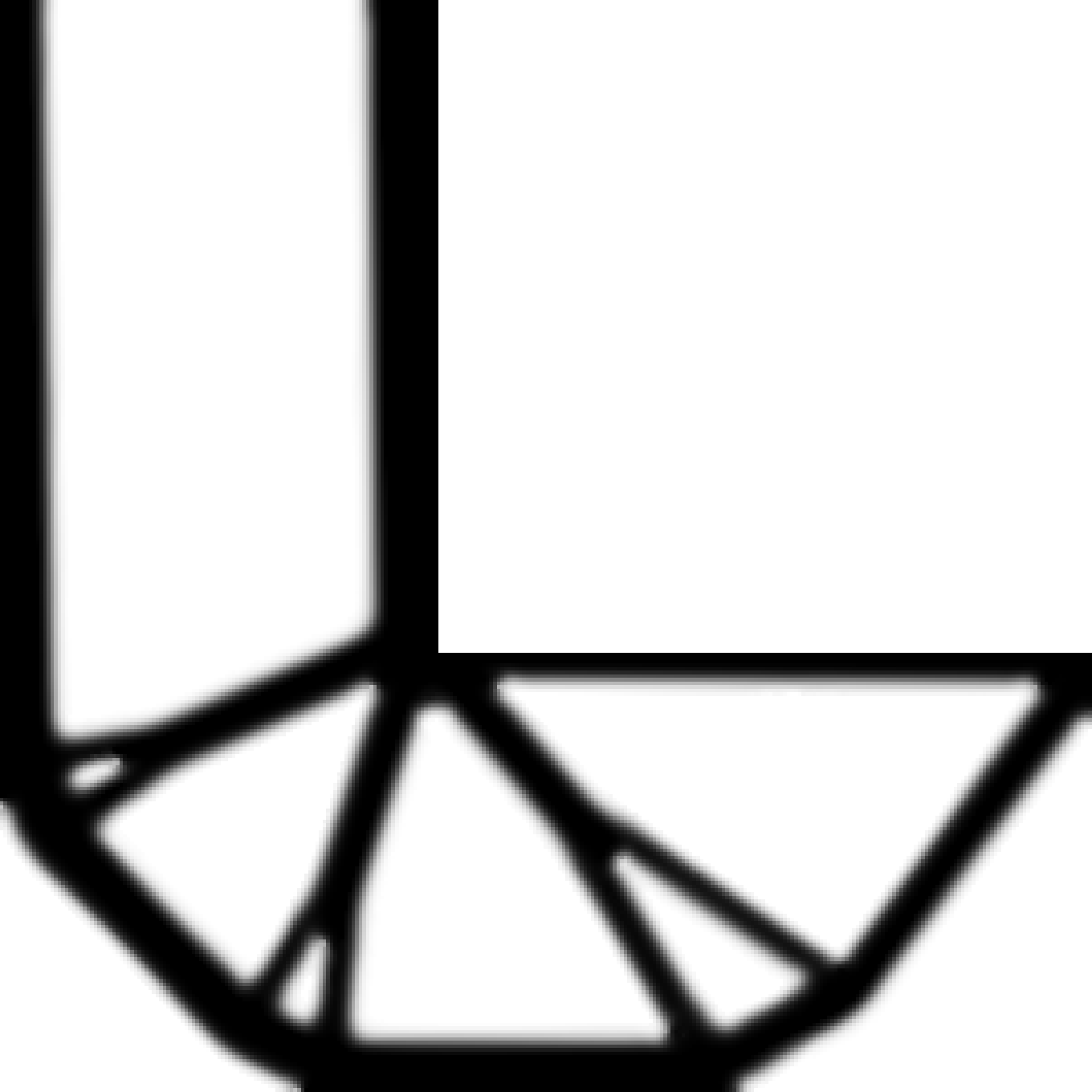} \\
        \midrule
        \multirow{2}{*}[-11.5pt]{0.4} & SIMP &
        \includegraphics[width=1.5cm,valign=m]{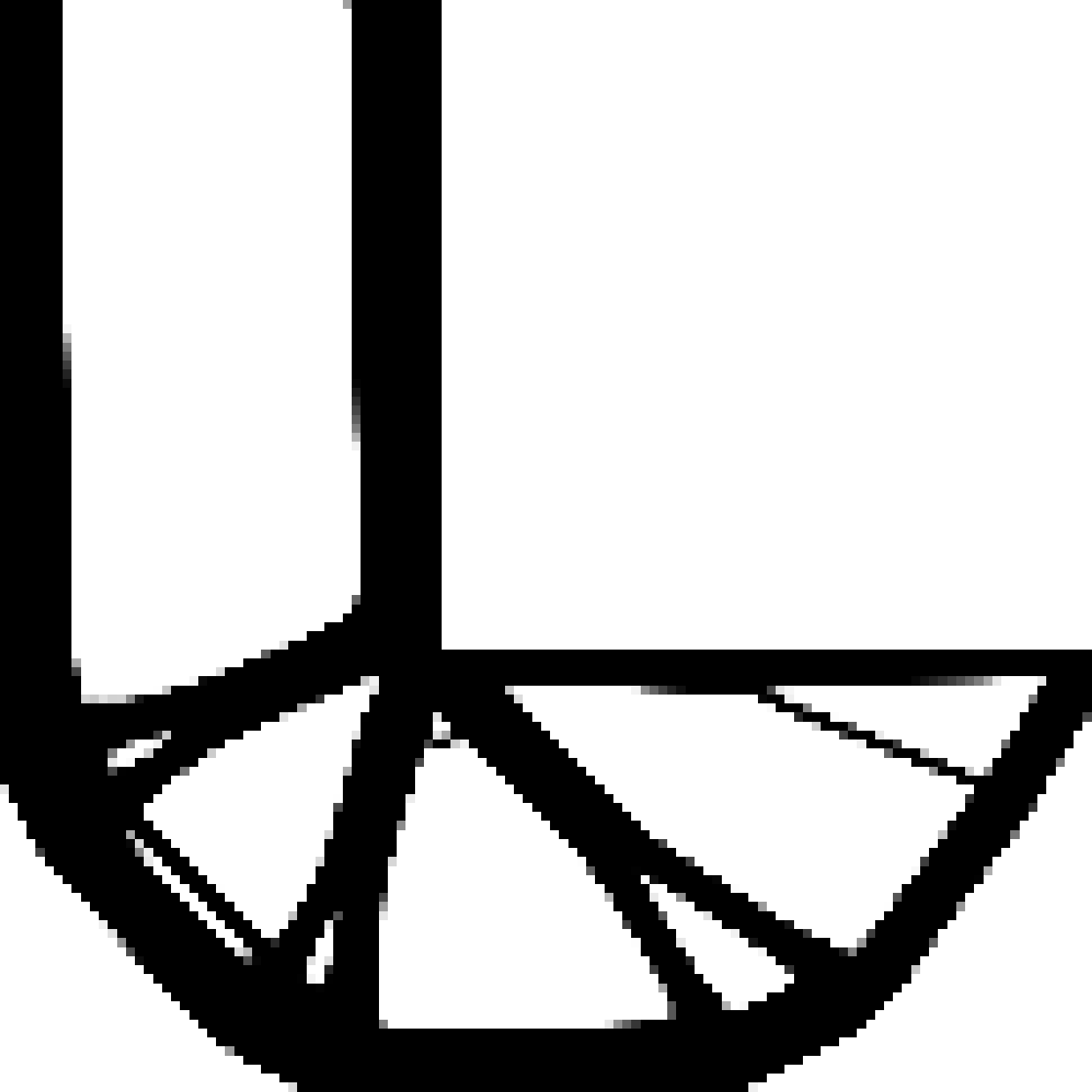} &
        \includegraphics[width=1.5cm,valign=m]{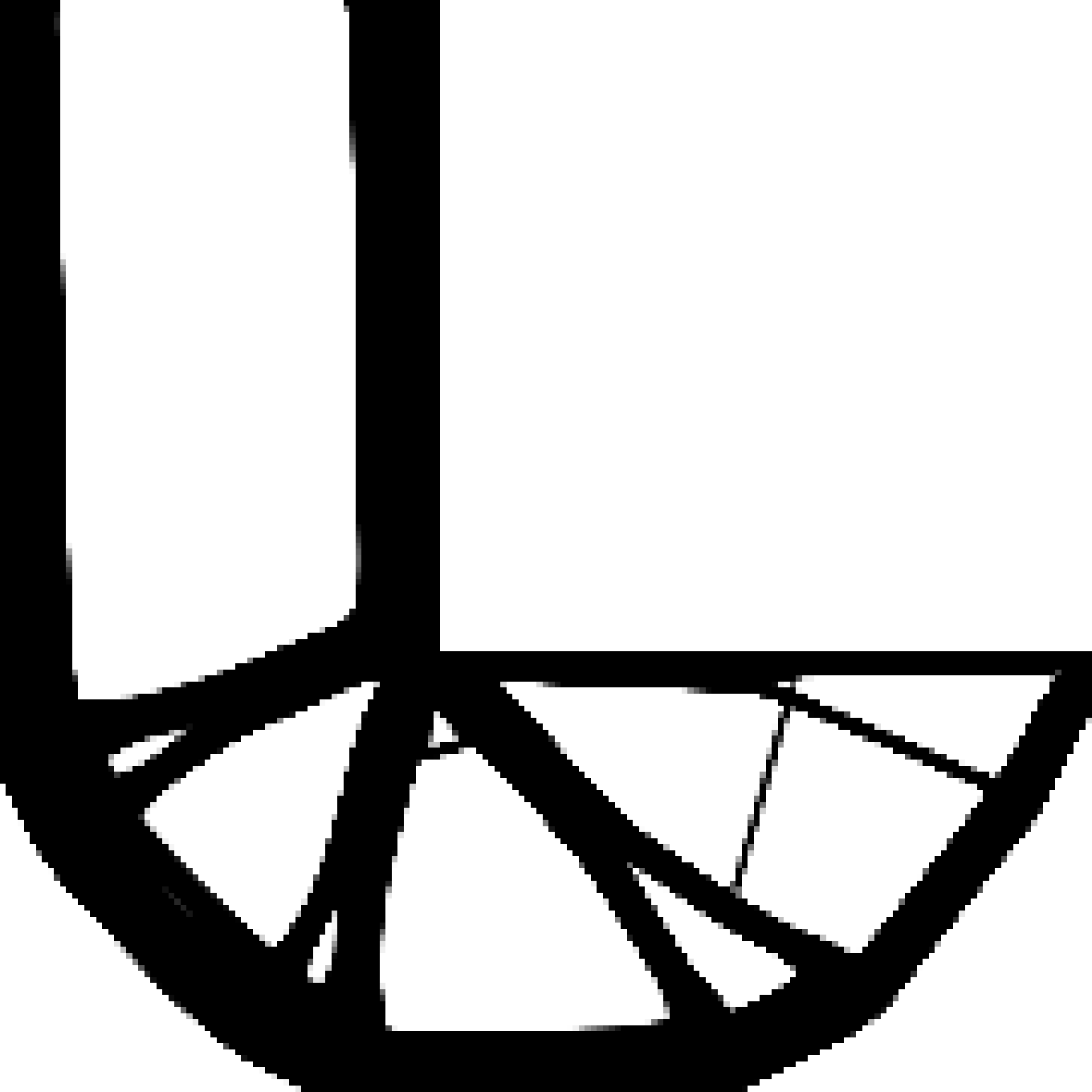} &
        \includegraphics[width=1.5cm,valign=m]{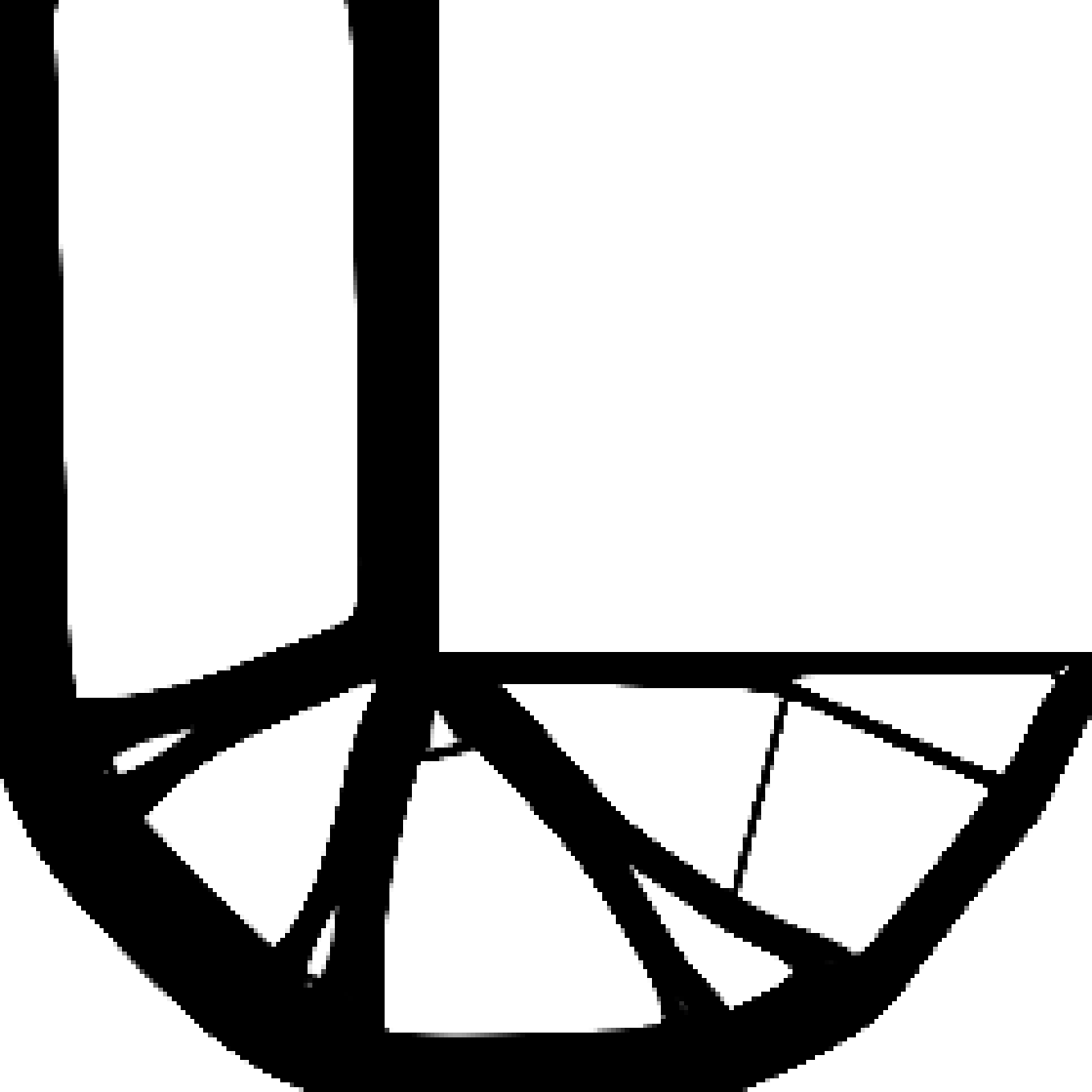} &
        \includegraphics[width=1.5cm,valign=m]{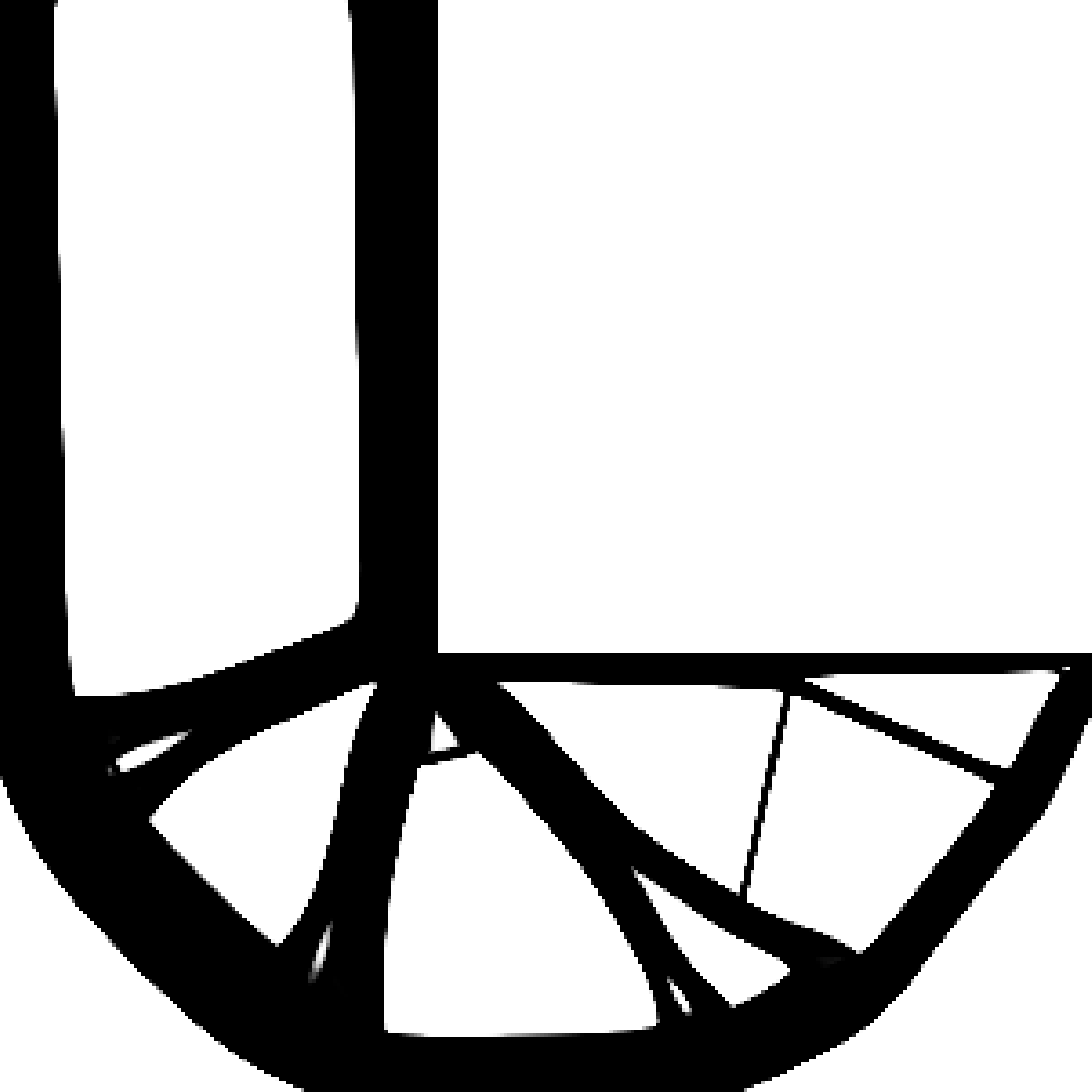} \\ \addlinespace[12pt]
                             & AC-ALM &
        \includegraphics[width=1.5cm,valign=m]{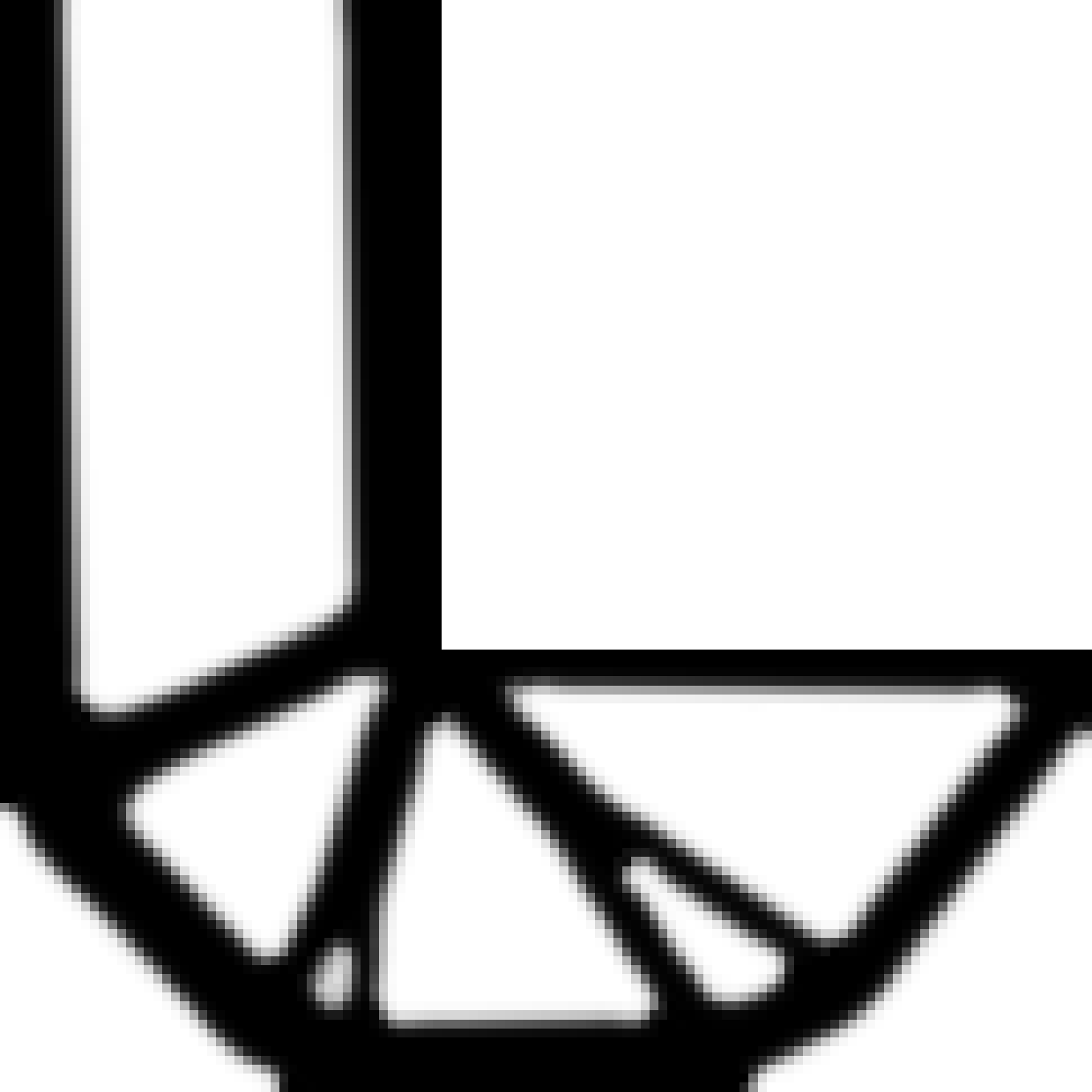} &
        \includegraphics[width=1.5cm,valign=m]{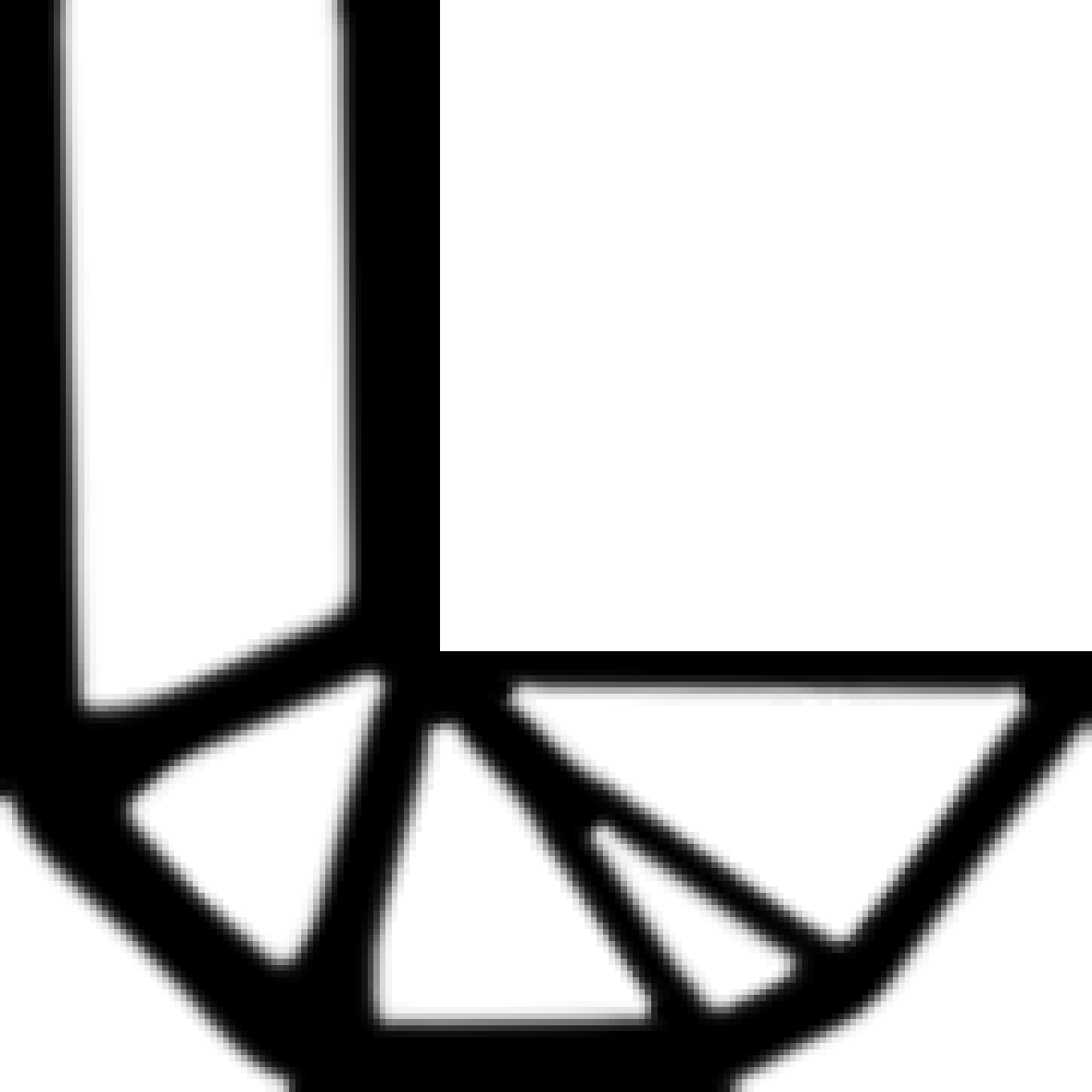} &
        \includegraphics[width=1.5cm,valign=m]{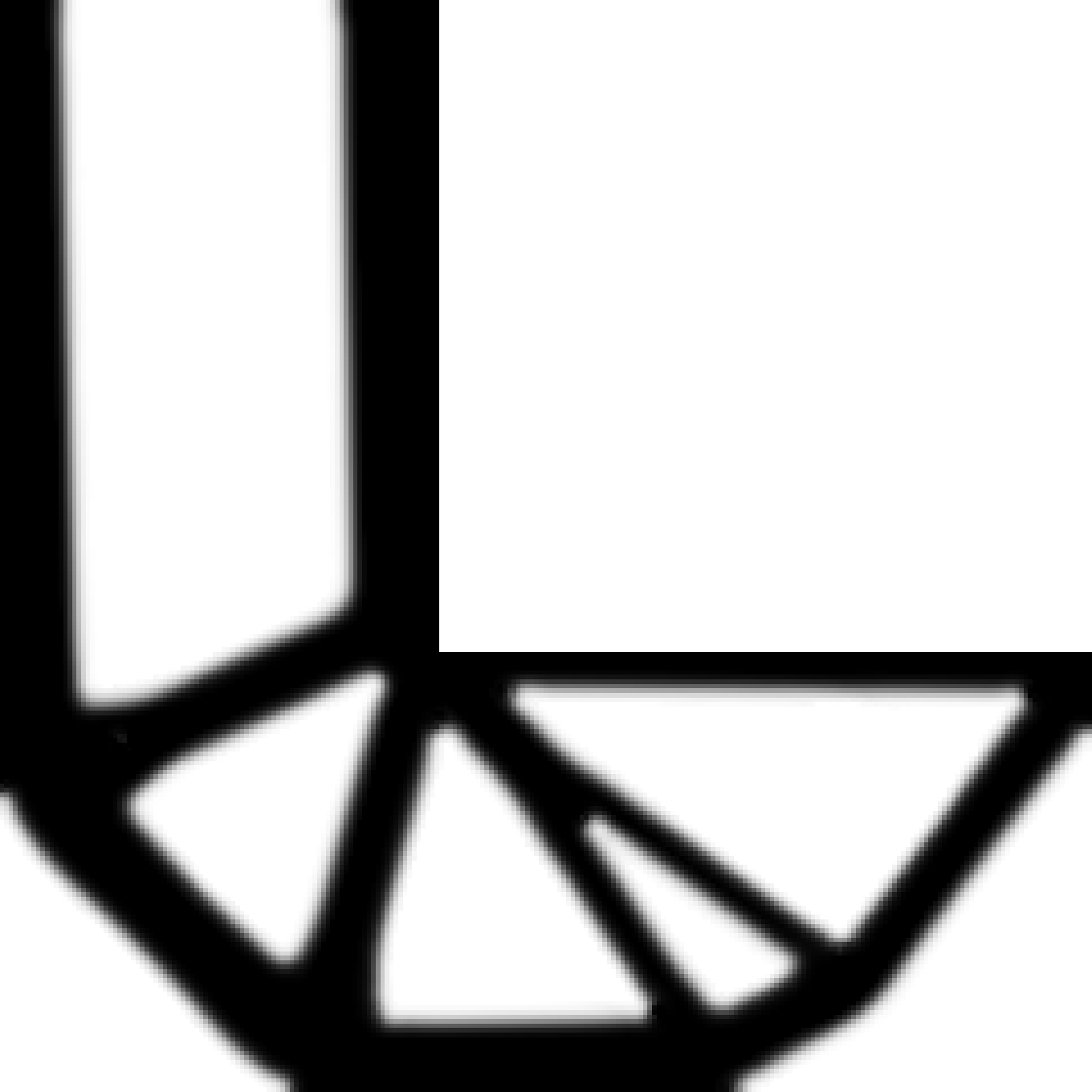} &
        \includegraphics[width=1.5cm,valign=m]{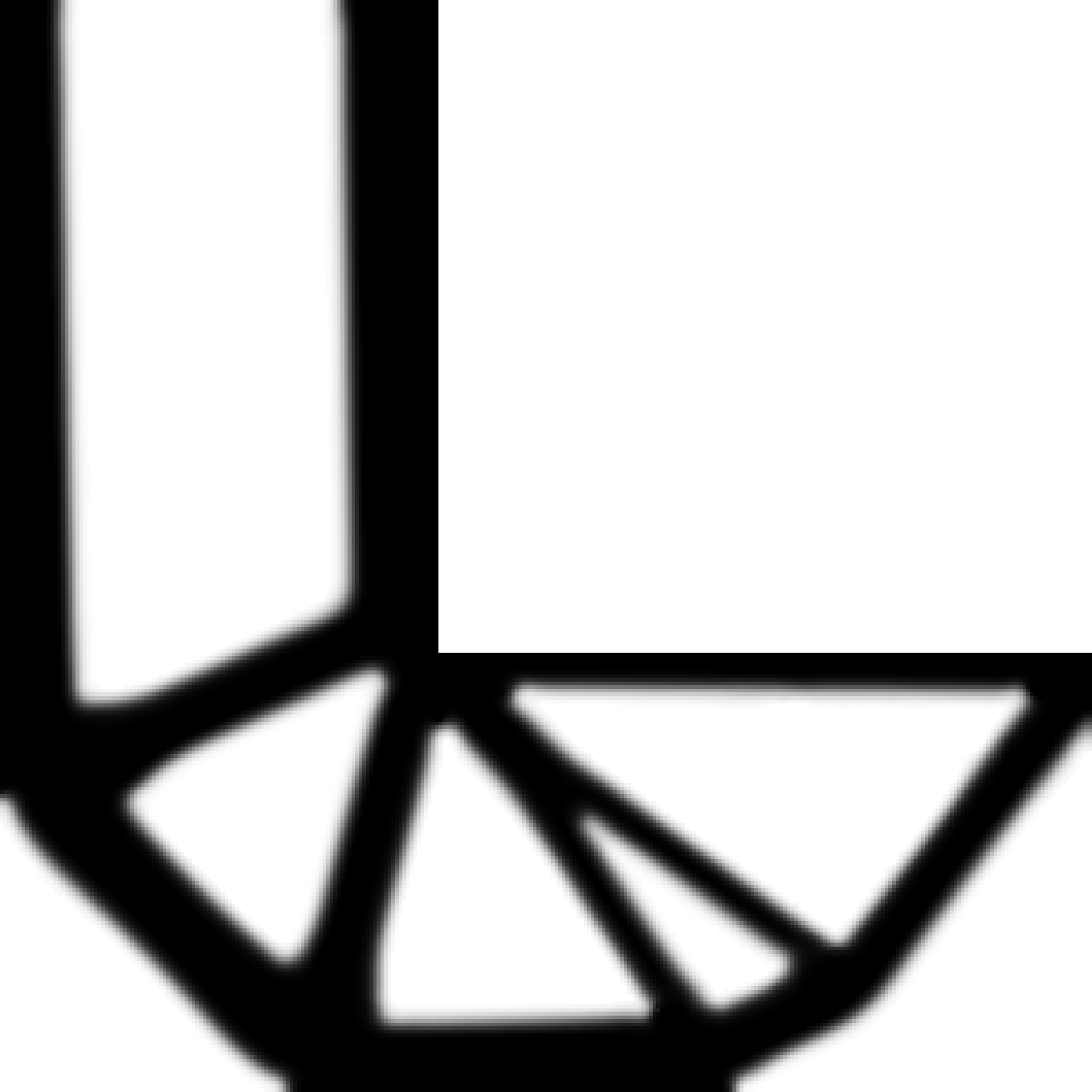} \\
        \bottomrule
    \end{tabular}
\end{table}

\subsubsection{Ablation Study: Necessity of the Lagrange Multiplier}\label{subsec:ablation_multiplier}

Having established benchmark compliance and mesh-independent regularity across three 
canonical geometries, we utilize the reference MBB beam configuration 
($180\times 120$, $V=0.3$, $r_{\min}=0.1$) as a standardized testbed. 
This isolates a key algorithmic question: is the first-order Augmented Lagrangian 
multiplier $\boldsymbol{\lambda}_\text{h}$ strictly necessary, or could the AC constraint 
be enforced via a simpler penalty-only formulation?

\begin{figure}[ht]
\centering
\includegraphics{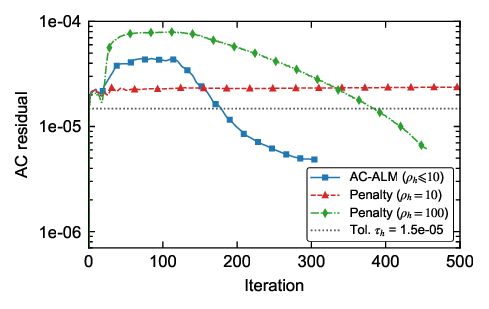}
\caption{Residual $\|\mathbf{h}\|_\text{norm}$ history: AC-ALM vs.\ penalty-only
(MBB, $180\times 120$). The dashed line indicates the target tolerance
$\tau_h \approx 1.5\times 10^{-5}$ evaluated from Table~\ref{tab:algorithm_parameters}
($\tau_h = 5\times 10^{-5}(\Delta x)^3$ at $\Delta x = 2/3$).}\label{fig:mbb_penalty_conv}
\end{figure}

To address this question, we compare the reference AC-ALM run against penalty-only variants 
with fixed $\rho_\text{h} = 10$ (matching the ALM upper bound $\rho_\text{h}^{\max}$) 
and an escalated penalty of $\rho_\text{h} = 100$. 
As shown in Fig.~\ref{fig:mbb_penalty_conv} and Table~\ref{tab:penalty_ablation}, 
the penalty-only formulation at $\rho_\text{h} = 10$ stagnates at 
$\|\mathbf{h}\|_\text{norm} \approx 2.40\times 10^{-5}$ ($1.62\,\tau_h$). 
It fails to achieve feasibility within the target tolerance $\tau_h \approx 1.48\times 10^{-5}$. 
This stagnation occurs because the penalty gradient $\rho_\text{h}\mat{H}^{\top}\mathbf{h}$ 
balances the objective gradient at a finite non-zero residual. 

In contrast, the accumulated multiplier $\boldsymbol{\lambda}_\text{h}$ supplies a persistent 
first-order correction, driving the residual norm down to $0.33\,\tau_h$ at a modest penalty of 
$\rho_\text{h} = 1.9$. 
Forcing feasibility with a pure penalty requires escalating $\rho_\text{h}$ to $100$ 
(Table~\ref{tab:penalty_ablation}). 
However, this severely inflates early residuals (Fig.~\ref{fig:mbb_penalty_conv}) and degrades 
subproblem conditioning, requiring $50\%$ more iterations and over four times the compute time. 
The ALM scheme thus achieves strict feasibility rapidly at well-conditioned penalty levels. 
Furthermore, across all benchmark configurations in this study, 
the maximum multiplier magnitude remained below $0.35$ 
($\max_i |\lambda_{h,i}| \leqslant 0.35 \ll \lambda_{\max} = 10$). 
Because the multiplier updates inherit the exact algebraic zero-mean property of the residual 
($\mathbf{m}^\top \mat{M}^{-1}\mathbf{h} = \mathbf{1}^\top \mathbf{h} \equiv 0$) 
and the safeguarding clamp remained inactive throughout, 
the multiplier gauge condition $\mathbf{m}^\top \boldsymbol{\lambda}_\text{h} \equiv 0$ 
was preserved to machine precision ($< 10^{-15}$) without requiring corrective projection.

\begin{table}[ht]
\centering
\footnotesize
\setlength{\tabcolsep}{3.5pt}
\caption{Effect of Lagrange multiplier on feasibility and convergence (MBB, $180\times 120$, $V=0.3$, $r_{\min}=0.1$).}\label{tab:penalty_ablation}
\begin{tabular}{lccccccr}
\toprule
\textbf{Method} & $\rho_\text{h}$ & $\mathcal{J}_\text{final}$ & $\|\mathbf{h}\|_{\text{norm}}$ & $\|\mathbf{h}\|_\text{norm}/\tau_h$ & \textbf{Iter.} & \textbf{Time (s)} & \textbf{Feasible} \\
\midrule
AC-ALM  & $\leqslant 10$ (final 1.9) & 80.64 & $4.85\times10^{-6}$ & 0.33 & 305 & 60.1 & Yes \\
Penalty & 10                         & 76.11 & $2.40\times10^{-5}$ & 1.60 & 500 & 367.8 & No  \\
Penalty & 100                        & 80.17 & $6.12\times10^{-6}$ & 0.41 & 455 & 268.8 & Yes \\
\bottomrule
\end{tabular}
\end{table}

\subsubsection{Quantitative Compliance and Computational Cost Comparison}

Table~\ref{tab:compliance_gap} consolidates the numerical performance
across all 48 two-dimensional benchmark pairs, reporting the final compliance,
relative compliance gap
$(\mathcal{J}_{\text{AC-ALM}} - \mathcal{J}_{\text{SIMP}})/\mathcal{J}_{\text{SIMP}}$, 
iteration count, and elapsed wall-clock time. 
For the MBB beam, the compliance gap ranges from $1.35\%$ to $6.54\%$ (mean $3.19\%$). 
For the cantilever beam, the gap spans $4.04\%$ to $12.30\%$ (mean $7.77\%$), 
and for the L-beam it ranges from $2.59\%$ to $9.60\%$ (mean $5.60\%$). 
In all three benchmarks, the gap peaks at $V = 0.3$ with $r_{\min} = 0.1$, 
where the unconstrained SIMP baseline is free to exploit arbitrarily thin,
mesh-dependent members. 
This quantifiable compliance difference represents the structural price paid
for enforcing diffuse-interface regularity and finite feature resolution. 
A comprehensive analysis and thresholding evaluation of this trade-off 
are presented in Section~\ref{discussion}.

In terms of computational effort, AC-ALM requires moderately more iterations than SIMP, 
with an average wall-clock time ratio of $1.37\times$. 
The overhead is modest because the AC residual operations
(sparse matrix-vector products and lumped solves) scale linearly with mesh size,
while the linear elasticity solve remains the primary bottleneck. 
Notably, on the finest meshes with $r_{\min} = 0.1$,
the relative computational time difference drops to between $-5\%$ and $+6\%$
in five of the six cases.
This occurs because the unregularized baseline itself suffers from slow convergence 
in the presence of fine-scale topological oscillations.
Across the complete set of 48 two-dimensional benchmark pairs, every case achieves residual feasibility 
prior to reaching $k_{\max}=500$, with a global peak ratio of $\|\mathbf{h}\|_\text{norm}/\tau_h = 0.67$. 
This ratio exhibits no systematic growth under mesh refinement 
(averaging $0.24$--$0.46$ across the meshes), 
demonstrating the mesh independence of the mass-weighted formulation. 
Numerical feasibility is certified strictly by the stopping criterion $\|\mathbf{h}\|_\text{norm} \leqslant \tau_h$ 
(corresponding to $\|\mathbf{h}\|_* \leqslant \delta$ via Eq.~\eqref{eq:delta_equiv}), 
while the peak junction metric $\|\mathbf{h}\|_{\infty,\text{norm}}$ enters exclusively
as a diagnostic for adaptive penalty updates.

\begingroup
\footnotesize
\setlength{\tabcolsep}{1.5pt}
\begin{longtable}{llrrrrrrrrr}
\caption{Compliance gap between SIMP and AC-ALM across 2D benchmarks (continuous designs). Relative gap: $(\mathcal{J}_\text{AC-ALM}-\mathcal{J}_\text{SIMP})/\mathcal{J}_\text{SIMP}\times 100\%$ (positive values indicate continuous compliance elevation in AC-ALM due to diffuse interfaces).}\label{tab:compliance_gap}\\
\toprule
\textbf{Problem} & \textbf{Mesh} & $\boldsymbol{V}$ & $\boldsymbol{r_{\min}}$ & $\boldsymbol{\mathcal{J}_\text{SIMP}}$ & $\boldsymbol{\mathcal{J}_\text{AC-ALM}}$ & \textbf{Gap (\%)} & \multicolumn{2}{c}{\textbf{Iter.}} & \multicolumn{2}{c}{\textbf{Time (s)}} \\
\cmidrule(lr){8-9}\cmidrule(lr){10-11}
 & & & & & & & \textbf{SIMP} & \textbf{AC-ALM} & \textbf{SIMP} & \textbf{AC-ALM} \\
\midrule
\endfirsthead
\caption[]{Compliance gap between SIMP and AC-ALM (continuous designs, continued).}\\
\toprule
\textbf{Problem} & \textbf{Mesh} & $\boldsymbol{V}$ & $\boldsymbol{r_{\min}}$ & $\boldsymbol{\mathcal{J}_\text{SIMP}}$ & $\boldsymbol{\mathcal{J}_\text{AC-ALM}}$ & \textbf{Gap (\%)} & \multicolumn{2}{c}{\textbf{Iter.}} & \multicolumn{2}{c}{\textbf{Time (s)}} \\
\cmidrule(lr){8-9}\cmidrule(lr){10-11}
 & & & & & & & \textbf{SIMP} & \textbf{AC-ALM} & \textbf{SIMP} & \textbf{AC-ALM} \\
\midrule
\endhead
\midrule
\multicolumn{11}{r}{Continued on next page}\\
\endfoot
\bottomrule
\endlastfoot
MBB & $120\times80$ & $0.3$ & $2$ & 79.041 & 81.612 & 3.25 & 224 & 335 & 19.44 & 28.39 \\
MBB & $180\times120$ & $0.3$ & $2$ & 79.772 & 82.261 & 3.12 & 212 & 327 & 43.25 & 61.93 \\
MBB & $240\times160$ & $0.3$ & $2$ & 80.880 & 83.456 & 3.18 & 216 & 289 & 83.11 & 102.49 \\
MBB & $300\times200$ & $0.3$ & $2$ & 82.113 & 84.278 & 2.64 & 220 & 267 & 145.24 & 169.12 \\
\addlinespace[2pt]
MBB & $120\times80$ & $0.4$ & $2$ & 59.650 & 60.969 & 2.21 & 215 & 345 & 17.32 & 27.68 \\
MBB & $180\times120$ & $0.4$ & $2$ & 60.736 & 62.074 & 2.20 & 222 & 330 & 38.23 & 63.52 \\
MBB & $240\times160$ & $0.4$ & $2$ & 61.823 & 62.951 & 1.82 & 219 & 307 & 78.94 & 105.53 \\
MBB & $300\times200$ & $0.4$ & $2$ & 62.759 & 63.608 & 1.35 & 222 & 264 & 142.78 & 163.22 \\
\addlinespace[2pt]
MBB & $120\times80$ & $0.3$ & $0.1$ & 77.306 & 80.747 & 4.45 & 219 & 363 & 19.67 & 33.96 \\
MBB & $180\times120$ & $0.3$ & $0.1$ & 76.894 & 80.644 & 4.88 & 232 & 305 & 43.11 & 60.05 \\
MBB & $240\times160$ & $0.3$ & $0.1$ & 77.334 & 81.462 & 5.34 & 242 & 323 & 95.48 & 120.27 \\
MBB & $300\times200$ & $0.3$ & $0.1$ & 77.574 & 82.650 & 6.54 & 245 & 321 & 177.77 & 187.20 \\
\addlinespace[2pt]
MBB & $120\times80$ & $0.4$ & $0.1$ & 59.101 & 60.480 & 2.33 & 258 & 357 & 20.66 & 33.63 \\
MBB & $180\times120$ & $0.4$ & $0.1$ & 59.480 & 61.238 & 2.95 & 230 & 334 & 40.87 & 67.60 \\
MBB & $240\times160$ & $0.4$ & $0.1$ & 60.337 & 61.787 & 2.40 & 259 & 337 & 93.48 & 121.64 \\
MBB & $300\times200$ & $0.4$ & $0.1$ & 61.149 & 62.542 & 2.28 & 241 & 284 & 162.46 & 163.49 \\
\addlinespace[2pt]
Cantilever & $120\times80$ & $0.3$ & $2$ & 59.566 & 65.432 & 9.85 & 190 & 294 & 14.84 & 28.11 \\
Cantilever & $180\times120$ & $0.3$ & $2$ & 59.563 & 64.806 & 8.80 & 212 & 265 & 39.24 & 53.24 \\
Cantilever & $240\times160$ & $0.3$ & $2$ & 59.845 & 65.146 & 8.86 & 214 & 293 & 82.16 & 109.58 \\
Cantilever & $300\times200$ & $0.3$ & $2$ & 60.155 & 64.522 & 7.26 & 216 & 248 & 149.86 & 151.42 \\
\addlinespace[2pt]
Cantilever & $120\times80$ & $0.4$ & $2$ & 43.294 & 45.779 & 5.74 & 191 & 329 & 16.39 & 31.59 \\
Cantilever & $180\times120$ & $0.4$ & $2$ & 43.478 & 45.821 & 5.39 & 206 & 302 & 38.00 & 63.59 \\
Cantilever & $240\times160$ & $0.4$ & $2$ & 43.696 & 45.988 & 5.24 & 221 & 293 & 78.02 & 115.12 \\
Cantilever & $300\times200$ & $0.4$ & $2$ & 43.885 & 45.658 & 4.04 & 221 & 241 & 142.41 & 158.07 \\
\addlinespace[2pt]
Cantilever & $120\times80$ & $0.3$ & $0.1$ & 57.637 & 63.752 & 10.61 & 203 & 297 & 16.34 & 27.93 \\
Cantilever & $180\times120$ & $0.3$ & $0.1$ & 55.961 & 62.363 & 11.44 & 226 & 276 & 42.44 & 56.10 \\
Cantilever & $240\times160$ & $0.3$ & $0.1$ & 55.609 & 61.804 & 11.14 & 227 & 266 & 93.68 & 102.07 \\
Cantilever & $300\times200$ & $0.3$ & $0.1$ & 55.732 & 62.588 & 12.30 & 239 & 272 & 188.23 & 178.67 \\
\addlinespace[2pt]
Cantilever & $120\times80$ & $0.4$ & $0.1$ & 42.463 & 45.307 & 6.70 & 214 & 346 & 16.40 & 32.66 \\
Cantilever & $180\times120$ & $0.4$ & $0.1$ & 42.242 & 44.791 & 6.04 & 225 & 352 & 39.94 & 77.01 \\
Cantilever & $240\times160$ & $0.4$ & $0.1$ & 42.191 & 44.560 & 5.61 & 235 & 297 & 85.01 & 115.70 \\
Cantilever & $300\times200$ & $0.4$ & $0.1$ & 42.389 & 44.630 & 5.29 & 242 & 284 & 162.38 & 193.99 \\
\addlinespace[2pt]
L-beam & $120\times120$ & $0.3$ & $2$ & 297.386 & 318.173 & 6.99 & 227 & 280 & 23.81 & 30.47 \\
L-beam & $180\times180$ & $0.3$ & $2$ & 302.678 & 319.270 & 5.48 & 228 & 269 & 50.21 & 64.55 \\
L-beam & $240\times240$ & $0.3$ & $2$ & 308.421 & 322.321 & 4.51 & 213 & 280 & 86.58 & 155.32 \\
L-beam & $300\times300$ & $0.3$ & $2$ & 312.354 & 326.566 & 4.55 & 212 & 264 & 177.30 & 218.85 \\
\addlinespace[2pt]
L-beam & $120\times120$ & $0.4$ & $2$ & 224.714 & 233.431 & 3.88 & 208 & 282 & 19.25 & 28.46 \\
L-beam & $180\times180$ & $0.4$ & $2$ & 229.009 & 236.941 & 3.46 & 212 & 297 & 48.25 & 69.12 \\
L-beam & $240\times240$ & $0.4$ & $2$ & 232.398 & 239.239 & 2.94 & 207 & 269 & 87.10 & 120.96 \\
L-beam & $300\times300$ & $0.4$ & $2$ & 235.757 & 241.852 & 2.59 & 211 & 257 & 160.38 & 209.46 \\
\addlinespace[2pt]
L-beam & $120\times120$ & $0.3$ & $0.1$ & 286.489 & 308.818 & 7.79 & 226 & 266 & 23.47 & 25.31 \\
L-beam & $180\times180$ & $0.3$ & $0.1$ & 281.919 & 302.619 & 7.34 & 243 & 280 & 58.02 & 63.71 \\
L-beam & $240\times240$ & $0.3$ & $0.1$ & 278.923 & 305.693 & 9.60 & 250 & 307 & 111.70 & 132.47 \\
L-beam & $300\times300$ & $0.3$ & $0.1$ & 277.649 & 303.527 & 9.32 & 259 & 261 & 206.32 & 204.82 \\
\addlinespace[2pt]
L-beam & $120\times120$ & $0.4$ & $0.1$ & 219.661 & 230.386 & 4.88 & 231 & 352 & 22.67 & 33.01 \\
L-beam & $180\times180$ & $0.4$ & $0.1$ & 215.955 & 227.052 & 5.14 & 232 & 298 & 52.44 & 65.26 \\
L-beam & $240\times240$ & $0.4$ & $0.1$ & 215.009 & 227.005 & 5.58 & 235 & 297 & 94.34 & 125.52 \\
L-beam & $300\times300$ & $0.4$ & $0.1$ & 215.198 & 227.239 & 5.60 & 248 & 252 & 183.97 & 195.21 \\
\end{longtable}
\endgroup

\subsection{3D Examples}\label{sec:num_examples_3d}

To demonstrate that the AC-ALM framework extends directly to spatial structures, 
we investigate two three-dimensional benchmarks: the 3D MBB and 3D L-beam. 
These examples evaluate the conserved residual formulation under spatial member 
curvature, combined bending-torsion loads, and geometric stress concentrations. 
Both cases are optimized under the intrinsic AC regime 
($r_{\min} = 0.1 \leqslant \frac{\sqrt{d}}{2}\Delta x$) with volume target $V = 0.15$, 
following the algorithmic parameters in Table~\ref{tab:algorithm_parameters}.

Both geometries employ uniform hexahedral meshes with element size $\Delta x = 1.0$. 
At this discretization, the core transition width 
$\xi_{0.1\to 0.9} = 3.11\,\ell_\text{AC} \approx 2.13$ 
spans $\approx 2.1$ elements, representing a compact transition layer at the lower resolution threshold 
of phase-field models. 
Here, $\Delta x = 1.0$ is deliberately selected to prioritize macroscale spatial truss 
formation within desktop computational bounds (approximately 153,000--158,000 DOFs) 
while maintaining numerical stability. 
Resolving the $\tanh$ interface across $4$--$5$ elements 
(as in 2D fine meshes where $\Delta x \leqslant 0.5$) would require $\Delta x = 0.5$ in 3D 
(exceeding $1.2\times 10^6$ DOFs), shifting into high-performance parallel computing.

\subsubsection{3D MBB Beam}\label{sec:num_examples_3d_mbb}

We evaluate the scalability of the framework on the three-dimensional MBB beam 
sketched in Fig.~\ref{fig:sketch_MBB_3d}. 
The domain is discretized into 48,000 hexahedral elements (157,563 DOFs), 
representing a fivefold increase in problem scale over the 2D benchmarks. 
A slightly gentler continuation rate of $\alpha_w = 0.04$ per iteration is adopted 
to accommodate spatial topological restructuring (Table~\ref{tab:algorithm_parameters}).

\begin{figure}[!ht]
\centering
\includegraphics{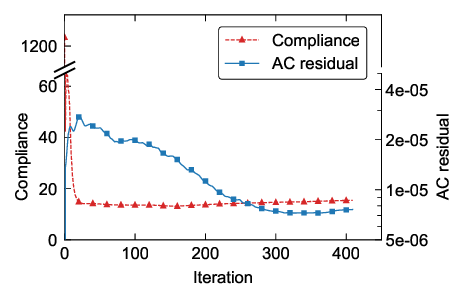}
\caption{Convergence history for 3D MBB under AC-ALM ($V=0.15$, $r_{\min}=0.1$).}\label{fig:3d_mbb_conv}
\end{figure}

\begin{figure}[!ht]
\centering
\includegraphics[width=1.0\textwidth]{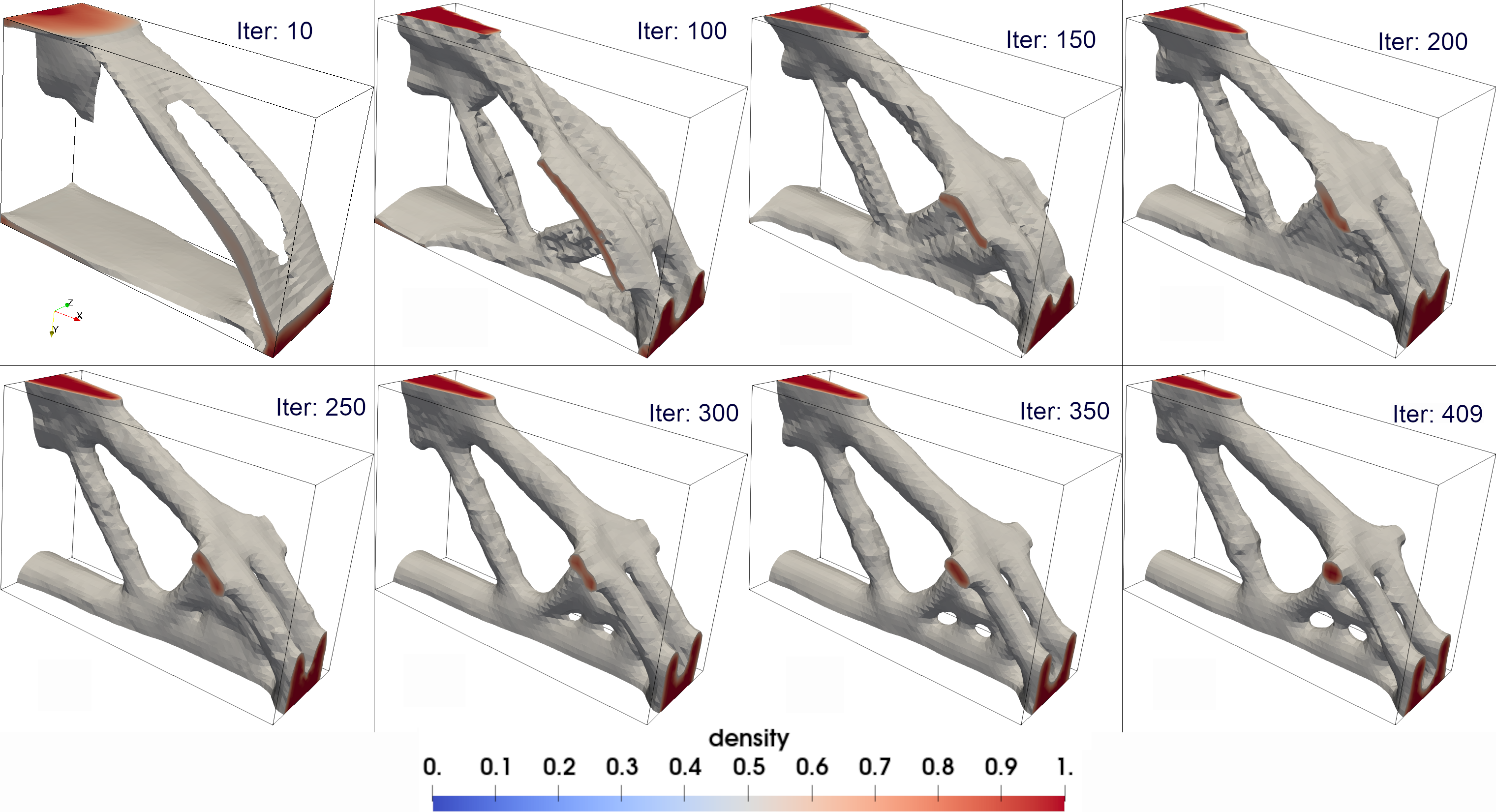}
\caption{Evolution and final topology of 3D MBB under AC-ALM ($V=0.15$, isosurface $\phi = 0.5$).}\label{fig:3d_mbb_iterations}
\end{figure}

The convergence history in Fig.~\ref{fig:3d_mbb_conv} confirms that the AC-ALM algorithm 
operates smoothly in 3D without architectural modifications. 
Because the uniform initial field is machine-precision feasible 
under the conserved formulation ($\mathbf{1}^{\top}\mathbf{h} \equiv 0$), 
the residual begins near zero, rises as 3D structural boundaries emerge, 
and is subsequently driven down by a factor of 3.6 
to stabilize safely below the tolerance threshold. 
The structural compliance reaches a minimum of $13.10$ at iteration 155 
before stabilizing at $15.45$ after 409 iterations. 
Figure~\ref{fig:3d_mbb_iterations} visualizes the 3D topology evolution, 
showing the progressive emergence of a spatial truss system 
that efficiently resists combined bending and torsion.

\begin{figure}[!ht]
\centering
\includegraphics{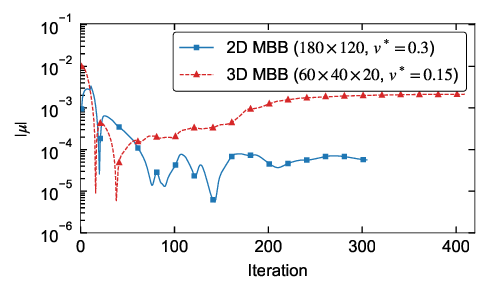}
\caption{Chemical potential history of 2D and 3D MBBs under conserved AC-ALM.}
\label{fig:mu_history}
\end{figure}

The 3D MBB example highlights a fundamental geometric distinction from 2D: 
the chemical potential converges to a positive value $\mu \approx +2.1\times 10^{-3}$ 
($|\mu|/\beta \approx 3.3\times 10^{-2}$), 
more than an order of magnitude larger relative to the double-well scale than in 2D. 
Figure~\ref{fig:mu_history} compares the full iteration histories of $\mu$ 
for the 2D and 3D MBB runs. 
Both start from the uniform-field state $\mu = W'(V)$. 
While the 2D history flattens near zero ($|\mu| \sim 7\times 10^{-5}$) 
once straight strut boundaries form, 
the 3D history climbs steadily to $+2.1\times 10^{-3}$. 
This dimensional contrast is a direct manifestation of interface curvature 
in the Allen--Cahn operator. 
Specifically, cylindrical 3D members possess a positive mean curvature ($\kappa > 0$), 
requiring a positive chemical potential $\mu \propto \kappa$ 
to maintain interface stationarity across the diffuse transition band. 
Retaining $\mu$ via the mean-zero projection is therefore essential in 3D 
to guarantee rigorous mathematical compatibility with volume conservation.

\subsubsection{3D L-Beam}\label{sec:num_examples_3d_lbeam}

As a final demonstration on a complex spatial geometry, 
we examine the 3D L-beam (Fig.~\ref{fig:sketch_Lbeam_3d}). 
The domain is discretized into 46,080 hexahedral elements 
(152,775 active displacement DOFs) to evaluate the ability of the optimizer 
to handle stress concentrations along a spatial re-entrant edge.

\begin{figure}[!ht]
\centering
    \includegraphics{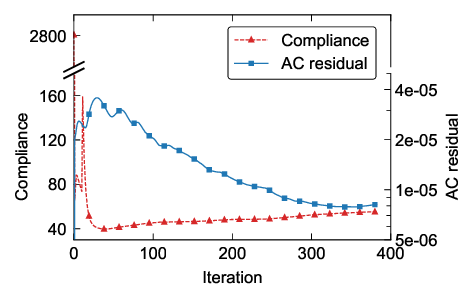}%
\caption{Convergence history for 3D L-beam under AC-ALM ($V=0.15$, $r_{\min}=0.1$).}\label{fig:3d_lbeam_conv}
\end{figure}

\begin{figure}[!ht]
\centering
    \includegraphics[width=1.0\textwidth]{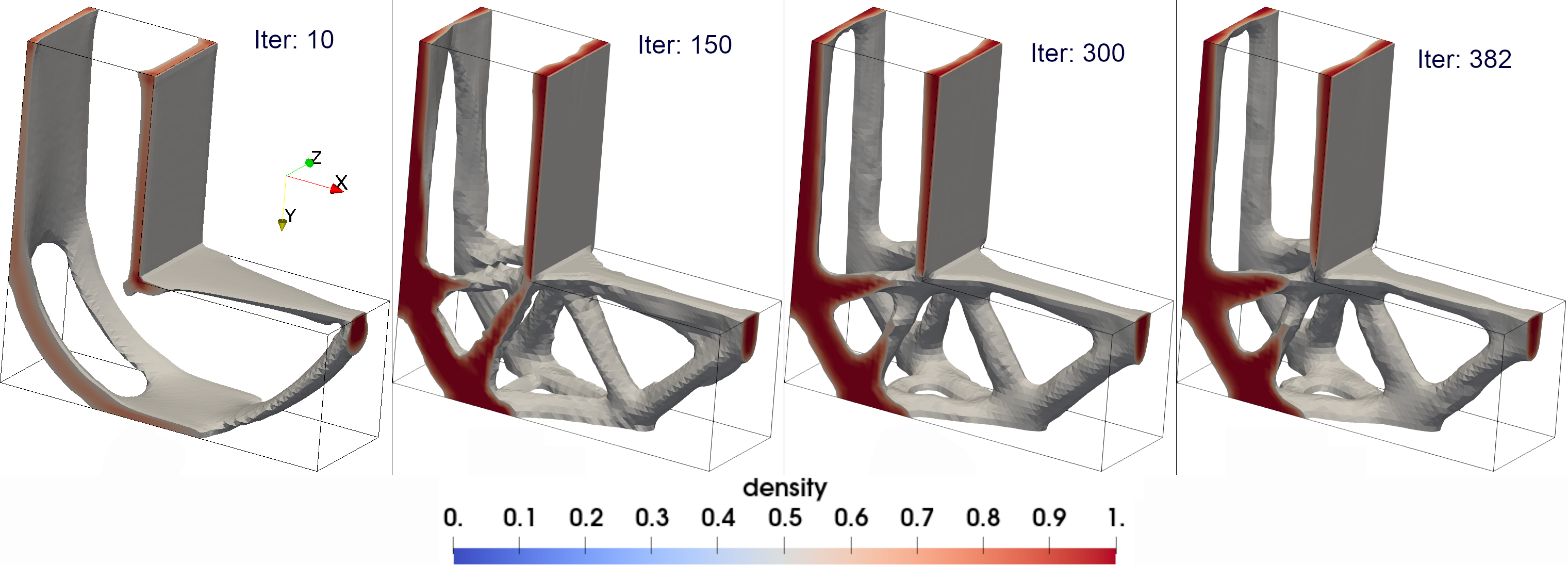}%
\caption{Evolution and final topology of 3D L-beam under AC-ALM ($V=0.15$, isosurface $\phi = 0.5$).}\label{fig:3d_lbeam_iterations}
\end{figure}

Figure~\ref{fig:3d_lbeam_conv} shows that the algorithm converges stably within 382 iterations, 
achieving a compliance of $55.46$ and a normalized residual of 
$\|\mathbf{h}\|_\text{norm} = 8.16\times 10^{-6}$, safely below the prescribed tolerance. 
The chemical potential again converges to a positive value ($\mu \approx +1.95\times 10^{-3}$), 
confirming the consistency of the mean-curvature effect across 3D geometries. 
As illustrated in Fig.~\ref{fig:3d_lbeam_iterations}, 
material smoothly concentrates into continuous, load-bearing arches 
wrapping around the re-entrant edge, 
yielding a well-connected spatial structure free from member fragmentation or numerical artifacts.

\section{Discussion}\label{discussion}

The numerical benchmarks demonstrate that the conserved AC-ALM framework
generates stable, well-connected topologies with controlled diffuse interfaces
across diverse meshes and volume fractions.
This section analyzes the underlying mechanisms across four themes:
(i) regularization mechanics and performance trade-offs;
(ii) parameter sensitivity, scaling equivalence, and penalty robustness;
(iii) discrete representations and box-constraint interactions; and
(iv) computational efficiency and cost breakdown.

\subsection{Regularization Mechanism and Performance Trade-offs}\label{sec:mechanism_tradeoffs}

\subsubsection{Compliance--Interface-Profile Trade-off and 0/1 Projection}\label{sec:compliance_tradeoff}

A primary observation from Table~\ref{tab:compliance_gap} is that
enforcing the Allen--Cahn constraint entails a measurable compliance cost.
Continuous AC-ALM compliances are $1.4\%$--$12.3\%$ higher than
the unconstrained SIMP baseline.
During optimization, compliance drops rapidly while the residual is large.
As ALM updates drive the residual norm below tolerance $\tau_h$
(equivalently $\delta$ in Eq.~\eqref{eq:delta_equiv}),
compliance rises slightly to a stable plateau.
A key question is whether this continuous compliance gap reflects structural deficiency
or an accounting artifact of the diffuse interface.

Under SIMP interpolation ($E(\phi) \propto \phi^p$ with $p=3$),
intermediate densities carry disproportionately low stiffness relative to volume.
In AC-ALM designs, $23\%$--$46\%$ (mean one-third) of all material resides
within intermediate elements ($0.05 < \bar{\phi}_e < 0.95$).
In contrast, SIMP exhibits only $6\%$--$21\%$ (mean $12\%$) intermediate material
at $r_{\min}=0.1$, and $17\%$--$36\%$ (mean $25\%$) at $r_{\min}=2.0$.
Because the residual constraint enforces a diffuse transition band
($\ell_{\text{AC}} \approx 0.686$, core width $\xi_{0.1\to 0.9} \approx 2.13$),
a substantial material fraction suffers the SIMP power-law penalty.
This inherently elevates the continuous compliance metric.

To isolate layout quality from intermediate-density stiffness penalties,
each of the 48 continuous design pairs in Table~\ref{tab:compliance_gap}
was projected onto a crisp 0/1 design at strictly identical volume.
The $\lfloor V N_e \rceil$ elements with highest densities are set to solid ($\phi=1$)
and the remainder to void ($\phi=0$).
We then re-evaluate compliance on the same mesh and boundary conditions.
Table~\ref{tab:threshold} summarizes the results,
reporting the range and mean over four meshes and two volume fractions.
Negative percentages indicate stiffer AC-ALM designs.

\begin{table}[!ht]
\centering
\caption{Post-processing audit: relative compliance difference $(\mathcal{J}_\text{AC}-\mathcal{J}_\text{SIMP})/\mathcal{J}_\text{SIMP}$ between continuous and volume-preserving thresholded designs (reporting range and mean across the eight cases in each problem--$r_{\min}$ group; negative values indicate that AC-ALM has lower compliance / is stiffer than SIMP).}
\label{tab:threshold}
\begin{tabular}{llccc}
\toprule
Problem & $r_{\min}$ & Continuous (\%) & Thresholded (\%) & AC-ALM stiffer / 8 cases \\
\midrule
MBB        & 2   & $+1.4$ to $+3.3$ ($+2.5$)  & $-2.1$ to $-1.3$ ($-1.6$) & 8/8 \\
MBB        & 0.1 & $+2.3$ to $+6.5$ ($+3.9$)  & $-3.4$ to $-1.8$ ($-2.6$) & 8/8 \\
Cantilever & 2   & $+4.0$ to $+9.8$ ($+6.9$)  & $-0.7$ to $+5.8$ ($+0.4$) & 7/8 \\
Cantilever & 0.1 & $+5.3$ to $+12.3$ ($+8.6$) & $-3.6$ to $+0.2$ ($-1.4$) & 6/8 \\
L-beam     & 2   & $+2.6$ to $+7.0$ ($+4.3$)  & $-1.5$ to $-0.2$ ($-0.6$) & 8/8 \\
L-beam     & 0.1 & $+4.9$ to $+9.6$ ($+6.9$)  & $-2.2$ to $-0.1$ ($-1.3$) & 8/8 \\
\bottomrule
\end{tabular}
\end{table}

In this volume-preserving thresholding audit,
the compliance of the AC-ALM designs drops by an average of $12.6\%$
(compared to $6.7\%$ for SIMP).
As a result, the continuous compliance gap vanishes.
The thresholded AC-ALM compliance differs from the baseline
by $-3.6\%$ to $+5.8\%$ (with a mean difference of $-1.2\%$).
Moreover, AC-ALM yields the stiffer layout in 45 of the 48 cases.
This includes all 24 cases on the two finest meshes,
which achieve $0.3\%$ to $3.6\%$ lower compliance.

Disaggregating the comparison demonstrates consistent performance across regimes.
Against the filtered SIMP baseline ($r_{\min}=2.0$),
AC-ALM yields lower compliance in 23 of 24 cases (averaging $0.6\%$ stiffer).
Against the unregularized baseline ($r_{\min}=0.1$),
AC-ALM yields lower compliance in 22 of 24 cases (averaging $1.8\%$ stiffer).
Across all 48 cases, the only three exceptions 
(one against filtered SIMP and two against unregularized SIMP) 
occur for the cantilever on the coarsest mesh ($120\times 80$).
There, thin diagonal members approach the transition width ($\xi \approx 2.13$)
and are barely resolved on $\Delta x = 1.0$.
Post-thresholding on this coarse grid occasionally produces isolated diagonal pixel contacts,
which share only a single corner vertex under 4-neighborhood connectivity.
On refined meshes, unregularized SIMP at $r_{\min}=0.1$ frequently retains
disconnected struts or single-node hinges that carry minimal load after thresholding.
In contrast, AC-ALM maintains well-connected, continuous structural chords.

These findings confirm that the continuous compliance gap reflects
the energy penalty of diffuse-interface elements rather than morphological deficiency.
Because AC-ALM maintains a symmetric, monotonic $\tanh$ profile,
thresholding cleanly promotes inner interface elements ($\phi \in [0.5, 0.95]$)
to full solid.
This fully recovers their load-bearing capacity.
Conversely, unregularized SIMP at $r_{\min}=0.1$ forms fragile
single-element-wide struts that volume-preserving rank thresholding can sever.

We emphasize that element-level volume-preserving thresholding serves strictly as a diagnostic audit on the identical mesh to isolate intrinsic layout quality from SIMP intermediate-density penalties.
Unlike explicit boundary-projection schemes that enforce crisp 0/1 interfaces directly within the optimization loop \citep{Guest2004a,WangFengwen2011a}, AC-ALM preserves a regularized continuous phase field throughout the optimization.
For downstream CAD modeling and manufacturing workflows, this discrete element thresholding is naturally distinguished from continuous boundary extraction: the nodal design field enables direct isosurface extraction (e.g., Marching Cubes at $\phi = 0.5$, Figs.~\ref{fig:3d_mbb_iterations} and \ref{fig:3d_lbeam_iterations}), producing smooth, watertight geometric reconstructions free from pixelated discretization artifacts.

\subsubsection{Interface Regularization: Filtering Comparison and Distinction from Perimeter Penalization}\label{sec:cmp_filter}

The regularizing action of the Allen--Cahn constraint operates
through the local reaction--diffusion balance $-\epsilon^2 \Delta \phi + W'(\phi) = \mu$.
In bulk regions, spatial gradients vanish and $W'(\phi) \approx 0$.
This confines constraint enforcement to the diffuse transition band.
Here, the double-well reaction promotes phase separation,
while the diffusion operator penalizes high-frequency spatial variations.
This intrinsic balance organizes intermediate densities into smooth,
residual-controlled diffuse profiles.
It naturally suppresses single-node hinges and mesh-dependent branch proliferation
without requiring external density or sensitivity filtering.

Comparing the filter-assisted ($r_{\min}=2.0$) and intrinsic ($r_{\min}=0.1$) regimes
(Section~\ref{subsec:2d_benchmarks}) highlights this behavior.
While the projection filter dictates feature size at $r_{\min}=2.0$,
AC-ALM remains stable and mesh-consistent at $r_{\min}=0.1$.
In contrast, conventional SIMP fractures into mesh-dependent artifacts at $r_{\min}=0.1$.

\begin{figure}[!ht]
\centering
\begin{subfigure}[b]{0.29\textwidth}\centering
  \includegraphics[width=\textwidth]{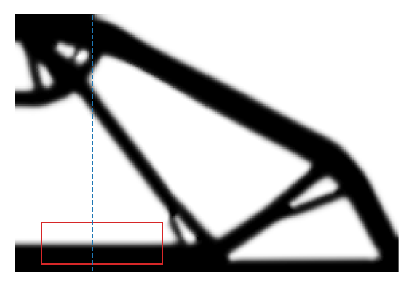}
  \caption{Topology}\label{fig:ifzoom_a}\end{subfigure}\hfill
\begin{subfigure}[b]{0.38\textwidth}\centering
  \includegraphics[width=\textwidth]{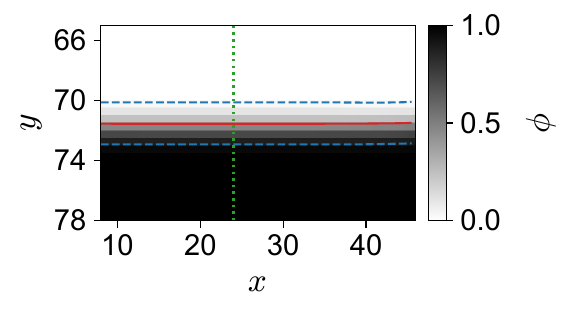}
  \caption{Lower-chord density $\phi$}\label{fig:ifzoom_b}\end{subfigure}\hfill
\begin{subfigure}[b]{0.29\textwidth}\centering
  \includegraphics[width=\textwidth]{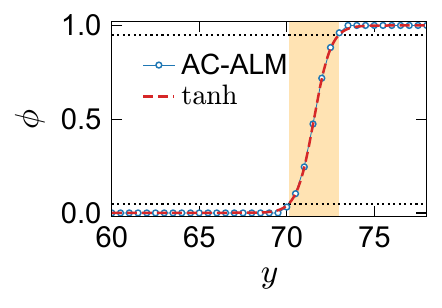}
  \caption{Profile vs.\ $\tanh$}\label{fig:ifzoom_c}\end{subfigure}
\caption{Diffuse-interface profile of AC-ALM MBB design ($240\times160$, $V=0.3$, $r_{\min}=0.1$).}
\label{fig:interface_zoom}
\end{figure}

In Fig.~\ref{fig:interface_zoom}(a), the optimized MBB topology is sampled
along $x=L/5$ (dashed line) and magnified at the lower chord.
The density field $\phi$ exhibits monotone contours (Fig.~\ref{fig:interface_zoom}b).
Figure~\ref{fig:interface_zoom}(c) compares the nodal profile
with the analytical $\tanh$ equilibrium across the transition band ($\phi=0.05$--$0.95$).
The measured width spans $5.6$ elements ($\approx 2.8$,
matching theoretical $\xi_{0.05\to 0.95} \approx 2.86$ on $\Delta x = 0.5$).
The core width $\xi_{0.1\to 0.9}$ spans $4.3$ elements ($\approx 2.13$,
in exact agreement with $3.11\,\ell_\text{AC} = 2.13$).
The interface width is thus governed intrinsically by $\ell_\text{AC} = \epsilon/\sqrt{\beta}$
rather than external filtering.

This mechanism also differs fundamentally from perimeter regularization~\citep{Haber1996a}.
Perimeter regularization penalizes total interface area
$\int_\Omega |\nabla\phi|\,\mathrm{d}\Omega$.
This inadvertently suppresses slender load-bearing struts in favor of bulky shapes.
In contrast, the AC constraint regulates the local transition profile
($\tanh$ shape and bandwidth $\propto \epsilon/\sqrt{\beta}$)
within tolerance $\|\mathbf{h}\|_* \leqslant \delta$.
It does not penalize total perimeter as an objective term.
Exact phase-field stationarity in the sharp-interface limit implies
constant-mean-curvature surfaces \citep{Modica1987a,Luckhaus1989a}.
However, the tolerance-regularized formulation allows the optimizer
to establish complex load-bearing trusses.
Residual concentrations localize naturally at member junctions,
where mechanical load paths require deviations from ideal phase-field equilibrium.

\subsubsection{Relationship with Classical Phase-Field Theory}\label{sec:phase_field_theory}

The proposed framework shares the formal structure
of classical mass-constrained phase-field models \citep{Rubinstein1992a,Blank2014a,Garcke2024a}.
However, it departs fundamentally from standard pseudo-time gradient-flow integrations
\citep{Takezawa2010a}.
Instead, it treats the conserved Allen--Cahn stationarity condition
as an equality constraint within an Augmented Lagrangian programming framework.
Departing from pseudo-time integration confers three operational advantages
over conventional phase-field gradient flows:
(i) it decouples optimization updates from stiff time-stepping trajectories,
allowing MMA to traverse infeasible intermediate states and explore design space
without time-step stability restrictions;
(ii) it enforces global volume directly as an explicit constraint in MMA
within a strict tolerance ($\tau_g = 2\times 10^{-3}$) rather than through
empirical multiplier or penalty tuning, while evaluating the chemical potential
$\mu$ in closed form (Eq.~\eqref{eq:mu_def}) to ensure algebraic residual compatibility; and
(iii) the observed sign and magnitude of $\mu$ are qualitatively consistent
with the formal Gibbs--Thomson asymptotic limit,
converging to near-zero for straight 2D struts while settling at a positive plateau in 3D
due to the positive mean curvature of spatial trusses (Section~\ref{num_boxresidual}).

\subsection{Parameter Sensitivity and Algorithmic Robustness}\label{sec:param_robustness}

\subsubsection{Sensitivity to Interface Parameters}\label{num_sensitivity}

To examine how Allen--Cahn parameters regulate structural layout,
we conduct a sensitivity study over $\epsilon \in \{0.15, 0.175, 0.20\}$
and $\beta \in \{0.05, 0.065, 0.08, 0.10\}$ on the reference MBB benchmark
($180\times 120$, $r_{\min}=0.1$) at volume fraction target $V=0.5$.
This symmetric midpoint of the double-well potential ($V=0.5$)
maximizes topological variation and interface area.
It tests the solver's ability to drive phase separation
via the AC residual rather than material scarcity.
Topologies are presented in Table~\ref{tab:mbb-sense-rmin0.1}.
Compliances and converged chemical potentials $\mu$
are listed in Table~\ref{tab:mbb-compliance-rmin0.1}.

\begin{table}[!htbp]
    \centering
    \small
    \setlength{\tabcolsep}{2pt}
    \caption{Topological sensitivity to interface parameters $\epsilon$ and $\beta$ (MBB, $180\times 120$, $V=0.5$, $r_{\min}=0.1$).}
    \label{tab:mbb-sense-rmin0.1}
    \begin{tabular}{ccccc}
        \toprule
        $\epsilon$\textbackslash{}$\beta$ & 0.05 & 0.065 & 0.08 & 0.1 \\
        \midrule
        0.15 &
        \includegraphics[width=1.5cm,valign=m]{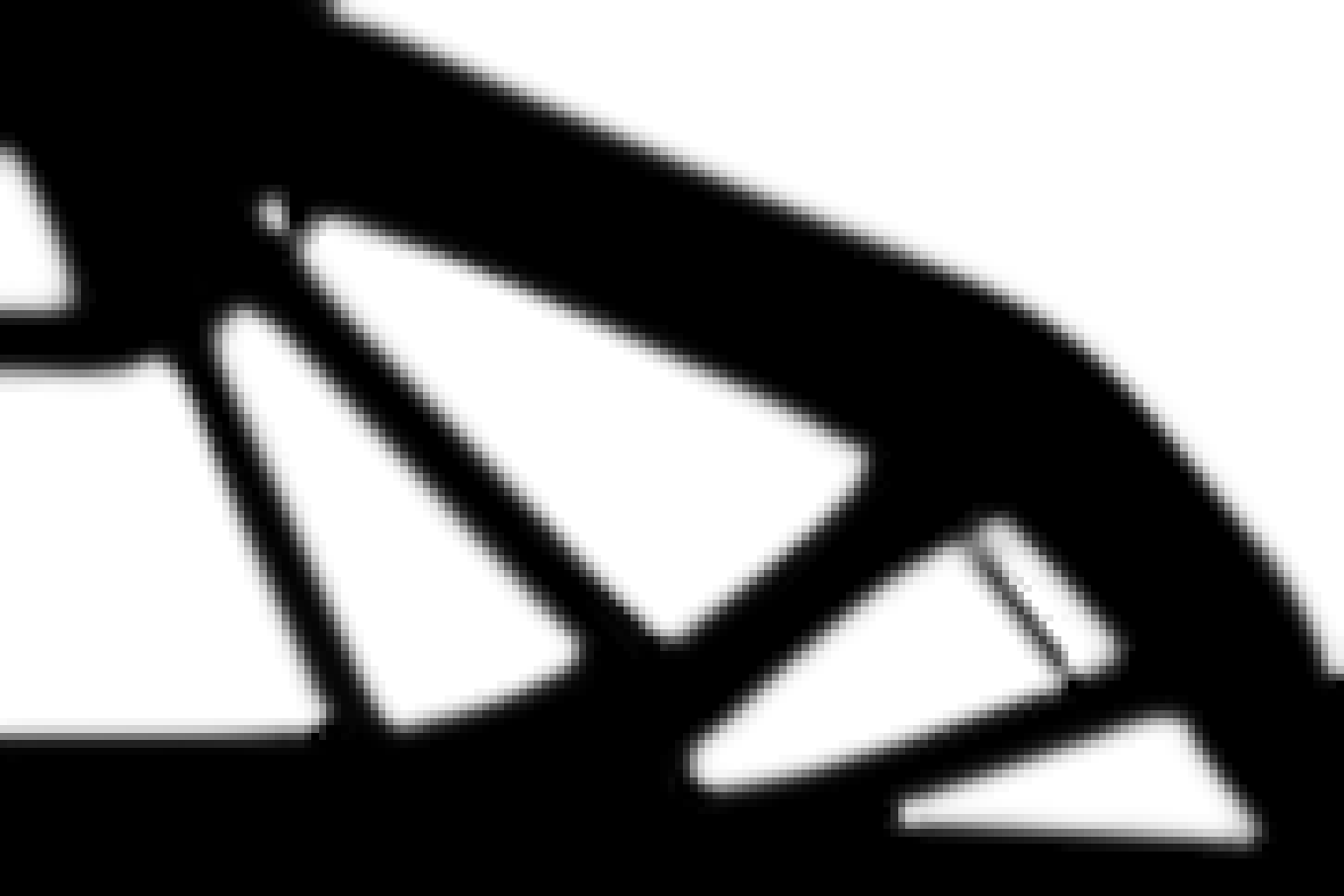} &
        \includegraphics[width=1.5cm,valign=m]{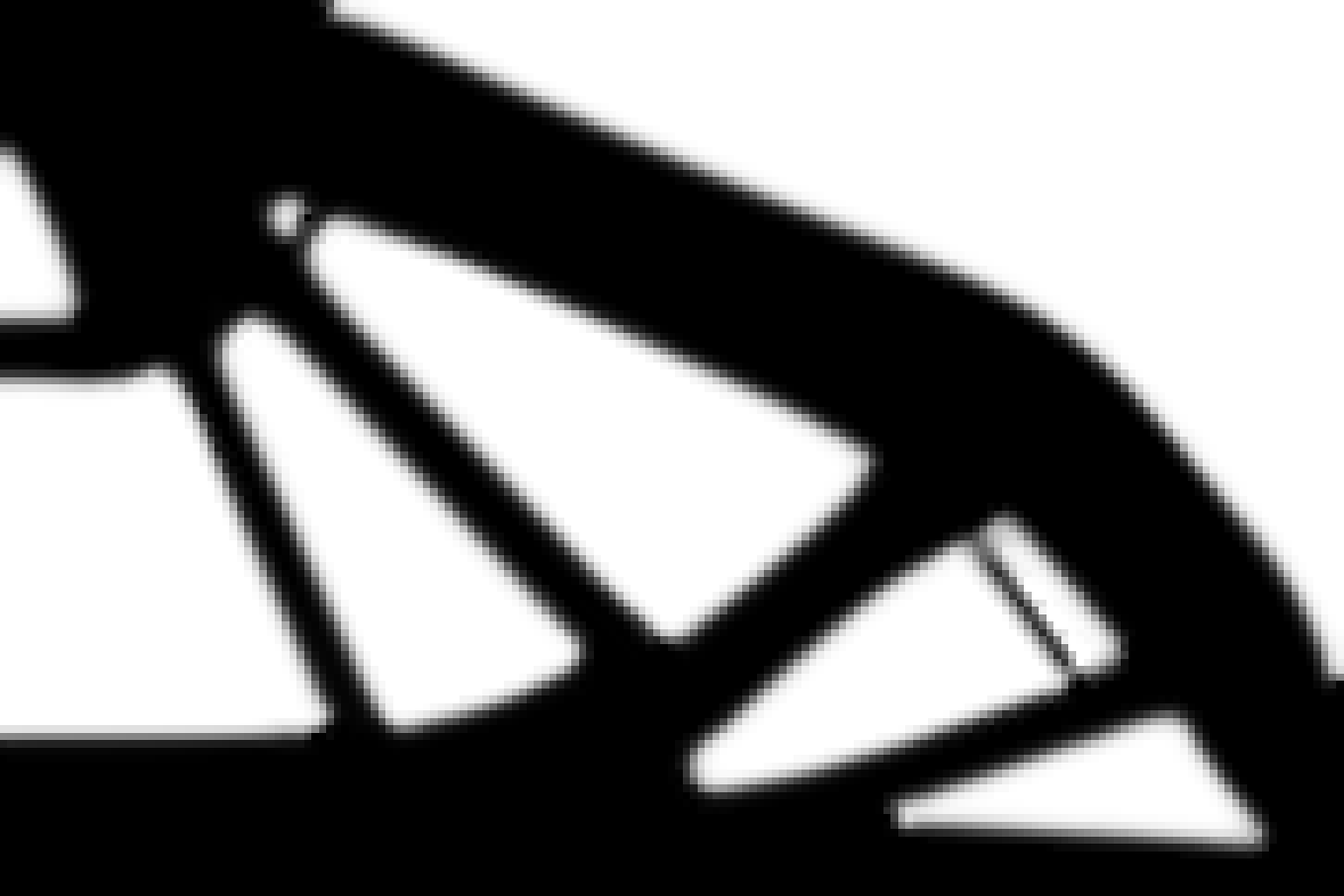} &
        \includegraphics[width=1.5cm,valign=m]{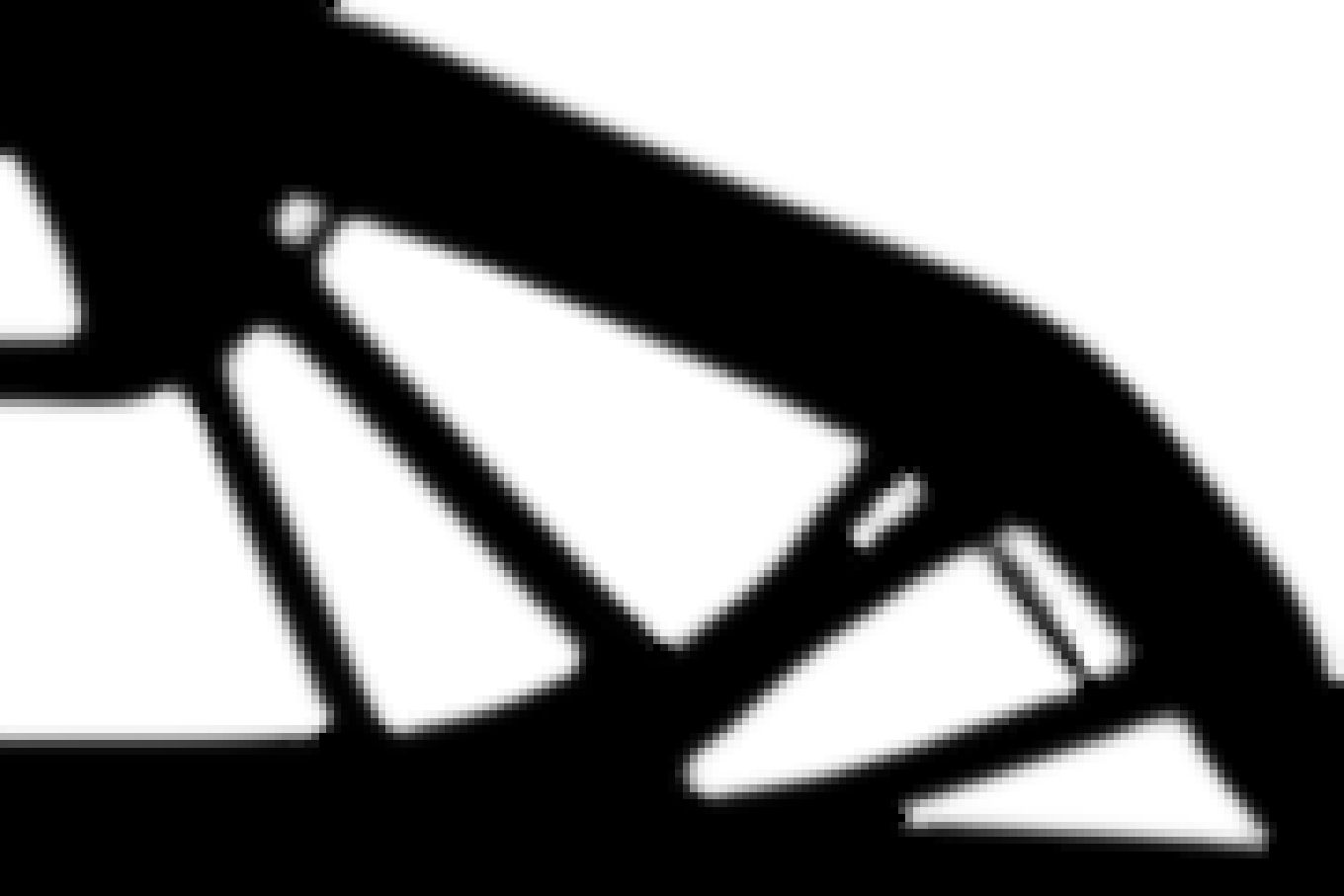} &
        \includegraphics[width=1.5cm,valign=m]{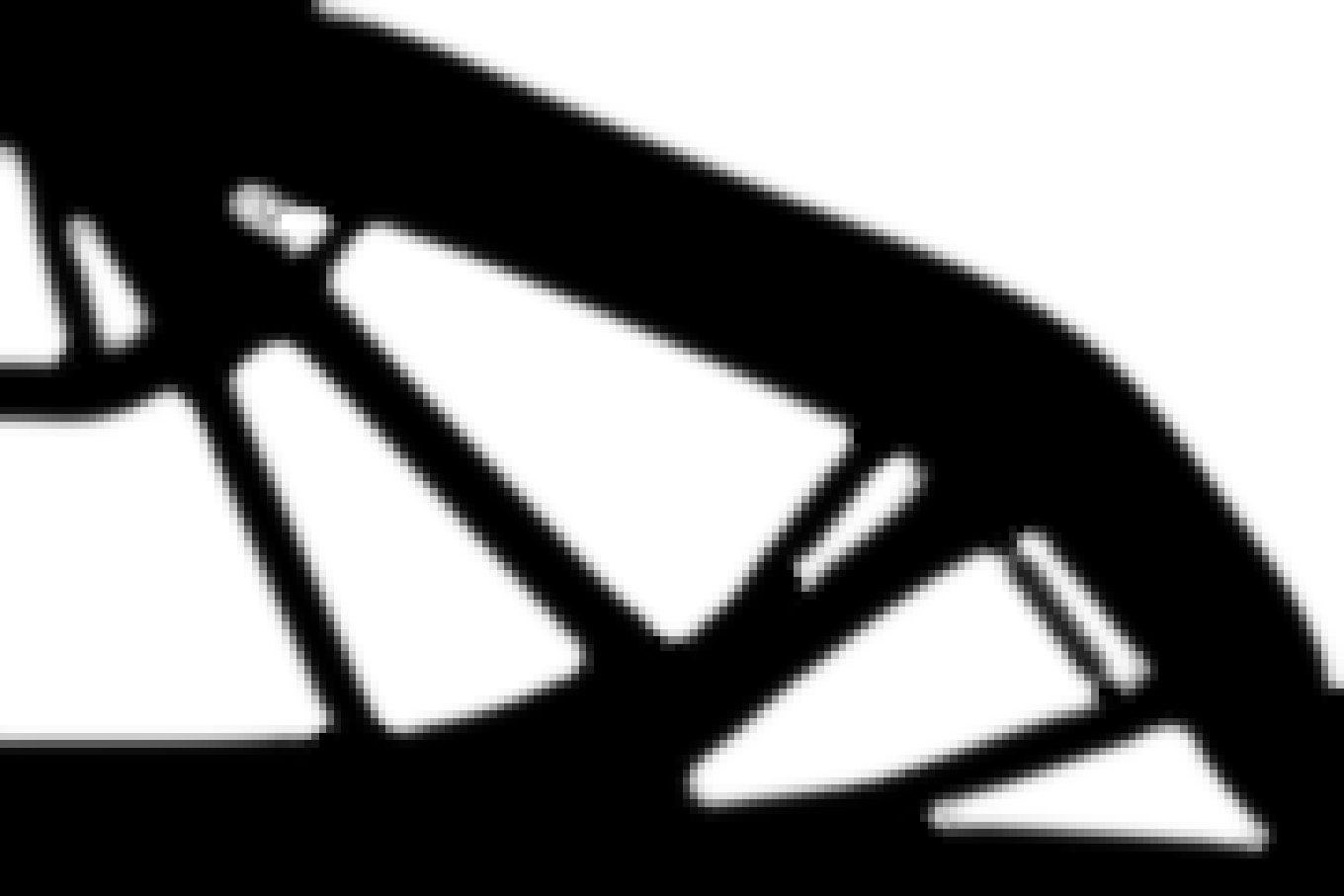} \\ \addlinespace[5pt]
        0.175 &
        \includegraphics[width=1.5cm,valign=m]{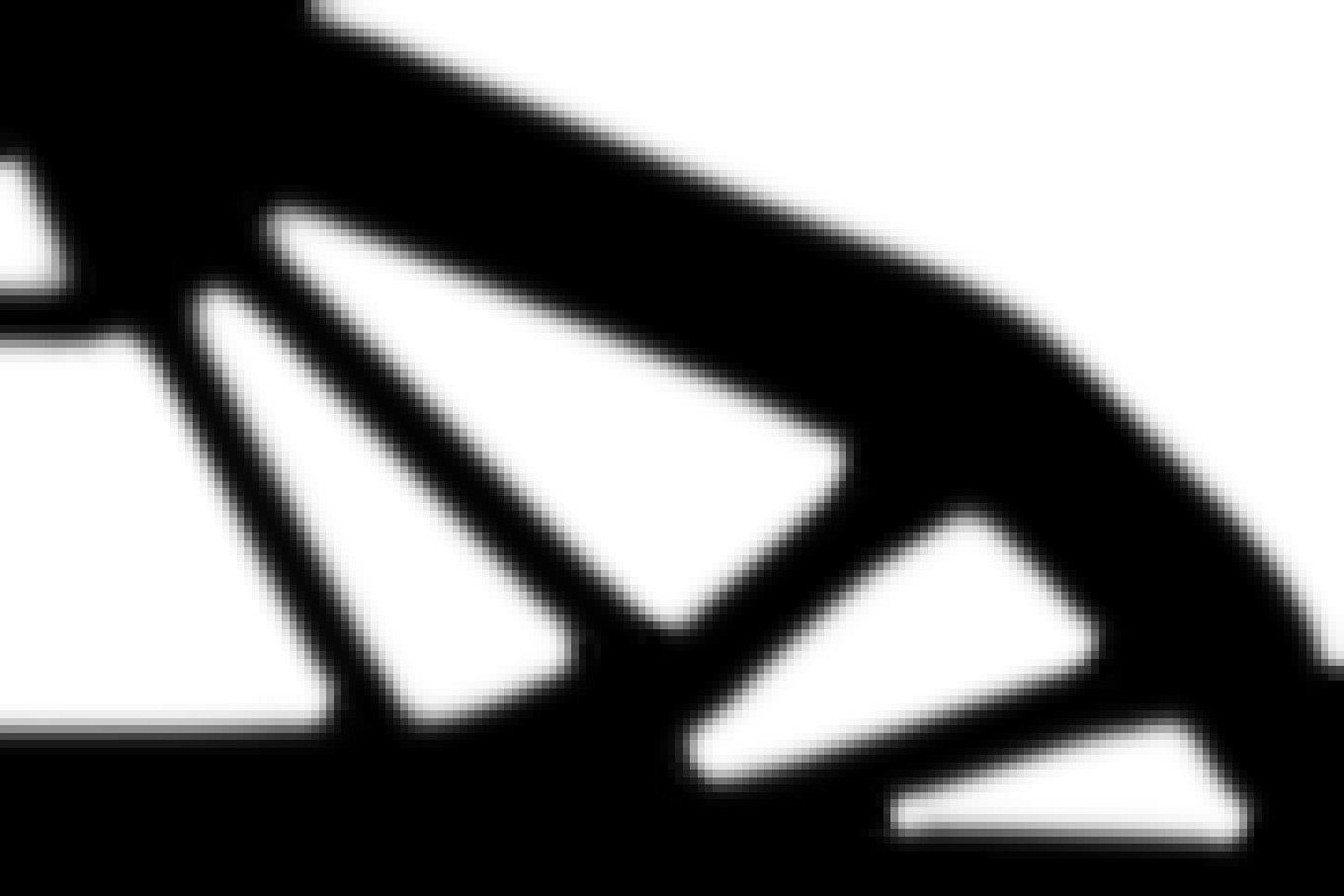} &
        \includegraphics[width=1.5cm,valign=m]{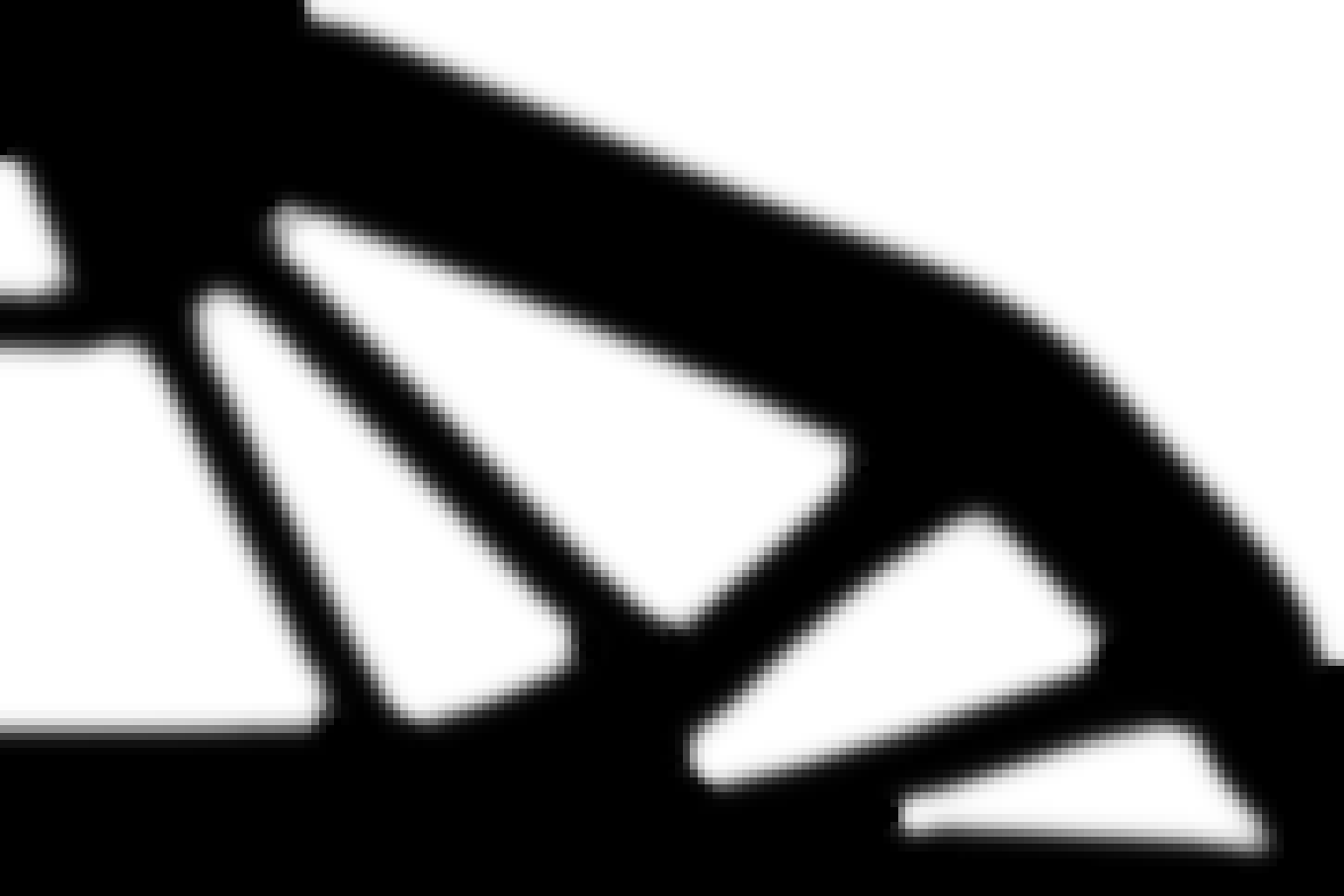} &
        \includegraphics[width=1.5cm,valign=m]{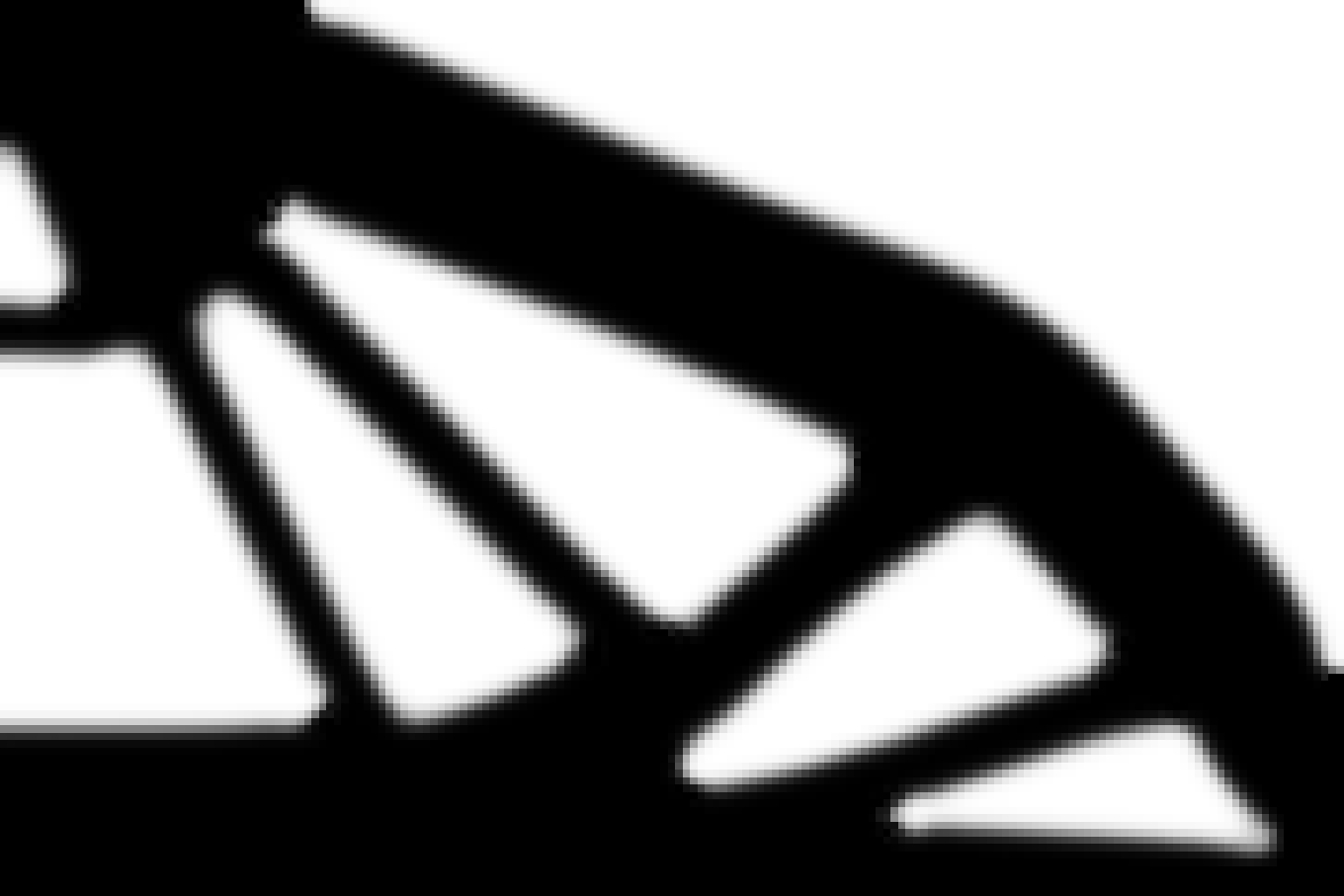} &
        \includegraphics[width=1.5cm,valign=m]{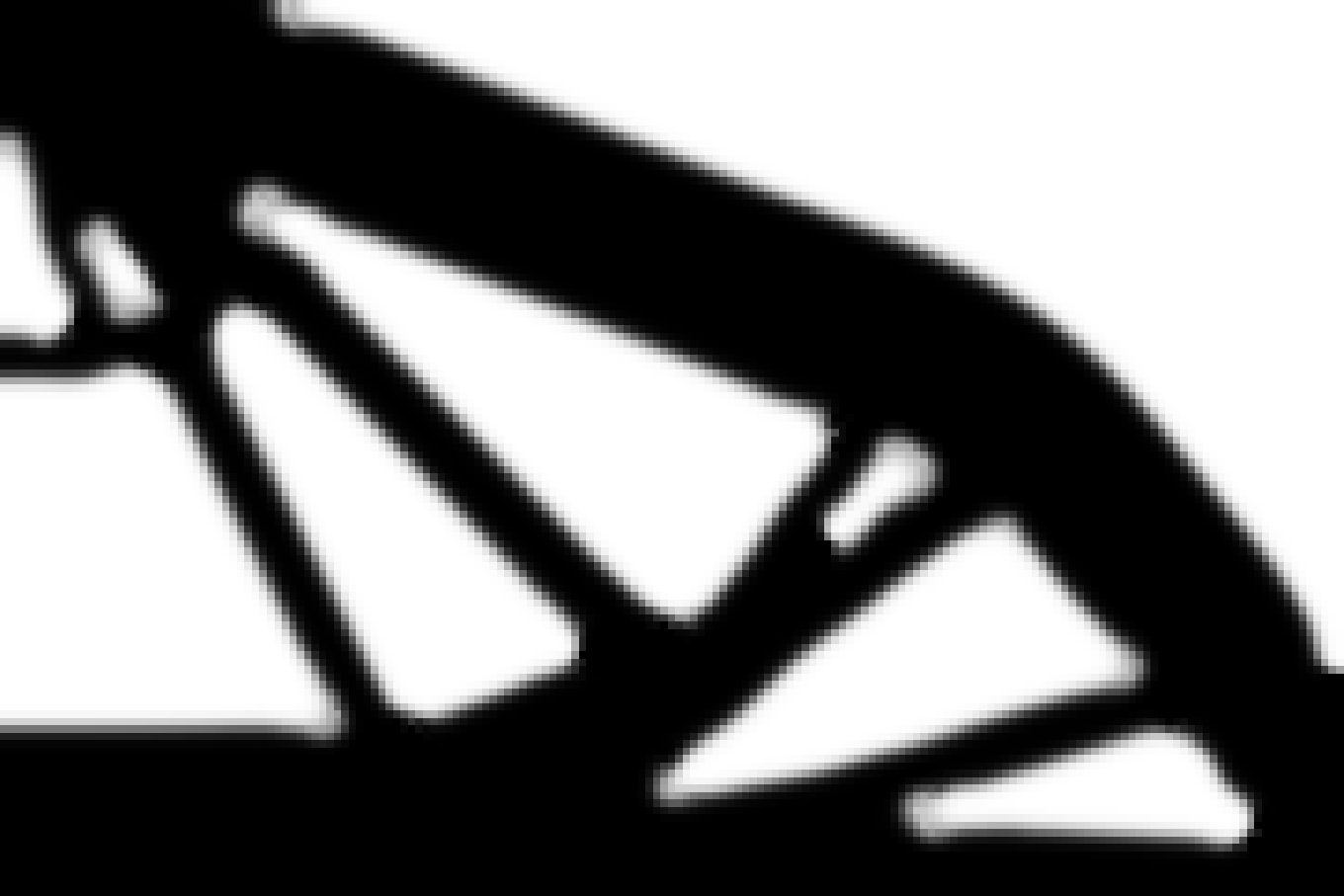} \\ \addlinespace[5pt]
        0.2 &
        \includegraphics[width=1.5cm,valign=m]{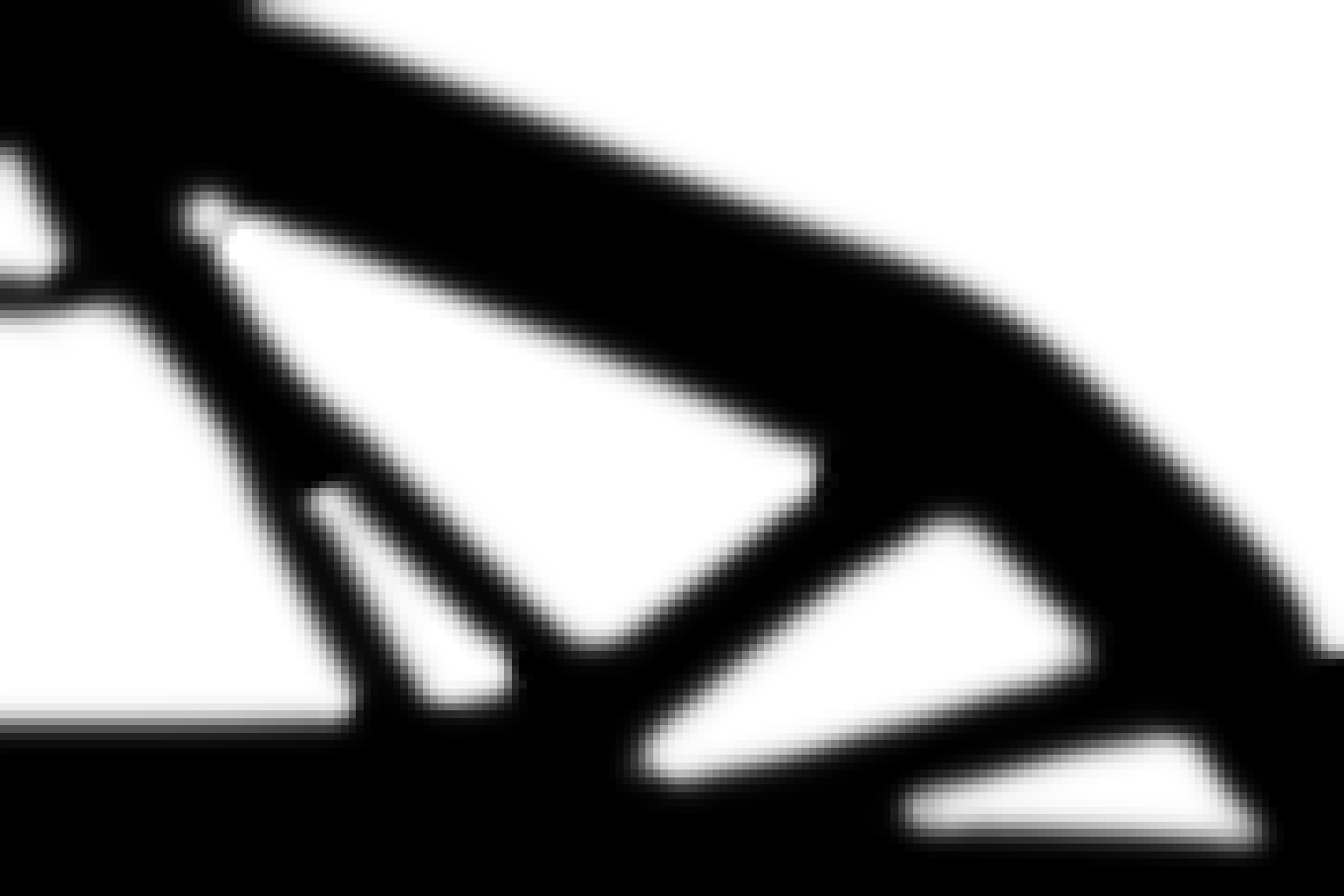} &
        \includegraphics[width=1.5cm,valign=m]{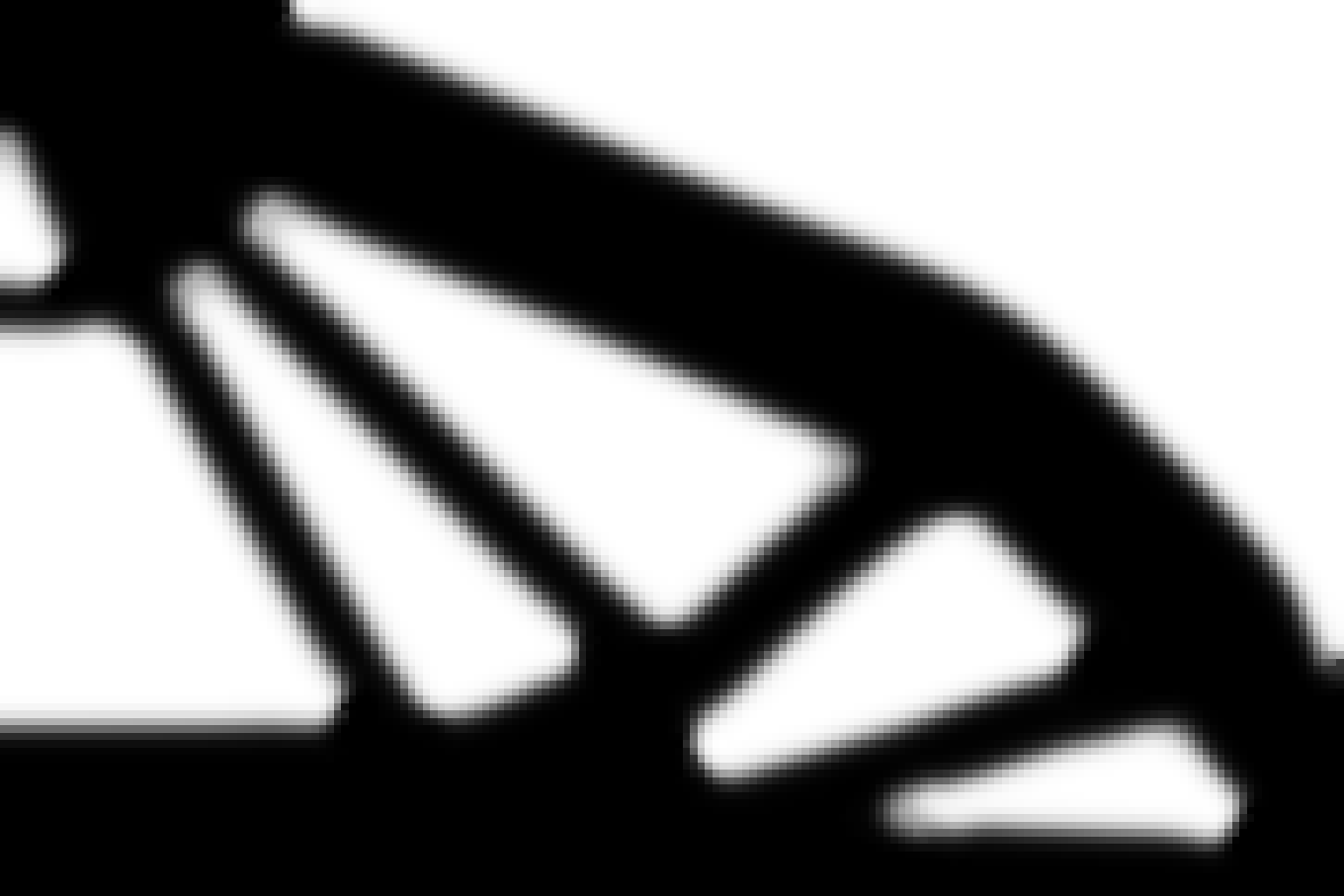} &
        \includegraphics[width=1.5cm,valign=m]{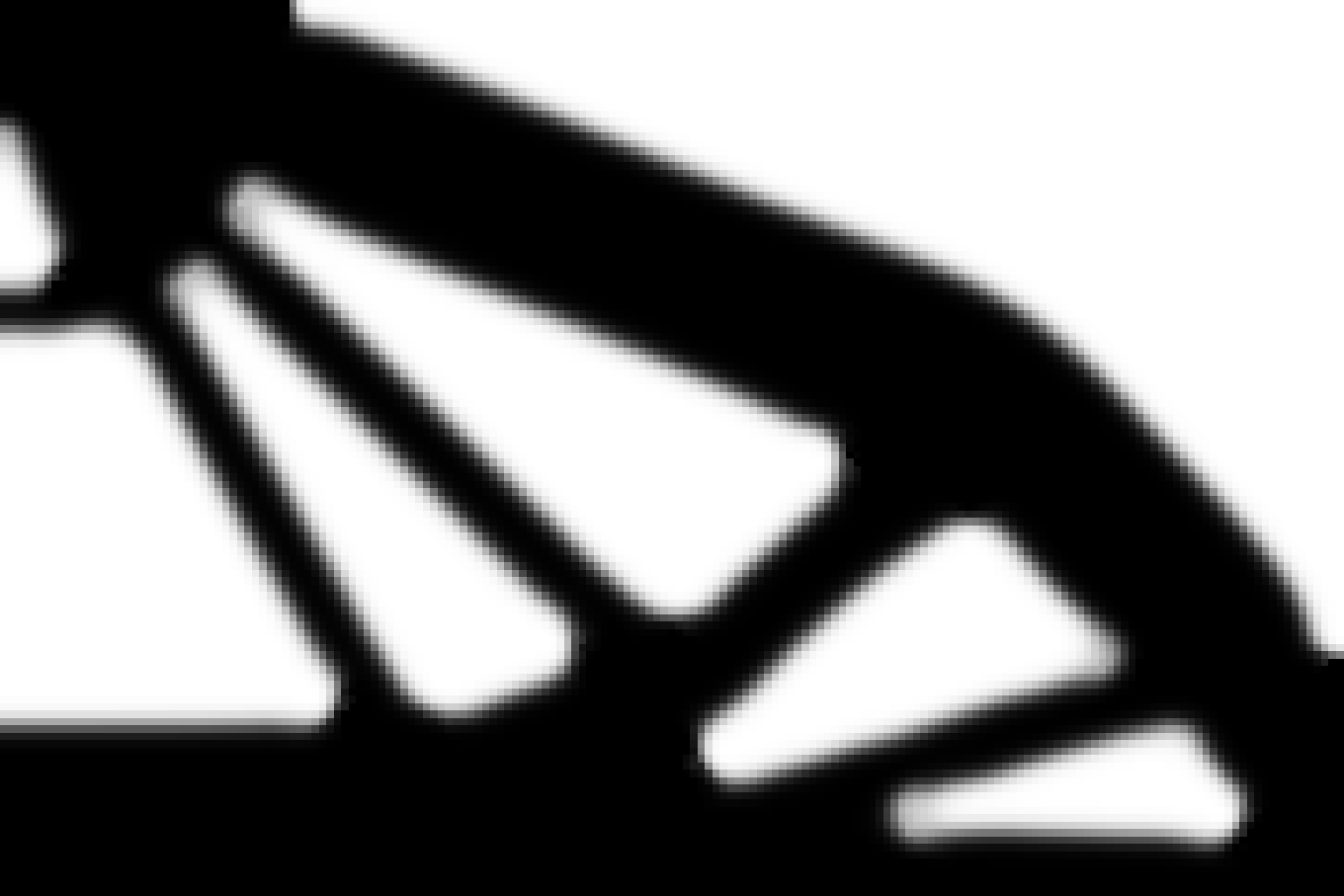} &
        \includegraphics[width=1.5cm,valign=m]{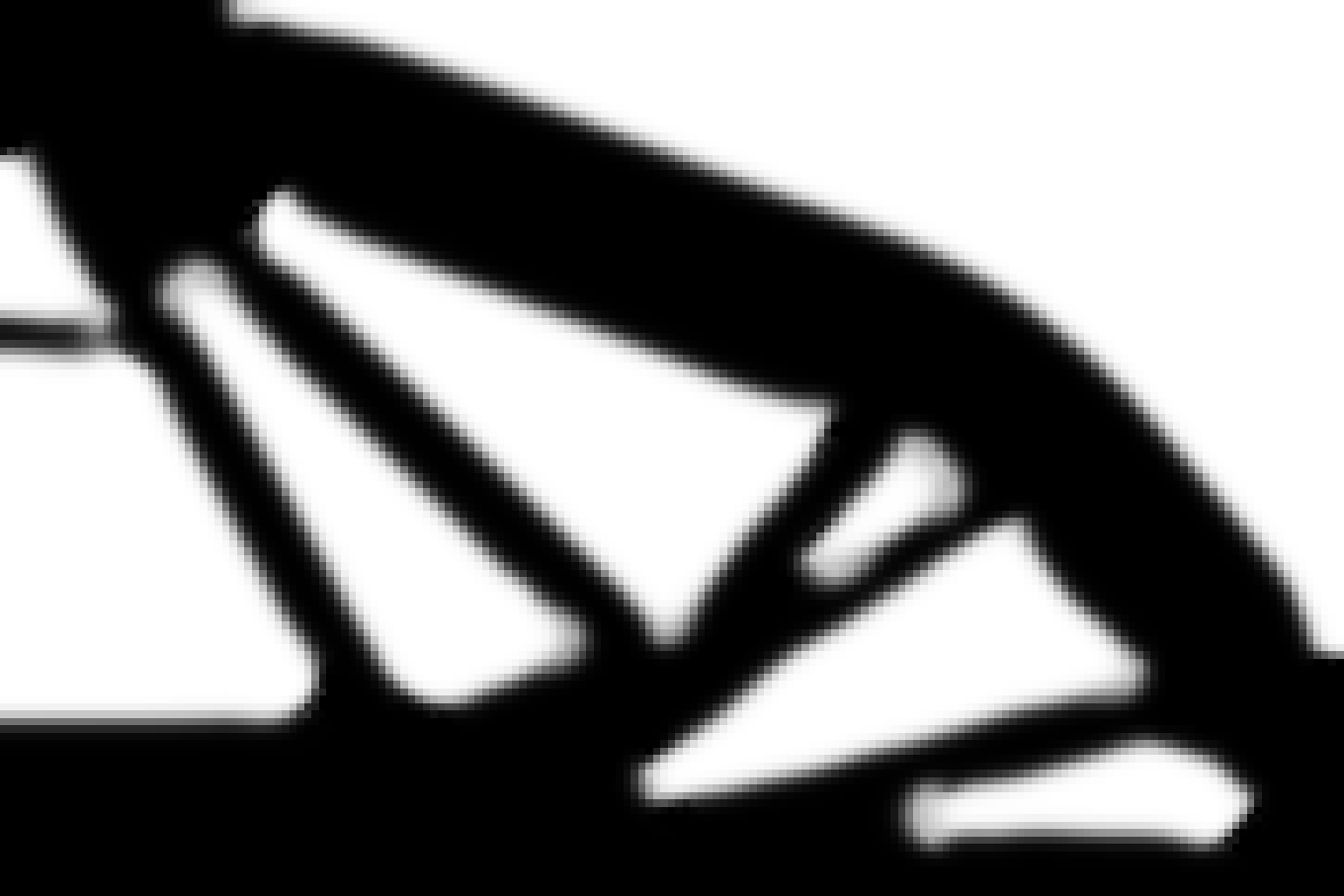} \\
        \bottomrule
    \end{tabular}
\end{table}
\begin{table}[!htbp]
    \centering
    \small
    \setlength{\tabcolsep}{5pt}
    \caption{Sensitivity of compliance $\mathcal{J}$ and chemical potential $\mu$ (MBB, $180\times 120$, $V=0.5$, $r_{\min}=0.1$).}
    \label{tab:mbb-compliance-rmin0.1}
    \begin{tabular}{lcccc}
        \toprule
        $\epsilon$\textbackslash{}$\beta$ & 0.05 & 0.065 & 0.08 & 0.1 \\
        \midrule
        \multicolumn{5}{l}{\textit{Final compliance} $\mathcal{J}$} \\
        0.15 & 50.655 & 50.392 & 50.326 & 50.345 \\
        0.175 & 51.194 & 50.825 & 50.532 & 50.950 \\
        0.2 & 51.677 & 51.477 & 51.204 & 51.759 \\
        \midrule
        \multicolumn{5}{l}{\textit{Converged chemical potential} $\mu$ ($\times 10^{-5}$)} \\
        0.15 & $-2.87$ & $-3.14$ & $-4.62$ & $-7.57$ \\
        0.175 & $-6.22$ & $-6.49$ & $-6.87$ & $-8.06$ \\
        0.2 & $-8.49$ & $-8.91$ & $-8.29$ & $-10.61$ \\
        \bottomrule
    \end{tabular}
\end{table}

\begin{figure}[!ht]
\centering
\includegraphics{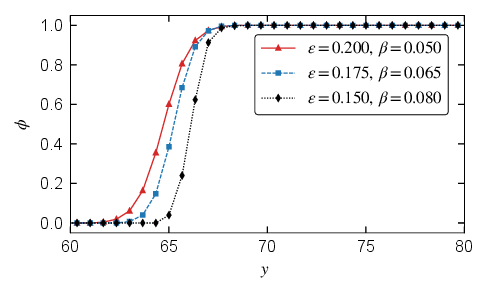}
\caption{1D interface profiles ($\phi$ vs.\ distance) along $x=L/5$ for combinations of $\epsilon$ and $\beta$ ($V=0.5$).}
\label{fig:mbb_sensitivity_profile}
\end{figure}

The sweep confirms that the interface length scale $\ell_\text{AC} = \epsilon/\sqrt{\beta}$
primarily dictates resolvable feature size.
At $\epsilon=0.15$, a narrow interface resolves fine secondary trusses
in the lower bay across all $\beta$.
As $\epsilon$ increases to $0.175$ and $0.20$, the interface widens,
merging fine struts into thicker structural members.
Across all combinations, the converged chemical potential remains small
($-1.1\times 10^{-4} \leqslant \mu \leqslant -2.9\times 10^{-5}$,
$|\mu|/\beta \sim 10^{-3}$).
This confirms that the projected constant mode is well-behaved.
Sampled profiles at $x=L/5$ (Fig.~\ref{fig:mbb_sensitivity_profile})
corroborate the theoretical scaling $\xi \propto \epsilon/\sqrt{\beta}$~\citep{Allen1979}.

Compliance exhibits distinct trends.
At fixed $\beta$, increasing $\epsilon$ monotonically increases compliance
(e.g., from $50.655$ to $51.677$ at $\beta=0.05$),
because more intermediate material is penalized by SIMP.
At fixed $\epsilon$, varying $\beta$ produces a subtle non-monotone compliance variation.
For example, at $\epsilon=0.15$, compliance drops from $50.655$ to $50.326$
before rising to $50.345$ at $\beta=0.10$.
This reflects competing mechanisms.
Larger $\beta$ narrows the interface ($\ell_{\text{AC}} \propto 1/\sqrt{\beta}$),
reducing intermediate material.
Simultaneously, it scales the effective ALM penalty by $\beta^2$
(Section~\ref{num_betascaling}).
Sizing $\epsilon$ and $\beta$ to span $2$--$4$ elements provides
a practical operating guideline.

\subsubsection{Scaling Equivalence of the Double-Well Coefficient}\label{num_betascaling}
To verify that $\beta$ acts as an ALM scale factor
rather than an independent shape parameter,
we apply the simultaneous scaling $(\beta, \epsilon) \to (s\beta, \sqrt{s}\epsilon)$
to the reference MBB benchmark ($180\times 120$, $V=0.3$, $r_{\min}=0.1$).
This transformation preserves $\ell_{\text{AC}} = \epsilon/\sqrt{\beta}$.
We compare two protocols over a sixteen-fold range $s \in \{0.25, 0.5, 2, 4\}$:
(A) uncompensated, with fixed ALM parameters; and
(B) compensated, where ALM quantities are rescaled as
$\rho_\text{h}^0, \rho_\text{h}^{\max} \to \cdot/s^2$,
$\lambda_{\max} \to \lambda_{\max}/s$, and
$\tau_h, \tau_{h,\max} \to s\cdot$.
Table~\ref{tab:beta_scaling} compiles the results.
Here, $\xi_{0.1\to0.9}$ is the nodal interface width along $x=L/5$, and
$\max|\Delta\phi|$ is the maximum nodal discrepancy
from the reference design ($s=1$).

\begin{table}[!ht]
\centering
\caption{Scaling study of $\beta$ at fixed $\ell_\text{AC} = \epsilon/\sqrt{\beta}$ (MBB, $180\times120$).}
\label{tab:beta_scaling}
\begin{tabular}{llccccc}
\toprule
Setting & $s$ & $\mathcal{J}$ & Iter. & $\|\mathbf{h}\|_{\ast}$ & $\xi_{0.1\to0.9}$ & $\max|\Delta\phi|$ \\
\midrule
Reference & 1 & 80.644 & 305 & 0.025 & 2.13 & -- \\
\midrule
(A) uncompensated & 0.25 & 76.467 & 239 & 0.249 & 1.03 & 1.0 \\
 & 0.5  & 78.186 & 289 & 0.096 & 1.38 & 1.0 \\
 & 2    & 83.058 & 460 & 0.021 & 2.16 & 1.0 \\
 & 4    & 86.357 & 500 & 0.024 & 2.00 & 1.0 \\
\midrule
(B) compensated & 0.25 & 80.644 & 305 & 0.025 & 2.13 & $0$ \\
 & 0.5  & 80.644 & 305 & 0.025 & 2.13 & $9.3\times10^{-10}$ \\
 & 2    & 80.644 & 305 & 0.025 & 2.13 & $9.3\times10^{-10}$ \\
 & 4    & 80.644 & 305 & 0.025 & 2.13 & $0$ \\
\bottomrule
\end{tabular}
\end{table}

Without compensation (protocol A), the effective penalty scales as $s^2$.
For $s < 1$, the penalty is insufficient: the residual exceeds tolerance,
and the interface compresses artificially.
For $s > 1$, over-penalization unnecessarily inflates iteration counts.
Conversely, with compensated parameters (protocol B),
the optimized design, iteration count, and residual norm are reproduced
to numerical round-off across the sixteen-fold range.
The measured core interface width $\xi_{0.1\to 0.9} = 2.13$
matches the theoretical prediction $3.11\,\ell_{\text{AC}} = 2.13$ exactly.
This confirms that the feasible set $\{\mathbf{h}=\mathbf{0}\}$
depends on $\beta$ only via $\epsilon/\sqrt{\beta}$.
Thus, $\beta$ is mathematically equivalent to an ALM parameter rescaling.

\subsubsection{Robustness to the Initial Penalty Parameter}\label{num_rho}

\begin{figure}[!ht]
\centering
\includegraphics{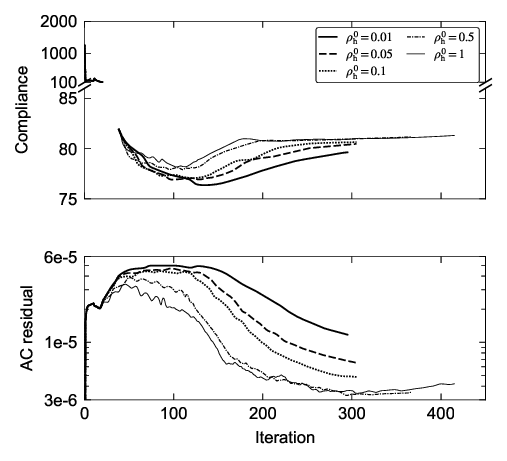}
\caption{Convergence histories for $\rho_\text{h}^0$ sweep (MBB, $180\times 120$, $V=0.3$). Top: compliance $\mathcal{J}$; bottom: normalized AC residual $\|\mathbf{h}\|_\text{norm}$.}
\label{fig:rho_sweep_history}
\end{figure}

\begin{figure}[!ht]
\centering
\begin{subfigure}[b]{0.19\textwidth}
    \centering\includegraphics[width=\textwidth]{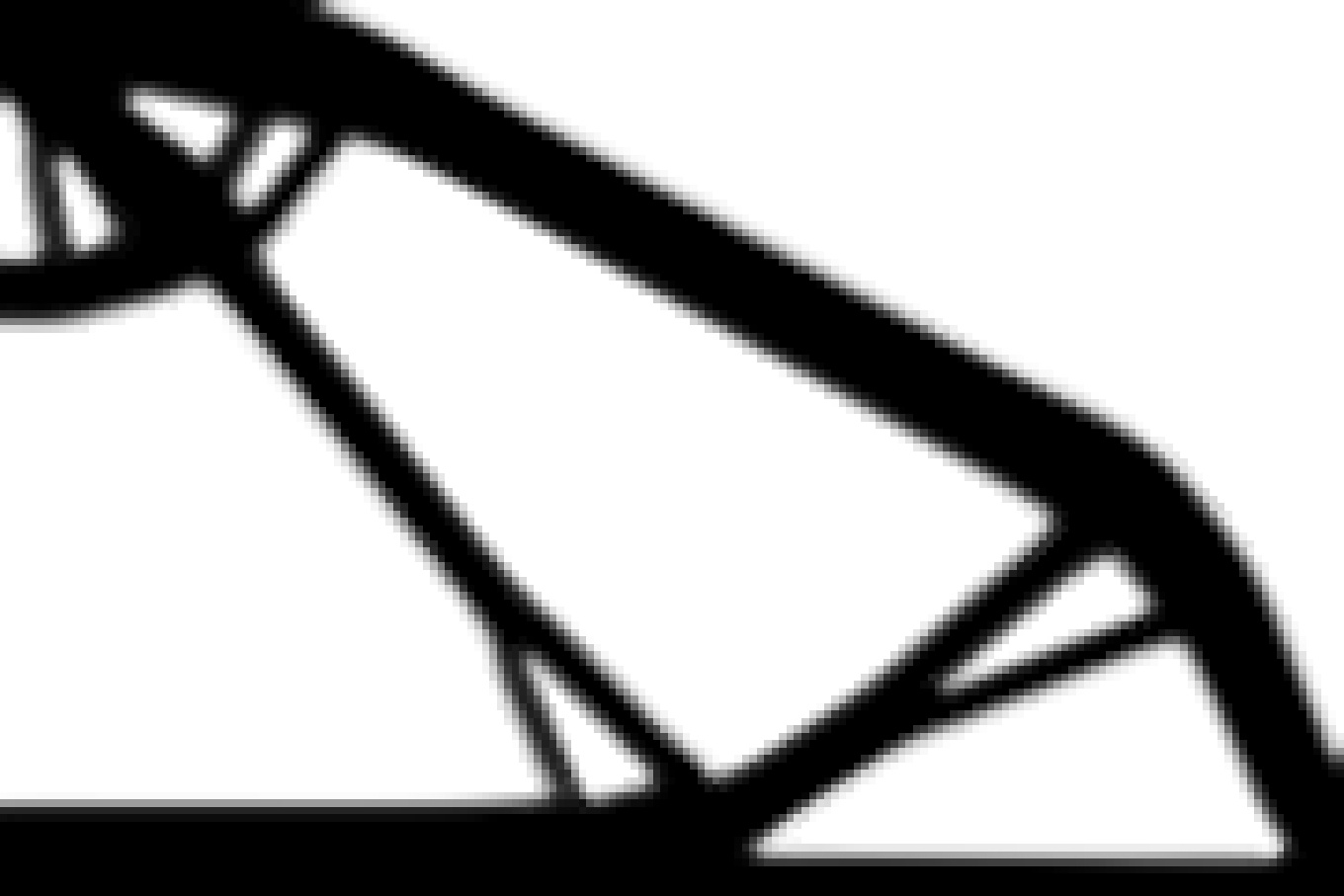}
    \caption{$\rho_\text{h}^0=0.01$}
\end{subfigure}\hfill
\begin{subfigure}[b]{0.19\textwidth}
    \centering\includegraphics[width=\textwidth]{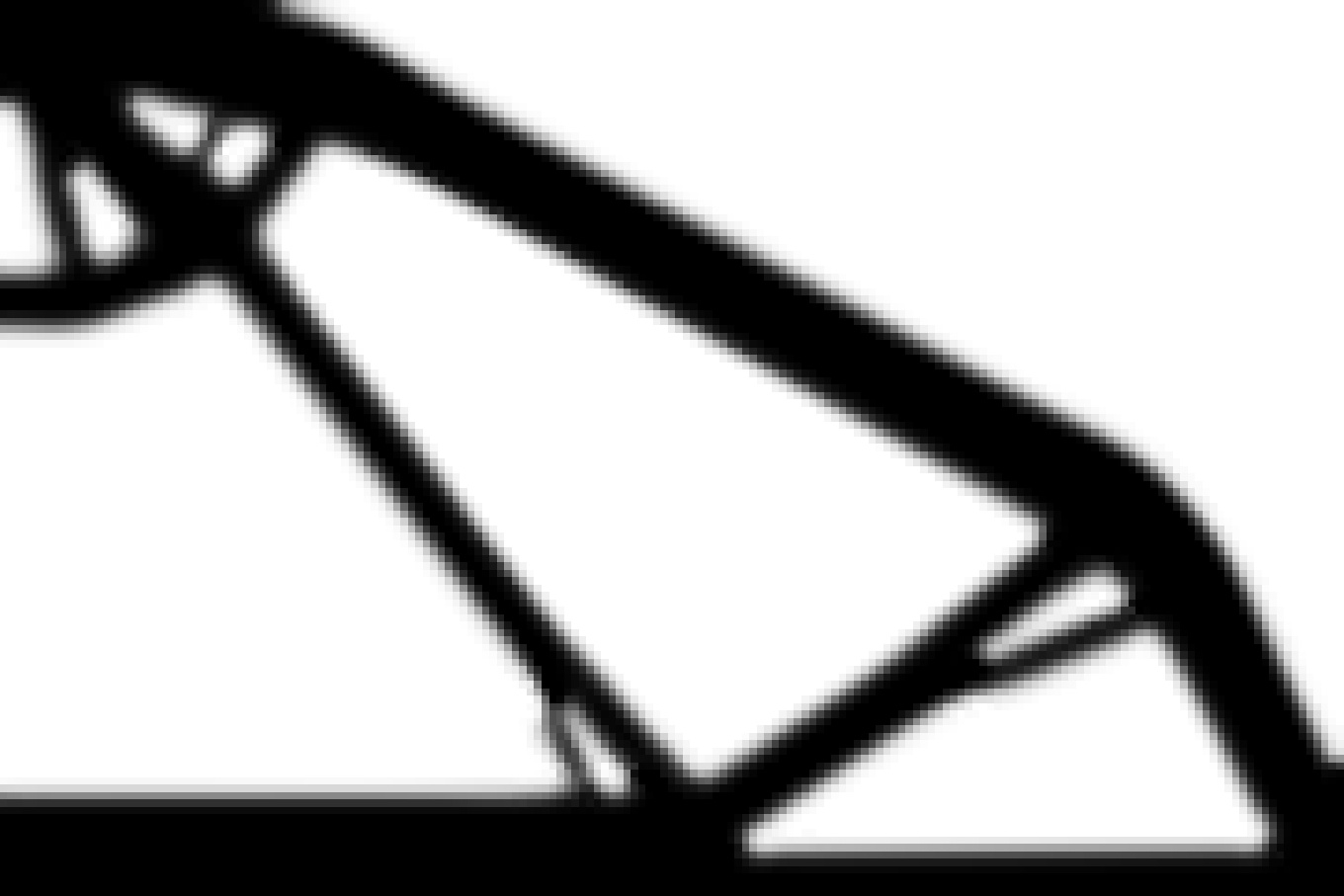}
    \caption{$\rho_\text{h}^0=0.05$}
\end{subfigure}\hfill
\begin{subfigure}[b]{0.19\textwidth}
    \centering\includegraphics[width=\textwidth]{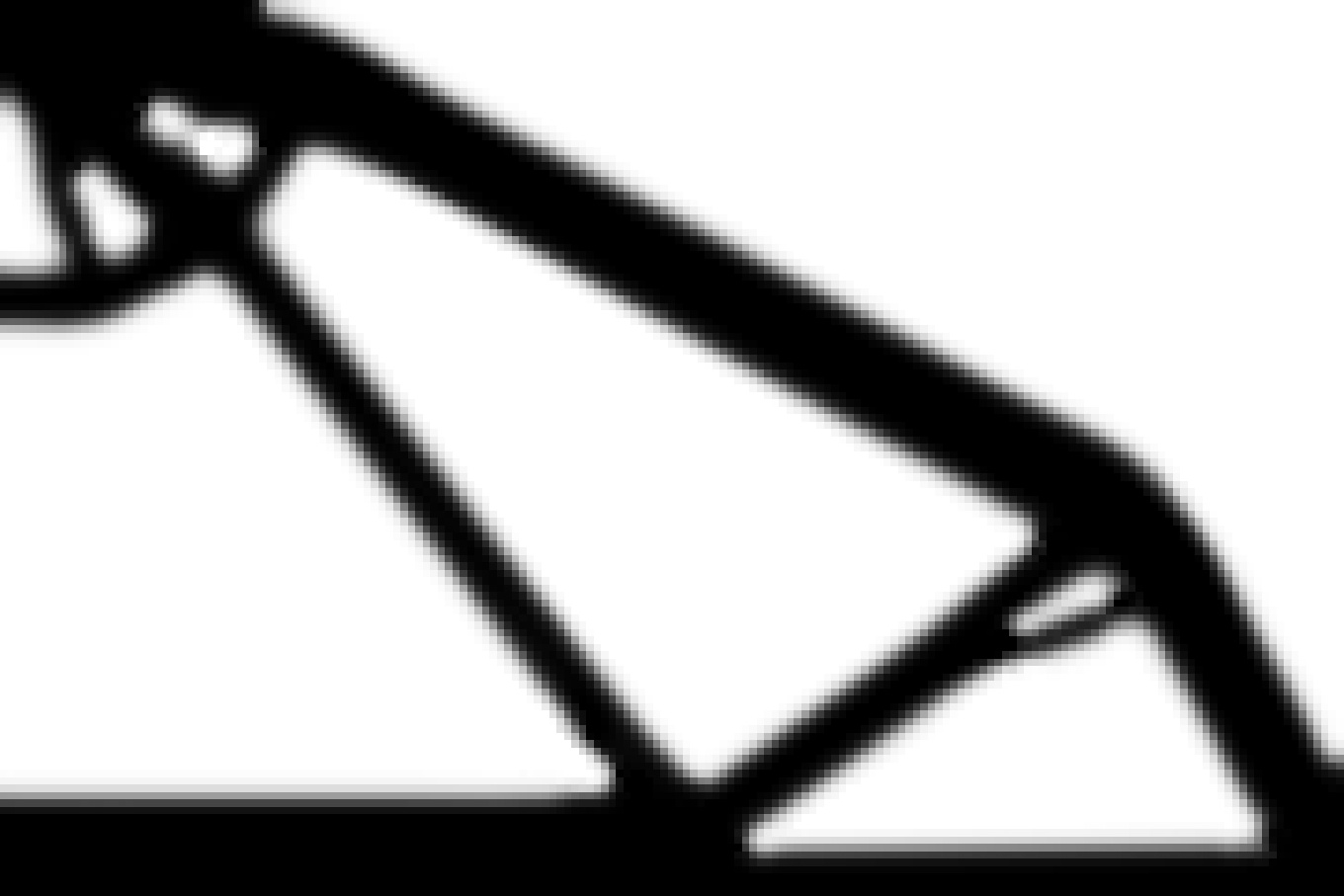}
    \caption{$\rho_\text{h}^0=0.1$}
\end{subfigure}\hfill
\begin{subfigure}[b]{0.19\textwidth}
    \centering\includegraphics[width=\textwidth]{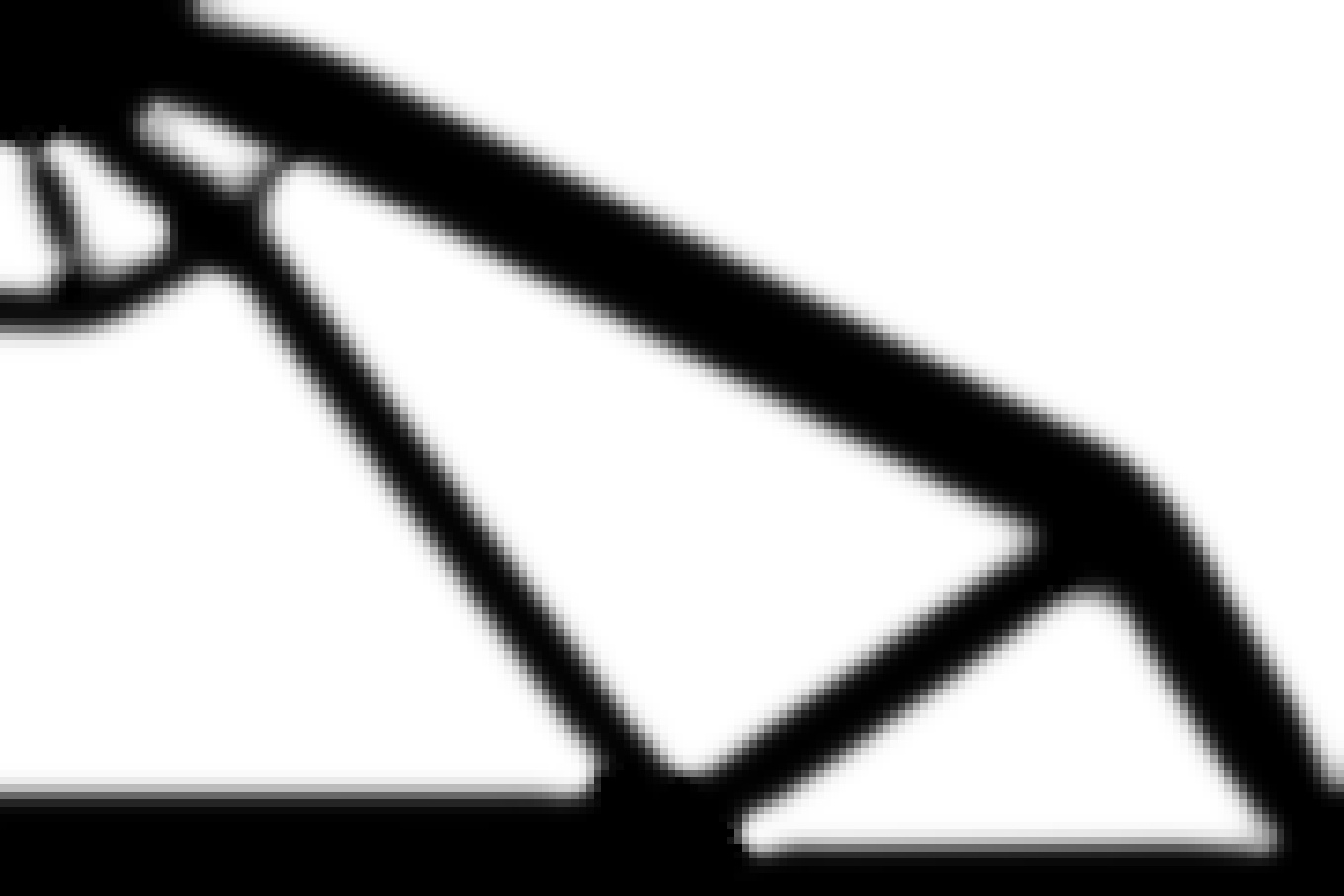}
    \caption{$\rho_\text{h}^0=0.5$}
\end{subfigure}\hfill
\begin{subfigure}[b]{0.19\textwidth}
    \centering\includegraphics[width=\textwidth]{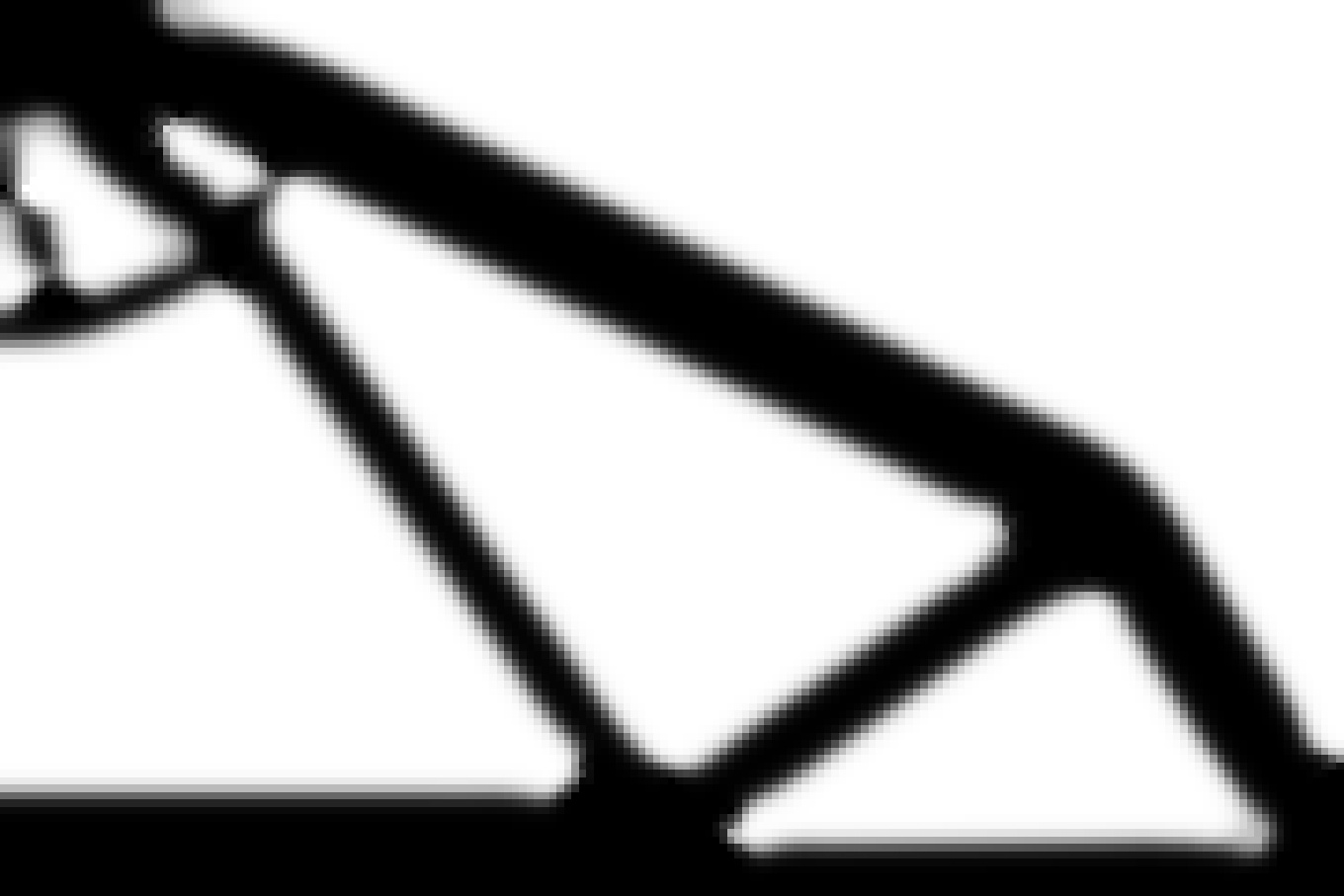}
    \caption{$\rho_\text{h}^0=1.0$}
\end{subfigure}
\caption{Optimized MBB topologies for $\rho_\text{h}^0$ sweep ($180\times 120$, $V=0.3$).}
\label{fig:rho_sweep_struct}
\end{figure}

The initial penalty $\rho_\text{h}^0 = 0.1$ is common to all benchmarks.
It is calibrated via the early gradient-scale balancing rule
(Section~\ref{subsec:algo_params}),
keeping the residual penalty gradient two to three orders of magnitude
below compliance sensitivity during early iterations ($k \in [1, 20]$).
To verify robustness, we sweep $\rho_\text{h}^0 \in \{0.01, 0.05, 0.1, 0.5, 1.0\}$
over two orders of magnitude on the MBB beam
($180\times 120$, $V=0.3$, $r_{\min}=0.1$) with all other parameters fixed.
Table~\ref{tab:rho_sensitivity} summarizes the results.
Figures~\ref{fig:rho_sweep_history} and~\ref{fig:rho_sweep_struct} plot
the convergence histories and final topologies.
While demonstrated on the MBB benchmark, this insensitivity supports
the gradient-scale rule as a reliable practical guideline for broader problem classes.

\begin{table}[!htbp]
    \centering
    \small
    \setlength{\tabcolsep}{5.5pt}
    \caption{Robustness to initial penalty $\rho_\text{h}^0$ (MBB, $180\times 120$).}
    \label{tab:rho_sensitivity}
    \begin{tabular}{lcccc}
        \toprule
        $\rho_\text{h}^0$ & Final $\mathcal{J}$ & Final $\|\mathbf{h}\|_\text{norm}$ & Final $\|\mathbf{h}\|_*$ & Iterations \\
        \midrule
        0.01 & 79.635 & $1.17\times 10^{-5}$ & 0.060 & 295 \\
        0.05 & 80.474 & $6.53\times 10^{-6}$ & 0.033 & 305 \\
        0.1  & 80.644 & $4.85\times 10^{-6}$ & 0.025 & 305 \\
        0.5  & 81.182 & $3.53\times 10^{-6}$ & 0.018 & 367 \\
        1.0  & 81.312 & $4.20\times 10^{-6}$ & 0.021 & 415 \\
        \bottomrule
    \end{tabular}
\end{table}

Across two orders of magnitude in $\rho_\text{h}^0$,
the primary topology is strictly preserved.
All designs share identical main chords, load-bearing diagonals,
and overall truss layouts (Fig.~\ref{fig:rho_sweep_struct}).
Minor variations are confined to secondary struts in the lower bay,
which thin slightly under stricter residual enforcement.
Final compliance varies within a narrow $2.1\%$ window
($79.64$ at $\rho_\text{h}^0=0.01$ to $81.31$ at $\rho_\text{h}^0=1.0$).
The lower compliance at $\rho_\text{h}^0=0.01$ reflects looser feasibility
($\|\mathbf{h}\|_\text{norm} = 1.17\times 10^{-5}$, $\|\mathbf{h}\|_* \approx 0.060$,
compared to $4.85\times 10^{-6}$ and $0.025$ at $\rho_\text{h}^0=0.1$;
Table~\ref{tab:rho_sensitivity}).
Moderate values $\rho_\text{h}^0 \in [0.1, 0.5]$ ensure tight feasibility
at comparable stiffness.

Convergence histories reveal a two-stage dynamic.
During iterations 1--20, all trajectories coincide
($\mathcal{J} \approx 103$, initial $\|\mathbf{h}\|_* \approx 0.17$).
This confirms that early layout formation is uninhibited by the penalty.
In the second stage, larger $\rho_\text{h}^0$ suppresses and advances
the peak residual excursion.
Peak $\|\mathbf{h}\|_\text{norm}$ drops from $5.0\times 10^{-5}$ at iteration 95
for $\rho_\text{h}^0=0.01$ to $3.4\times 10^{-5}$ at iteration 46 for $\rho_\text{h}^0=1.0$.
Thus, $\rho_\text{h}^0$ governs the convergence rate
and diffuse-interface engagement timing rather than the final layout.
This confirms that the gradient-scale rule requires no empirical fine-tuning.

\subsection{Discrete Formulations and Box-Constraint Interaction}\label{sec:discrete_box}

\subsubsection{Mesh Dependence of the Effective Penalty: Euclidean vs. Mass-Weighted ALM}\label{num_meshpenalty}

As shown in Eq.~\eqref{eq:penalty_equiv}, an unweighted Euclidean penalty
$\frac{\rho_\text{E}}{2}\|\mathbf{h}\|^2 = \frac{\rho_\text{E}}{2}\mathbf{h}^\top\mathbf{h}$
is asymptotically equivalent to an effective physical penalty
$\rho_\text{E} v_e$, where $v_e = (\Delta x)^2$
is the element area.
Consequently, if the algorithmic penalty parameters $\rho_\text{E}^0$ and $\rho_\text{E}^{\max}$
are held constant across mesh refinement (matching the mass-weighted $\rho_\text{h}$ values),
the effective physical penalty weakens as $(\Delta x)^2$.
Across the four MBB meshes,
the effective initial penalty $\rho_\text{E}^0 v_e$ decreases from $0.100$ on $120\times 80$ to $0.016$ on $300\times 200$.

To demonstrate why the mass-weighted ALM formulation is essential,
we repeat the MBB refinement series ($V=0.3$, $r_{\min}=0.1$)
using the unweighted Euclidean ALM form,
keeping all other parameters unchanged.
The comparative performance is presented in Table~\ref{tab:mesh_penalty}.
Here, $\|\mathbf{h}\|_{\ast}$ is the mesh-independent residual
of Eq.~\eqref{eq:tol_sense} ($\delta \approx 0.075$,
corresponding to tolerance $\tau_h$ via Eq.~\eqref{eq:delta_equiv}),
and $\xi_{0.1\to0.9}$ is the nodal interface width measured along $x=L/5$,
compared against the one-dimensional prediction $3.11\,\epsilon/\sqrt{\beta} = 2.13$.

\begin{table}[!ht]
\centering
\caption{MBB mesh refinement: Euclidean vs.\ mass-weighted ALM penalties ($V=0.3$, $r_{\min}=0.1$).}
\label{tab:mesh_penalty}
\begin{tabular}{lcccccccc}
\toprule
 & \multicolumn{4}{c}{Euclidean penalty} & \multicolumn{4}{c}{Mass-weighted penalty} \\
\cmidrule(lr){2-5}\cmidrule(lr){6-9}
Mesh & $\mathcal{J}$ & Iter. & $\|\mathbf{h}\|_{\ast}$ & $\xi_{0.1\to0.9}$ & $\mathcal{J}$ & Iter. & $\|\mathbf{h}\|_{\ast}$ & $\xi_{0.1\to0.9}$ \\
\midrule
$120\times80$  & 80.67 & 347 & 0.016 & 2.39 & 80.75 & 363 & 0.017 & 2.38 \\
$180\times120$ & 80.43 & 319 & 0.034 & 2.00 & 80.64 & 305 & 0.025 & 2.13 \\
$240\times160$ & 80.92 & 288 & 0.048 & 1.77 & 81.46 & 323 & 0.031 & 2.13 \\
$300\times200$ & 81.46 & 271 & 0.055 & 1.60 & 82.65 & 321 & 0.028 & 2.12 \\
\bottomrule
\end{tabular}
\end{table}

With the Euclidean penalty, the residual norm $\|\mathbf{h}\|_*$ grows
by a factor of $3.4$ under mesh refinement.
Concurrently, the interface width shrinks from $2.39$ to $1.60$ ($75\%$ of target)
on the finest mesh, because the weakening penalty allows SIMP
to artificially sharpen the boundary.
In contrast, under the mass-weighted penalty,
$\|\mathbf{h}\|_*$ stabilizes across refined meshes.
Across the three sufficiently resolved meshes ($180\times 120$, $240\times 160$, 
and $300\times 200$), the measured core interface width matches 
the theoretical prediction ($2.13$) to within $0.5\%$ ($2.13$, $2.13$, and $2.12$). 
On the coarsest mesh ($120\times 80$), $\xi_{0.1\to0.9} = 2.38$ 
deviates by $\approx 11.7\%$. 
This occurs because the interface spans only $\approx 2.1$ elements at $\Delta x = 1.0$, 
leaving the steep transition under-resolved by bilinear shape functions. 
Once adequately resolved ($\Delta x \leqslant 0.67$, $\geqslant 3.2$ elements), 
the theoretical width is accurately recovered. 
The mass-weighted formulation thus ensures consistent 
feature regularization across discretizations.

\subsubsection{Residual Decomposition and Curvature-Induced Background Density}\label{num_boxresidual}

To investigate the pointwise interaction between the Allen--Cahn residual
and the design bounds $0 \leqslant \phi_i \leqslant 1$,
nodes of a converged structure are partitioned into three disjoint sets:
(i) inactive nodes ($10^{-3} < \phi_i < 1 - 10^{-3}$) in the interface interior;
(ii) bulk-active nodes lying on a bound alongside all immediate neighbors; and
(iii) edge-active nodes on a bound adjacent to the diffuse transition band.
Because $\|\mathbf{h}\|_*^2$ sums over nodes, the contributions add in quadrature.
Table~\ref{tab:box_residual} reports this decomposition
for the 2D MBB reference design ($180\times120$, $V=0.3$)
and both 3D benchmarks ($V=0.15$).
The table lists nodal percentages and individual contributions to $\|\mathbf{h}\|_{\ast}$.

\begin{table}[!ht]
\centering
\caption{Decomposition of converged residual $\|\mathbf{h}\|_{\ast}$ across node sets.}
\label{tab:box_residual}
\begin{tabular}{lccccc}
\toprule
Case & $\mu/\beta$ & $\|\mathbf{h}\|_{\ast}$ & Inactive & Bulk-active & Edge-active \\
\midrule
2D MBB & $-1.3\times10^{-3}$ & 0.025 & 26\% / 0.022 & 62\% / 0.001 & 12\% / 0.010 \\
3D MBB & $+3.3\times10^{-2}$ & 0.024 & 96\% / 0.023 & 0.5\% / 0.002 & 3.2\% / 0.007 \\
3D L-beam & $+3.0\times10^{-2}$ & 0.027 & 97\% / 0.026 & 0.2\% / 0.001 & 3.2\% / 0.006 \\
\bottomrule
\end{tabular}
\end{table}

On bulk-active nodes, the pointwise residual equals $-\mu$ within precision
(mean $\tilde{h}/\beta = +1.31\times 10^{-3}$ vs.\ $-\mu/\beta = +1.30\times 10^{-3}$ in 2D).
Because $|\mu|/\beta$ is small,
bulk-active contributions to the total residual norm are negligible ($\leqslant 0.002$).
Edge-active nodes account for $7\%$--$18\%$ of $\|\mathbf{h}\|_*^2$.
This reflects the truncation of exponential $\tanh$ tails by the box bounds.
The dominant residual norm resides on inactive nodes within the transition band,
localizing naturally at structural member junctions.

The 3D benchmarks reveal an intriguing consequence of interface geometry.
The positive chemical potential $\mu > 0$ ($\mu/\beta \approx +3.3\times 10^{-2}$)
is qualitatively consistent with the predominant positive mean curvature
of spatial trusses ($\kappa > 0$) via the asymptotic Gibbs--Thomson relation.
At $\mu > 0$, the unconstrained stationarity state shifts slightly
to the positive root of $W'(\phi_b) = \mu$ ($\phi_b \approx 0.016$--$0.017$).
This leaves $>96\%$ of nodes mathematically inactive.
This diffuse background is mechanically inert under SIMP
($E(\phi_b) \propto \phi_b^3 \approx 4.9\times 10^{-6}$, contributing negligible stiffness).
It is fully reclaimed upon 0/1 thresholding
or $\phi=0.5$ isosurface extraction ($V_\text{eff} \approx 0.136$ of $V=0.15$).
These observations justify why the tolerance-regularized early-stopping formulation
is solved instead of idealized equality.
Enforcing $\|\mathbf{h}\|_{\ast} \leqslant \delta$ cleanly absorbs localized bulk-shift
and boundary-truncation residuals without requiring non-smooth obstacle solvers.
At the same time, variational inequalities remain a natural avenue
for strict $\phi_i \equiv 0$ bounds.

\subsection{Computational Efficiency and Cost Breakdown}\label{sec:comp_efficiency}

Table~\ref{tab:comp_cost} details the average wall-clock time fraction
per iteration for the 2D MBB ($180 \times 120$)
and 3D MBB benchmarks at $r_{\min}=0.1$.
In both cases, state equation FEA ($\mathbf{K}\mathbf{U}=\mathbf{F}$)
dominates computational cost:
$99.58\%$ of the $65$\,s/iteration in 3D, and $48.44\%$ in 2D
(where MMA subproblem solving requires $42.05\%$, totaling $0.16$\,s/iteration).

\begin{table}[!ht]
\centering
\caption{Breakdown of average computational cost per iteration (2D and 3D MBB).}
\label{tab:comp_cost}
\begin{tabular}{llrr}
\toprule
 & & \multicolumn{2}{c}{\textbf{Time Fraction (\%)}} \\
\textbf{Module} & \textbf{Operation} & \textbf{2D ($180 \times 120$)} & \textbf{3D MBB} \\
\midrule
\quad State Equation & FEA ($\mathbf{K}\mathbf{U}=\mathbf{F}$) & 48.44 & 99.58 \\
\quad Sensitivity Analysis & Compliance Sensitivity & 0.04 & 0.01 \\
\quad & AC Residual \& Gradient & 5.91 & 0.11 \\
\quad Optimization & MMA Solving & 42.05 & 0.25 \\
\quad Other & Assembly, I/O, etc. & 3.56 & 0.05 \\
\bottomrule
\end{tabular}
\end{table}

Unlike transient phase-field formulations requiring implicit time-stepping,
evaluating the AC residual and its sensitivity involves only sparse
matrix-vector multiplications on the scalar design field.
This drops the measured per-iteration residual-and-sensitivity overhead
to only $0.11\%$ of the linear elasticity solve time
in the reported 3D benchmarks (Table~\ref{tab:comp_cost}).
While direct Cholesky factorization dominates the FEA solve here,
iterative solvers would reduce state solve times.
Nonetheless, because AC residual and gradient evaluations are strictly matrix-free,
their computational burden remains negligible.
Scalability is therefore governed entirely by the state elasticity solver.

\section{Conclusions}
\label{sec:conclusions}

This paper establishes a residual-constrained phase-field framework 
for compliance topology optimization, enforcing the finite-element residual 
of the steady-state \emph{conserved} Allen--Cahn condition, 
$-\epsilon^2\Delta\phi + W^\prime(\phi) = \mu$, 
with an Augmented Lagrangian Method.
The principal methodological contributions and findings are summarized as follows:
\begin{enumerate}
    \item \textbf{Volume-compatible formulation and matrix-free sensitivity without residual adjoint solves:} 
    Retaining the chemical potential $\mu$ and evaluating it in closed form
    as the spatial domain average of $W^\prime(\phi)$ removes the constant mode
    from the residual.
    This guarantees algebraic zero-mean compatibility with external volume targets
    via a zero-mean residual ($\mathbf{1}^\top\mathbf{h} \equiv 0$).
    The resulting symmetric-plus-rank-one residual Jacobian allows the ALM sensitivities
    to be evaluated matrix-free without solving additional adjoint PDEs
    for the AC residual constraint.
    This reduces the measured per-iteration residual-and-sensitivity overhead
    to only $0.11\%$ of the linear elasticity solve time in the 3D MBB benchmark (Table~\ref{tab:comp_cost}).

    \item \textbf{Reconciliation of the compliance gap via 0/1 projection:} 
    Across 48 benchmark configurations, the continuous compliance elevation
    ($1.4\%$--$12.3\%$) in AC-ALM stems from diffuse-interface elements
    under SIMP penalization ($E \propto \phi^3$). 
    In a volume-preserving 0/1 thresholding post-processing audit,
    this gap largely vanishes.
    The binarized AC-ALM topologies achieve equal or superior stiffness
    in 45 of 48 cases (averaging $1.2\%$ stiffer: 23/24 against filtered
    and 22/24 against unregularized SIMP),
    while substantially suppressing single-node hinges and strut fragmentation
    compared to unregularized baselines.

    \item \textbf{Consistent feature regularization via mass-weighted ALM:} 
    Conventional Euclidean penalty formulations weaken under mesh refinement,
    causing residual drift and interface shrinkage. 
    In contrast, the mass-weighted ALM maintains consistent feature regularization. 
    Across sufficiently resolved meshes ($\geqslant 3.2$ elements), 
    it matches the analytical bandwidth $3.11\,\ell_\text{AC} \approx 2.13$ 
    to within $0.5\%$ along representative straight members
    (with an expected $11.7\%$ numerical broadening 
    on the coarsest under-resolved mesh). 
    Furthermore, scaling analyses confirm that $\beta$ acts mathematically
    as an ALM penalty scale factor rather than an independent geometric parameter.

    \item \textbf{Three-dimensional spatial viability and methodological scope:} 
    The converged chemical potential $\mu$ agrees qualitatively with the Gibbs--Thomson limit,
    settling near zero for 2D straight struts while remaining positive for 3D curved members.
    This explains the small curvature-induced background density. 
    While demonstrating 3D viability on MBB and L-beam examples,
    the discretization ($\Delta x = 1.0$) yields a compact interface ($\approx 2.1$ elements).
    Multi-mesh investigations at $\Delta x \leqslant 0.5$ ($\geqslant 1.2\times 10^6$ DOFs)
    and variational-inequality extensions remain promising avenues.
\end{enumerate}

This framework provides an effective foundation for PDE-constrained topology optimization,
with extensions to multi-material designs, manufacturing constraints,
and multiphysics problems.

\backmatter

\section*{Statements and Declarations}

\bmhead{Author Contributions}
R.D.: Conceptualization, methodology, software, simulation, analysis,
writing (original draft, review \& editing), funding acquisition.
X.X.: Software, simulation, analysis, writing (original draft, review \& editing).
W.C.: Analysis, review \& editing, funding acquisition.
Z.Z.: Analysis, review \& editing, funding acquisition.

\bmhead{Funding}
Financial support from the National Natural Science Foundation of China
(Grant Nos. 52375268, 51405397, 51675450),
Sichuan Science and Technology Program (2023YFG0182),
and Southwest Jiaotong University Multidisciplinary Research Fund (2682025ZD005)
is gratefully acknowledged.

\bmhead{Data availability}
The datasets generated and analyzed during the current study are available
within the article and supplementary files;
additional data are available from the corresponding author upon reasonable request.

\bmhead{Code availability}
The code are available from the corresponding author upon reasonable request.

\bmhead{Replication of results}
The data can be reproduced by Algorithm 1 with the parameters listed in Table 1.

\bmhead{Conflict of interest}
The authors declare no competing financial or nonfinancial interests.

\bmhead{Ethical approval and Consent to participate}
This article does not contain any studies with human or animal participants.


\setlength{\bibsep}{0.0pt plus 0.2pt}


\end{document}